# A novel methodology for antenna design and optimization: *Variable* $Z_0$ (ver. 2)

**Richard A. Formato**

Registered Patent Attorney & Consulting Engineer
Cataldo & Fisher, LLC, 400 TradeCenter, Suite 5900
Woburn, MA 01801 USA
rf2@ieee.org, rformato@cataldofisher.com

**Abstract —** This paper describes "Variable $Z_0$," a novel and proprietary approach to antenna design and optimization. The new methodology is illustrated by applying it to the design of a resistively-loaded bowtie antenna and two broadband Yagi-Uda arrays. *Variable* $Z_0$ is applicable to any antenna design or optimization methodology. Using it will result in generally better antenna designs across any user-specified set of performance objectives.

## 1. Introduction

This paper describes what the author believes to be a novel and proprietary antenna design methodology called "Variable $Z_0$," $Z_0$ being the antenna feed system characteristic impedance. *Design* refers to the process of specifying a complete set of parameters that define an antenna meeting specific performance objectives. *Optimization* refers to specifying a complete set of parameters that define the antenna that *best* meets specific performance objectives. While *Variable* $Z_0$ may be particularly useful for improving impedance bandwidth (IBW), the methodology can be used to meet any performance objectives, even ones that do not consider IBW at all.

Traditional antenna design and optimization methods, processes, or procedures ("methodology") view $Z_0$ as a *fixed* design *parameter* with a constant value specified at the start of the methodology. $Z_0$ therefore is not a *variable* quantity whose value is determined *by* the methodology. This distinction is quite important, because traditional methodology excludes from the outset all designs that could provide better performance by using some other value of $Z_0$. *Variable* $Z_0$ improves on traditional methodologies by treating $Z_0$ as another design *variable* in the set of variable antenna parameters to be determined by the methodology. *Variable* $Z_0$ is applied to the design of a numerically optimized ultra wideband (UWB) bowtie antenna and to two broadband Yagi-Uda arrays to illustrate its use.

## 2. $Z_0$ as a design *variable*

IBW is defined as the frequency band or bands within which $Z_{in}$, the antenna input impedance, is matched within specified limits to the feed system characteristic impedance, $Z_0$. The required degree of matching can be specified in terms of the antenna's actual input impedance (resistance, $R_{in}$, and reactance, $X_{in}$) as a function of frequency, or, as is more often the case, in terms of a maximum voltage standing wave ratio (VSWR). IBW typically is specified as VSWR // $Z_0 \le$ 2:1 (// denoting "relative to"), which is equivalent to a return loss (scattering parameter $S_{11}$) approximately less than $-10$ dB. Other

VSWR thresholds can be used instead, and frequently are. Zehforoosh *et al.* [1] describes the design of an UWB microstrip antenna and provides a good summary of $Z_0$'s significance as a design parameter in the context of IBW. As is usual in the traditional methodology, $Z_0$ in [1] is a fixed parameter, not a variable quantity as it is in *Variable* $Z_0$.

While *Variable* $Z_0$ methodology is useful in any antenna design, its widest range of applicability likely will be in improving IBW, which consequently is emphasized in this paper. The literature is replete with examples of various design and optimization approaches to improving IBW. Hoorfar's review of Evolutionary Programming (EP) in electromagnetic optimization [2] provides several antenna design examples involving IBW, all of which employ fixed $Z_0$. An optimized dipole array design using fixed $Z_0$ is described in [3]. An extensive study of the impact of various antenna parameters on down-link performance appears in [4], but $Z_0$ is not one of those parameters. Boryssenko [5] presents an extensive discussion of methodologies for designing wideband antennas, yet $Z_0$ is not included as a design *variable*. The multiband and UWB printed bowties described in [6] and [7], respectively, both utilize fixed $Z_0$ as a design parameter. Other UWB bowtie configurations (slot [8], UWB balun-fed [9], and dual-band slot [10]) are all designed using constant values for $Z_0$. Of course, the traditional practice of specifying $Z_0$ as a fixed parameter is not restricted to bowties. It is the approach used in all antenna design and optimization methodologies, as, for example, in [11] (broad-band array optimization) and [12] (Particle Swarm Optimized Vivaldi antenna). The state-of-the-art, which fixes $Z_0$'s value, limits achievable IBW (as well as meeting other performance objectives) because better results are obtainable when $Z_0$ is considered to be a design *variable* whose value is determined *by* the design or optimization process, instead of being a user-supplied constant at the outset of that process.

## 3. Design example: resistively loaded broadband bowtie

As discussed above, the bowtie antenna has been studied extensively, and it provides a simple structure to illustrate how better IBW can be achieved using *Variable* $Z_0$ instead of traditional methodology. Its structure is shown schematically in Fig. 1. The antenna comprises two symmetrical wire triangles connected by a feed wire. The perfectly electrically conducting (PEC) wire has diameter $2a$, and the arm and feed wire lengths are $L_{arm}$ and $L_{feed}$, respectively. The bowtie's flare angle (vertex angle) is $2\alpha$. The antenna is impedance-loaded with four symmetrically placed loading resistors of equal-value, $R_{load}$, that increase IBW at the expense of radiation efficiency (see [13] for a discussion of impedance loading). The resistors are inserted in series with the antenna wire a distance $L_{load}$ from the vertex. Excitation is provided by an RF (radio frequency) source at the center of the feed wire.

The bowtie structure was modeled in free space with the Numerical Electromagnetics Code Ver 4.1D (NEC4, double precision) [14,15], the "modeling engine." NEC4 is a widely used Method of Moments (MoM) modeling code for wire structures developed at Lawrence Livermore National Laboratory. The RF source is located at the origin of NEC's standard right-handed spherical polar coordinate system with the antenna the Y-Z plane and the feed wire along the Y-axis. For readers interested in NEC MoM modeling who do not have access to NEC4, freeware versions of NEC2 are available [16] (source code and executables). While NEC4 is particularly useful for wire structures like the bowtie, any suitable modeling engine, numerical, analytical or hybrid, can be used instead. The engine must be capable of accurately calculating all performance parameters that are used in evaluating how good a particular antenna design is.

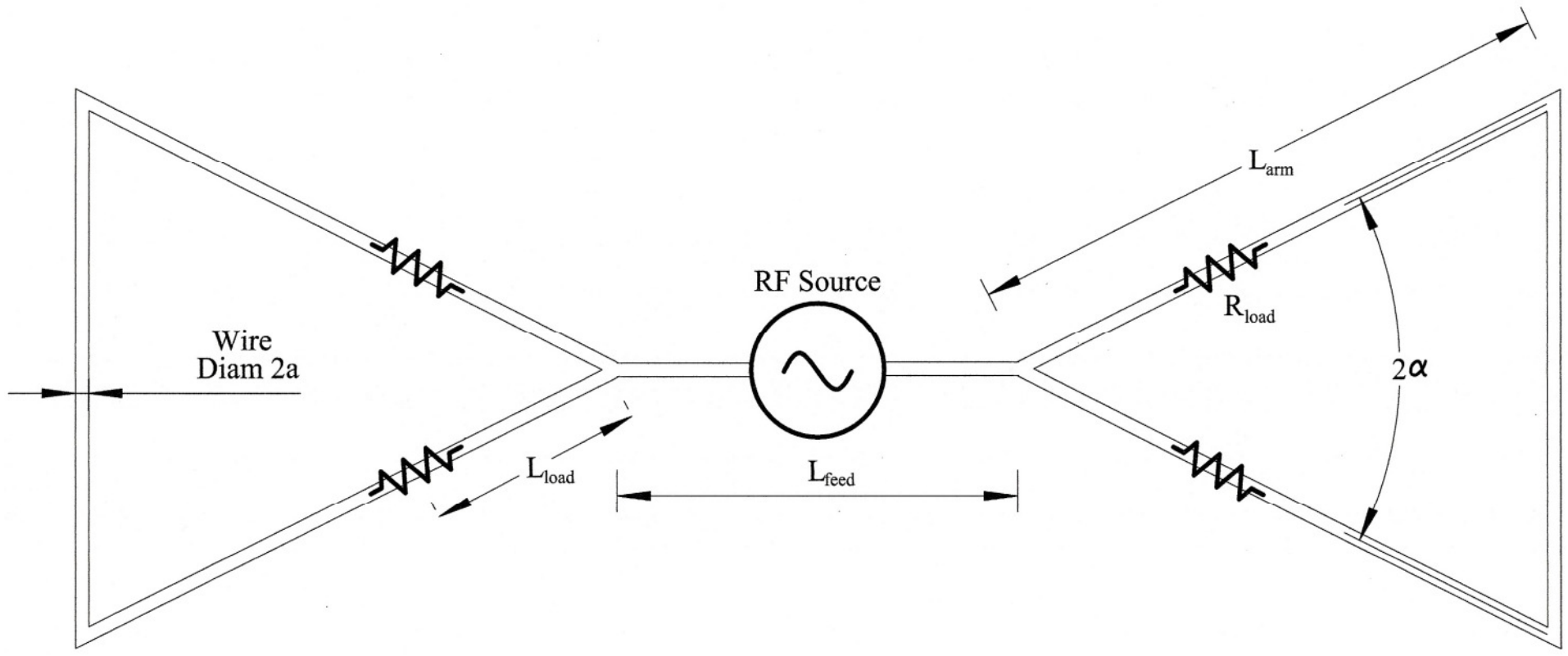

Fig. 1. Schematic diagram of the resistively loaded bowtie antenna.

## 4. Bowtie design variables

There are five bowtie design *variables*: $L_{arm}$, $\alpha$, $L_{load}$, $R_{load}$, and, importantly and differently from traditional methodology, $Z_0$. Each one of these quantities is allowed to vary in the design or optimization process. Their values are determined by meeting specific performance goals or by achieving the best performance using a designer-specified "fitness" or "objective" function that measures how well a particular set of values meets the desired performance goals. The key idea in *Variable* $Z_0$ is that the value of $Z_0$ be allowed to change during the design or optimization methodology because different values result in different antenna performance. It is likely that specific performance objectives can be met with multiple values of $Z_0$, especially in the design setting. Even in the optimization setting there may be multiple values for a "best" design corresponding to multiple global optima, or that result from using a stochastic optimization algorithm or one that has converged to a local optimum.

Note that the wire radius $a$ and the feed wire length $L_{feed}$ are *not* included in the design variable list because these quantities are *fixed parameters*. As in the traditional design methodology, their values are assigned at the beginning of the design or optimization process, and they do not change as a result of that process. $Z_0$ is treated that way in traditional methodology, that is, $Z_0$ is assigned a value that does not change and is not determined by the methodology. Of course, neither $a$ nor $L_{feed}$ need be fixed. They could be allowed to vary with their values determined by how well the performance goals are met, because different values of $a$ or $L_{feed}$ might improve performance. However, in the bowtie design discussed here these quantities are held constant as examples of the fixed design parameters used in traditional methodology.

## 5. Optimization methodology

Any methodology can be used to design an antenna using *Variable* $Z_0$ (it is applicable regardless of the methodology). For the bowtie design example, CFO (Central Force Optimization) [17] was used. But any number of other commonly employed algorithms, such as Particle Swarm (PSO) [18], Ant Colony (ACO) [19], Group Search Optimizer (GSO) [20], or Differential Evolution (DE) [21], could be used instead. This "product by process" approach applies to any methodology, deterministic ones like CFO; stochastic metaheuristics like PSO, ACO, GSO or DE; analytic approaches such as extended Wu-King impedance loading [13]; or even "seat of the pants" optimization based on experience, intuition, or a "best guess." The specific optimization methodology is irrelevant to the novelty of treating $Z_0$ as a design variable instead of a fixed parameter. Thus, *Variable* $Z_0$ can be used advantageously with *any* design or optimization methodology.

The bowtie fitness function (to be maximized by CFO) is defined as

$$F(Z_0, R_{load}, L_{load}, L_{arm}, \alpha) = \frac{5 \cdot \min(G_{\max}) + \min(\varepsilon)}{\left|\max(R_{in}) - Z_0\right| \cdot \{\max(VSWR // Z_0) - \min(VSWR // Z_0\} \cdot \{\max(X_{in}) - \min(X_{in})\}}.$$

$\varepsilon$ is the bowtie's radiation efficiency, $G_{\max}$ its maximum gain, and $Z_{in} = R_{in} + jX_{in}$ ($j = \sqrt{-1}$) its input impedance, all as functions of frequency. The first term in the denominator drives the input resistance toward $Z_0$ (assumed purely resistive), while the second and third terms, respectively, minimize the VSWR and input reactance variability with frequency. A particular bowtie's fitness increases with increasing radiation efficiency and minimum maximum gain. It decreases with increasing mismatch between $Z_0$ and $R_{in}$ and increasing variability in VSWR and input reactance. The coefficient 5 in the gain term was determined empirically. Of course, the antenna designer is free to specify any desired fitness function, with different fitnesses creating different decision space landscapes resulting in different antenna designs.

A parameter-free CFO implementation was used as described in [22-24] (only the fitness function is supplied as input to the algorithm; Appendix I contains a complete source code listing). Because there are five design variables, the decision space is 5-dimensional ($N_d = 5$) with the following hardwired parameters: $N_t = 250$, $N_\gamma = 11$, $\max\left(\frac{N_p}{N_d}\right) = 8$, $F_{rep}^{init} = 0.5$, $\Delta F_{rep} = 0.1$, $F_{rep}^{\min} = 0.05$, $\alpha = 1$, $\beta = 1$, $G = 2$, $\Delta t = 0.5$. Decision space adaptation was applied every 20th time step with an early termination criterion of fitness saturation for 25 consecutive steps (variation $\le 10^{-6}$). The best fitness returned by CFO was $1.5806 \times 10^{-4}$ (probe #1) with $\frac{N_p}{N_d} = 2$ and $\gamma = 1$ at step 35.

The decision space was defined as follows: $0.01 \le L_{arm} \le 0.08$ m, $10° \le \alpha \le 80°$ (bowtie variables); $1 \le \left\lfloor \frac{L_{load}}{L_{arm}} \right\rfloor \le 9$, $1 \le R_{load} \le 1000\ \Omega$ (loading variables); $50 \le Z_0 \le 1000\ \Omega$ (*variable* feed system characteristic impedance). The *fixed* design parameters were $L_{feed} = 0.02$ m and wire radius $a = 0.0005$ m (note all dimensions in meters). These values, which are specified by the antenna designer, were determined based on experience.

The bowtie's performance was computed every $100\,\text{MHz}$ across the band $800 \le f_{mhz} \le 12{,}000$ ($f_{mhz}$ being the frequency in megahertz) and the resulting data used to compute the fitness function. The NEC-4 model employed 9 segments in each of the six wires forming the bowtie triangles, and three segments in the feed wire. Gain was computed every 5 degrees from $0 \le \theta \le 90°$ at $\phi = 0°$ (see [14,15] for coordinate system details). Note that the range for $L_{load}$ is specified as an integer ratio to $L_{arm}$ because NEC loads a wire by segment *number* ("LD" cards), not by coordinate. The distance $L_{arm}$ consequently must be converted to an integer for use in NEC4.

## 5. Bowtie results

Design values for the *Variable* $Z_0$ CFO-optimized loaded bowtie appear in Fig. 2. Its computed performance from 800 to 12,000 MHz is summarized in Fig. 3. Post-processing extended the frequency range from 200 to 15,000 MHz with data computed every 15 MHz (Appendices II and III contain the computed data). Figures 8 and 11-15 plot the results, and Fig. 16 shows the NEC4 input file. Fig. 17 shows the antenna geometry as visualized using 4nec2 [16(i)] (rectangles are loading resistors). In order to compare *Variable* $Z_0$ to the traditional methodology, another CFO run was made holding $Z_0$ constant. The same run parameters and decision space boundaries were employed, except $50 \le Z_0 \le 50\ \Omega$ to fix its value. Results of the comparison run appear in Figs. 4 and 5.

As a preliminary matter, the NEC4 model must be validated using the *unloaded* antenna's average free space power gain test (AGT) as discussed at [15, Part I, p. 101]. Radiation patterns were computed from 200 to 15,000 MHz at $5°$ and $10°$ increments, respectively, for $0 \le \theta \le 90°$ and $0 \le \phi \le 180°$ with power gain averaged over $\pi$ steradians. The AGT plot for the *Variable* $Z_0$ bowtie appears in Fig. 6, and in Fig. 7 for the fixed $Z_0$ design.

The average power gain for the *Variable* $Z_0$ bowtie is between about 1.05 and 1.1 over most of the band with a peak near 1.18. Values within 20% of unity usually are considered adequate for model validation, so that in this case there is confidence in the bowtie data throughout the entire frequency range. By comparison, the average gain for the fixed $Z_0 = 50\ \Omega$ design shows a sharp increase above 1.2 for frequencies beyond about 14,000 MHz, so that confidence in those data is considerably less.

The primary objective of this design example is minimizing VSWR, which is plotted for the optimized *Variable* $Z_0$ antenna in Fig. 8. The bowtie structure computed by CFO was modeled with and without resistive loading and its VSWR computed relative to the optimum CFO-computed feed system $Z_0$ of $715\ \Omega$. Interestingly, this value for $Z_0$, which was determined as a result of its being *variable* in the optimization process, is unusual and quite far from the "standard" values that are much lower (mostly $\sim 50\ \Omega$). Of course, impedance-matching the optimized value of $Z_0$ to a "standard" value, say, $50\ \Omega$, requires a sufficiently broadband matching network, for example, a broadband transformer, lumped or distributed matching network, tapered transmission line, and so on, with some matching devices being better than others depending on the frequency range.

| | |
|---|---|
| $L_{arm}$ | 0.051 m |
| $\alpha$ | 39.4° |
| $\left\lfloor \frac{L_{load}}{L_{arm}} \right\rfloor$ | 6 |
| $L_{load}$ | 0.03117 m |
| $R_{load}$ | 166.93 Ω |
| $Z_0$ | 715 Ω |

Fig. 2. *Variable*-$Z_0$ CFO-optimized bowtie.

| *Parameter* | *Minimum* | *Maximum* |
|---|---|---|
| $VSWR // 715\ \Omega$ | 1.06 | 6.43 |
| $R_{in}$ | 111.51 Ω | 729.32 Ω |
| $X_{in}$ | −286 Ω | 335.73 Ω |
| $\varepsilon$ | 24.25 % | 98.88 % |
| $G_{max}$ | −3.34 dBi | 4.71 dBi |

Fig. 3. *Variable*-$Z_0$ bowtie performance 800-12,000 MHz.

| | |
|---|---|
| $L_{arm}$ | 0.08 m |
| $\alpha$ | 30° |
| $\left\lfloor \frac{L_{load}}{L_{arm}} \right\rfloor$ | 9 |
| $L_{load}$ | 0.07555 |
| $R_{load}$ | 1000 Ω |
| $Z_0$ | 50 Ω *fixed* |

Fig. 4. CFO-optimized fixed $Z_0 = 50\ \Omega$ bowtie.

Fig. 9 plots the fixed $Z_0$ bowtie's VSWR. which is much higher than it is for the *Variable* $Z_0$ antenna across the entire band. At no point is $VSWR \leq 2:1$, and for the most part it is $> 5:1$. Since IBW is the primary performance measure in this case, this example shows how the *Variable* $Z_0$ methodology can lead to much better antenna designs because $Z_0$ then is determined *by* the design or optimization process instead of being assigned a value that cannot change.

| *Parameter* | *Minimum* | *Maximum* |
|---|---|---|
| $VSWR // 50\ \Omega$ | 2.67 | 13.58 |
| $R_{in}$ | 119.38 Ω | 676.71 Ω |
| $X_{in}$ | − 285.46 Ω | 294.05 Ω |
| $\varepsilon$ | 43.76 % | 91.85 % |
| $G_{max}$ | −1.37 dBi | 5.14 dBi |

Fig. 5. Fixed $Z_0 = 50\ \Omega$ bowtie performance 800-12,000 MHz.

Figure 10 summarizes the *Variable* $Z_0$ bowtie's IBW using the commonly employed threshold $VSWR // Z_0 \leq 2:1$. The antenna has several low-VSWR bands, two of which would be considered UWB because the fractional bandwidth is greater than 25% of the band's center frequency, $f_c$. The loading resistance has the desired effect of increasing IBW, but it does so at the expense of radiation efficiency as shown in Figure 11. At low frequencies $\varepsilon$ is quite low, and it decreases very quickly with decreasing frequency. But near 3000 MHz $\varepsilon$ exhibits a sharp increase, with very reasonable values at higher frequencies. The unloaded PEC bowtie's efficiency is 100 % and consequently is not plotted.

Figs. 12 and 13, respectively, plot the *Variable* $Z_0$ antenna's input resistance and reactance. As expected, the effect of adding resistive loading is to reduce fluctuations in these variables. Figures 14 and 15 plot the maximum and minimum gains, respectively. As expected, adding resistive loading generally reduces maximum gain as expected, but, interestingly, it actually increases minimum gain in certain frequency ranges.

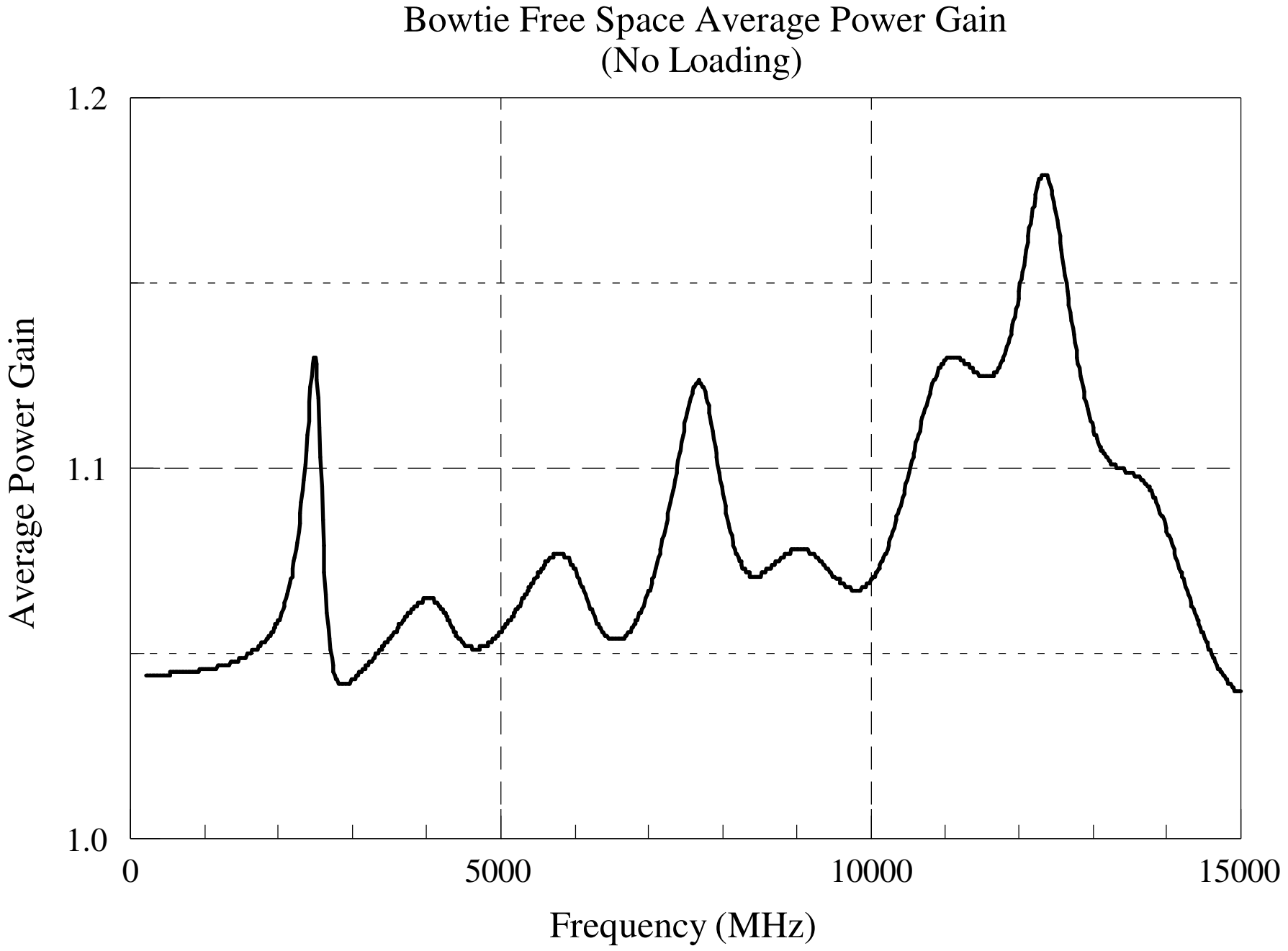

Fig.6. *Variable*-$Z_0$ bowtie NEC4 AGT plot ($Z_0 = 715\ \Omega$).

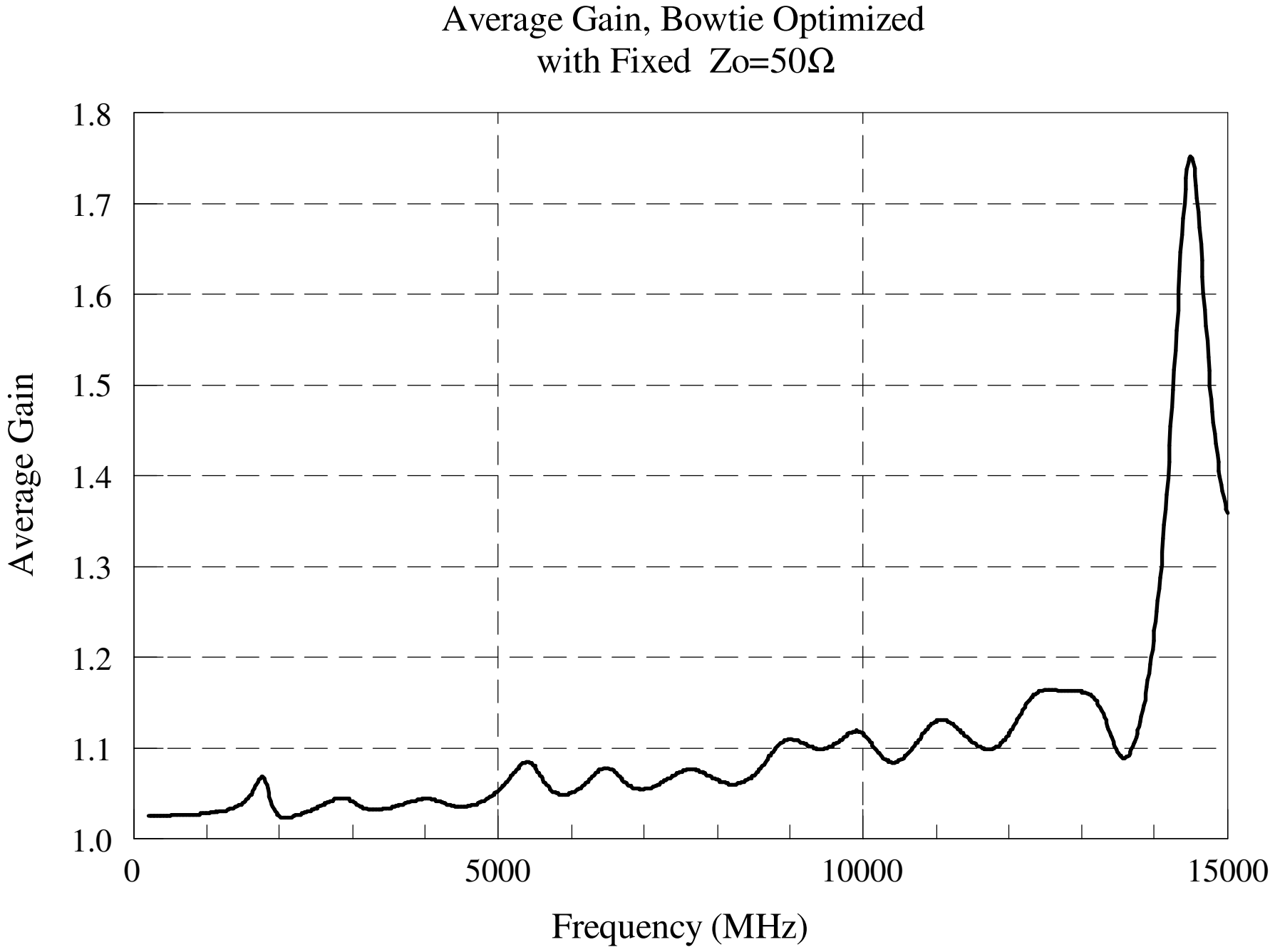

Fig.7. Fixed-$Z_0$ bowtie NEC4 AGT plot ($Z_0 = 50\ \Omega$ ).

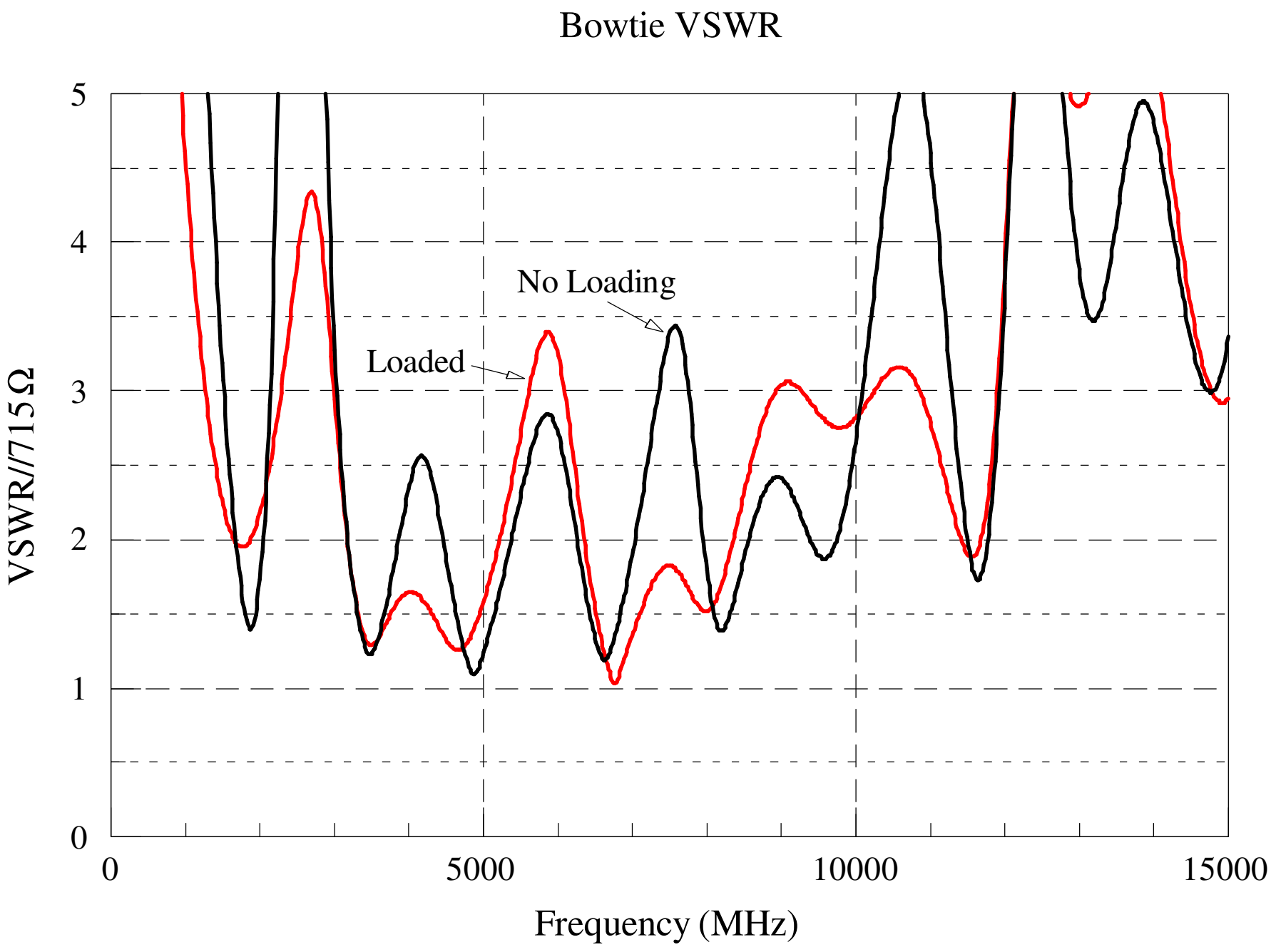

Fig.8. *Variable*-$Z_0$ bowtie VSWR ( $Z_0 = 715\ \Omega$ ).

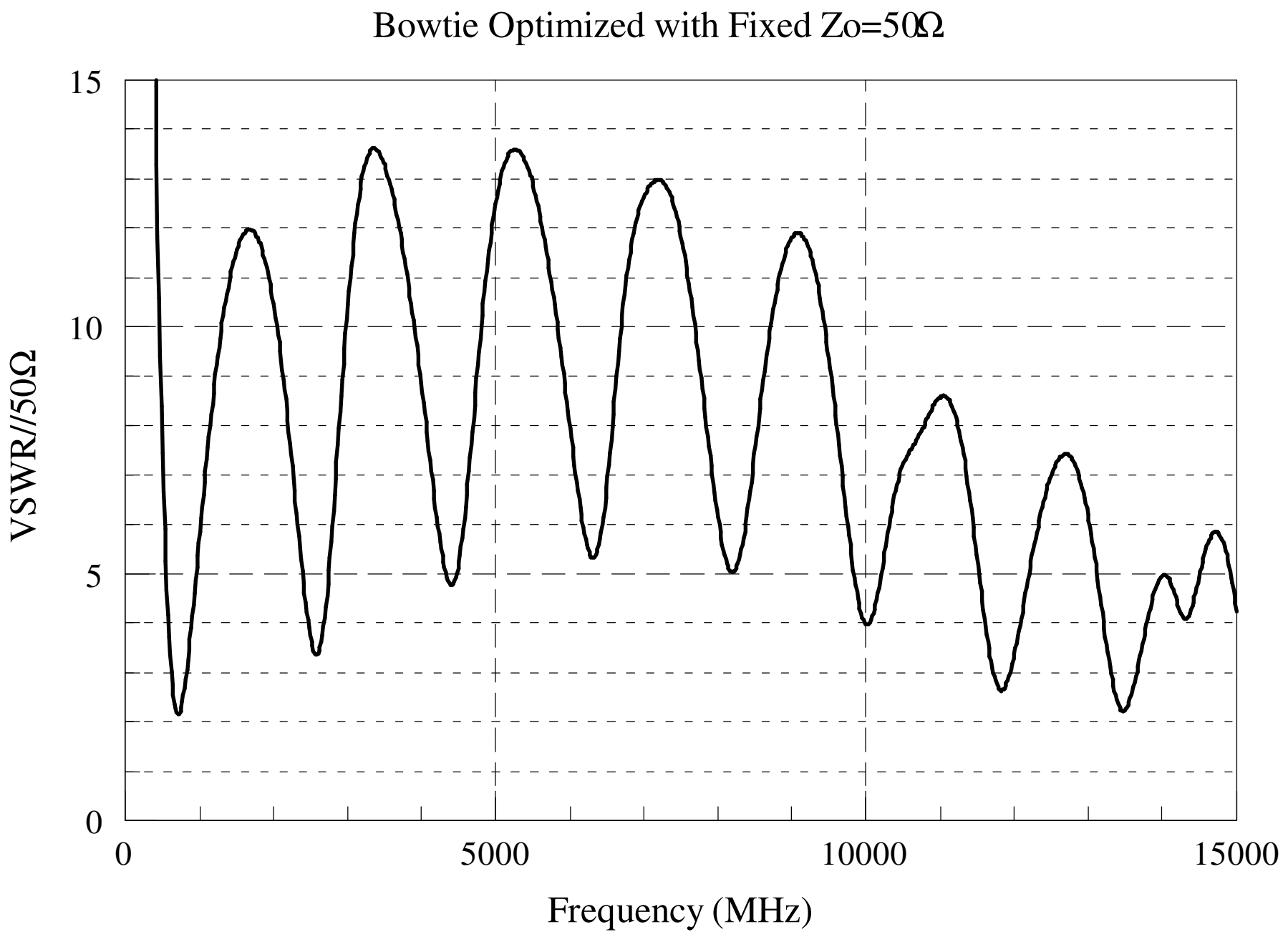

Fig.9. Fixed-$Z_0$ bowtie VSWR ( $Z_0 = 50\ \Omega$ ).

| *Resistively Loaded* | | | *No Loading* | | |
|---|---|---|---|---|---|
| Band Limits (MHz) | Width (MHz) | % of $f_c$ | Band Limits (MHz) | Width (MHz) | % of $f_c$ |
| 1655-1865 | 210 | 11.9% | 1685-2030 | 345 | 18.6% |
| 3185-5210 | 2025 | 48.2% | 3185-3860 | 675 | 18.9% |
| 6365-8345 | 1980 | 26.9% | 4460-5375 | 915 | 18.6% |
| 11420-11675 | 255 | 2.21% | 6260-7025 | 765 | 11.52% |
| - | - | - | 7955-8555 | 600 | 7.27% |
| - | - | - | 9395-9740 | 345 | 3.6% |
| - | - | - | 11510-11765 | 255 | 2.19% |

Fig.10. *Variable-* $Z_0$ bowtie $VSWR // 715\ \Omega \leq 2:1$ bands ($f_c$=band center).

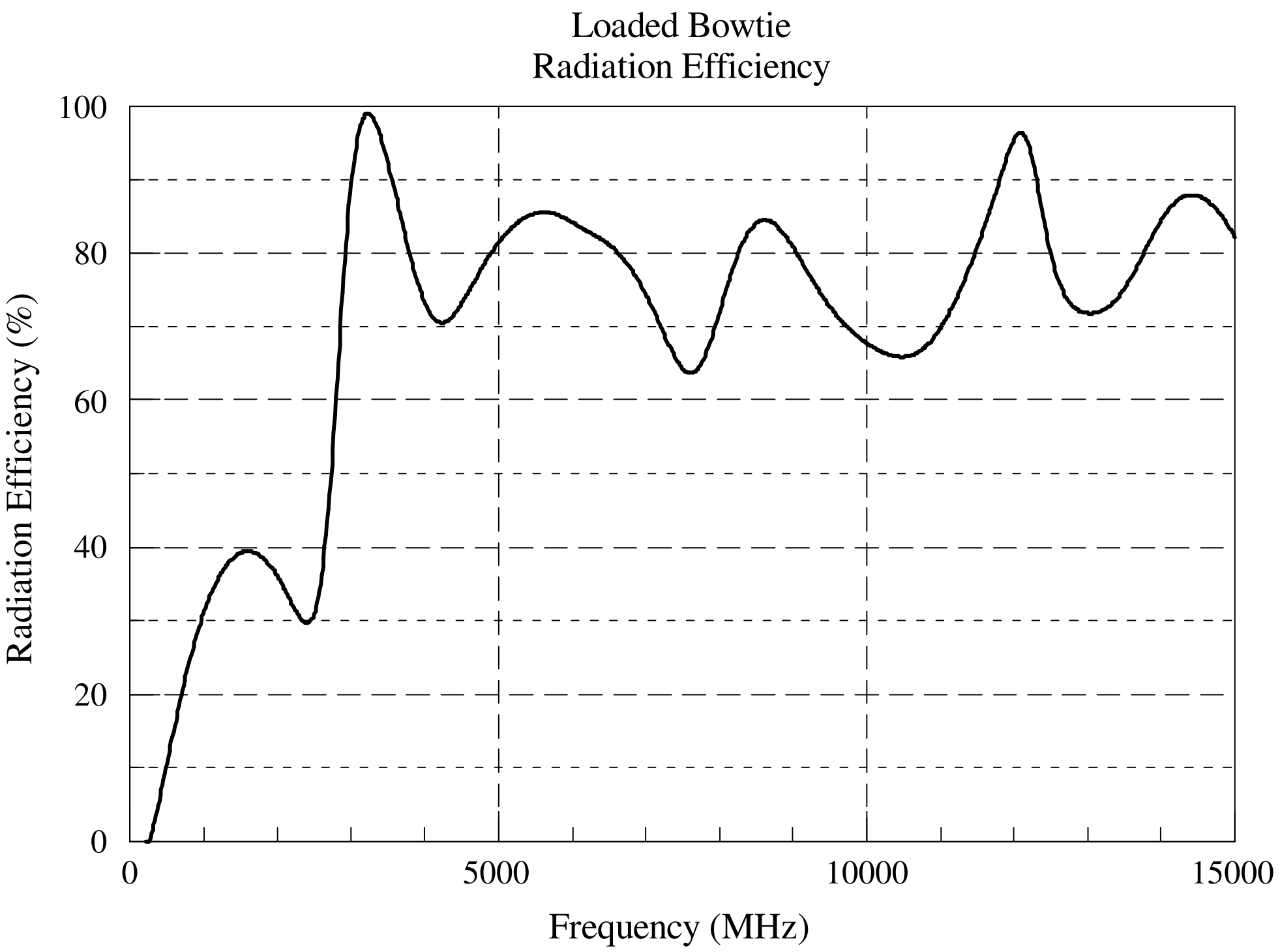

Fig.11. *Variable-* $Z_0$ Resistively loaded bowtie radiation efficiency.

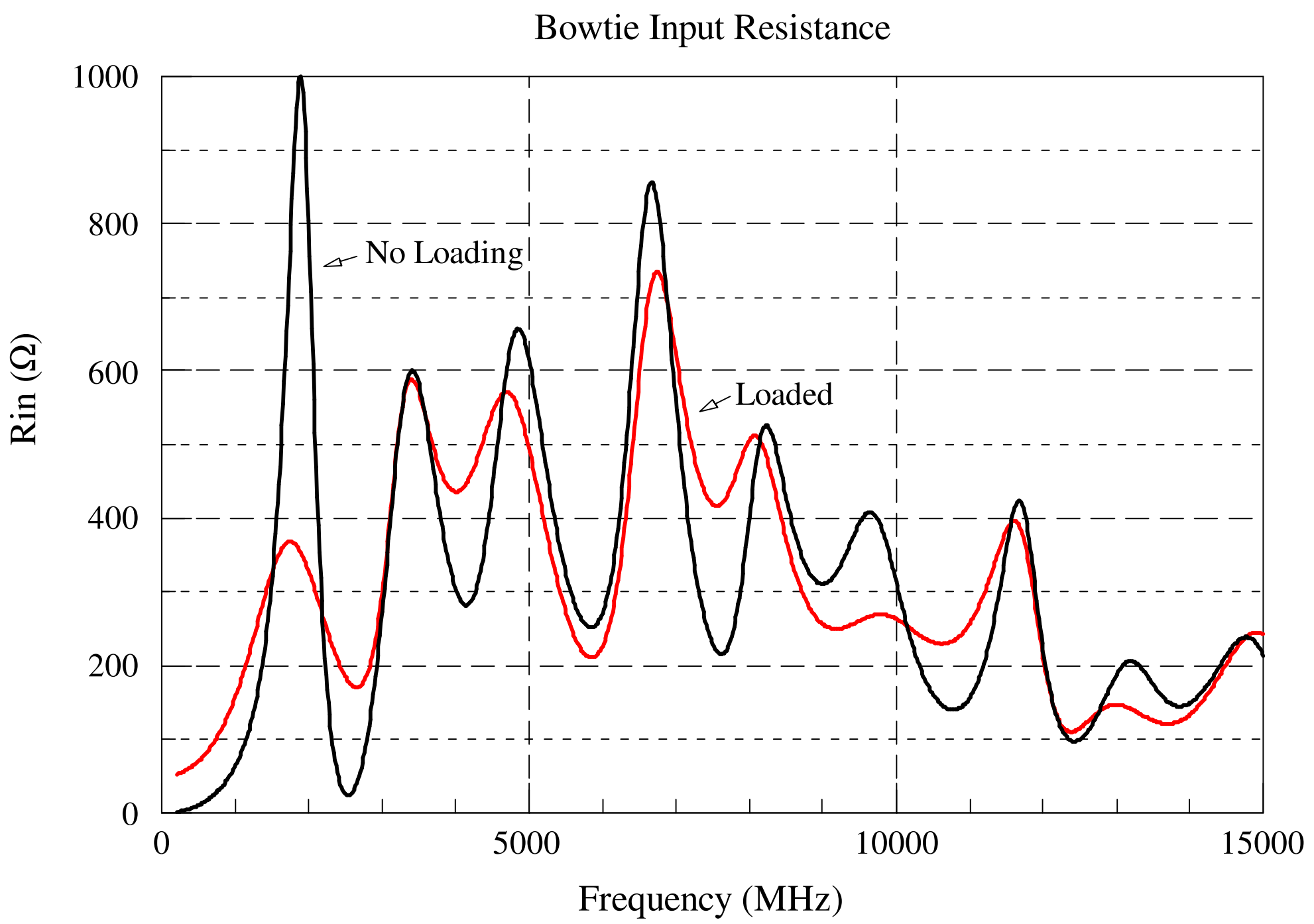

Fig.12. *Variable-* $Z_0$ bowtie input resistance.

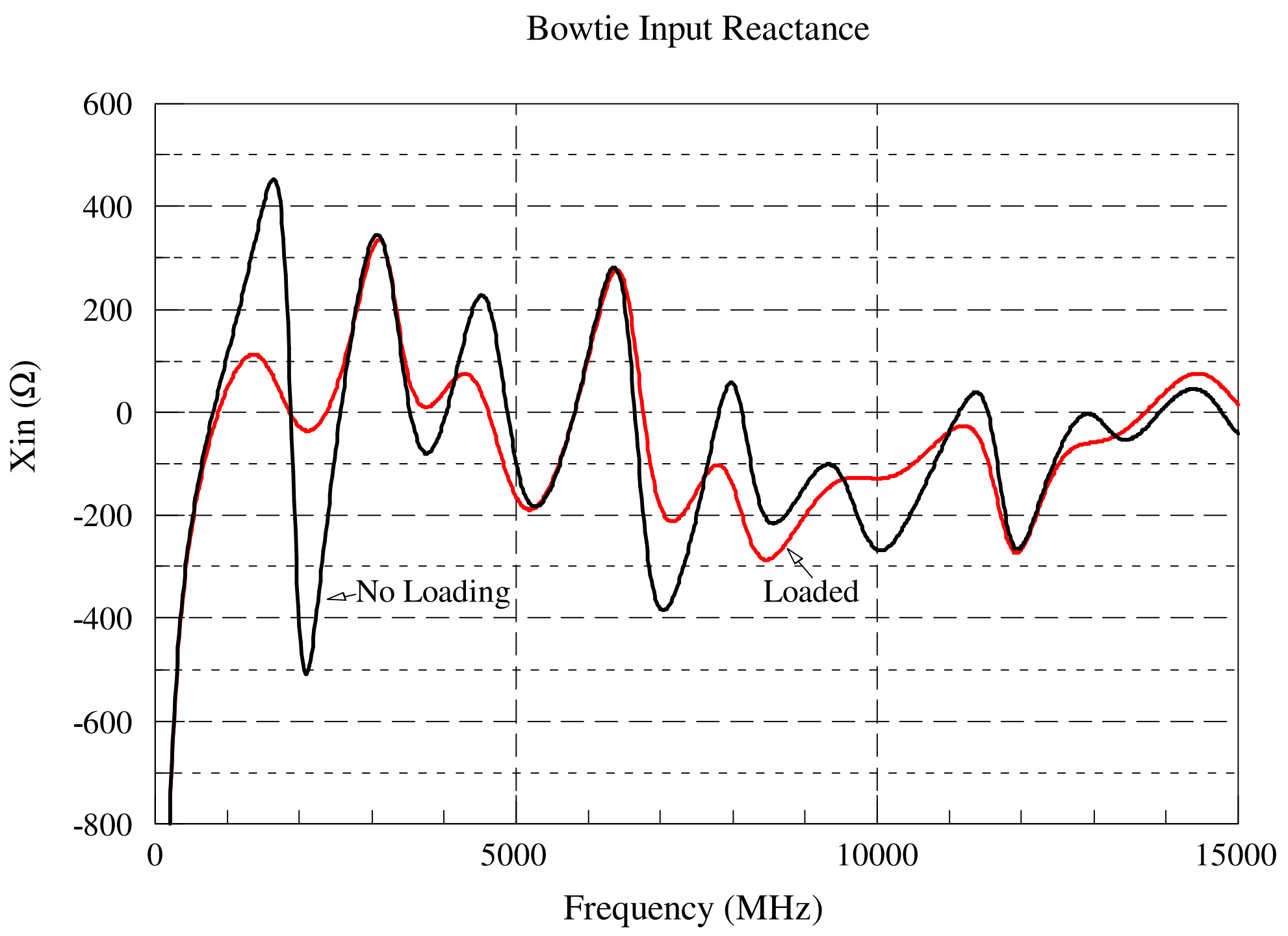

Fig. 13. *Variable-* $Z_0$ bowtie input reactance.

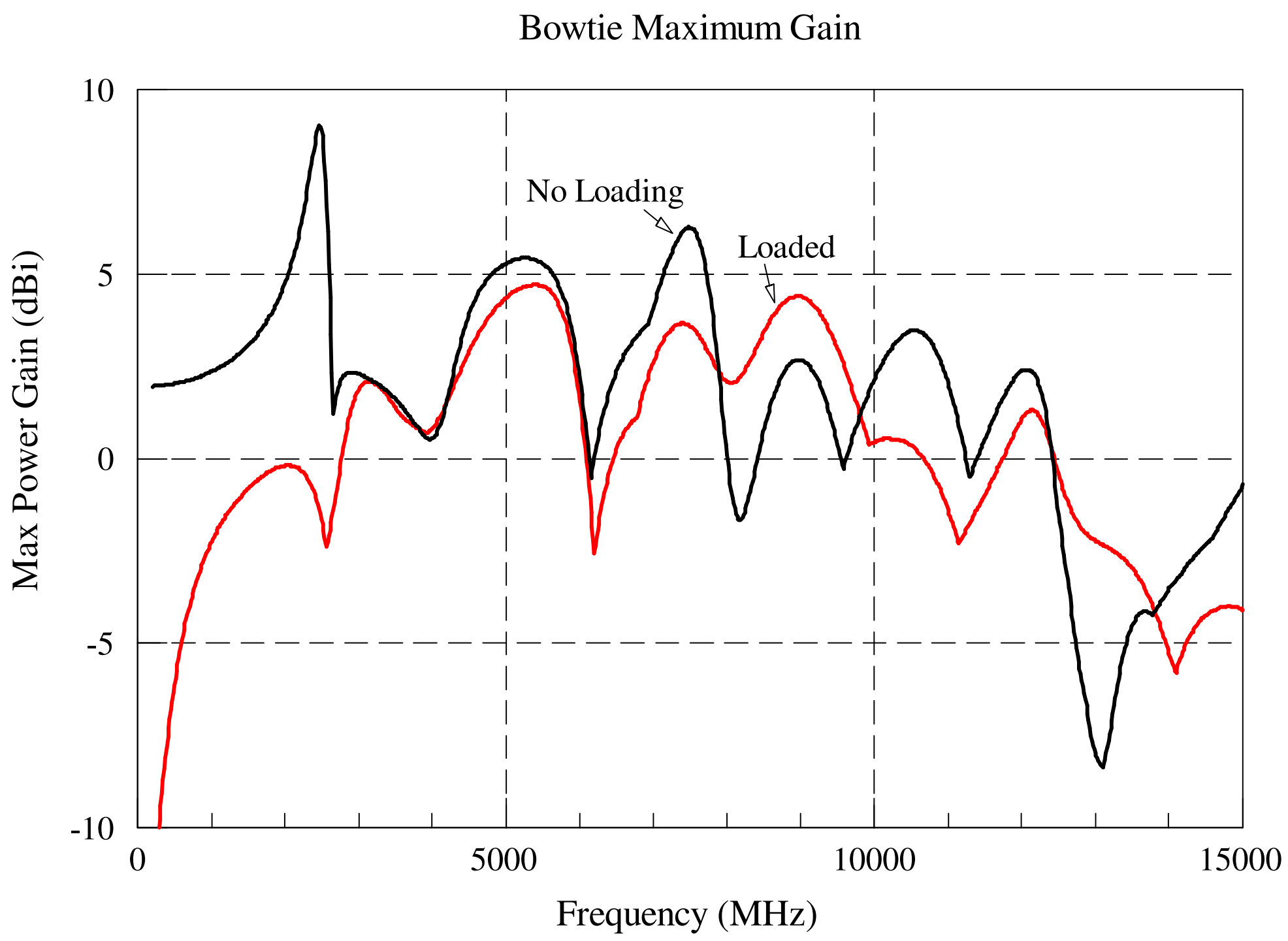

Fig. 14. *Variable-* $Z_0$ bowtie maximum power gain.

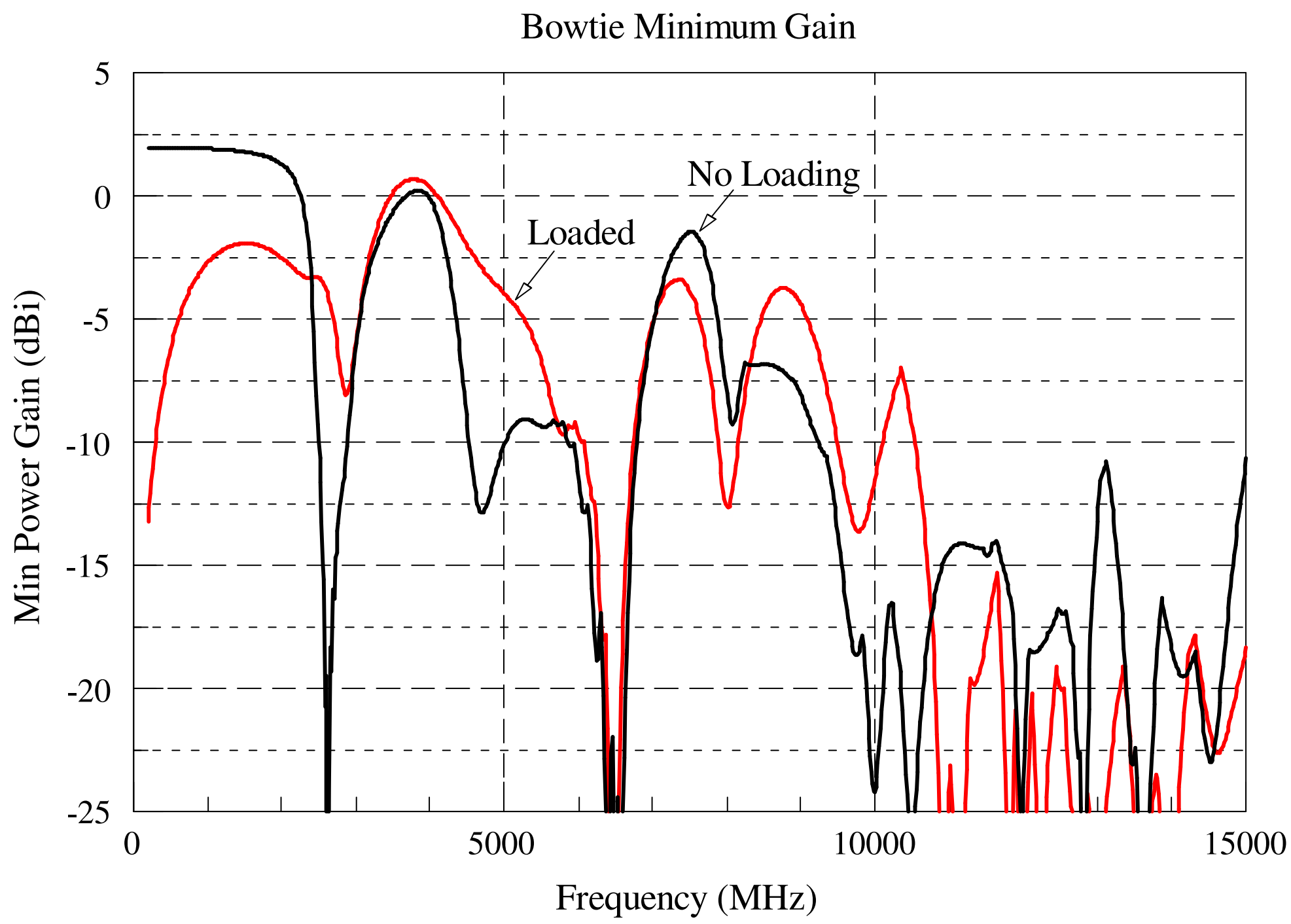

Fig. 15. *Variable-* $Z_0$ bowtie minimum power gain.

```
CM File: BESTBOWTIE.NEC
CM R-LOADED BOWTIE IN FREE SPACE WITH
CM Zo AS AN OPTIMIZATION PARAMETER.
CM Antenna in Y-Z plane.
CM Run ID: 07022011_221747
CM Fitness function:
CM [Min(Eff)+5*Min(Gmax)]/[|Zo-MaxRin|*(MaxVSWR-MinVSWR)*(MaxXin-MinXin)]
CM Arm Length = .051 meters
CM Bowtie HALF Angle = 39.4 degrees
CM Zo = 715 ohms
CM Rload = 166.93 ohms
CM Loaded Seg # = 6/9
CM File ID 07032011225840
CM Nd = 5, p = 1, j = 35
CE
GW1,3,0.,-.01,0.,0.,.01,0.,.0005
GW2,9,0.,.01,0.,0.,.049,.032,.0005
GW3,9,0.,.01,0.,0.,.049,-.032,.0005
GW4,9,0.,.049,.032,0.,.049,-.032,.0005
GW5,9,0.,-.01,0.,0.,-.049,.032,.0005
GW6,9,0.,-.01,0.,0.,-.049,-.032,.0005
GW7,9,0.,-.049,.032,0.,-.049,-.032,.0005
GE
LD0,2,6,6,166.93,0.,0.
LD0,3,6,6,166.93,0.,0.
LD0,5,6,6,166.93,0.,0.
LD0,6,6,6,166.93,0.,0.
FR 0,1001,0,0,200.,15.
EX 0,1,2,1,1.,0.
RP 0,19,1,1001,0.,0.,5.,0.,100000.
EN
```

Fig. 16. *Variable-* $Z_0$ bowtie NEC4 input file.

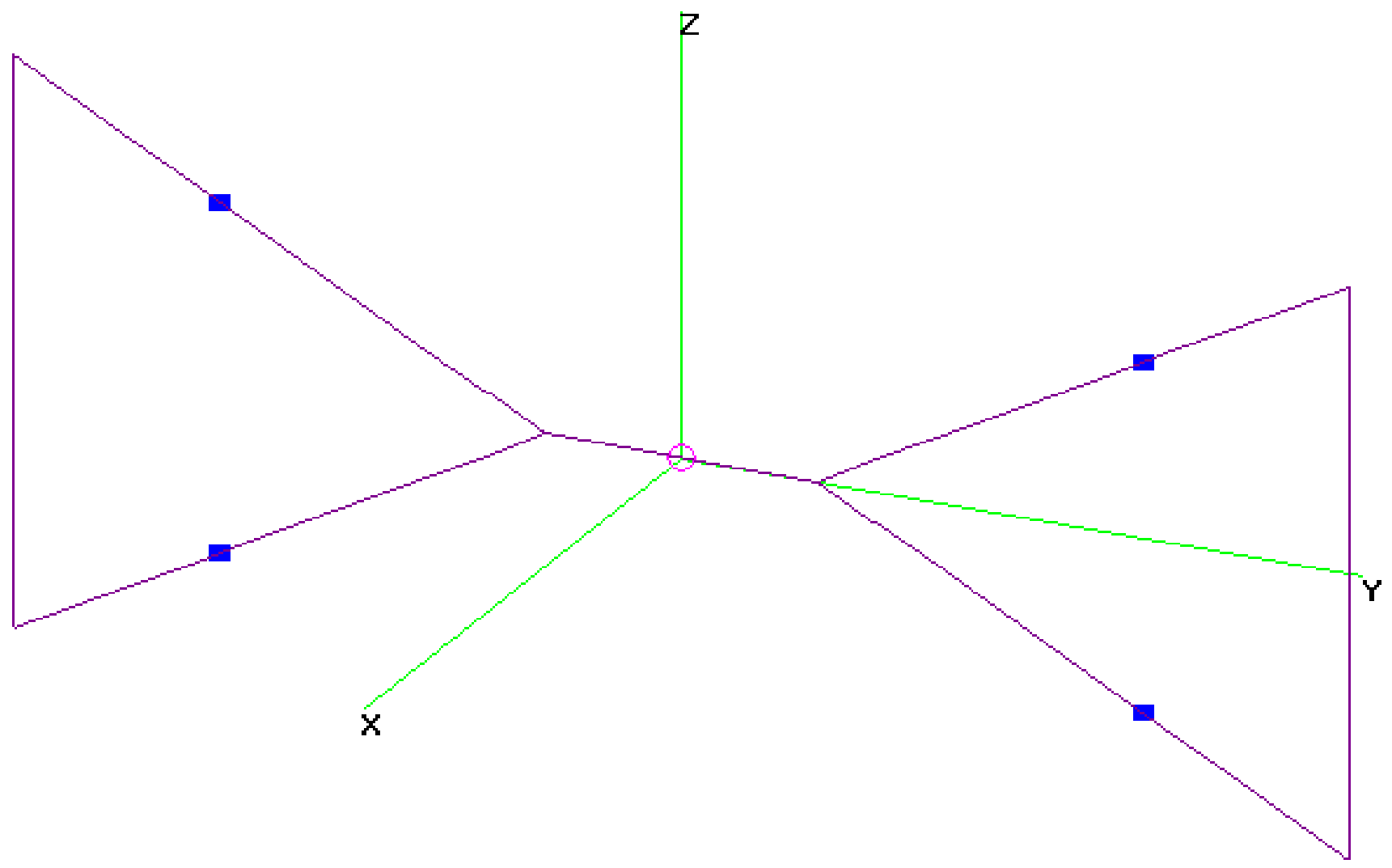

Fig. 17. *Variable-* $Z_0$ bowtie 4nec2 geometry display.

## 6. Design example: broadband Yagi-Uda arrays

The second *Variable* $Z_0$ design example is two 6-element Yagi-Uda ("Yagi") arrays. A Yagi comprises a single center-fed dipole driven element (DE), one or more reflectors (REF) on one side of DE (usually only one), and a group of directors on the other side (often many). DE is excited by the RF source, and all other elements are parasitic. Fig. 18 shows the array geometry as visualized by 4nec2 for Yagi design #1 (red circle is the RF source). All elements are parallel to the y-axis with the reflector on the axis. If the element radius is fixed (usually the same value for all elements), there are 11 geometric design variables, the 5 element spacings and their 6 lengths. Of course, in this case there is a twelfth design variable, *viz.*, $Z_0$. Fixing the element radius is an example of the traditional methodology in which certain design parameters are fixed at the outset, so that their values are not determined by the methodology. Fig. 19 shows the NEC4 input file for this antenna.

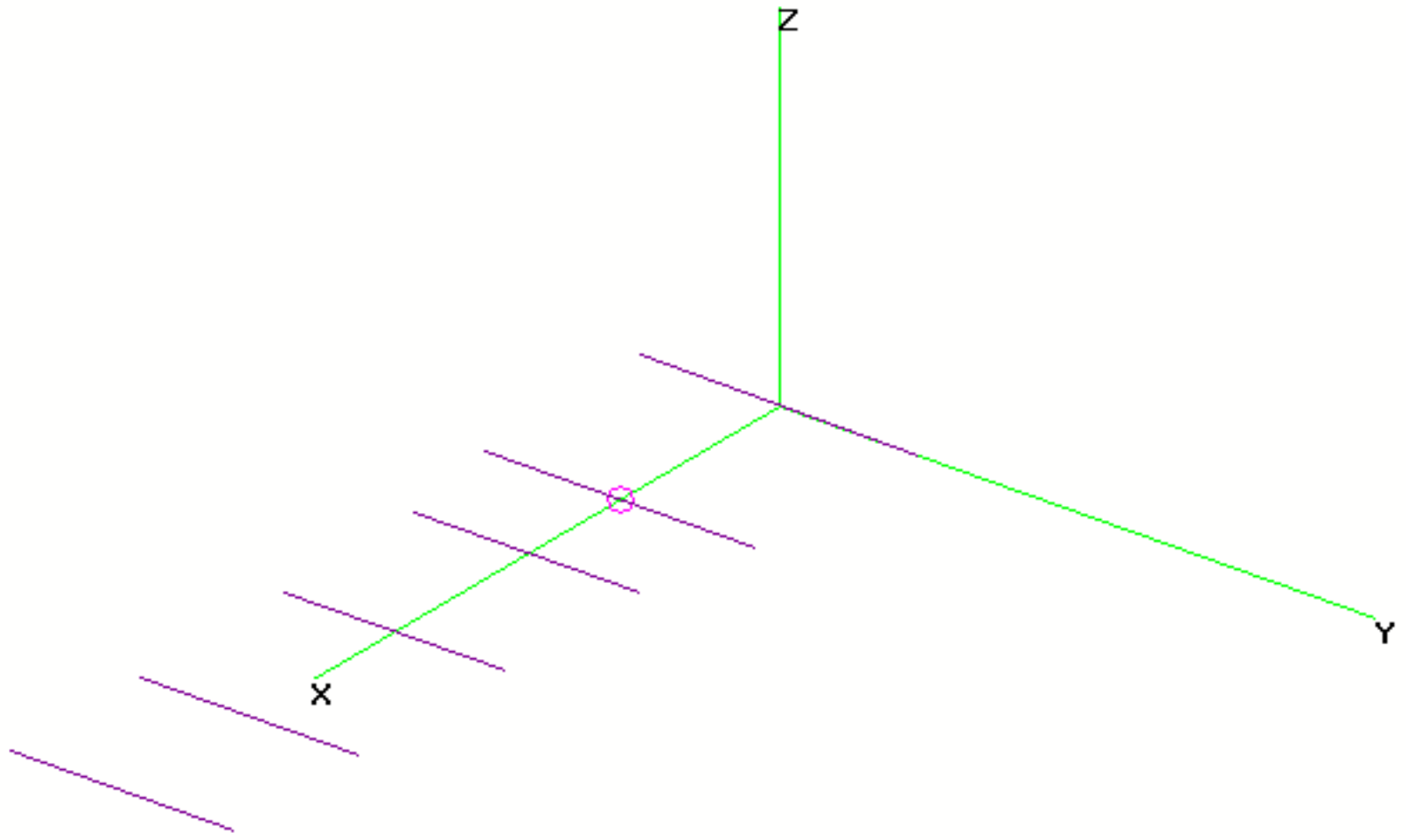

Fig. 18. 4nec2 geometry display for *Variable*-$Z_0$ 6-element Yagi design #1.

```
CM File: YAGI.NEC
CM YAGI ARRAY IN FREE SPACE
CM Band center frequency, Fc = 299.8 MHz
CM Run ID: 07102011_205004
CM Fitness function:
CM Gfwd(L)/5-2*VSWR(L)+Gfwd(M)-4*VSWR(M)+Gfwd(U)-2*VSWR(U)/5
CM where L,M,U are lower/mid/upper frequencies
CM Zo=65.75 ohms
CM File ID 07122011125139
CM Nd= 13, p= 20, j= 160
CE
GW1,9,0.,-.234,0.,0.,.234,0.,.00635
GW2,9,.343,-.228,0.,.343,.228,0.,.00635
GW3,9,.54,-.19,0.,.54,.19,0.,.00635
GW4,9,.827,-.186,0.,.827,.186,0.,.00635
GW5,9,1.137,-.184,0.,1.137,.184,0.,.00635
GW6,9,1.41,-.189,0.,1.41,.189,0.,.00635
GE
FR 0,1501,0,0,200.,.1
EX 0,2,5,1,1.,0.
RP 0,19,19,1001,0.,0.,5.,10.,100000.
EN
```

Fig. 19. *Variable*-$Z_0$ Yagi design #1 NEC4 input file.

The antenna was CFO-optimized against the following fitness function (to be maximized):

$$F(Z_0, L_i, S_j) = c_1 \cdot G_{fwd}(f_L) - c_2 \cdot VSWR(f_L) + c_3 \cdot G_{fwd}(f_C) - c_4 \cdot VSWR(f_C) + c_5 \cdot G_{fwd}(f_U) - c_6 \cdot VSWR(f_U).$$

$L_i$, $i = 1,..,6$ and $S_j$, $j = 1,..,5$ are the element lengths and spacings, respectively. The constants $c_i$ are empirically determined weighting coefficients; $G_{fwd}(f)$ the Yagi's forward directivity (along the +X-axis, $\theta = 90°, \phi = 0°$); and $f_L$, $f_C$, and $f_U$, respectively, the lower, center, and upper frequencies defining the band within which the Yagi is optimized.

The same CFO parameters that were used for the loaded bowtie were used for the two Yagi designs, but with $\max\left(\frac{N_p}{N_d}\right) = 6$ for both antennas, and with $N_t = 250$, $N_t = 200$ for the first and second designs, respectively. Note that, although the Yagi problem contains 12 design *variables*, as a matter of programming convenience in dealing with array dimensions, CFO treats the problem as 13-D. The Yagi element REF is assigned a fixed "spacing" of zero because it is placed on the Y-axis, and this is the 13th "variable" (see the source code listing in Appendix I for details). For Yagi design #1, CFO's best fitness was 14.62534041 (CFO probe #20) with $\frac{N_p}{N_d} = 4$ and $\gamma = 0.6$ at step 160, while for design #2 it was 0.93193733 (probe #28) with $\frac{N_p}{N_d} = 4$ and $\gamma = 0.2$ at step 168.

The decision space was $0.2 \le L_i \le 0.6\lambda$, $0.1 \le S_j \le 0.5\lambda$, where $\lambda$ is the wavelength corresponding to the band center frequency $f_C$. The *variable* feed system impedance was bounded by $5 \le Z_0 \le 600\,\Omega$. The first Yagi was optimized over the band 275-325 MHz and the second over 250-350 MHz (in both cases $\lambda \approx 1\text{m}$ @ $f_C$). In post processing, performance was computed from 200-350 MHz every 0.1 MHz. All array elements have the same radius $0.00635\lambda$ (0.5-inch diameter elements at 299.8 MHz). In the NEC4 model, lengths were rounded to three decimal places (except radius) and $Z_0$ to two. The fitness weighting coefficients are tabulated in Fig. 20.

| *Design #* | $c_1$ | $c_2$ | $c_3$ | $c_4$ | $c_5$ | $c_6$ |
|---|---|---|---|---|---|---|
| 1 | 0.2 | 2 | 1 | 4 | 1 | 0.4 |
| 2 | 0.2 | 4 | 1 | 8 | 1 | 0.8 |

Fig. 20. *Variable-*$Z_0$ Yagi fitness coefficients.

Fig. 21 plots the evolution of CFO's best fitness for design #1, while Fig. 22 shows the convergence of CFO's probes onto the maximum ($D_{avg}$ is the average distance between the probe with the best fitness and all other probes). The best probe number as a function of time step appears in Fig. 23. These results are typical of CFO's behavior, both on antenna problems and on established benchmark functions (see, for example, [25-27]).

The optimum feed system impedance computed by CFO was $Z_0 = 65.75\,\Omega$ for design #1 and $Z_0 = 89.88\,\Omega$ for design #2. Although Yagi arrays are generally considered "low" impedance antennas (typically $Z_{in} \approx 10 - 30\,\Omega$), the optimized feed system $Z_0$'s are relatively high. These values, however, are more readily matched to the "standard" values of $50 - 75\,\Omega$, which is a fortuitous outcome. The optimized array geometry is tabulated in Fig. 24. The spacings have been added to specify each element's position ($X$ − coordinate, "boom dist") along the Yagi's boom (+$X$ − axis).

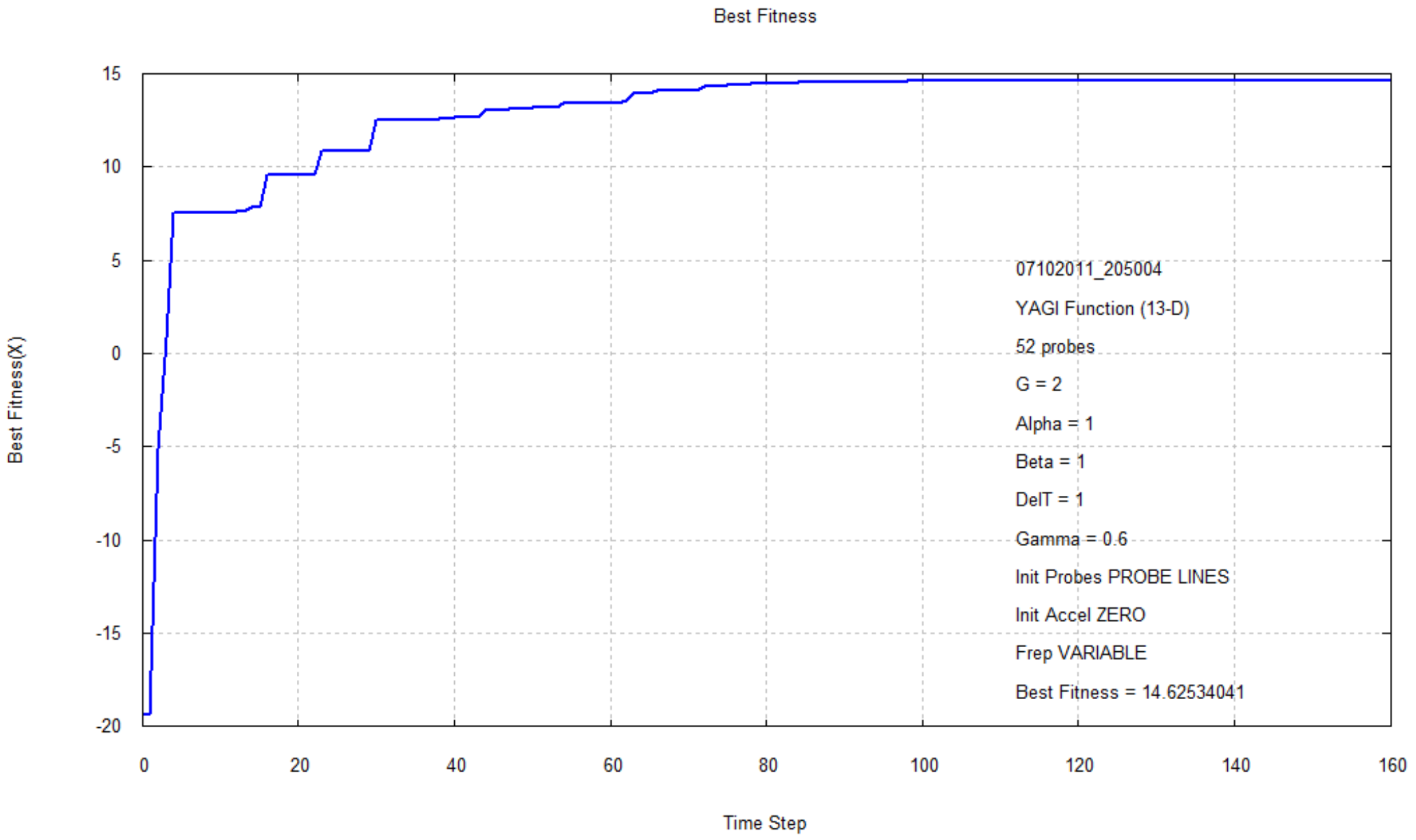

Fig. 21. *Variable-* $Z_0$ CFO fitness evolution for Yagi design #1.

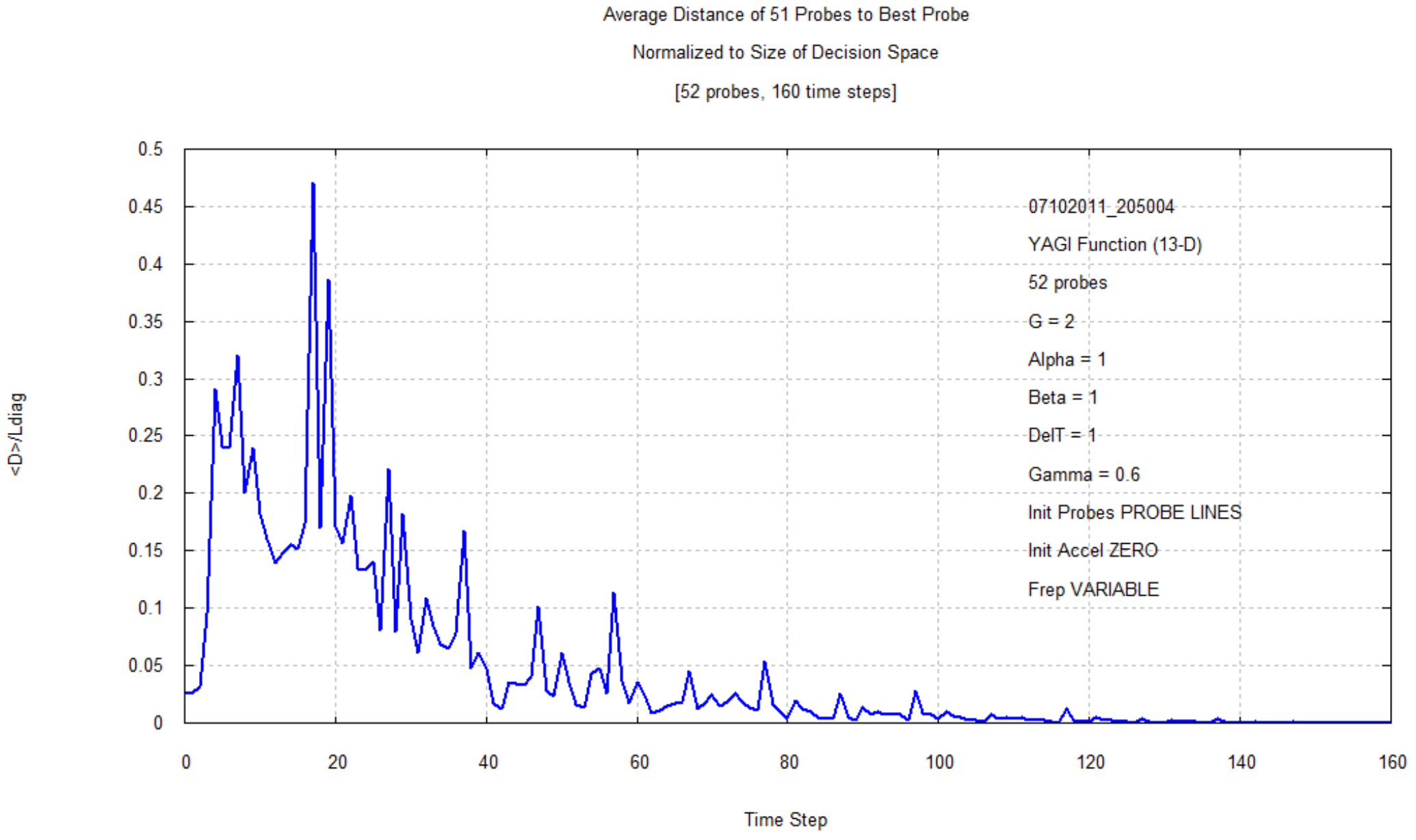

Fig. 22. *Variable-* $Z_0$ CFO $D_{avg}$ evolution for Yagi design #1.

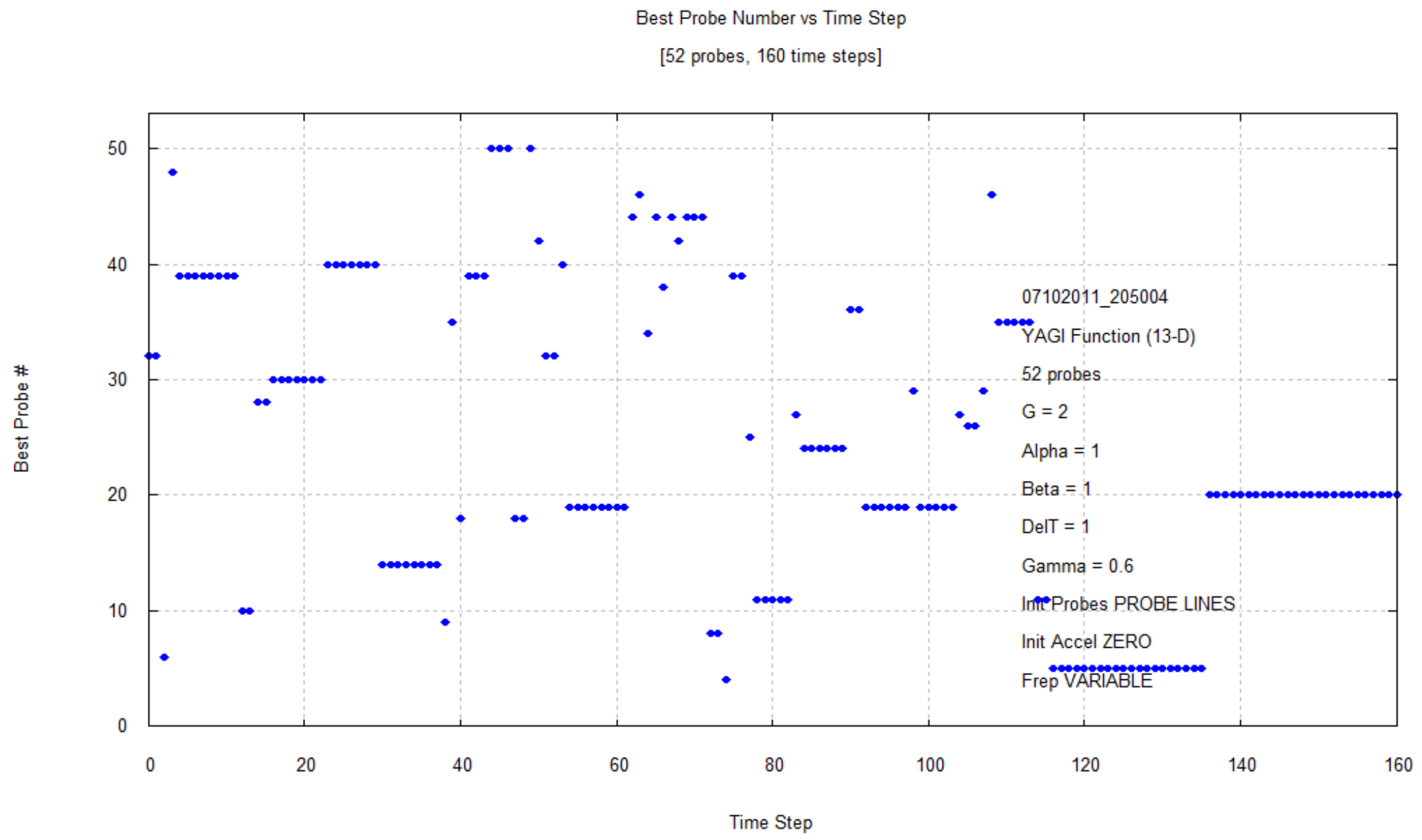

Fig. 23. *Variable-* $Z_0$ CFO best probe evolution for Yagi design #1.

| *element #* | *design* #1 ($Z_0 = 65.75\,\Omega$) | | *design* #2 ($Z_0 = 89.88\,\Omega$) | |
|---|---|---|---|---|
| | *length* ($\lambda$) | *boom dist* ($\lambda$) | *length* ($\lambda$) | *boom dist* ($\lambda$) |
| 1(REF) | 0.468 | 0 | 0.564 | 0 |
| 2 (DE) | 0.456 | 0.343 | 0.500 | 0.305 |
| 3 | 0.380 | 0.540 | 0.370 | 0.434 |
| 4 | 0.372 | 0.827 | 0.360 | 0.715 |
| 5 | 0.368 | 1.137 | 0.364 | 0.921 |
| 6 | 0.378 | 1.410 | 0.344 | 1.238 |

Fig. 24. Geometry of *Variable-* $Z_0$ CFO-optimized Yagi arrays.

VSWR curves for the *Variable* $Z_0$ optimized Yagis appear in Fig. 25 (note, $0.994 \le \text{AGT} \le 0.998$ not plotted). For design #1, $VSWR // 65.75\Omega \le 2:1$ from 273.5 MHz to 319.6 MHz and from 329.3 MHz to 333.8 MHz with a peak of 2.2:1 between these bands at 324.8 MHz. As a practical matter, the 2:1 VSWR IBW can be reasonably considered to be 273.5-333.8 MHz (60.3 MHz) centered on 303.65 MHz, resulting in an IBW of 19.86%. This bandwidth is quite large for a Yagi, which usually is thought of as "narrowband" structure (typically, IBW of a few percent).

The performance of design # 2 is even better. Its VSWR is less than 2:1 from 249.7 to 322.7 MHz (73 MHz) centered at 286.2 MHz for an IBW of 25.5%, which is quite remarkable because this erstwhile "narrowband" device actually exhibits ***UWB*** performance without any impedance loading or other measures to increase bandwidth. Of course, for these two designs, matching the optimized feed system impedances of $Z_0 = 65.75\ \Omega$ and $Z_0 = 89.88\ \Omega$ to the usual $50\ \Omega$ requires 1.32:1 and 1.8:1 broadband transformers, respectively (or other suitable matching device).

These interesting results highlight another consideration in antenna design, the effect of different fitness functions on the resulting antenna. Any change in the objective function, however slight, changes the decision space landscape, and, consequently, the "optimum" antenna in that new landscape. The question is, How to specify an objective function that actually meets specific performance objectives?, a conundrum not easily answered. The best way to address it is with a *deterministic* optimizer instead of a *stochastic* one. Deterministic algorithms like CFO return the same result for every run with the same setup parameters, while stochastic metaheuristics, ACO or PSO, for example, yield different results with every run because they rely on variables whose values are unknowable in advance of their being calculated from a probability distribution.

The only way the antenna designer can know that changing the fitness function actually produced a better antenna, instead of being the result of an optimizer's inherent stochasticity, is by using a deterministic algorithm. If this run's Yagi is better than the last, is it because of those new coefficients, or, instead, luck of the draw? There simply is no way to know. The flip side is that a genuinely better objective function might produce a worse antenna if it is used with a stochastic optimizer.

Deterministic metaheuristics such as CFO are much better suited to real world problems like designing antennas, an issue discussed at greater length in [25] by way of an example (resistively loaded monopole). In the Yagi designs considered here, changing the fitness coefficients slightly produced significantly different antennas. Yet another set of coefficients will yield yet another optimal design, and a different fitness function altogether will result in possibly quite different antennas. In this regard, the *Variable* $Z_0$ methodology is advantageous because it provides additional flexibility in meeting specific performance goals by creating a decision landscape with one more degree of freedom, a landscape that should be searched deterministically.

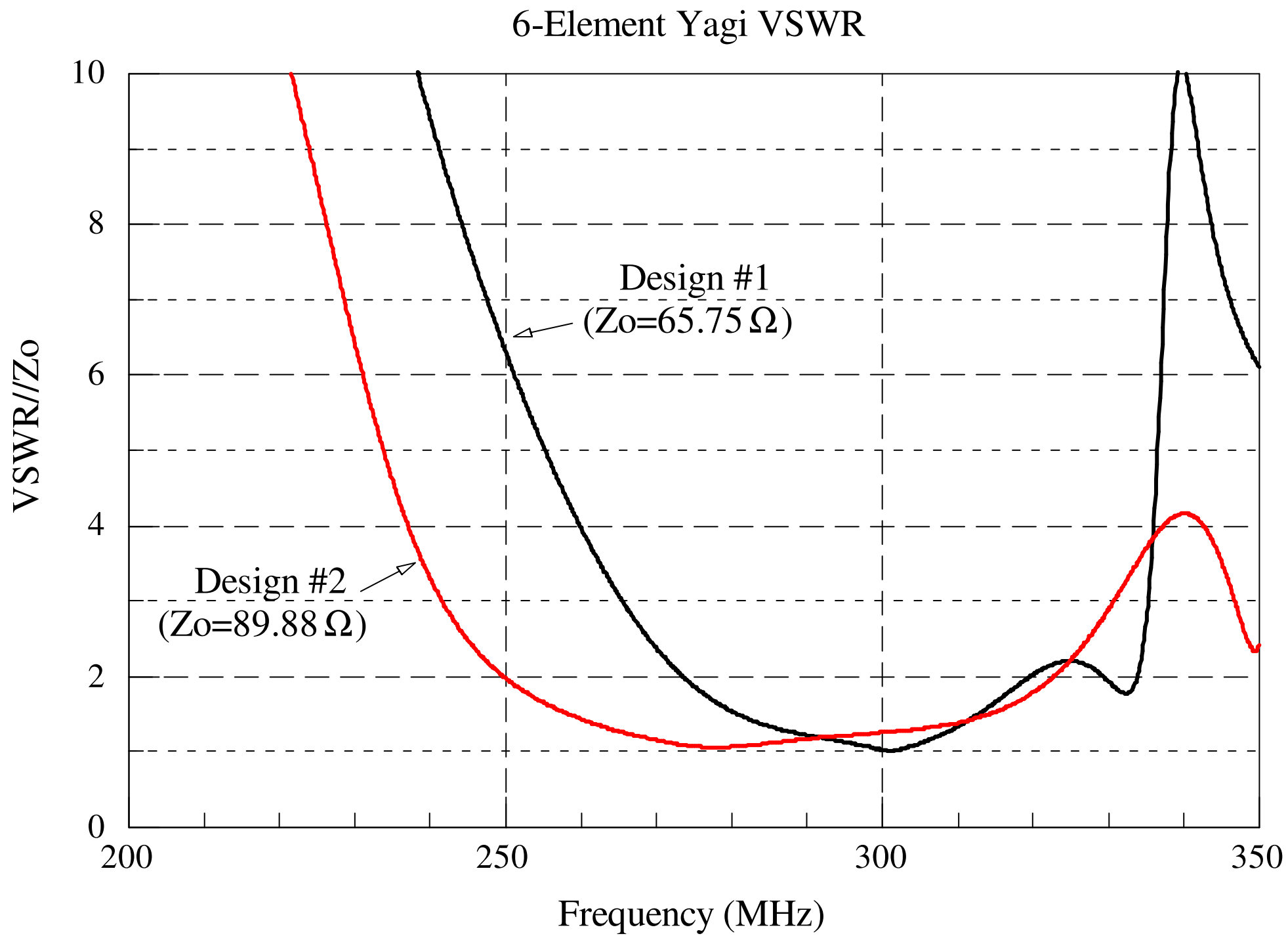

Fig. 25. *Variable*-$Z_0$ CFO-optimized Yagi $VSWR//Z_0$.

Figs. 26-28, respectively, plot the two Yagis' $G_{fwd}$, $R_{in}$, and $X_{in}$. The gain for design #1 increases monotonically from 8.29 dBi at 273.5MHz to 11.96 dBi at 333.8 MHz with a band-center gain of 9.73 dBi at 303.15 MHz. The corresponding values for design #2 are 7.66 dBi at 249.7 MHz, 9.5 dBi at 322.7 MHz, and 7.39 dBi at 286.2 MHz; and the gain is flatter than for design #1. These figures are quite good for a Yagi array with such large IBW. For example, the "A3" wideband Yagi design reported in [28] has $VSWR//50\ \Omega \leq 2:1$ from 262.5-309.8 MHz centered at 286.15 MHz (IBW 16.53%) with a maximum

in-band $G_{fwd}$ of 10.24 dBi as computed by NEC4. This antenna was optimized using the traditional approach of *fixing* $Z_0$ at $50\ \Omega$ with a deterministic optimizer (DCLS). While the *Variable* $Z_0$ Yagis and the DCLS-optimized array cannot be compared "head-to-head" because different optimization objectives were used, their performances can be compared as being representative of the state-of-the-art for broadband array design. This comparison clearly shows that the *Variable* $Z_0$ approach can produce better antennas.

Turning to the input impedance, $R_{in}$ is moderate across the band for both Yagis, but large fluctuations begin to appear at the high end. $X_{in}$ also is moderate and increases monotonically across the band, almost linearly. Each Yagi exhibits a single resonance, design #1 at 301.05 MHz and design #2 at 274.7 MHz.

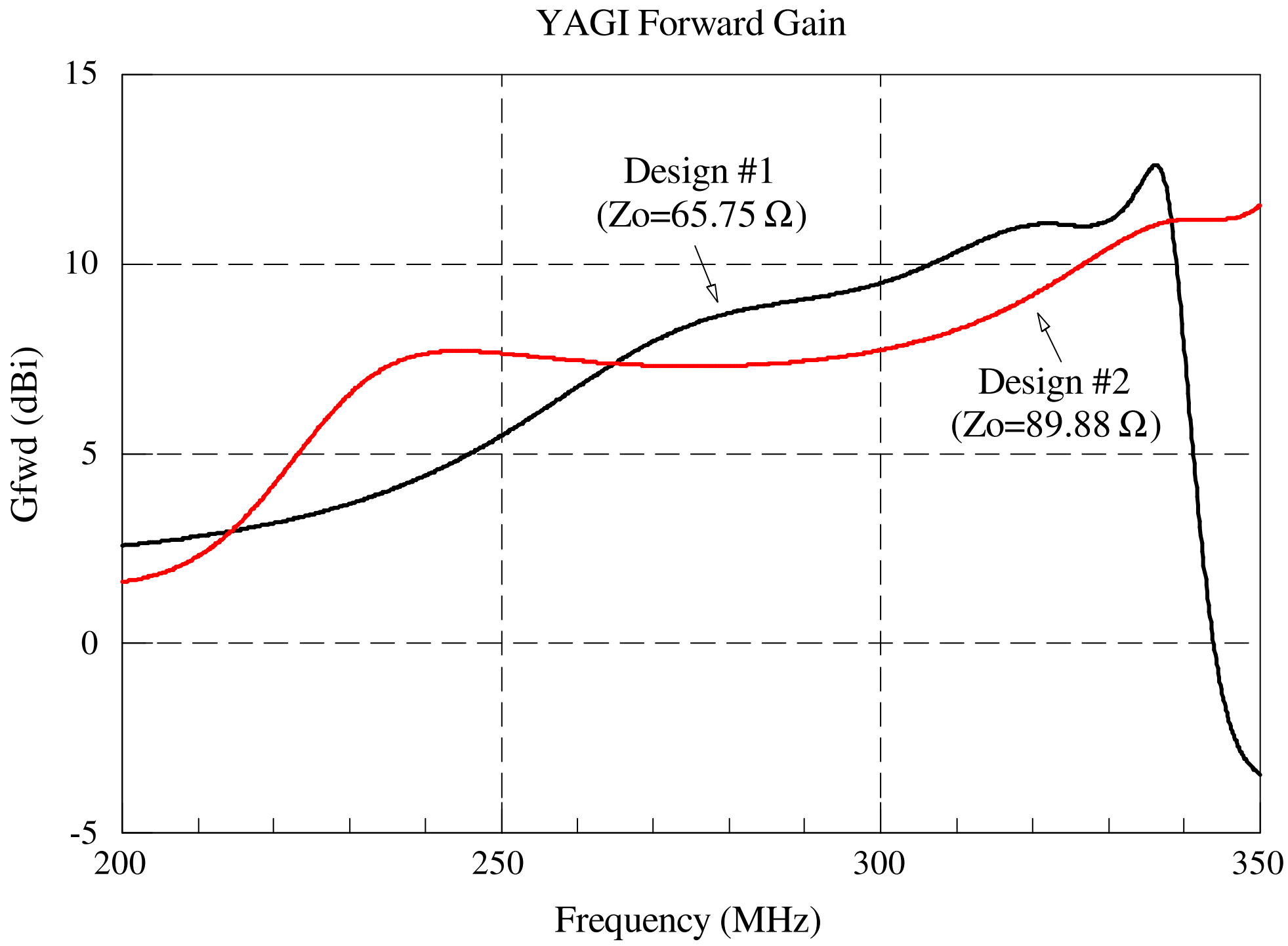

Fig. 26. *Variable-* $Z_0$ CFO-optimized Yagi $G_{fwd}$.

## 7. Conclusion

This paper describes *Variable* $Z_0$, which, to the author's knowledge, is a novel and proprietary methodology for antenna design and optimization. As an example, *Variable* $Z_0$ is applied to designing a CFO-optimized resistively-loaded bowtie antenna and two broadband Yagi-Uda arrays, both with good results. The resulting bowtie exhibits UWB properties in two bands useful for UWB systems, and one of the Yagis is UWB without any broadbanding mechanism such as impedance-loading. The performance of the *Variable* $Z_0$ Yagis compares very favorably to another state-of-the-art deterministically optimized array.

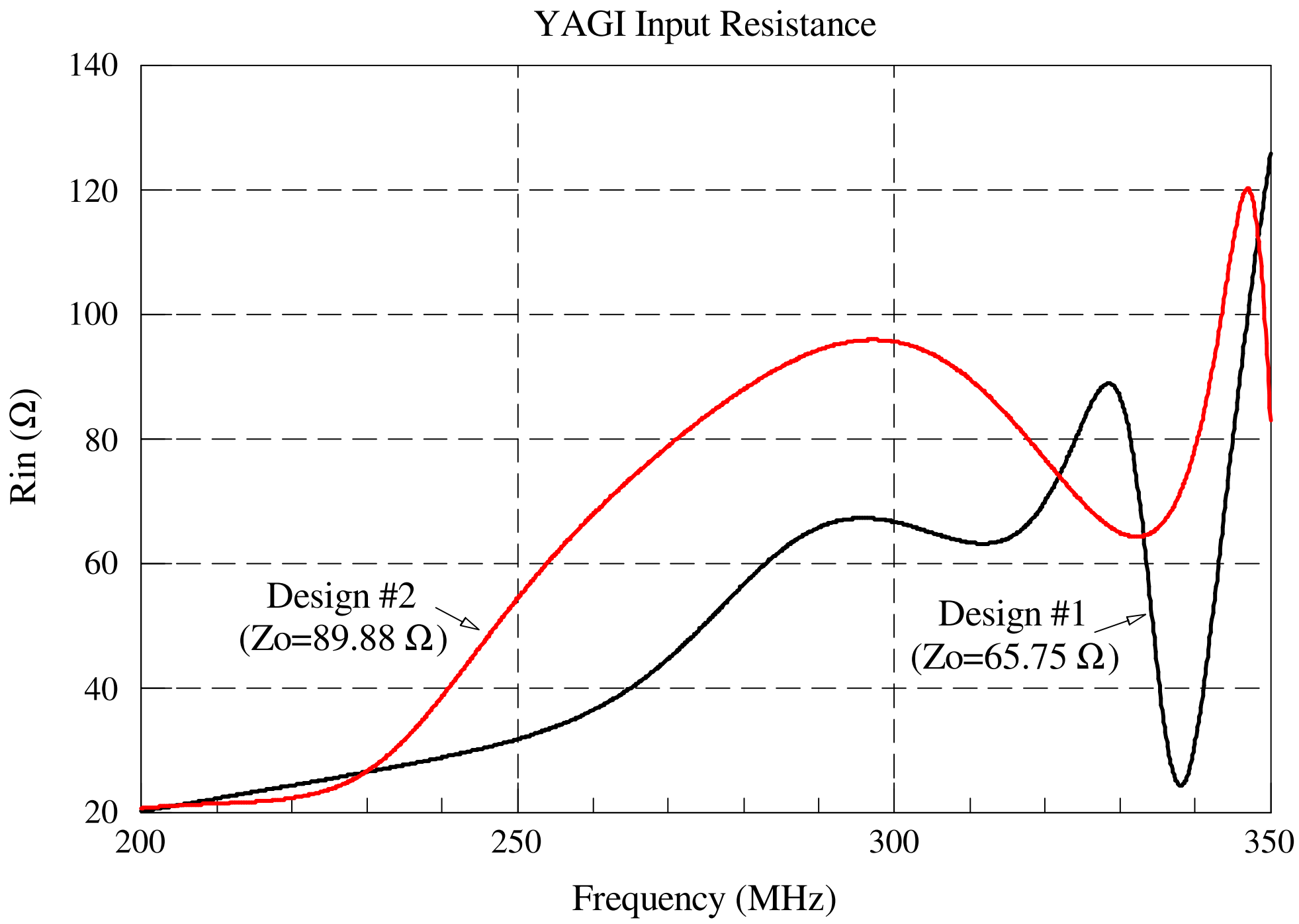

Fig. 27. *Variable*-$Z_0$ CFO-optimized Yagi $R_{in}$.

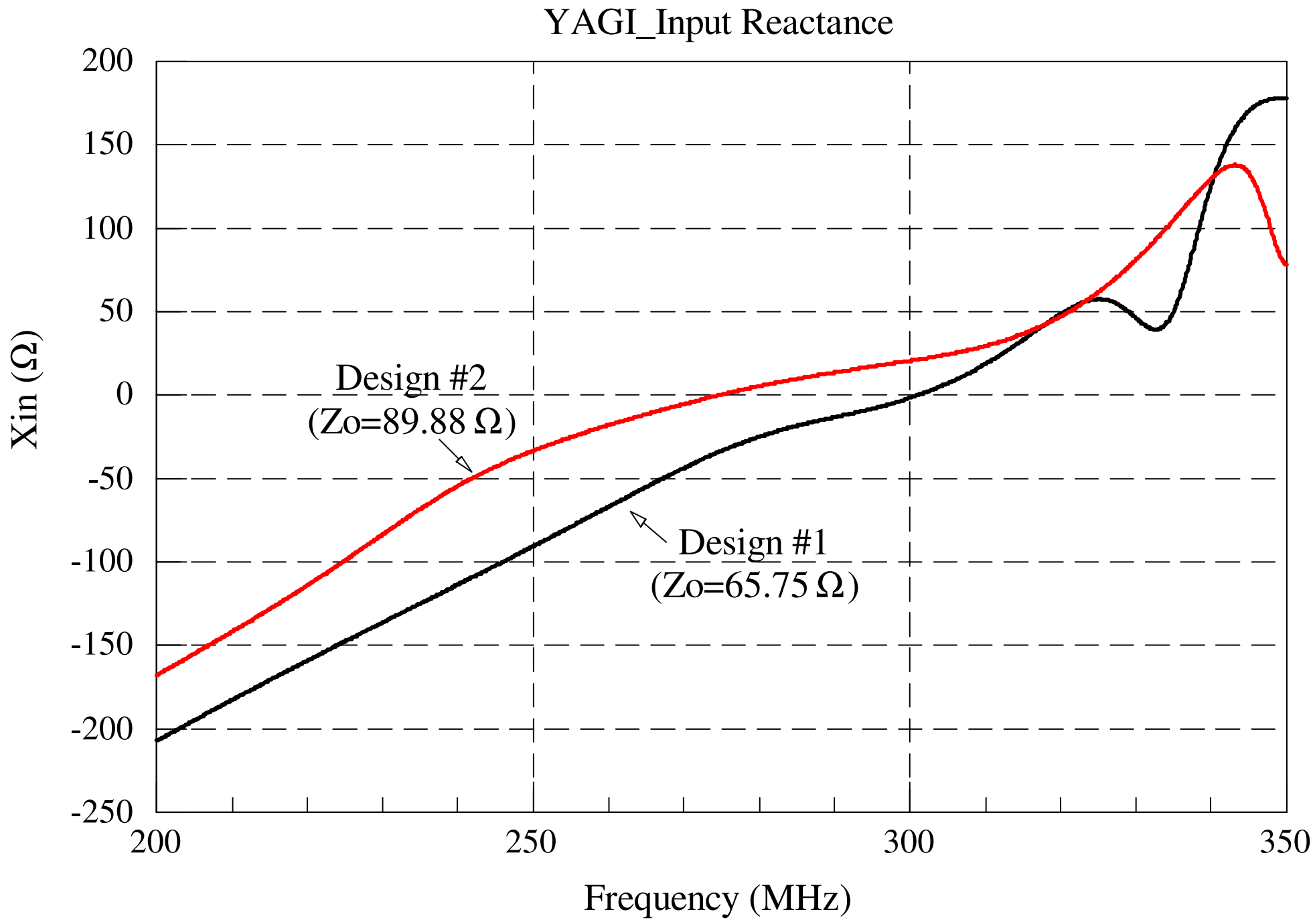

Fig. 28. *Variable*-$Z_0$ CFO-optimized Yagi $X_{in}$.

***14 July 2011***
***Brewster, Massachusetts***

# Appendix I

```
'Program 'CFO_R_ONLY_LOADED_BOWTIE_YAGI_07-10-2011.BAS'
'Compiled with Power Basic/Windows Compiler 9.04.0122 (www.PowerBasic.com).

'CFO-OPTIMIZED R-LOADED (1 RESISTOR/ARM) BOWTIE IN FREE SPACE AND
'FS YAGI-UDA ARRAY BOTH WITH Zo AS AN OPTIMIZATION <<<VARIABLE>>>.

'LAST MOD 07-10-2011 ~2049 HRS EDT

'****************** YAGI FITNESS FUNCTION ***********************
'THE USER CAN USE THE FOLLOWING FITNESS FUNCTION:
'        F=[A*FwdGain(DBi)-B*|Zo-Max(Rin)|-
'           C*Max|Xin|-D*(MaxVSWR-MinVSWR)]/(A+B+C+D)
'WHERE THE COEFFICIENTS A, B, and C are USER-SPECIFIED IN THE FILE
'YagiCoeff.TXT.  THIS FILE MUST CONTAIN FOUR LINES WITH A, B, C,
'AND D, RESPECTIVELY, ON ONE LINE EACH.  IF THE FILE DDOES NOT
'EXIST, A DEFAULT FILE IS CREATED WITH A=20, B=2, C=3, D=4.
'OR, SOME OTHER FITNESS MAY BE HARDWIRED IN THE YAGI FUNCTION.
'CHECK THE SOURCE LISTING TO SEE WHICH IS BEING USED.

'****************** YAGI CFO SETUP PARAMETERS ******************
'THE TEXT FILE YAGI_CFO.CFG CONTAINS THREE CFO SETUP PARAMETERS:
'   Nt& (# time steps)
'   NumGammas%(# gamma values)
'   MaxProbesPerDimension% (max # probes/dim on probe lines)
'EACH ONE MUST BE ENTERED AS A NUMBER ON A SEPARATE LINE.  IF THE
'FILE YAGI_CFO.CFG DOES NOT EXIST, A DEFAULT VERSION IS CREATED.
'THIS APPLIES ONLY TO THE YAGI MODEL.  OTHER BUILT-IN BENCHMARK
'FUNCTIONS USE THE HARDWIRED INTERNAL VALUES FOR THESE PARAMETERS.

'****************** IMPORTANT MODELING NOTE ********************
'THE YAGI ELEMENT SEGMENT LENGTH IN NEC CAN BE VARIABLE OR FIXED.
'IF FIXED, THEN THE NUMBER OF SEGMENTS VARIES FROM ONE ELEMENT TO
'THE NEXT, BUT ALL SEGMENTS ARE PERFECTLY ALIGNED, WHICH, AS A
'GENERAL RULE IS RECOMMENDED IN THE NEC MODELING GUIDELINES.  BUT
'EXPERIENCE SUGGESTS THAT THIS IS NOT NECESSARILY THE BEST
'APPROACH UNDER ALL CIRCUMSTANCES, SO VARIABLE LENGTH SEGMENTS
'ARE ALLOWED AS WELL.  SEGMENTATION MUST BE VALIDATED USING THE
'NEC-RECOMMENDED AGT TEST.  THE SETUP FILE 'YagiSeg.TXT' CONTAINS
'TWO LINES.  THE FIRST IS EITHER THE WORD 'VARIABLE' (NO QUOTES)
'OR THE WORD 'FIXED,' WHICH DETERMINES WHETHER VARIABLE LENGTH OR
'FIXED LENGTH SEGMENTS ARE USED.  THE NEXT LINE IS THE SEGMENT
'LENGTH IN WAVELENGTHS (RANGE 0.02-0.2).  THIS FILE WILL BE
'CREATED IF IT DOES NOT EXIST USING VARIABLE SEGMENT LENGTH AS
'THE DEFAULT WITH 9 SEGMENTS PER ARRAY ELEMENT.  NOTE THAT IF
'FIXED SEGMENT LENGTH IS USED, EACH ELEMENT'S PHYSICAL LENGTH
'IS ADJUSTED TO BE AN INTEGER NUMBER OF SEGMENTS IN ORDER TO
'ALIGN EVERY SEGMENT.
'***************************************************************

'****************** BOWTIE FITNESS FUNCTION ********************
'THE HARDWIRED BOWTIE FITNESS FUNCTION IS:
'F=[Min(Eff)+5*Min(Gmax)]/{|Zo-Max(Rin)|*[Max(VSWR)-Min(VSWR)]*
'                                        [Max(Xin)-Min(Xin)]}

'**************** BOWTIE CFO SETUP PARAMETERS *******************
'THE TEXT FILE BOWTIE_CFO.CFG CONTAINS THREE CFO SETUP PARAMETERS:
'   Nt& (# time steps)
'   NumGammas%(# gamma values)
'   MaxProbesPerDimension% (max # probes/dim on probe lines)
'EACH ONE MUST BE ENTERED AS A NUMBER ON A SEPARATE LINE.  IF THE
'FILE BOWTIE_CFO.CFG DOES NOT EXIST, A DEFAULT VERSION IS CREATED.
'THIS APPLIES ONLY TO THE BOWTIE MODEL.  OTHER BUILT-IN BENCHMARK
'FUNCTIONS USE THE HARDWIRED INTERNAL VALUES FOR THESE PARAMETERS.

'==============================================================
'NOTE: ALL PBM FUNCTIONS HAVE WIRE RADIUS SET TO 0.00001 LAMBDA
'==============================================================

'THIS PROGRAM IMPLEMENTS A SIMPLE VERSION OF "CENTRAL
'FORCE OPTIMIZATION."  IT IS DISTRIBUTED FREE OF CHARGE
'TO INCREASE AWARENESS OF CFO AND TO ENCOURAGE EXPERI-
'MENTATION WITH THE ALGORITHM.

'CFO IS A MULTIDIMENSIONAL SEARCH AND OPTIMIZATION
'ALGORITHM THAT LOCATES THE GLOBAL MAXIMA OF A FUNCTION.
'UNLIKE MOST OTHER ALGORITHMS, CFO IS COMPLETELY DETERMIN-
'ISTIC, SO THAT EVERY RUN WITH THE SAME SETUP PRODUCES
'THE SAME RESULTS.

'(c) 2006-2011 Richard A. Formato

'ALL RIGHTS RESERVED WORLDWIDE

'THIS PROGRAM IS FREEWARE.  IT MAY BE COPIED AND
'DISTRIBUTED WITHOUT LIMITATION AS LONG AS THIS
'COPYRIGHT NOTICE AND THE GNUPLOT AND REFERENCE
'INFORMATION BELOW ARE INCLUDED WITHOUT MODIFICATION,
'AND AS LONG AS NO FEE OR COMPENSATION IS CHARGED,
'INCLUDING "TIE-IN" OR "BUNDLING" FEES CHARGED FOR
'OTHER PRODUCTS.

'==================================================================

'THIS PROGRAM REQUIRES wgnuplot.exe TO DISPLAY PLOTS.
'Gnuplot is a copyrighted freeware plotting program
'available at http://www.gnuplot.info/index.html.

'IT ALSO REQUIRES A VERSION OF THE Numerical Electromagnetics
'Code (NEC) in order to run the PBM benchmarks, Bowtie and Yagi
'antenna models.  If this file is not present, a runtime error
'occurs.  Remove the code that checks for the NEC EXE is there
'is no interest in the these functions.

'==================================================================
'CFO REFERENCES (author: Richard A. Formato unless otherwise noted)
'------------------------------------------------------------------

'"A Novel Methodology for Antenna Design and Optimization: Variable Zo," July 2011, http://arXiv.org/abs/1107.1437.

'"Central Force Optimization with Variable Initial Probes and Adaptive Decision Space," Applied Mathematics and Computation,
' Vol. 217, 2011, pp. 8866-8872, 2011.

'"New Techniques for Increasing Antenna Bandwidth with Impedance Loading," Progress In Electromagnetics Research B, vol. 29, pp. 269-288, 2011,
'http://www.jpier.org/pierb/pier.php?paper=11021904.

'"Parameter-Free Deterministic Global Search with Simplified Central Force Optimization," in Advanced Intelligent Computing Theories and
'Applications (ICIC2010), Lecture Notes in Computer Science (D.-S. Huang, Z. Zhao, V. Bevilacqua and J, C, Figueroa,Eds.), LNCS 6215, pp. 309-318,
'Springer-Verlag Berlin Heidelberg, 2010 [book chapter].

'"Antenna Benchmark Performance and Array Synthesis using Central Force Optimization," IET (U.K.) Microwaves, Antennas & Propagation vol. 4, no. 5,
'pp. 583-592, 2010.  (doi: 10.1049/iet-map.2009.0147). [with G. M. Qubati and N. I. Dib].

'"Improved CFO Algorithm for Antenna Optimization," Prog. Electromagnetics Research B, pp. 405-425, 2010
'http://www.jpier.org/pierb/pier.php?paper=09112309  (doi:10.2528/PIERB09112309).
```

```
'"Central Force Optimization and NEOs - First Cousins?," Journal of Multiple-Valued Logic and Soft Computing, vol. 16, pp. 547-565, 2010.

'"Convergence Analysis and Performance of the Extended Artificial Physics Optimization Algorithm," Applied Mathematics and Computation (in press)
'[with L. Xie and J. Zeng].

'"Central Force Optimization with Variable Initial Probes and Adaptive Decision Space," Applied Mathematics and Computation, July 2011.

'"High-Performance Indoor VHF-UHF Antennas: Technology Update Report," National Association of Broadcasters (NAB), FASTROAD (Flexible Advanced
'Services for Television and Radio On All Devices), Technology Advocacy Program, 15 May 2010 [with M. W. Cross and E. Merulla]
'http://www.nabfastroad.org/NABHighperformanceIndoorTVantennaRpt.pdf

'"Central Force Optimisation: a New Gradient-Like Metaheuristic for Multidimensional Search and Optimisation," Int. J. Bio-Inspired Computation,
'vol. 1, no. 4, pp. 217-238, 2009.  (doi: 10.1504/IJBIC.2009.024721).

'"Central Force Optimization: A New Deterministic Gradient-Like Optimization Metaheuristic," OPSEARCH, Journal of the Operations Research Society of India,
'vol. 46, no. 1, pp. 25-51, 2009.  (doi: 10.1007/s12597-009-0003-4).

'"Central Force Optimization: A New Computational Framework for Multidimensional Search and Optimization," in Nature Inspired Cooperative Strategies for
'Optimization (NICSO 2007), Studies in Computational Intelligence 129 (N. Krasnogor, G. Nicosia, M. Pavone, and D. Pelta, Eds.), vol. 129, Springer-Verlag,
'Heidelberg, 2008 [book chapter].

'"Central Force Optimization: A New Metaheuristic with Applications in Applied Electromagnetics," Prog. Electromagnetics Research, vol. PIER 77,
'pp. 425-491, 2007: http://www.jpier.org/pier/pier.php?paper=07082403 (doi:10.2528/PIER07082403).

'"Are Near Earth Objects the Key to Optimization Theory?," Dec., 2009.  http://arxiv.org/abs/0912.1394.

'"Pseudorandomness in Central Force Optimization," Jan., 2010.  http://arxiv.org/abs/1001.0317.

'"Central Force Optimization Applied to the PBM Suite of Antenna Benchmarks," Feb., 2010, http://arxiv.org/abs/1003.0221.

'"On the Utility of Directional Information for Repositioning Errant Probes in Central Force Optimization," May, 2010, http://arxiv.org/abs/1005.5490.

'"Issues in Antenna Optimization - A Monopole Case Study," Mar., 2011.  http://arXiv.org/abs/1103.5629.

'"Central Force Optimization with Variable Initial Probes and Adaptive Decision Space," Applied Mathematics and Computation, Vol. 217, 2011, pp. 8866-8872.

'===========================================================================================================================================
========

#COMPILE EXE

#DIM ALL

%USEMACROS = 1

#INCLUDE "Win32API.inc"

DEFEXT A-Z

'------ EQUATES -----

%IDC_FRAME1     = 101
%IDC_FRAME2     = 102

%IDC_Function_Number1  = 121
%IDC_Function_Number2  = 122
%IDC_Function_Number3  = 123
%IDC_Function_Number4  = 124
%IDC_Function_Number5  = 125
%IDC_Function_Number6  = 126
%IDC_Function_Number7  = 127
%IDC_Function_Number8  = 128
%IDC_Function_Number9  = 129
%IDC_Function_Number10 = 130
%IDC_Function_Number11 = 131
%IDC_Function_Number12 = 132
%IDC_Function_Number13 = 133
%IDC_Function_Number14 = 134
%IDC_Function_Number15 = 135
%IDC_Function_Number16 = 136
%IDC_Function_Number17 = 137
%IDC_Function_Number18 = 138
%IDC_Function_Number19 = 139
%IDC_Function_Number20 = 140
%IDC_Function_Number21 = 141
%IDC_Function_Number22 = 142
%IDC_Function_Number23 = 143
%IDC_Function_Number24 = 144
%IDC_Function_Number25 = 145
%IDC_Function_Number26 = 146
%IDC_Function_Number27 = 147
%IDC_Function_Number28 = 148
%IDC_Function_Number29 = 149
%IDC_Function_Number30 = 150
%IDC_Function_Number31 = 151
%IDC_Function_Number32 = 152
%IDC_Function_Number33 = 153
%IDC_Function_Number34 = 154
%IDC_Function_Number35 = 155
%IDC_Function_Number36 = 156
%IDC_Function_Number37 = 157
%IDC_Function_Number38 = 158
%IDC_Function_Number39 = 159
%IDC_Function_Number40 = 160
%IDC_Function_Number41 = 161
%IDC_Function_Number42 = 162
%IDC_Function_Number43 = 163
%IDC_Function_Number44 = 164
%IDC_Function_Number45 = 165
%IDC_Function_Number46 = 166
%IDC_Function_Number47 = 167
%IDC_Function_Number48 = 168
%IDC_Function_Number49 = 169
%IDC_Function_Number50 = 170

'------------------------------ GLOBAL CONSTANTS & SYMBOLS ---------------------------

GLOBAL YagiSegmentLength$, YagiCoefficients$

GLOBAL NumYagiElements% 'added 05/12/2011 instead of passing in function call

GLOBAL YAGIsecsPerRun, YagiSegmentLengthWvln, YagiFitnessCoefficients() AS EXT

GLOBAL BowtieSegmentLength$, BowtieCoefficients$

GLOBAL NumBowtieElements% 'added 05/12/2011 instead of passing in function call

GLOBAL BOWTIEsecsPerRun, BowtieSegmentLengthWvln, BowtieFitnessCoefficients() AS EXT

GLOBAL XiOffset() AS EXT 'offset array for Rosenbrock F6 function

GLOBAL XiMin(), XiMax(), DiagLength, StartingXiMin(), StartingXiMax() AS EXT 'decision space boundaries, length of diagonal

GLOBAL Aij() AS EXT 'array for Shekel's Foxholes function

GLOBAL EulerConst, Pi, Pi2, Pi4, TwoPi, FourPi, e, Root2 AS EXT 'mathematical constants

GLOBAL Alphabet$, Digits$, RunID$  'upper/lower case alphabet, digits 0-9 & Run ID
```

```
GLOBAL Quote$, SpecialCharacters$  'quotation mark & special symbols
GLOBAL Mu0, Eps0, c, eta0 AS EXT   'E&M constants
GLOBAL Rad2Deg, Deg2Rad, Feet2Meters, Meters2Feet, Inches2Meters, Meters2Inches AS EXT 'conversion factors
GLOBAL Miles2Meters, Meters2Miles, NautMi2Meters, Meters2NautMi AS EXT                 'conversion factors
GLOBAL ScreenWidth&, ScreenHeight& 'screen width & height
GLOBAL xOffset&, yOffset&          'offsets for probe plot windows
GLOBAL FunctionNumber%
GLOBAL AddNoiseToPBM2$
'------------------------------ TEST FUNCTION DECLARATIONS --------------------------------
DECLARE FUNCTION F1(R(),Nd%,p%,j&)              'F1 (n-D)
DECLARE FUNCTION F2(R(),Nd%,p%,j&)              'F2(n-D)
DECLARE FUNCTION F3(R(),Nd%,p%,j&)              'F3 (n-D)
DECLARE FUNCTION F4(R(),Nd%,p%,j&)              'F4 (n-D)
DECLARE FUNCTION F5(R(),Nd%,p%,j&)              'F5 (n-D)
DECLARE FUNCTION F6(R(),Nd%,p%,j&)              'F6 (n-D)
DECLARE FUNCTION F7(R(),Nd%,p%,j&)              'F7 (n-D)
DECLARE FUNCTION F8(R(),Nd%,p%,j&)              'F8 (n-D)
DECLARE FUNCTION F9(R(),Nd%,p%,j&)              'F9 (n-D)
DECLARE FUNCTION F10(R(),Nd%,p%,j&)             'F10 (n-D)
DECLARE FUNCTION F11(R(),Nd%,p%,j&)             'F11 (n-D)
DECLARE FUNCTION F12(R(),Nd%,p%,j&)             'F12 (n-D)
DECLARE FUNCTION u(Xi,a,k,m)                    'Auxiliary function for F12 & F13
DECLARE FUNCTION F13(R(),Nd%,p%,j&)             'F13 (n-D)
DECLARE FUNCTION F14(R(),Nd%,p%,j&)             'F14 (n-D)
DECLARE FUNCTION F15(R(),Nd%,p%,j&)             'F15 (n-D)
DECLARE FUNCTION F16(R(),Nd%,p%,j&)             'F16 (n-D)
DECLARE FUNCTION F17(R(),Nd%,p%,j&)             'F17 (n-D)
DECLARE FUNCTION F18(R(),Nd%,p%,j&)             'F18 (n-D)
DECLARE FUNCTION F19(R(),Nd%,p%,j&)             'F19 (n-D)
DECLARE FUNCTION F20(R(),Nd%,p%,j&)             'F20 (n-D)
DECLARE FUNCTION F21(R(),Nd%,p%,j&)             'F21 (n-D)
DECLARE FUNCTION F22(R(),Nd%,p%,j&)             'F22 (n-D)
DECLARE FUNCTION F23(R(),Nd%,p%,j&)             'F23 (n-D)
DECLARE FUNCTION F24(R(),Nd%,p%,j&)             'F24 (n-D)
DECLARE FUNCTION F25(R(),Nd%,p%,j&)             'F25 (n-D)
DECLARE FUNCTION F26(R(),Nd%,p%,j&)             'F26 (n-D)
DECLARE FUNCTION F27(R(),Nd%,p%,j&)             'F27 (n-D)
DECLARE FUNCTION ParrottF4(R(),Nd%,p%,j&)       'Parrott F4 (1-D)
DECLARE FUNCTION SGO(R(),Nd%,p%,j&)             'SGO Function (2-D)
DECLARE FUNCTION GoldsteinPrice(R(),Nd%,p%,j&)  'Goldstein-Price Function (2-D)
DECLARE FUNCTION StepFunction(R(),Nd%,p%,j&)    'Step Function (n-D)
DECLARE FUNCTION Schwefel226(R(),Nd%,p%,j&)     'Schwefel Prob. 2.26 (n-D)
DECLARE FUNCTION Colville(R(),Nd%,p%,j&)        'Colville Function (4-D)
DECLARE FUNCTION Griewank(R(),Nd%,p%,j&)        'Griewank (n-D)
DECLARE FUNCTION Himmelblau(R(),Nd%,p%,j&)      'Himmelblau (2-D)
DECLARE FUNCTION Rosenbrock(R(),Nd%,p%,j&)      'Rosenbrock (n-D)
DECLARE FUNCTION Sphere(R(),Nd%,p%,j&)          'Sphere (n-D)
DECLARE FUNCTION HimmelblauNLO(R(),Nd%,p%,j&)   'Himmelblau NLO (5-D)
DECLARE FUNCTION Tripod(R(),Nd%,p%,j&)          'Tripod (2-D)
DECLARE FUNCTION Sign(X)                        'Auxiliary function for Tripod
DECLARE FUNCTION RosenbrockF6(R(),Nd%,p%,j&)    'Rosenbrock F6 (10-D)
DECLARE FUNCTION CompressionSpring(R(),Nd%,p%,j&)'Compression Spring (3-D)
DECLARE FUNCTION GearTrain(R(),Nd%,p%,j&)       'Gear Train (4-D)
DECLARE FUNCTION PBM_1(R(),Nd%,p%,j&)           'PBM Benchmark #1
DECLARE FUNCTION PBM_2(R(),Nd%,p%,j&)           'PBM Benchmark #2
DECLARE FUNCTION PBM_3(R(),Nd%,p%,j&)           'PBM Benchmark #3
DECLARE FUNCTION PBM_4(R(),Nd%,p%,j&)           'PBM Benchmark #4
DECLARE FUNCTION PBM_5(R(),Nd%,p%,j&)           'PBM Benchmark #5
DECLARE FUNCTION BOWTIE(R(),Nd%,p%,j&)          'FREE-SPACE RLC-LOADED BOWTIE
DECLARE FUNCTION YAGI_ARRAY(R(),Nd%,p%,j&)      'FREE-SPACE YAGI ARRAY
'----------------------------------- SUB DECLARATIONS -------------------------------------
DECLARE SUB SieveOfEratosthenes(N&&,Primes&&(),NumPrimes&&)
DECLARE SUB IPD_Halton(Np%,Nd%,Nt&,R(),Gamma)
DECLARE SUB ReplaceCommentCard(NECfile$)
DECLARE SUB CopyBestMatrices(Np%,Nd%,Nt&,R(),M(),Rbest(),Mbest())
```

```
DECLARE SUB CheckNECFiles(NECfileError$)

DECLARE SUB GetTestFunctionNumber(FunctionName$)

DECLARE SUB FillArrayAij

DECLARE SUB Plot3DbestProbeTrajectories(NumTrajectories%,M(),R(),Np%,Nd%,LastStep&,FunctionName$)

DECLARE SUB Plot2DbestProbeTrajectories(NumTrajectories%,M(),R(),Np%,Nd%,LastStep&,FunctionName$)

DECLARE SUB Plot2DindividualProbeTrajectories(NumTrajectories%,M(),R(),Np%,Nd%,LastStep&,FunctionName$)

DECLARE SUB Show2Dprobes(R(),Np%,Nt&,j&,Frep,BestFitness,BestProbeNumber%,BestTimeStep&,FunctionName$,RepositionFactor$,Gamma)

DECLARE SUB Show3Dprobes(R(),Np%,Nd%,Nt&,j&,Frep,BestFitness,BestProbeNumber%,BestTimeStep&,FunctionName$,RepositionFactor$,Gamma)

DECLARE SUB StatusWindow(FunctionName$,StatusWindowHandle???)

DECLARE                                                                                                                        SUB
PlotResults(FunctionName$,Nd%,Np%,BestFitnessOverall,BestNpNd%,BestGamma,Neval&&,Rbest(),Mbest(),BestProbeNumberOverall%,BestTimeStepOverall&,LastStepBestR
un&,Alpha,Beta)

DECLARE                                                                                                                        SUB
DisplayRunParameters(FunctionName$,Nd%,Np%,Nt&,G,DeltaT,Alpha,Beta,Frep,R(),A(),M(),PlaceInitialProbes$,InitialAcceleration$,RepositionFactor$,RunCFO$,Shri
nkDS$,CheckForEarlyTermination$)

DECLARE SUB GetBestFitness(M(),Np%,StepNumber&,BestFitness,BestProbeNumber%,BestTimeStep&)

DECLARE                                                                                                                        SUB
Tabulate1DprobeCoordinates(Max1DprobesPlotted%,Nd%,Np%,LastStep&,G,DeltaT,Alpha,Beta,Frep,R(),M(),PlaceInitialProbes$,InitialAcceleration$,RepositionFactor
$,FunctionName$,Gamma)

DECLARE                                                                                                                        SUB
GetPlotAnnotation(PlotAnnotation$,Nd%,Np%,Nt&,G,DeltaT,Alpha,Beta,Frep,M(),PlaceInitialProbes$,InitialAcceleration$,RepositionFactor$,FunctionName$,Gamma)

DECLARE                                                                                                                        SUB
ChangeRunParameters(NumProbesPerDimension%,Np%,Nd%,Nt&,G,Alpha,Beta,DeltaT,Frep,PlaceInitialProbes$,InitialAcceleration$,RepositionFactor$,FunctionName$)

DECLARE SUB CLEANUP

DECLARE                                                                                                                        SUB
Plot1DprobePositions(Max1DprobesPlotted%,Nd%,Np%,LastStep&,G,DeltaT,Alpha,Beta,Frep,R(),M(),PlaceInitialProbes$,InitialAcceleration$,RepositionFactor$,Func
tionName$,Gamma)

DECLARE SUB DisplayMmatrix(Np%,Nt&,M())

DECLARE SUB DisplayMbestMatrix(Np%,Nt&,Mbest())

DECLARE SUB DisplayMmatrixThisTimeStep(Np%,j&,M())

DECLARE SUB DisplayAmatrix(Np%,Nd%,Nt&,A())

DECLARE SUB DisplayAmatrixThisTimeStep(Np%,Nd%,j&,A())

DECLARE SUB DisplayRmatrix(Np%,Nd%,Nt&,R())

DECLARE SUB DisplayRmatrixThisTimeStep(Np%,Nd%,j&,R(),Gamma)

DECLARE SUB DisplayXiMinMax(Nd%,XiMin(),XiMax())

DECLARE SUB DisplayRunParameters2(FunctionName$,Nd%,Np%,Nt&,G,DeltaT,Alpha,Beta,Frep,PlaceInitialProbes$,InitialAcceleration$,RepositionFactor$)

DECLARE                                                                                                                        SUB
PlotBestProbeVsTimeStep(Nd%,Np%,LastStep&,G,DeltaT,Alpha,Beta,Frep,M(),PlaceInitialProbes$,InitialAcceleration$,RepositionFactor$,FunctionName$,Gamma)

DECLARE                                                                                                                        SUB
PlotBestFitnessEvolution(Nd%,Np%,LastStep&,G,DeltaT,Alpha,Beta,Frep,M(),PlaceInitialProbes$,InitialAcceleration$,RepositionFactor$,FunctionName$,Gamma)

DECLARE                                                                                                                        SUB
PlotAverageDistance(Nd%,Np%,LastStep&,G,DeltaT,Alpha,Beta,Frep,M(),PlaceInitialProbes$,InitialAcceleration$,RepositionFactor$,FunctionName$,R(),DiagLength,
Gamma)

DECLARE SUB Plot2Dfunction(FunctionName$,R())

DECLARE SUB Plot1Dfunction(FunctionName$,R())

DECLARE                                                                                                                        SUB
GetFunctionRunParameters(FunctionName$,Nd%,Np%,Nt&,G,DeltaT,Alpha,Beta,Frep,R(),A(),M(),DiagLength,PlaceInitialProbes$,InitialAcceleration$,RepositionFacto
r$)

DECLARE SUB InitialProbeDistribution(Np%,Nd%,Nt&,R(),PlaceInitialProbes$,Gamma)

DECLARE SUB RetrieveErrantProbes(Np%,Nd%,j&,R(),Frep)

DECLARE SUB Retrieveerrantprobes2(Np%,Nd%,j&,R(),A(),Frep)

DECLARE SUB CFO(FunctionName$,Nd%,Nt&,R(),A(),M(),DiagLength,BestFitnessOverall,BestNpNd%,BestGamma,Neval&&,Rbest(),_
          Mbest(),BestProbeNumberOverall%,BestTimeStepOverall&,LastStepBestRun&,Alpha,Beta)   'Self-contained   CFO   routine   ->   NO   USER-SPECIFIED
PARAMETERS

DECLARE SUB IPD(Np%,Nd%,Nt&,R(),Gamma)

DECLARE SUB ResetDecisionSpaceBoundaries(Nd%)

DECLARE SUB ThreeDplot(PlotFileName$,PlotTitle$,Annotation$,xCoord$,yCoord$,zCoord$, _
                       XaxisLabel$,YaxisLabel$,ZaxisLabel$,zMin$,zMax$,GnuPlotEXE$,A$)

DECLARE SUB ThreeDplot2(PlotFileName$,PlotTitle$,Annotation$,xCoord$,yCoord$,zCoord$,XaxisLabel$,_
                        YaxisLabel$,ZaxisLabel$,zMin$,zMax$,GnuPlotEXE$,A$,xStart$,xStop$,yStart$,yStop$)

DECLARE SUB ThreeDplot3(PlotFileName$,PlotTitle$,Annotation$,xCoord$,yCoord$,zCoord$, _
                        XaxisLabel$,YaxisLabel$,ZaxisLabel$,zMin$,zMax$,GnuPlotEXE$,xStart$,xStop$,yStart$,yStop$)

DECLARE SUB TwoDplot(PlotFileName$,PlotTitle$,xCoord$,yCoord$,XaxisLabel$,YaxisLabel$, _
                     LogXaxis$,LogYaxis$,xMin$,xMax$,yMin$,yMax$,xTics$,yTics$,GnuPlotEXE$,LineType$,Annotation$)

DECLARE SUB TwoDplot2Curves(PlotFileName1$,PlotFileName2$,PlotTitle$,Annotation$,xCoord$,yCoord$,XaxisLabel$,YaxisLabel$, _
                            LogXaxis$,LogYaxis$,xMin$,xMax$,yMin$,yMax$,xTics$,yTics$,GnuPlotEXE$,LineSize)

DECLARE SUB TwoDplot3curves(NumCurves%,PlotFileName1$,PlotFileName2$,PlotFileName3$,PlotTitle$,Annotation$,xCoord$,yCoord$,XaxisLabel$,YaxisLabel$, _
                            LogXaxis$,LogYaxis$,xMin$,xMax$,yMin$,yMax$,xTics$,yTics$,GnuPlotEXE$)

DECLARE SUB CreateGNUplotINIfile(PlotWindowULC_X%,PlotWindowULC_Y%,PlotWindowWidth%,PlotWindowHeight%)

DECLARE SUB Delay(NumSecs)

DECLARE SUB MathematicalConstants

DECLARE SUB AlphabetAndDigits

DECLARE SUB SpecialSymbols

DECLARE SUB EMconstants

DECLARE SUB ConversionFactors

DECLARE SUB ShowConstants
```

```
DECLARE SUB GetNECdata(NECoutputFile$,NumFreqs%,NumRadPattAngles%,Zo,FrequencyMHZ(),RadEfficiencyPCT(),MaxGainDBI(),MinGainDBI(),RinOhms(),XinOhms(),VSWR(),ForwardGainDBi(),FileStatus$,FileID$)

DECLARE SUB ComplexMultiply(ReA,ImA,ReB,ImB,ReC,ImC)

DECLARE SUB ComplexDivide(ReA,ImA,ReB,ImB,ReC,ImC)

'------ FUNCTION DECLARATIONS -------

DECLARE FUNCTION DecimalToVanDerCorputBaseN(N&&,Nbase%)

DECLARE FUNCTION ProbeWeight2(Nd%,Np%,R(),M(),p%,j&)

DECLARE FUNCTION ProbeWeight(Nd%,R(),p%,j&) 'computes a 'weighting factor' based on probe's position (greater weight if closer to decision space boundary)

DECLARE FUNCTION SlopeRatio(M(),Np%,StepNumber&)

DECLARE CALLBACK FUNCTION DlgProc

DECLARE FUNCTION HasFITNESSsaturated$(Nsteps&,j&,Np%,Nd%,M(),R(),DiagLength)

DECLARE FUNCTION HasDAVGsaturated$(Nsteps&,j&,Np%,Nd%,M(),R(),DiagLength)

DECLARE FUNCTION OscillationInDavg$(j&,Np%,Nd%,M(),R(),DiagLength)

DECLARE FUNCTION DavgThisStep(j&,Np%,Nd%,M(),R(),DiagLength)

DECLARE FUNCTION NoSpaces$(X,NumDigits%)

DECLARE FUNCTION FormatFP$(X,Ndigits%)

DECLARE FUNCTION FormatInteger$(M%)

DECLARE FUNCTION TerminateNowForSaturation$(j&,Nd%,Np%,Nt&,G,DeltaT,Alpha,Beta,R(),A(),M())

DECLARE FUNCTION MagVector(V(),N%)

DECLARE FUNCTION UniformDeviate(u&&)

DECLARE FUNCTION RandomNum(a,b)

DECLARE FUNCTION GaussianDeviate(Mu,Sigma)

DECLARE FUNCTION UnitStep(X)

DECLARE FUNCTION Fibonacci&&(N%)

DECLARE FUNCTION ObjectiveFunction(R(),Nd%,p%,j&,FunctionName$)

DECLARE FUNCTION UnitStep(X)

DECLARE FUNCTION FP2String2$(X!)

DECLARE FUNCTION FP2String$(X)

DECLARE FUNCTION Int2String$(X%)

DECLARE FUNCTION StandingWaveRatio(Zo,ReZ,ImZ)

'====================================================================================================

'----- MAIN PROGRAM ------

FUNCTION PBMAIN () AS LONG

'   ------ CFO Parameters -----

    LOCAL Nd%, Np%, Nt&

    LOCAL G, DeltaT, Alpha, Beta, Frep AS EXT

    LOCAL PlaceInitialProbes$, InitialAcceleration$, RepositionFactor$

    LOCAL R(), A(), M(), Rbest(), Mbest() AS EXT    'position, acceleration & fitness matrices

    LOCAL FunctionName$           'name of objective function

'   ------------------------ Miscellaneous Setup Parameters ----------------------

    LOCAL N%, O%, P%, i%, YN&, Neval&&, NevalTotal&&, BestNpNd%, NumTrajectories%, Max1DprobesPlotted%, LastStepBestRun&, Pass%

    LOCAL A$, RunCFO$, CFOversion$, NECfileError$, RunStart$, RunStop$

    LOCAL BestGamma, BestFitnessThisRun, BestFitnessOverall, StartTime, StopTime, OptimumFitness AS EXT

    LOCAL BestProbeNum%, BestTimeStep&, BestProbeNumberOverall%, BestTimeStepOverall&, StatusWindowHandle???

'   -------------------- Global Constants --------------------

    REDIM Aij(1 TO 2, 1 TO 25) '(GLOBAL array for Shekel's Foxholes function)

    CALL FillArrayAij

    CALL MathematicalConstants 'NOTE: Calling order is important!!

    CALL AlphabetAndDigits 'NOTE: GLOBAL VARIABLE RunID$ IS SET HERE

    CALL SpecialSymbols

    CALL EMconstants

    CALL ConversionFactors        ': CALL ShowConstants 'to verify constants have been set

    BOWTIEsecsPerRun = 3##

'   -------------------------- General Setup ---------------------------

    CFOversion$ = "CFO Ver. 06-25-2011"

    RANDOMIZE TIMER  'seed random number generator with program start time

    DESKTOP GET SIZE TO ScreenWidth&, ScreenHeight&  'get screen size (global variables)

    IF DIR$("wgnuplot.exe") = "" THEN
        MSGBOX("WARNING!  'wgnuplot.exe' not found.  Run terminated.") : EXIT FUNCTION
    END IF

    IF DIR$("Fitness")    <> "" THEN KILL "Fitness"
    IF DIR$("Davg")       <> "" THEN KILL "Davg"
    IF DIR$("Best Probe") <> "" THEN KILL "Best Probe"

'   ---------------------------------------------------------------------- CFO RUN PARAMETERS -------------------------------------------------------------------------

    CALL GetTestFunctionNumber(FunctionName$)' : exit function 'DEBUG
```

```
    CALL
GetFunctionRunParameters(FunctionName$,Nd%,Np%,Nt&,G,DeltaT,Alpha,Beta,Frep,R(),A(),M(),DiagLength,PlaceInitialProbes$,InitialAcceleration$,RepositionFacto
r$) 'NOTE: Parameters returned but not used in this version!!

    REDIM R(1 TO Np%, 1 TO Nd%, 0 TO Nt&), A(1 TO Np%, 1 TO Nd%, 0 TO Nt&), M(1 TO Np%, 0 TO Nt&) 'position, acceleration & fitness matrices

    REDIM Rbest(1 TO Np%, 1 TO Nd%, 0 TO Nt&), Mbest(1 TO Np%, 0 TO Nt&) 'overall best position & fitness matrices

'   -------- PLOT 1D and 2D FUNCTIONS ON-SCREEN FOR VISUALIZATION --------

    IF Nd% = 2 AND INSTR(FunctionName$,"PBM_") > 0 THEN

        CALL CheckNECFiles(NECfileError$)

        IF NECfileError$ = "YES" THEN
            EXIT FUNCTION
        ELSE
            MSGBOX("Begin computing plot of function "+FunctionName$+"?  May take a while - be patient...")
        END IF

    END IF

    SELECT CASE Nd%
        CASE 1 : CALL Plot1Dfunction(FunctionName$,R()) : REDIM R(1 TO Np%, 1 TO Nd%, 0 TO Nt&) 'erases coordinate data in R()used to plot function
        CASE 2 : CALL Plot2Dfunction(FunctionName$,R()) : REDIM R(1 TO Np%, 1 TO Nd%, 0 TO Nt&) 'ditto
    END SELECT

'   ------------------------------------------------------------------------------  RUN CFO ---------------------------------------------------------------
--------------------

    YN& = MSGBOX("RUN CFO ON FUNCTION " + FunctionName$ + "?"+CHR$(13)+CHR$(13)+"Get some coffee & sit back...",%MB_YESNO,"CONFIRM RUN") : IF YN& = %IDYES
THEN RunCFO$ = "YES"

    IF RunCFO$ = "YES" THEN

        StartTime = TIMER : RunStart$ = "Started at "+TIME$+", "+DATE$


'       ------------- BOWTIE -----------
        IF FunctionName$ = "BOWTIE" THEN
            N% = FREEFILE : OPEN "RunTime_BOWTIE.DAT" FOR OUTPUT AS #N% : PRINT #N%, RunStart$ : CLOSE #N%

            IF DIR$("BOWTIE_CFO.CFG") = "" THEN
                N% = FREEFILE
                OPEN "BOWTIE_CFO.CFG" FOR OUTPUT AS #N%
                PRINT #N%, "150" : PRINT #N%, "11" : PRINT #N%, "4" 'CFO parameters Nt&, NumGammas%, MaxProbesPerDimension%, respectively
                CLOSE #N%
            END IF

        END IF

'       ------------- YAGI -----------
        IF FunctionName$ = "YAGI" THEN
            N% = FREEFILE : OPEN "RunTime_YAGI.DAT" FOR OUTPUT AS #N% : PRINT #N%, RunStart$ : CLOSE #N%

            IF DIR$("YAGI_CFO.CFG") = "" THEN
                N% = FREEFILE
                OPEN "YAGI_CFO.CFG" FOR OUTPUT AS #N%
                PRINT #N%, "150" : PRINT #N%, "11" : PRINT #N%, "4" 'CFO parameters Nt&, NumGammas%, MaxProbesPerDimension%, respectively
                CLOSE #N%
            END IF

        END IF

        CALL
CFO(FunctionName$,Nd%,Nt&,R(),A(),M(),DiagLength,BestFitnessOverall,BestNpNd%,BestGamma,Neval&&,Rbest(),Mbest(),BestProbeNumberOverall%,BestTimeStepOverall
&,LastStepBestRun&,Alpha,Beta)

        StopTime = TIMER : RunStop$ = "Ended at "+TIME$+", "+DATE$

        Np% = BestNpNd%*Nd%

'       -------------- Add Best Run Results to CFO Setup Parameter File ---------------

        N% = FREEFILE
        OPEN "CFO_"+RunID$+".PAR" FOR APPEND AS #N%
            PRINT #N%,""
            PRINT #N%,"Best Run Data:"
            PRINT #N%,"--------------"
            PRINT #N%,"Fitness    = "+STR$(ROUND(BestFitnessOverall,8))
            PRINT #N%,"Np/Nd      = "+STR$(BestNpNd%)
            PRINT #N%,"Nd         = "+STR$(Nd%)
            PRINT #N%,"Np         = "+STR$(Np%)
            PRINT #N%,"Gamma      = "+STR$(BestGamma)
            PRINT #N%,"Probe #    = "+STR$(BestProbeNumberOverall%)
            PRINT #N%,"#Time Step = "+STR$(BestTimeStepOverall&)
            PRINT #N%,"#Last Step = "+STR$(LastStepBestRun&)
        CLOSE #N%

        CALL
PlotResults(FunctionName$,Nd%,Np%,BestFitnessOverall,BestNpNd%,BestGamma,Neval&&,Rbest(),Mbest(),BestProbeNumberOverall%,BestTimeStepOverall&,LastStepBestR
un&,Alpha,Beta)

'       ------------- BOWTIE -----------
        IF FunctionName$ = "BOWTIE" THEN

            N% = FREEFILE
            OPEN "RunTime_BOWTIE.DAT" FOR APPEND AS #N%
                PRINT #N%,RunStop$
                PRINT #N%,"Runtime = "+STR$(ROUND((StopTime-StartTime)/3600##,2))+" hrs"
                PRINT #N%,"   Note: runtime will be incorrect if start/stop times transition midnight..."
                PRINT #N%,FunctionName$+", Total Function Evaluations = "+STR$(Neval&&)
                PRINT #N%,"Avg Time/Run = " +STR$(ROUND((StopTime-StartTime)/Neval&&,6))
            CLOSE #N%

            IF DIR$("BestBOWTIE.NEC") <> "" THEN KILL "BestBOWTIE.NEC"
            IF DIR$("BestBOWTIE.OUT") <> "" THEN KILL "BestBOWTIE.OUT"
            OptimumFitness  = BOWTIE(Rbest(),Nd%,BestProbeNumberOverall%,BestTimeStepOverall&)
            CALL ReplaceCommentCard("BOWTIE.NEC")
            NAME "BOWTIE.NEC" AS "BestBOWTIE.NEC"
            CALL ReplaceCommentCard("BOWTIE.OUT")
            NAME "BOWTIE.OUT" AS "BestBOWTIE.OUT"

            O% = FREEFILE
            OPEN "BestBOWTIE.NEC" FOR APPEND AS #O%
                PRINT #O%,"RLC-LOADED BOWTIE"
                PRINT #O%,"" : PRINT #O%,"OPTIMUM FITNESS = "+STR$(ROUND(OptimumFitness,3))+" @ Fo = 299.8 MHz, "+DATE$+", "+TIME$
                PRINT  #O%,""  :  PRINT  #O%,USING$("Best  Gamma:  ##.###         BestNp/Nd:  ###         Nt:  #####         Neval:  #######         LastStep:
#####",BestGamma,BestNpNd,Nt&,Neval&&,LastStepBestRun&)
                PRINT #O%, ""
                PRINT #O%,RunStart$
                PRINT #O%,RunStop$
            CLOSE #O%

            P% = FREEFILE
            OPEN "BestBOWTIE.OUT" FOR APPEND AS #P%
                PRINT #P%,"RLC-LOADED BOWTIE"
                PRINT #P%,"" : PRINT #P%,"OPTIMUM FITNESS = "+STR$(ROUND(OptimumFitness,3))+" @ Fo = 299.8 MHz, "+DATE$+", "+TIME$
```

```
                PRINT  #P%,""  :  PRINT  #P%,USING$("Best  Gamma:  ##.###         BestNp/Nd:  ###         Nt:  #####         Neval:  #######         LastStep:
#####",BestGamma,BestNpNd,Nt&,Neval&&,LastStepBestRun&)
                PRINT #P%, ""
                PRINT #P%,RunStart$
                PRINT #P%,RunStop$
            CLOSE #P%

            SHELL "Read_NEC_Output_File_BOWTIE.exe"

        END IF 'FunctionName$ = "BOWTIE"

'       ---------- YAGI ARRAY --------
        IF FunctionName$ = "YAGI" THEN

            N% = FREEFILE
            OPEN "RunTime_YAGI.DAT" FOR APPEND AS #N%
                PRINT #N%,RunStop$
                PRINT #N%,"Runtime = "+STR$(ROUND((StopTime-StartTime)/3600##,2))+" hrs"
                PRINT #N%,"   Note: runtime will be incorrect if start/stop times transition midnight..."
                PRINT #N%,FunctionName$+", Total Function Evaluations = "+STR$(Neval&&)
                PRINT #N%,"Avg Time/Run = " +STR$(ROUND((StopTime-StartTime)/Neval&&,6))
            CLOSE #N%

            IF DIR$("BestYAGI.NEC") <> "" THEN KILL "BestYAGI.NEC"
            IF DIR$("BestYAGI.OUT") <> "" THEN KILL "BestYAGI.OUT"
            OptimumFitness  = YAGI_ARRAY(Rbest(),Nd%,BestProbeNumberOverall%,BestTimeStepOverall&)
            CALL ReplaceCommentCard("YAGI.NEC")
            NAME "YAGI.NEC" AS "BestYAGI.NEC"
            CALL ReplaceCommentCard("YAGI.OUT")
            NAME "YAGI.OUT" AS "BestYAGI.OUT"

            O% = FREEFILE
            OPEN "BestYAGI.NEC" FOR APPEND AS #O%
                PRINT #O%,"YAGI ARRAY"
                PRINT #O%,"" : PRINT #O%,"OPTIMUM FITNESS = "+STR$(ROUND(OptimumFitness,3))+" @ Fo = 299.8 MHz, "+DATE$+", "+TIME$
                PRINT  #O%,""  :  PRINT  #O%,USING$("Best  Gamma:  ##.###         BestNp/Nd:  ###         Nt:  #####         Neval:  #######         LastStep:
#####",BestGamma,BestNpNd,Nt&,Neval&&,LastStepBestRun&)
                PRINT #O%, ""
                PRINT #O%,RunStart$
                PRINT #O%,RunStop$
            CLOSE #O%

            P% = FREEFILE
            OPEN "BestYAGI.OUT" FOR APPEND AS #P%
                PRINT #P%,"YAGI ARRAY"
                PRINT #P%,"" : PRINT #P%,"OPTIMUM FITNESS = "+STR$(ROUND(OptimumFitness,3))+" @ Fo = 299.8 MHz, "+DATE$+", "+TIME$
                PRINT  #P%,""  :  PRINT  #P%,USING$("Best  Gamma:  ##.###         BestNp/Nd:  ###         Nt:  #####         Neval:  #######         LastStep:
#####",BestGamma,BestNpNd,Nt&,Neval&&,LastStepBestRun&)
                PRINT #P%, ""
                PRINT #P%,RunStart$
                PRINT #P%,RunStop$
            CLOSE #P%

            SHELL "Read_NEC_Output_File_YAGI_ARRAY.exe"

        END IF 'FunctionName$ = "YAGI"

        MSGBOX(FunctionName$+CHR$(13)+"Total Function Evaluations = "+STR$(Neval&&)+CHR$(13)+"Runtime = "+STR$(ROUND((StopTime-StartTime)/3600##,2))+_
                                      " hrs, Avg Time/Run = " +STR$(ROUND((StopTime-StartTime)/Neval&&,6)))

        NAME "Fitness" AS "Fitness"+RunID$+".DAT" : NAME "Davg" AS "Davg"+RunID$+".DAT" : NAME "Best Probe" AS "BestProbe"+RunID$+".DAT" 'append Run ID and
'DAT' to data file names to identify file type

'       ------------ Cleanup Temp Files --------------
        IF DIR$("cmd2d.gp") <> "" THEN KILL "cmd2d.gp"
        IF DIR$("cmd3d.gp") <> "" THEN KILL "cmd3d.gp"
        IF DIR$("NECtemp")  <> "" THEN KILL "NECtemp"

    END IF 'RunCFO$ = "YES"

ExitPBMAIN:

END FUNCTION 'PBMAIN()

'=====================================================================================                                     CFO                    SUBROUTINE
========================================================================================

SUB
CFO(FunctionName$,Nd%,Nt&,R(),A(),M(),DiagLength,BestFitnessOverall,BestNpNd%,BestGamma,Neval&&,Rbest(),Mbest(),BestProbeNumberOverall%,BestTimeStepOverall
&,LastStepBestRun&,Alpha,Beta)

LOCAL p%, i%, j& 'Standard Indices: Probe #, Coordinate #, Time Step #

LOCAL Np% 'Number of Probes

LOCAL MaxProbesPerDimension% 'Maximum # probes per dimension (depends on Nd%)

LOCAL k%, L% 'Dummy summation indices

LOCAL M% 'file number

LOCAL NumProbesPerDimension%, GammaNumber%, NumGammas% 'Probes/dimension on each probe line; gamm point #; # gamma points

LOCAL SumSQ, Denom, Numerator, Gamma, Frep, DeltaFrep, FrepInit, MinFrep AS EXT

LOCAL BestProbeNumber%, BestTimeStep&, LastStep&, BestProbeNumberThisRun%, BestTimeStepThisRun&

LOCAL BestFitness, BestFitnessThisRun, eta AS EXT

LOCAL FitnessSaturation$

'<<<<<<<<<<<<<< NOTE: THIS IS A 'PARAMETER FREE' CFO IMPLEMENTATION.  SEE THE ICIC2010 REFERENCE ABOVE FOR DETAILS. >>>>>>>>>>>>>>

'---------------- Initial Parameter Values ------------------

NumGammas% = 3'11 'default value for functions other than BOWTIE

Alpha = 1## : Beta = 1## 'CFO EXPONENTS Alpha & Beta

Nt& = 1000 'set to a large value expecting early termination

IF FunctionName$ = "F7" THEN Nt& = 100 'to reduce runtime because this function contains random noise

Neval&& = 0

FrepInit  = 0.5##

DeltaFrep = 0.1##

MinFrep   = 0.05##

SELECT CASE Nd% 'set Np%/Nd% based on Nd% to avoid excessive runtimes

    CASE   1 TO   6 : MaxProbesPerDimension% = 14
    CASE   7 TO  10 : MaxProbesPerDimension% = 12
    CASE  11 TO  15 : MaxProbesPerDimension% = 10
    CASE  16 TO  20 : MaxProbesPerDimension% =  8
    CASE  21 TO  30 : MaxProbesPerDimension% =  6
    CASE ELSE       : MaxProbesPerDimension% =  4
```

```
END SELECT

'------------ BOWTIE -----------
IF FunctionName$ = "BOWTIE" THEN
    M% = FREEFILE
    OPEN "BOWTIE_CFO.CFG" FOR INPUT AS #M%
        INPUT #M%, Nt& : INPUT #M%, NumGammas% : INPUT #M%, MaxProbesPerDimension%
    CLOSE #M%
    IF Nt&                    < 5 OR Nt&                    > 500 THEN Nt& = 150
    IF NumGammas%             < 2 OR NumGammas%             > 31  THEN NumGammas% = 11
    IF MaxProbesPerDimension% < 2 OR MaxProbesPerDimension% > 10  THEN MaxProbesPerDimension% = 6
END IF

'--------- YAGI ARRAY --------
IF FunctionName$ = "YAGI" THEN
    M% = FREEFILE
    OPEN "YAGI_CFO.CFG" FOR INPUT AS #M%
        INPUT #M%, Nt& : INPUT #M%, NumGammas% : INPUT #M%, MaxProbesPerDimension%
    CLOSE #M%
    IF Nt&                    < 5 OR Nt&                    > 500 THEN Nt& = 150
    IF NumGammas%             < 2 OR NumGammas%             > 31  THEN NumGammas% = 11
    IF MaxProbesPerDimension% < 2 OR MaxProbesPerDimension% > 10  THEN MaxProbesPerDimension% = 6
END IF

'--------------

LastStep& = Nt&

'IF FunctionName$ = "BOWTIE" THEN
'    MaxProbesPerDimension% =  6 'to reduce runtime for LOADED BOWTIE
'    MSGBOX("Approximate runtime: "+STR$(ROUND(BOWTIEsecsPerRun*NumGammas*Nd%*MaxProbesPerDimension%*Nt&/40##,1))+" minutes")
'END IF

'   ------------------------- Save Setup Parameters for This Run -----------------------------

    M% = FREEFILE

    OPEN "CFO_"+RunID$+".PAR" FOR OUTPUT AS #M%
        PRINT #M%,"CFO RUN PARAMETERS, Run ID: "+RunID$
        PRINT #M%,"----------------------------"+STRING$(LEN(RunID$),"-")
        PRINT #M%,"Alpha           = "+STR$(Alpha)
        PRINT #M%,"Beta            = "+STR$(Beta)
        PRINT #M%,"Nt              = "+STR$(Nt&)
        PRINT #M%,"Initial Frep    = "+STR$(FrepInit)
        PRINT #M%,"Delta Frep      = "+STR$(DeltaFrep)
        PRINT #M%,"Min Frep        = "+STR$(MinFrep)
        PRINT #M%,"# Gammas        = "+STR$(NumGammas%)
        PRINT #M%,"Nt              = "+STR$(Nt&)
        PRINT #M%,"Max Probes/Dim  = "+STR$(MaxProbesPerDimension%)
    CLOSE #M%

'   -------------------------- Np/Nd LOOP ---------------------

BestFitnessOverall = -1E4200 'very large negative number...

FOR NumProbesPerDimension% = 2 TO MaxProbesPerDimension% STEP 2

'FOR NumProbesPerDimension% = 4 TO MaxProbesPerDimension% STEP 2 'original code

Np% = NumProbesPerDimension%*Nd%

'   -------------------------- GAMMA LOOP ---------------------

FOR GammaNumber% = 1 TO NumGammas%

Gamma = (GammaNumber%-1)/(NumGammas%-1)

REDIM R(1 TO Np%, 1 TO Nd%, 0 TO Nt&), A(1 TO Np%, 1 TO Nd%, 0 TO Nt&), M(1 TO Np%, 0 TO Nt&) 're-initializes Position Vector/Acceleration/Fitness matrices
to zero

'STEP (A1) ------------- Compute Initial Probe Distribution (Step 0)----------------

    CALL IPD(Np%,Nd%,Nt&,R(),Gamma) 'Probe Line IPD intersecting on diagonal at a point determined by Gamma

'STEP (A2) ------------- Compute Initial Fitness Matrix (Step 0) ------------------------

    FOR p% = 1 TO Np% : M(p%,0) = ObjectiveFunction(R(),Nd%,p%,0,FunctionName$) : INCR Neval&& : NEXT p%

'STEP (A3) ------------- Set Initial Probe Accelerations to ZERO (Step 0)----------------

    FOR p% = 1 TO Np% : FOR i% = 1 TO Nd% : A(p%,i%,0) = 0## : NEXT i% : NEXT p%

'STEP (A4) ------------- Initialize Frep ----------------

    Frep = FrepInit '0.5##

'   ====================================== LOOP ON TIME STEPS STARTING AT STEP #1 ===========================================

    BestFitnessThisRun = M(1,0)

    FOR j& = 1 TO Nt&

'STEP (B) ----------- Compute Probe Position Vectors for this Time Step --------

        FOR p% = 1 TO Np% : FOR i% = 1 TO Nd% : R(p%,i%,j&) = R(p%,i%,j&-1) + A(p%,i%,j&-1) : NEXT i% : NEXT p% 'note: factor of 1/2 combined with G=2 to
produce unity coefficient

'STEP (C) ----------- Retrieve Errant Probes ---------------

       CALL RetrieveErrantProbes(Np%,Nd%,j&,R(),Frep)

'         CALL RetrieveErrantprobes2(Np%,Nd%,j&,R(),A(),Frep) 'added 04-01-10

'STEP (D) ----------- Compute Fitness Matrix for Current Probe Distribution ---------

       FOR p% = 1 TO Np% : M(p%,j&) = ObjectiveFunction(R(),Nd%,p%,j&,FunctionName$) : INCR Neval&& : NEXT p%

'STEP (E) ----------- Compute Accelerations Based on Current Probe Distribution & Fitnesses ----------------

        FOR p% = 1 TO Np%

            FOR i% = 1 TO Nd%

                A(p%,i%,j&) = 0

                FOR k% = 1 TO Np%

                    IF k% <> p% THEN

                        SumSQ = 0## : FOR L% = 1 TO Nd%  : SumSQ = SumSQ + (R(k%,L%,j&)-R(p%,L%,j&))^2 : NEXT L% 'dummy index

                        IF SumSQ <> 0## THEN 'to avoid zero denominator (added 03-20-10) [NOTE THAT POWER BASIC DOES NOT FAULT ON DIVIDE-BY-ZERO...]

                            Denom = SQR(SumSQ) : Numerator = UnitStep((M(k%,j&)-M(p%,j&)))*(M(k%,j&)-M(p%,j&))

                            A(p%,i%,j&) = A(p%,i%,j&) + (R(k%,i%,j&)-R(p%,i%,j&))*Numerator^Alpha/Denom^Beta 'ORIGINAL VERSION WITH VARIABLE Alpha & Beta

                        END IF 'added 03-20-10
```

```
                END IF
            NEXT k% 'dummy index
        NEXT i% 'coord (dimension) #
    NEXT p% 'probe #
'   --------- Get Best Fitness & Corresponding Probe # and Time Step ---------
    CALL GetBestFitness(M(),Np%,j&,BestFitness,BestProbeNumber%,BestTimeStep&)
    IF BestFitness >= BestFitnessThisRun THEN
        BestFitnessThisRun = BestFitness : BestProbeNumberThisRun% = BestProbeNumber% : BestTimeStepThisRun& = BestTimeStep&
    END IF
'   ----- Increment Frep -----
    Frep = Frep + DeltaFrep
    IF Frep > 1## THEN Frep = MinFrep 'keep Frep in range [0.05,1]
'   ---------- Starting at Step #20 Shrink Decision Space Around Best Probe Every 20th Step -----------
'   IF j& MOD 10 = 0 AND j& >= 20 THEN
    IF j& MOD 20 = 0 AND j& >= 20 THEN
        FOR  i%  =  1  TO  Nd%  :  XiMin(i%)  =  XiMin(i%)+(R(BestProbeNumber%,i%,BestTimeStep&)-XiMin(i%))/2##  :  XiMax(i%)  =  XiMax(i%)-(XiMax(i%)-
R(BestProbeNumber%,i%,BestTimeStep&))/2## : NEXT i% 'shrink DS by 0.5
       CALL RetrieveErrantProbes(Np%,Nd%,j&,R(),Frep) 'TO RETRIEVE PROBES LYING OUTSIDE SHRUNKEN DS 'ADDED 02-07-2010
'        CALL RetrieveErrantprobes2(Np%,Nd%,j&,R(),A(),Frep) 'added 04-01-10
    END IF
'STEP (F) ----------- Check for Early Run Termination ---------
    IF HasFITNESSsaturated$(25,j&,Np%,Nd%,M(),R(),DiagLength) = "YES" THEN
        LastStep& = j&
        EXIT FOR
    END IF
    NEXT j& 'END TIME STEP LOOP
'---------------- Best Overall Fitness & Corresponding Parameters -------------------
IF BestFitnessThisRun >= BestFitnessOverall THEN
    BestFitnessOverall = BestFitnessThisRun : BestProbeNumberOverall% = BestProbeNumberThisRun% : BestTimeStepOverall& = BestTimeStepThisRun&
    BestNpNd% = NumProbesPerDimension%      : BestGamma = Gamma : LastStepBestRun& = LastStep&
    CALL CopyBestMatrices(Np%,Nd%,Nt&,R(),M(),Rbest(),Mbest())
END IF
'STEP (G) ----- Reset Decision Space Boundaries to Initial Values -----
CALL ResetDecisionSpaceBoundaries(Nd%)
NEXT GammaNumber% 'END GAMMA LOOP
NEXT NumProbesPerDimension% 'END Np/Nd LOOP
END SUB 'CFO()
'=====================================================================================================================================================
=======
SUB IPD_Halton(Np%,Nd%,Nt&,R(),Gamma) 'added 06-29-2011
LOCAL DeltaXi, DelX1, DelX2, Di AS EXT
LOCAL NumProbesPerDimension%, p%, i%, k%, NumX1points%, NumX2points%, x1pointNum%, x2pointNum%
            IF Nd% > 1 THEN
                NumProbesPerDimension% = Np%\Nd% 'even #
            ELSE
                NumProbesPerDimension% = Np%
            END IF
            FOR i% = 1 TO Nd%
                FOR p% = 1 TO Np%
                    R(p%,i%,0) = XiMin(i%) + Gamma*(XiMax(i%)-XiMin(i%))
                NEXT Np%
            NEXT i%
            FOR i% = 1 TO Nd% 'place probes probe line-by-probe line (i% is dimension number)
                DeltaXi = (XiMax(i%)-XiMin(i%))/(NumProbesPerDimension%-1)
                FOR k% = 1 TO NumProbesPerDimension%
                    p% = k% + NumProbesPerDimension%*(i%-1) 'probe #
                    R(p%,i%,0) = XiMin(i%) + (k%-1)*DeltaXi
                NEXT k%
            NEXT i%
END SUB 'IPD_Halton()
'--------------------
FUNCTION DecimalToVanDerCorputBaseN(N&&,Nbase%)
'Returns the decimal value of the VDC base Nbase% 'radical inverse function' to create a Low Discrepancy sequence of decimal values.
'Maps integer argument into a quasirandom real number on the interval [0,1).
'Refs: (1) "Improved Particle Swarm Optimization with Low-Discrepancy Sequences,"
'           Pant, Thangaraj, Grosan & Abraham, 2008 IEEE Congress on Evolutionary Computation (CEC2008), p. 3011.
'      (2) "Quasirandom Ramblings," Computing Science column in American Scientist Magazine, July-Augist 2011, p. 282.
```

```
'          (www.american scientist.org)

'CHECK VALUES FOR BASE 2
'-----------------------
'Integer(N&&)  Base 2 Representation   Digits Reversed    VDC Decimal Value
'------------  ---------------------   ---------------    -----------------
'      0                  0                   0            0.0
'      2                 10                  01            0.25
'      3                 11                  11            0.75
'      4                100                 001            0.125
'      6                110                 011            0.375
'      7                111                 111            0.875
'     15               1111                1111            0.9375
'    120            1111000             0001111            0.1171875
'    121            1111001             1001111            0.6171875
'    532         1000010100          0010100001            0.1572265625

'CHECK VALUES FOR BASE 3
'-----------------------
'Integer(N&&)  Base 3 Representation   Digits Reversed    VDC Decimal Value
'------------  ---------------------   ---------------    -----------------
'      0                  0                   0            0.0
'      2                  2                   2            0.666...
'      3                 10                  01            0.111...
'      7                 21                  12            0.555...
'    120              11110               01111            0.164609053497942
'    532             201201              102102            0.422496570544719

'CHECK VALUES FOR BASE 5
'-----------------------
'Integer(N&&)  Base 5 Representation   Digits Reversed    VDC Decimal Value
'------------  ---------------------   ---------------    -----------------
'      0                  0                   0            0.0
'      1                  1                   1            0.2
'      2                  2                   2            0.4
'      3                  3                   3            0.6
'      4                  4                   4            0.8
'      5                 10                  01            0.04
'      7                 12                  21            0.44
'     17                 32                  23            0.52
'    121                441                 144            0.392
'    532               4112                2114            0.4544

LOCAL S AS EXT, R AS EXT, I&&, D&&

    S = 0##

    IF N&& > 2^63-2 THEN
        MSGBOX("WARNING! VDC arg > 2^63-2.  Zero returned.") : GOTO ExitVDC
    END IF

    I&& = N&& : R = 1##/Nbase%

    DO WHILE I&& <> 0 : D&& = I&& MOD Nbase% : S = S + D&&*R : I&& = (I&&-D&&)/Nbase% : R = R/Nbase% : LOOP

ExitVDC:

DecimalToVanDerCorputBaseN = S

END FUNCTION

'-----------

SUB SieveOfEratosthenes(N&&,Primes&&(),NumPrimes&&)

'Returns primes between 2 and N&& and their number.

'NOTE: There are 303 primes between 2 and 2000. Accuracy of this routine has been checked against table of the first 1000 primes (# primes up to N&&=7919).

'Ref: Caldwell, C., "The Prime Pages," Univ. Tennessee at Martin, http://primes.utm.edu

LOCAL i&&, iStart&&, p&&, j&&, TempArray&&()

    REDIM TempArray&&(1 TO N&&)

    TempArray&&(1) = 0 : FOR i&& = 2 TO N&& : TempArray&&(i&&) = 1 : NEXT i&&

    p&& = 2 : WHILE p&&^2 =< N&& : j&& = p&&^2 : WHILE j&& =< N&& : TempArray&&(j&&) = 0 : j&& = j&& + p&& : WEND : p&& = p&& + 1 : WEND 'inserts 0 or 1
depending on whether or not # is a prime

    NumPrimes&& = 0

    FOR i&& = 1 TO N&&  'converts the 1's into their corresponding prime numbers & retrieves # primes between 2 and N&&
        IF TempArray&&(i&&) = 1 THEN INCR NumPrimes&&
        TempArray&&(i&&) = i&&*TempArray&&(i&&)
    NEXT i&&

    REDIM Primes&&(1 TO NumPrimes&&)

    iStart&& = 1
    FOR j&& = 1 TO NumPrimes&&
        FOR i&& = iStart&& TO N&&
            IF TempArray&&(i&&) <> 0 THEN
                Primes&&(j&&) = TempArray&&(i&&) : iStart&& = i&&+1 : EXIT FOR
            END IF
        NEXT i&&
    NEXT j&&

END SUB 'SieveOfEratosthenes()

'-----------------------------

SUB IPD(Np%,Nd%,Nt&,R(),Gamma)

LOCAL DeltaXi, DelX1, DelX2, Di AS EXT

LOCAL NumProbesPerDimension%, p%, i%, k%, NumX1points%, NumX2points%, x1pointNum%, x2pointNum%

            IF Nd% > 1 THEN

                NumProbesPerDimension% = Np%\Nd% 'even #

            ELSE

                NumProbesPerDimension% = Np%

            END IF

            FOR i% = 1 TO Nd%

                FOR p% = 1 TO Np%

                    R(p%,i%,0) = XiMin(i%) + Gamma*(XiMax(i%)-XiMin(i%))

                NEXT Np%

            NEXT i%

            FOR i% = 1 TO Nd% 'place probes probe line-by-probe line (i% is dimension number)
```

```
                DeltaXi = (XiMax(i%)-XiMin(i%))/(NumProbesPerDimension%-1)
                FOR k% = 1 TO NumProbesPerDimension%
                    p% = k% + NumProbesPerDimension%*(i%-1) 'probe #
                    R(p%,i%,0) = XiMin(i%) + (k%-1)*DeltaXi
                NEXT k%
            NEXT i%
END SUB 'IPD()
'----
FUNCTION HasFITNESSsaturated$(Nsteps&,j&,Np%,Nd%,M(),R(),DiagLength)
LOCAL A$, B$
LOCAL k&, p%
LOCAL BestFitness, SumOfBestFitnesses, BestFitnessStepJ, FitnessSatTOL AS EXT
    A$ = "NO" : B$ = "j="+STR$(j&)+CHR$(13)
    FitnessSatTOL = 0.000001## 'tolerance for FITNESS saturation
    IF j& < Nsteps& + 10 THEN GOTO ExitHasFITNESSsaturated 'execute at least 10 steps after averaging interval before performing this check
    SumOfBestFitnesses = 0##
    FOR k& = j&-Nsteps&+1 TO j& 'GET BEST FITNESSES STEP-BY-STEP FOR Nsteps& INCLUDING THIS STEP j& AND COMPUTE AVERAGE VALUE.
'       BestFitness = M(k&,1) 'ORIG CODE 03-23-2010: THIS IS A MISTAKE!
        BestFitness = -1E4200 'THIS LINE CORRECTED 03-23-2010 PER DISCUSSION WITH ROB GREEN.
                              'INITIALIZE BEST FITNESS AT k&-th TIME STEP TO AN EXTREMELY LARGE NEGATIVE NUMBER.
        FOR p% = 1 TO Np% 'PROBE-BY-PROBE GET MAXIMUM FITNESS
            IF M(p%,k&) >= BestFitness THEN BestFitness = M(p%,k&)
        NEXT p%
        IF k& = j& THEN BestFitnessStepJ = BestFitness  'IF AT THE END OF AVERAGING INTERVAL, SAVE BEST FITNESS FOR CURRENT TIME STEP j&
        SumOfBestFitnesses = SumOfBestFitnesses + BestFitness
        B$ = B$ + "k="+STR$(k&)+"  BestFit="+STR$(BestFitness)+"   SumFit="+STR$(SumOfBestFitnesses)+CHR$(13)
    NEXT k&
    IF ABS(SumOfBestFitnesses/Nsteps&-BestFitnessStepJ) =< FitnessSatTOL THEN A$ = "YES" 'saturation if (avg value - last value) are within TOL
ExitHasFITNESSsaturated:
    HasFITNESSsaturated$ = A$
END FUNCTION 'HasFITNESSsaturated$()
'-----------------------------------
SUB RetrieveErrantprobes2(Np%,Nd%,j&,R(),A(),Frep) 'added 04-01-10
LOCAL ErrantProbe$
LOCAL p%, i%, k%
LOCAL Xik, dMax, Eta(), EtaStar, SumSQ, MagAj1 AS EXT
REDIM Eta(1 TO Nd%, 1 TO 2) 'Eta(i%,k%)
    FOR p% = 1 TO Np%
        ErrantProbe$ = "NO" 'presume each probe is inside DS
        FOR i% = 1 TO Nd% 'check to see if probe p lies outside DS (any coordinate exceeding a bounday)
            IF (R(p%,i%,j&) > XiMax(i%) OR R(p%,i%,j&) < XiMin(i%)) AND A(p%,i%,j&-1) <> 0## THEN 'probe lies outside DS
                ErrantProbe$ = "YES" : EXIT FOR
            END IF
        NEXT i%
        IF ErrantProbe$ = "YES" THEN 'reposition probe p inside DS with acceleration direction preserved
            FOR i% = 1 TO Nd% 'compute array of Eta values
                FOR k% = 1 TO 2
                    SELECT CASE k%
                        CASE 1 : Xik = XiMin(i%)
                        CASE 2 : Xik = XiMax(i%)
                    END SELECT
                    Eta(i%,k%) = (Xik-R(p%,i%,j&-1))/A(p%,i%,j&-1)
                NEXT k%
            NEXT i%
            EtaStar = 1E4200 'very large positive number
            FOR i% = 1 TO Nd% 'get min Eta value > 0
                FOR k% = 1 TO 2
                    IF Eta(i%,k%) =< EtaStar AND Eta(i%,k%) >= 0## THEN EtaStar = Eta(i%,k%)
                NEXT k%
            NEXT i%
            SumSQ = 0## : FOR i% = 1 TO Nd% : SumSQ = SumSQ + A(p%,i%,j&-1)^2 : NEXT i% : MagAj1 = SQR(SumSQ) 'magnitude of acceleration vector A(probe #p)
at step j&-1
            dMax = EtaStar*MagAj1 'distance to nearest boundary plane from position of probe p% at step j&-1
            FOR i% = 1 TO Nd% 'change probe p's i-th coordinate in proportion to the ratio of position vector lengths at steps j& and j&-1
                R(p%,i%,j&) = R(p%,i%,j&-1) + Frep*dMax*A(p%,i%,j&-1)/MagAj1 'preserves acceleration directional information by scaling dMax by Frep, which
is arbitrary but precisely known
            NEXT i%
```

```
        END IF 'ErrantProbe$ = "YES"
    NEXT p%
END SUB 'Retrieveerrantprobes2()
'------------------------------
SUB RetrieveErrantProbes(Np%,Nd%,j&,R(),Frep)
LOCAL p%, i%
    FOR p% = 1 TO Np%
        FOR i% = 1 TO Nd%
            IF R(p%,i%,j&) < XiMin(i%) THEN R(p%,i%,j&) = MAX(XiMin(i%) + Frep*(R(p%,i%,j&-1)-XiMin(i%)),XiMin(i%)) 'CHANGED 02-07-10
            IF R(p%,i%,j&) > XiMax(i%) THEN R(p%,i%,j&) = MIN(XiMax(i%) - Frep*(XiMax(i%)-R(p%,i%,j&-1)),XiMax(i%))
        NEXT i%
    NEXT p%
END SUB 'RetrieveErrantProbes()
'------------------------------
SUB ResetDecisionSpaceBoundaries(Nd%)
    LOCAL i%
    FOR i% = 1 TO Nd% : XiMin(i%) = StartingXiMin(i%) : XiMax(i%) = StartingXiMax(i%) : NEXT i%
END SUB
'------
SUB CopyBestMatrices(Np%,Nd%,Nt&,R(),M(),Rbest(),Mbest())
LOCAL p%, i%, j&
REDIM Rbest(1 TO Np%, 1 TO Nd%, 0 TO Nt&), Mbest(1 TO Np%, 0 TO Nt&) 're-initializes Best Position Vector/Fitness matrices to zero
    FOR p% = 1 TO Np%
        FOR i% = 1 TO Nd%
            FOR j& = 0 TO Nt&
                Rbest(p%,i%,j&) = R(p%,i%,j&) : Mbest(p%,j&) = M(p%,j&)
            NEXT j&
        NEXT i%
    NEXT p%
END SUB
'------
'FORGET THIS IDEA !!!!
FUNCTION ProbeWeight(Nd%,R(),p%,j&) 'computes a 'weighting factor' based on probe's position (greater weight if closer to decision space boundary)
LOCAL MinDistCoordinate%, i% 'Dimension number.  Remember, XiMin(), XiMax()& DiagLength are GLOBAL.
LOCAL MinDistance, dStar, MaxWeight AS EXT
    MinDistance = DiagLength 'largest dimension of decision space
    FOR i% = 1 TO Nd% 'compute distance to closest boundary
        IF ABS(R(p%,i%,j&)-XiMin(i%)) =< MinDistance THEN
            MinDistance = ABS(R(p%,i%,j&)-XiMin(i%)) : MinDistCoordinate% = i%
        END IF
        IF ABS(XiMax(i%)-R(p%,i%,j&)) =< MinDistance THEN
            MinDistance = ABS(XiMax(i%)-R(p%,i%,j&)) : MinDistCoordinate% = i%
        END IF
    NEXT i%
    dStar = MinDistance/(XiMax(MinDistCoordinate%)-XiMin(MinDistCoordinate%)) 'normalized minimum distance, [0-1]
    MaxWeight = 2##
'     ProbeWeight = 1## + 2##*MaxWeight*abs(dStar-0.5##)
    ProbeWeight = 1## + 4##*MaxWeight*(dStar-0.5##)^2
END FUNCTION 'ProbeWeight
'------------------------
'FORGET THIS IDEA !!!!
FUNCTION ProbeWeight2(Nd%,Np%,R(),M(),p%,j&) 'computes a 'weighting factor' based on probe's position (greater weight if closer to decision space boundary)
LOCAL MinDistCoordinate%, ProbeNum%, BestProbeThisStep%, i% 'Dimension number.  Remember, XiMin(), XiMax()& DiagLength are GLOBAL.
LOCAL Distance, SumSQ, dStar, MaxWeight, BestFitnessThisStep AS EXT
    BestFitnessThisStep = M(1,j&)
    FOR ProbeNum% = 1 TO Np% 'get number of best probe this step
        IF M(ProbeNum%,j&) >= BestFitnessThisStep THEN
            BestFitnessThisStep = M(ProbeNum%,j&) : BestProbeThisStep% = ProbeNum%
        END IF
    NEXT ProbeNum%
    SumSQ = 0##
    FOR i% = 1 TO Nd% 'compute distance from probe #p% to thes best probe this step
        SumSQ = (R(p%,i%,j&) - R(BestProbeThisStep%,i%,j&))^2
    NEXT i%
    Distance = SQR(SumSQ)
```

```
    dStar = Distance/DiagLength 'range [0-1]

'   ----------------- Compute weight Factor ---------------------

    MaxWeight = 0##

'    ProbeWeight2 = 1## + 2##*MaxWeight*abs(dStar-0.5##)

    ProbeWeight2 = 1## + 4##*MaxWeight*(dStar-0.5##)^2

END FUNCTION 'ProbeWeight2

'--------------------------

FUNCTION SlopeRatio(M(),Np%,StepNumber&)

LOCAL p% 'probe #

LOCAL NumSteps%

LOCAL BestFitnessAtStepNumber, BestFitnessAtStepNumberMinus1, BestFitnessAtStepNumberMinus2, Z AS EXT

    Z = 1## 'assumes no slope change

    IF StepNumber& < 10 THEN GOTO ExitSlopeRatio 'need at least 10 steps for this test

    NumSteps% = 2

    BestFitnessAtStepNumber        = M(1,StepNumber&)               : FOR p% = 1 TO Np% : IF M(p%,StepNumber&)               >= BestFitnessAtStepNumber
THEN BestFitnessAtStepNumber = M(p%,StepNumber&)                   : NEXT p%

    BestFitnessAtStepNumberMinus1 = M(1,StepNumber&-NumSteps%)    : FOR p% = 1 TO Np% : IF M(p%,StepNumber&-NumSteps%)    >= BestFitnessAtStepNumberMinus1
THEN BestFitnessAtStepNumberMinus1 = M(p%,StepNumber&-NumSteps%)   : NEXT p%

    BestFitnessAtStepNumberMinus2 = M(1,StepNumber&-2*NumSteps%)  : FOR p% = 1 TO Np% : IF M(p%,StepNumber&-2*NumSteps%) >= BestFitnessAtStepNumberMinus2
THEN BestFitnessAtStepNumberMinus2 = M(p%,StepNumber&-2*NumSteps%) : NEXT p%

    Z = (BestFitnessAtStepNumber-BestFitnessAtStepNumberMinus1)/(BestFitnessAtStepNumberMinus1-BestFitnessAtStepNumberMinus2)

ExitSlopeRatio:

    SlopeRatio = Z

END FUNCTION

'-----------

SUB
PlotResults(FunctionName$,Nd%,Np%,BestFitnessOverall,BestNpNd%,BestGamma,Neval&&,Rbest(),Mbest(),BestProbeNumberOverall%,BestTimeStepOverall&,LastStepBestR
un&,Alpha,Beta)

LOCAL LastStep&, BestFitnessProbeNumber%, BestFitnessTimeStep&, NumTrajectories%, Max1DprobesPlotted%, i%

LOCAL RepositionFactor$, PlaceInitialProbes$, InitialAcceleration$, A$, B$

LOCAL G, DeltaT, Frep AS EXT

'     G = 2## : DeltaT = 1## : Frep = 0.5## : RepositionFactor$ = "VARIABLE" : PlaceInitialProbes$ = "UNIFORM ON-AXIS " : InitialAcceleration$ = "FIXED"
'THESE ARE NOW HARDWIRED IN THE CFO EQUATIONS 'ORIG CODE 05/14/2011
    G = 2## : DeltaT = 1## : Frep = 0.5## : RepositionFactor$ = "VARIABLE" : PlaceInitialProbes$ = "PROBE LINES" : InitialAcceleration$ = "ZERO" 'MOD
05/14/2011

    B$ = "" : IF Nd% > 1 THEN B$ = "s"

    A$ = FunctionName$ + CHR$(13) +_
         "Best Fitness = " + REMOVE$(STR$(BestFitnessOverall),ANY" ")      + " returned by" + CHR$(13) +_
         "Probe # "        + REMOVE$(STR$(BestProbeNumberOverall%),ANY" ") +_
         " at Time Step "  + REMOVE$(STR$(BestTimeStepOverall&),ANY" ")    + CHR$(13) + CHR$(13) + "P" + REMOVE$(STR$(BestProbeNumberOverall%),ANY" ") + "
coordinate" + B$ + ":" + CHR$(13)

    FOR i% = 1 TO Nd% : A$ = A$ + STR$(i%)+"    "+REMOVE$(STR$(ROUND(Rbest(BestProbeNumberOverall%,i%,BestTimeStepOverall&),8)),ANY" ")+CHR$(13) : NEXT i%

    MSGBOX(A$)

'   ----------------------------------------------- PLOT EVOLUTION OF BEST FITNESS, AVG DISTANCE & BEST PROBE # -------------------------------------------
-------------

    CALL
PlotBestFitnessEvolution(Nd%,Np%,LastStepBestRun&,G,DeltaT,Alpha,Beta,Frep,Mbest(),PlaceInitialProbes$,InitialAcceleration$,RepositionFactor$,FunctionName$
,BestGamma)

'    MSGBOX("Enter for next plot...")

    CALL
PlotAverageDistance(Nd%,Np%,LastStepBestRun&,G,DeltaT,Alpha,Beta,Frep,Mbest(),PlaceInitialProbes$,InitialAcceleration$,RepositionFactor$,FunctionName$,Rbes
t(),DiagLength,BestGamma)

'    MSGBOX("Enter for next plot...")

    CALL
PlotBestProbeVsTimeStep(Nd%,Np%,LastStepBestRun&,G,DeltaT,Alpha,Beta,Frep,Mbest(),PlaceInitialProbes$,InitialAcceleration$,RepositionFactor$,FunctionName$,
BestGamma)

'   --------------------------------------------- PLOT TRAJECTORIES OF BEST PROBES FOR 2/3-D FUNCTIONS --------------------------------------------------

    IF Nd% = 2 THEN

        NumTrajectories% = 10 : CALL Plot2DbestProbeTrajectories(NumTrajectories%,Mbest(),Rbest(),Np%,Nd%,LastStepBestRun&,FunctionName$)

        NumTrajectories% = 16 : CALL Plot2DindividualProbeTrajectories(NumTrajectories%,Mbest(),Rbest(),Np%,Nd%,LastStepBestRun&,FunctionName$)

    END IF

    IF Nd% = 3 THEN

        NumTrajectories% = 4 : CALL Plot3DbestProbeTrajectories(NumTrajectories%,Mbest(),Rbest(),Np%,Nd%,LastStepBestRun&,FunctionName$)

    END IF

'   ----------- For 1-D Objective Functions, Tabulate Probe Coordinates & if Np% =< Max1DprobesPlotted% Plot Evolution of Probe Positions ------------

    IF Nd% = 1 THEN

        Max1DprobesPlotted% = 15

        CALL
Tabulate1DprobeCoordinates(Max1DprobesPlotted%,Nd%,Np%,LastStepBestRun&,G,DeltaT,Alpha,Beta,Frep,Rbest(),Mbest(),PlaceInitialProbes$,InitialAcceleration$,R
epositionFactor$,FunctionName$,BestGamma)

        IF Np% =< Max1DprobesPlotted% THEN _
        CALL
Plot1DprobePositions(Max1DprobesPlotted%,Nd%,Np%,LastStepBestRun&,G,DeltaT,Alpha,Beta,Frep,Rbest(),Mbest(),PlaceInitialProbes$,InitialAcceleration$,Reposit
ionFactor$,FunctionName$,BestGamma)

        CALL CLEANUP 'delete probe coordinate plot files, if any

    END IF
```

```
END SUB 'PlotResults()
'=====================================================================================================================================================================================================
SUB CheckNECFiles(NECfileError$)
LOCAL N%
    NECfileError$ = "NO"
'   ---------------------------- NEC Files Required for PBM Antenna Benchmarks -----------------------------
    IF DIR$("n41_2k1.exe") = "" THEN
            MSGBOX("WARNING!  'n41_2k1.exe' not found.  Run terminated.") : NECfileError$ = "YES" : EXIT SUB
    END IF
'   -------------- These Files are Required for NEC-4.1, Some for NEC-2 -------------------
    N% = FREEFILE : OPEN "ENDERR.DAT"   FOR OUTPUT AS #N% : PRINT #N%, "NO"      : CLOSE #N%
    N% = FREEFILE : OPEN "FILE_MSG.DAT" FOR OUTPUT AS #N% : PRINT #N%, "NO"      : CLOSE #N%
    N% = FREEFILE : OPEN "NHSCALE.DAT"  FOR OUTPUT AS #N% : PRINT #N%, "0.00001" : CLOSE #N%
END SUB
'------
SUB GetBestFitness(M(),Np%,StepNumber&,BestFitness,BestProbeNumber%,BestTimeStep&)
LOCAL p%, i&, A$
    BestFitness = M(1,0)
    FOR i& = 0 TO StepNumber&
        FOR p% = 1 TO Np%
            IF M(p%,i&) >= BestFitness THEN
                BestFitness = M(p%,i&) : BestProbeNumber% = p% : BestTimeStep& = i&
            END IF
        NEXT p%
    NEXT i&
END SUB
'============================================================= FUNCTION DEFINITIONS =============================================================
FUNCTION ObjectiveFunction(R(),Nd%,p%,j&,FunctionName$) 'Objective function to be MAXIMIZED is defined here
    SELECT CASE FunctionName$
        CASE "ParrottF4"     : ObjectiveFunction = ParrottF4(R(),Nd%,p%,j&)      'Parrott F4 (1-D)
        CASE "SGO"           : ObjectiveFunction = SGO(R(),Nd%,p%,j&)            'SGO Function (2-D)
        CASE "GP"            : ObjectiveFunction = GoldsteinPrice(R(),Nd%,p%,j&) 'Goldstein-Price Function (2-D)
        CASE "STEP"          : ObjectiveFunction = StepFunction(R(),Nd%,p%,j&)   'Step Function (n-D)
        CASE "SCHWEFEL_226"  : ObjectiveFunction = Schwefel226(R(),Nd%,p%,j&)    'Schwefel Prob. 2.26 (n-D)
        CASE "COLVILLE"      : ObjectiveFunction = Colville(R(),Nd%,p%,j&)       'Colville Function (4-D)
        CASE "GRIEWANK"      : ObjectiveFunction = Griewank(R(),Nd%,p%,j&)       'Griewank Function (n-D)
        CASE "HIMMELBLAU"    : ObjectiveFunction = Himmelblau(R(),Nd%,p%,j&)     'Himmelblau Function (2-D)
        CASE "ROSENBROCK"    : ObjectiveFunction = Rosenbrock(R(),Nd%,p%,j&)     'Rosenbrock Function (n-D)
        CASE "SPHERE"        : ObjectiveFunction = Sphere(R(),Nd%,p%,j&)         'Sphere Function (n-D)
        CASE "HIMMELBLAUNLO" : ObjectiveFunction = HIMMELBLAUNLO(R(),Nd%,p%,j&)  'Himmelblau NLO (5-D)
        CASE "TRIPOD"        : ObjectiveFunction = Tripod(R(),Nd%,p%,j&)         'Tripod (2-D)
        CASE "ROSENBROCKF6"  : ObjectiveFunction = RosenbrockF6(R(),Nd%,p%,j&)   'RosebrockF6 (10-D)
        CASE "COMPRESSIONSPRING"  : ObjectiveFunction = CompressionSpring(R(),Nd%,p%,j&)   'Compression Spring (3-D)
        CASE "GEARTRAIN"     : ObjectiveFunction = GearTrain(R(),Nd%,p%,j&)      'Gear Train (4-D)
'       ------------------------- GSO Paper Benchmark Functions ---------------------------
        CASE  "F1"           : ObjectiveFunction = F1(R(),Nd%,p%,j&)             'F1  (n-D)
        CASE  "F2"           : ObjectiveFunction = F2(R(),Nd%,p%,j&)             'F2  (n-D)
        CASE  "F3"           : ObjectiveFunction = F3(R(),Nd%,p%,j&)             'F3  (n-D)
        CASE  "F4"           : ObjectiveFunction = F4(R(),Nd%,p%,j&)             'F4  (n-D)
        CASE  "F5"           : ObjectiveFunction = F5(R(),Nd%,p%,j&)             'F5  (n-D)
        CASE  "F6"           : ObjectiveFunction = F6(R(),Nd%,p%,j&)             'F6  (n-D)
        CASE  "F7"           : ObjectiveFunction = F7(R(),Nd%,p%,j&)             'F7  (n-D)
        CASE  "F8"           : ObjectiveFunction = F8(R(),Nd%,p%,j&)             'F8  (n-D)
        CASE  "F9"           : ObjectiveFunction = F9(R(),Nd%,p%,j&)             'F9  (n-D)
        CASE "F10"           : ObjectiveFunction = F10(R(),Nd%,p%,j&)            'F10 (n-D)
        CASE "F11"           : ObjectiveFunction = F11(R(),Nd%,p%,j&)            'F11 (n-D)
        CASE "F12"           : ObjectiveFunction = F12(R(),Nd%,p%,j&)            'F12 (n-D)
        CASE "F13"           : ObjectiveFunction = F13(R(),Nd%,p%,j&)            'F13 (n-D)
        CASE "F14"           : ObjectiveFunction = F14(R(),Nd%,p%,j&)            'F14 (2-D)
        CASE "F15"           : ObjectiveFunction = F15(R(),Nd%,p%,j&)            'F15 (4-D)
        CASE "F16"           : ObjectiveFunction = F16(R(),Nd%,p%,j&)            'F16 (2-D)
        CASE "F17"           : ObjectiveFunction = F17(R(),Nd%,p%,j&)            'F17 (2-D)
        CASE "F18"           : ObjectiveFunction = F18(R(),Nd%,p%,j&)            'F18 (2-D)
        CASE "F19"           : ObjectiveFunction = F19(R(),Nd%,p%,j&)            'F19 (3-D)
        CASE "F20"           : ObjectiveFunction = F20(R(),Nd%,p%,j&)            'F20 (6-D)
        CASE "F21"           : ObjectiveFunction = F21(R(),Nd%,p%,j&)            'F21 (4-D)
        CASE "F22"           : ObjectiveFunction = F22(R(),Nd%,p%,j&)            'F22 (4-D)
        CASE "F23"           : ObjectiveFunction = F23(R(),Nd%,p%,j&)            'F23 (4-D)
'       ----------------------------- PBM Antenna Benchmarks -----------------------------
        CASE "PBM_1"         : ObjectiveFunction = PBM_1(R(),Nd%,p%,j&)          'PBM_1 (2-D)
        CASE "PBM_2"         : ObjectiveFunction = PBM_2(R(),Nd%,p%,j&)          'PBM_2 (2-D)
        CASE "PBM_3"         : ObjectiveFunction = PBM_3(R(),Nd%,p%,j&)          'PBM_3 (2-D)
        CASE "PBM_4"         : ObjectiveFunction = PBM_4(R(),Nd%,p%,j&)          'PBM_4 (2-D)
        CASE "PBM_5"         : ObjectiveFunction = PBM_5(R(),Nd%,p%,j&)          'PBM_5 (2-D)
'       --------------------------- LOADED BOWTIE IN FREE SPACE -------------------------
        CASE "BOWTIE"          : ObjectiveFunction = BOWTIE(R(),Nd%,p%,j&)       'FREE SPACE LOADED BOWTIE
'       --------------------------- YAGI ARRAY IN FREE SPACE -------------------------
        CASE "YAGI"            : ObjectiveFunction = YAGI_ARRAY(R(),Nd%,p%,j&)   'FREE SPACE YAGI ARRAY
```

```
    END SELECT
END FUNCTION 'ObjectiveFunction()
'------
SUB GetFunctionRunParameters(FunctionName$,Nd%,Np%,Nt&,G,DeltaT,Alpha,Beta,Frep,R(),A(),M(),_
                             DiagLength,PlaceInitialProbes$,InitialAcceleration$,RepositionFactor$)
LOCAL i%, NumProbesPerDimension%, NN%, NumCollinearElements%
LOCAL A$
    SELECT CASE FunctionName$
        CASE "ParrottF4"
            Nd% = 1 : Np% = 3
            REDIM XiMin(1 TO Nd%), XiMax(1 TO Nd%) : XiMin(1) = 0## : XiMax(1) = 1##
            REDIM StartingXiMin(1 TO Nd%), StartingXiMax(1 TO Nd%) : FOR i% = 1 TO Nd% : StartingXiMin(i%) = XiMin(i%) : StartingXiMax(i%) = XiMax(i%) :
NEXT i%
        CASE "SGO"
            Nd% = 2 : Np% = 8
            REDIM XiMin(1 TO Nd%), XiMax(1 TO Nd%) : FOR i% = 1 TO Nd% : XiMin(i%) = -50## : XiMax(i%) = 50## : NEXT i%
            REDIM StartingXiMin(1 TO Nd%), StartingXiMax(1 TO Nd%) : FOR i% = 1 TO Nd% : StartingXiMin(i%) = XiMin(i%) : StartingXiMax(i%) = XiMax(i%) :
NEXT i%
        CASE "GP"
            Nd% = 2 : Np% = 8
            REDIM XiMin(1 TO Nd%), XiMax(1 TO Nd%) : FOR i% = 1 TO Nd% : XiMin(i%) = -100## : XiMax(i%) = 100## : NEXT i%
            REDIM StartingXiMin(1 TO Nd%), StartingXiMax(1 TO Nd%) : FOR i% = 1 TO Nd% : StartingXiMin(i%) = XiMin(i%) : StartingXiMax(i%) = XiMax(i%) :
NEXT i%
        CASE "STEP"
            Nd% = 2 : Np% = 8
            REDIM XiMin(1 TO Nd%), XiMax(1 TO Nd%) : FOR i% = 1 TO Nd% : XiMin(i%) = -100## : XiMax(i%) = 100## : NEXT i%
'           REDIM XiMin(1 TO Nd%), XiMax(1 TO Nd%) : XiMin(1) = 72## : XiMax(1) = 78## : XiMin(2) = 27## : XiMax(2) = 33## 'use this to plot STEP detail
            REDIM StartingXiMin(1 TO Nd%), StartingXiMax(1 TO Nd%) : FOR i% = 1 TO Nd% : StartingXiMin(i%) = XiMin(i%) : StartingXiMax(i%) = XiMax(i%) :
NEXT i%
        CASE "SCHWEFEL_226"
            Nd% = 30 : Np% = 120
            REDIM XiMin(1 TO Nd%), XiMax(1 TO Nd%) : FOR i% = 1 TO Nd% : XiMin(i%) = -500## : XiMax(i%) = 500## : NEXT i%
            REDIM StartingXiMin(1 TO Nd%), StartingXiMax(1 TO Nd%) : FOR i% = 1 TO Nd% : StartingXiMin(i%) = XiMin(i%) : StartingXiMax(i%) = XiMax(i%) :
NEXT i%
        CASE "COLVILLE"
            Nd% = 4 : Np% = 16
            REDIM XiMin(1 TO Nd%), XiMax(1 TO Nd%) : FOR i% = 1 TO Nd% : XiMin(i%) = -10## : XiMax(i%) = 10## : NEXT i%
            REDIM StartingXiMin(1 TO Nd%), StartingXiMax(1 TO Nd%) : FOR i% = 1 TO Nd% : StartingXiMin(i%) = XiMin(i%) : StartingXiMax(i%) = XiMax(i%) :
NEXT i%
        CASE "GRIEWANK"
            Nd% = 2 : Np% = 8
            REDIM XiMin(1 TO Nd%), XiMax(1 TO Nd%) : FOR i% = 1 TO Nd% : XiMin(i%) = -600## : XiMax(i%) = 600## : NEXT i%
            REDIM StartingXiMin(1 TO Nd%), StartingXiMax(1 TO Nd%) : FOR i% = 1 TO Nd% : StartingXiMin(i%) = XiMin(i%) : StartingXiMax(i%) = XiMax(i%) :
NEXT i%
       CASE "HIMMELBLAU"
            Nd% = 2 : Np% = 8
            REDIM XiMin(1 TO Nd%), XiMax(1 TO Nd%) : FOR i% = 1 TO Nd% : XiMin(i%) = -6## : XiMax(i%) = 6## : NEXT i%
            REDIM StartingXiMin(1 TO Nd%), StartingXiMax(1 TO Nd%) : FOR i% = 1 TO Nd% : StartingXiMin(i%) = XiMin(i%) : StartingXiMax(i%) = XiMax(i%) :
NEXT i%
        CASE "ROSENBROCK" '(n-D)
            Nd% = 2 : Np% = 8
            REDIM XiMin(1 TO Nd%), XiMax(1 TO Nd%) : FOR i% = 1 TO Nd% : XiMin(i%) = -2## : XiMax(i%) = 2## : NEXT i% :'XiMin(i%) = -6## : XiMax(i%) = 6##
: NEXT i%
            REDIM StartingXiMin(1 TO Nd%), StartingXiMax(1 TO Nd%) : FOR i% = 1 TO Nd% : StartingXiMin(i%) = XiMin(i%) : StartingXiMax(i%) = XiMax(i%) :
NEXT i%
        CASE "SPHERE" '(n-D)
            Nd% = 2 : Np% = 8
            REDIM XiMin(1 TO Nd%), XiMax(1 TO Nd%) : FOR i% = 1 TO Nd% : XiMin(i%) = -100## : XiMax(i%) = 100## : NEXT i%
            REDIM StartingXiMin(1 TO Nd%), StartingXiMax(1 TO Nd%) : FOR i% = 1 TO Nd% : StartingXiMin(i%) = XiMin(i%) : StartingXiMax(i%) = XiMax(i%) :
NEXT i%
        CASE "HIMMELBLAUNLO" '(5-D)
            Nd% = 5 : Np% = 20
            REDIM XiMin(1 TO Nd%), XiMax(1 TO Nd%)
            XiMin(1) = 78## : XiMax(1) = 102##
            XiMin(2) = 33## : XiMax(2) = 45##
            XiMin(3) = 27## : XiMax(3) = 45##
            XiMin(4) = 27## : XiMax(4) = 45##
            XiMin(5) = 27## : XiMax(5) = 45##
            REDIM StartingXiMin(1 TO Nd%), StartingXiMax(1 TO Nd%) : FOR i% = 1 TO Nd% : StartingXiMin(i%) = XiMin(i%) : StartingXiMax(i%) = XiMax(i%) :
NEXT i%
        CASE "TRIPOD" '(2-D)
            Nd% = 2 : Np% = 8
            REDIM XiMin(1 TO Nd%), XiMax(1 TO Nd%) : FOR i% = 1 TO Nd% : XiMin(i%) = -100## : XiMax(i%) = 100## : NEXT i%
```

```
            REDIM StartingXiMin(1 TO Nd%), StartingXiMax(1 TO Nd%) : FOR i% = 1 TO Nd% : StartingXiMin(i%) = XiMin(i%) : StartingXiMax(i%) = XiMax(i%) :
NEXT i%

        CASE "ROSENBROCKF6" '(10-D)

            Nd% = 10
            Np% = 40

            REDIM XiOffset(1 TO Nd%)

'            XiOffset(1) =  81.0232##
'            XiOffset(2) = -48.3950##
'            XiOffset(3) =  19.2316##
'            XiOffset(4) =  -2.5231##
'            XiOffset(5) =  70.4338##
'            XiOffset(6) =  47.1774##
'            XiOffset(7) =  -7.8358##
'            XiOffset(8) = -86.6693##
'            XiOffset(9) =  57.8532##
'            XiOffset(10) = 0##

'            XiOffset(1)  =  80##
'            XiOffset(2)  = -50##
'            XiOffset(3)  =  20##
'            XiOffset(4)  =  -3##
'            XiOffset(5)  =  70##
'            XiOffset(6)  =  47##
'            XiOffset(7)  =  -8##
'            XiOffset(8)  = -87##
'            XiOffset(9)  =  58##
'            XiOffset(10) =   0##

            XiOffset(1)  =   5##
            XiOffset(2)  = -25##
            XiOffset(3)  =   5##
            XiOffset(4)  = -15##
            XiOffset(5)  =   5##
            XiOffset(6)  = -25##
            XiOffset(7)  =  25##
            XiOffset(8)  =  -5##
            XiOffset(9)  =   5##
            XiOffset(10) = -15##

'            XiOffset(1)  =  0##
'            XiOffset(2)  =  81.0232##
'            XiOffset(3)  = -48.3950##
'            XiOffset(4)  =  19.2316##
'            XiOffset(5)  =  -2.5231##
'            XiOffset(6)  =  70.4338##
'            XiOffset(7)  =  47.1774##
'            XiOffset(8)  =  -7.8358##
'            XiOffset(9)  = -86.6693##
'            XiOffset(10) =  57.8532##

            REDIM XiMin(1 TO Nd%), XiMax(1 TO Nd%) : FOR i% = 1 TO Nd% : XiMin(i%) = -100## : XiMax(i%) = 100## : NEXT i%

            REDIM StartingXiMin(1 TO Nd%), StartingXiMax(1 TO Nd%) : FOR i% = 1 TO Nd% : StartingXiMin(i%) = XiMin(i%) : StartingXiMax(i%) = XiMax(i%) :
NEXT i%

         CASE "COMPRESSIONSPRING" '(3-D)

            Nd% = 3 : Np% = 12

            REDIM XiMin(1 TO Nd%), XiMax(1 TO Nd%)

            XiMin(1) = 1##     : XiMax(1) =  70## 'integer values only!!
            XiMin(2) = 0.6##   : XiMax(2) =   3##
            XiMin(3) = 0.207## : XiMax(3) = 0.5##

            REDIM StartingXiMin(1 TO Nd%), StartingXiMax(1 TO Nd%) : FOR i% = 1 TO Nd% : StartingXiMin(i%) = XiMin(i%) : StartingXiMax(i%) = XiMax(i%) :
NEXT i%

         CASE "GEARTRAIN" '(4-D)

            Nd% = 4 : Np% = 16

            REDIM XiMin(1 TO Nd%), XiMax(1 TO Nd%) : FOR i% = 1 TO Nd% : XiMin(i%) = 12# : XiMax(i%) = 60## : NEXT i%

            REDIM StartingXiMin(1 TO Nd%), StartingXiMax(1 TO Nd%) : FOR i% = 1 TO Nd% : StartingXiMin(i%) = XiMin(i%) : StartingXiMax(i%) = XiMax(i%) :
NEXT i%

        CASE "F1" '(n-D)

            Nd% = 30 : Np% = 60

            REDIM XiMin(1 TO Nd%), XiMax(1 TO Nd%) : FOR i% = 1 TO Nd% : XiMin(i%) = -100## : XiMax(i%) = 100## : NEXT i%

            REDIM StartingXiMin(1 TO Nd%), StartingXiMax(1 TO Nd%) : FOR i% = 1 TO Nd% : StartingXiMin(i%) = XiMin(i%) : StartingXiMax(i%) = XiMax(i%) :
NEXT i%

       CASE "F2" '(n-D)

            Nd% = 30 : Np% = 60

            REDIM XiMin(1 TO Nd%), XiMax(1 TO Nd%) : FOR i% = 1 TO Nd% : XiMin(i%) = -10## : XiMax(i%) = 10## : NEXT i%

            REDIM StartingXiMin(1 TO Nd%), StartingXiMax(1 TO Nd%) : FOR i% = 1 TO Nd% : StartingXiMin(i%) = XiMin(i%) : StartingXiMax(i%) = XiMax(i%) :
NEXT i%

      CASE "F3" '(n-D)

            Nd% = 30 : Np% = 60

            REDIM XiMin(1 TO Nd%), XiMax(1 TO Nd%) : FOR i% = 1 TO Nd% : XiMin(i%) = -100## : XiMax(i%) = 100## : NEXT i%

            REDIM StartingXiMin(1 TO Nd%), StartingXiMax(1 TO Nd%) : FOR i% = 1 TO Nd% : StartingXiMin(i%) = XiMin(i%) : StartingXiMax(i%) = XiMax(i%) :
NEXT i%

      CASE "F4" '(n-D)

            Nd% = 30 : Np% = 60

            REDIM XiMin(1 TO Nd%), XiMax(1 TO Nd%) : FOR i% = 1 TO Nd% : XiMin(i%) = -100## : XiMax(i%) = 100## : NEXT i%

            REDIM StartingXiMin(1 TO Nd%), StartingXiMax(1 TO Nd%) : FOR i% = 1 TO Nd% : StartingXiMin(i%) = XiMin(i%) : StartingXiMax(i%) = XiMax(i%) :
NEXT i%

      CASE "F5" '(n-D)

            Nd% = 30 : Np% = 60

            REDIM XiMin(1 TO Nd%), XiMax(1 TO Nd%) : FOR i% = 1 TO Nd% : XiMin(i%) = -30## : XiMax(i%) = 30## : NEXT i%

            REDIM StartingXiMin(1 TO Nd%), StartingXiMax(1 TO Nd%) : FOR i% = 1 TO Nd% : StartingXiMin(i%) = XiMin(i%) : StartingXiMax(i%) = XiMax(i%) :
NEXT i%

       CASE "F6" '(n-D) STEP

            Nd% = 30 : Np% = 60
```

```
            REDIM XiMin(1 TO Nd%), XiMax(1 TO Nd%) : FOR i% = 1 TO Nd% : XiMin(i%) = -100## : XiMax(i%) = 100## : NEXT i%
            REDIM StartingXiMin(1 TO Nd%), StartingXiMax(1 TO Nd%) : FOR i% = 1 TO Nd% : StartingXiMin(i%) = XiMin(i%) : StartingXiMax(i%) = XiMax(i%) :
NEXT i%
        CASE "F7" '(n-D)
            Nd% = 30 : Np% = 60
            REDIM XiMin(1 TO Nd%), XiMax(1 TO Nd%) : FOR i% = 1 TO Nd% : XiMin(i%) = -1.28## : XiMax(i%) = 1.28## : NEXT i%
            REDIM StartingXiMin(1 TO Nd%), StartingXiMax(1 TO Nd%) : FOR i% = 1 TO Nd% : StartingXiMin(i%) = XiMin(i%) : StartingXiMax(i%) = XiMax(i%) :
NEXT i%
        CASE "F8" '(n-D)
            Nd% = 30 : Np% = 60
            REDIM XiMin(1 TO Nd%), XiMax(1 TO Nd%) : FOR i% = 1 TO Nd% : XiMin(i%) = -500## : XiMax(i%) = 500## : NEXT i%
            REDIM StartingXiMin(1 TO Nd%), StartingXiMax(1 TO Nd%) : FOR i% = 1 TO Nd% : StartingXiMin(i%) = XiMin(i%) : StartingXiMax(i%) = XiMax(i%) :
NEXT i%
        CASE "F9" '(n-D)
            Nd% = 30 : Np% = 60
            REDIM XiMin(1 TO Nd%), XiMax(1 TO Nd%) : FOR i% = 1 TO Nd% : XiMin(i%) = -5.12## : XiMax(i%) = 5.12## : NEXT i%
            REDIM StartingXiMin(1 TO Nd%), StartingXiMax(1 TO Nd%) : FOR i% = 1 TO Nd% : StartingXiMin(i%) = XiMin(i%) : StartingXiMax(i%) = XiMax(i%) :
NEXT i%
        CASE "F10" '(n-D) Ackley's Function
            Nd% = 30 : Np% = 60
            REDIM XiMin(1 TO Nd%), XiMax(1 TO Nd%) : FOR i% = 1 TO Nd% : XiMin(i%) = -32## : XiMax(i%) = 32## : NEXT i%
            REDIM StartingXiMin(1 TO Nd%), StartingXiMax(1 TO Nd%) : FOR i% = 1 TO Nd% : StartingXiMin(i%) = XiMin(i%) : StartingXiMax(i%) = XiMax(i%) :
NEXT i%
        CASE "F11" '(n-D)
            Nd% = 30 : Np% = 60
            REDIM XiMin(1 TO Nd%), XiMax(1 TO Nd%) : FOR i% = 1 TO Nd% : XiMin(i%) = -600## : XiMax(i%) = 600## : NEXT i%
            REDIM StartingXiMin(1 TO Nd%), StartingXiMax(1 TO Nd%) : FOR i% = 1 TO Nd% : StartingXiMin(i%) = XiMin(i%) : StartingXiMax(i%) = XiMax(i%) :
NEXT i%
         CASE "F12" '(n-D) Penalized #1
            Nd% = 30 : Np% = 60
            REDIM XiMin(1 TO Nd%), XiMax(1 TO Nd%) : FOR i% = 1 TO Nd% : XiMin(i%) = -50## : XiMax(i%) = 50## : NEXT i%
'            REDIM XiMin(1 TO Nd%), XiMax(1 TO Nd%) : FOR i% = 1 TO Nd% : XiMin(i%) = -5## : XiMax(i%) = 5## : NEXT i% 'use this interval for second run to
improve performance
            REDIM StartingXiMin(1 TO Nd%), StartingXiMax(1 TO Nd%) : FOR i% = 1 TO Nd% : StartingXiMin(i%) = XiMin(i%) : StartingXiMax(i%) = XiMax(i%) :
NEXT i%
        CASE "F13" '(n-D) Penalized #2
            Nd% = 30 : Np% = 60
            REDIM XiMin(1 TO Nd%), XiMax(1 TO Nd%) : FOR i% = 1 TO Nd% : XiMin(i%) = -50## : XiMax(i%) = 50## : NEXT i%
            REDIM StartingXiMin(1 TO Nd%), StartingXiMax(1 TO Nd%) : FOR i% = 1 TO Nd% : StartingXiMin(i%) = XiMin(i%) : StartingXiMax(i%) = XiMax(i%) :
NEXT i%
        CASE "F14" '(2-D) Shekel's Foxholes
            Nd% = 2 : Np% = 8
            REDIM XiMin(1 TO Nd%), XiMax(1 TO Nd%) : FOR i% = 1 TO Nd% : XiMin(i%) = -65.536## : XiMax(i%) = 65.536## : NEXT i%
            REDIM StartingXiMin(1 TO Nd%), StartingXiMax(1 TO Nd%) : FOR i% = 1 TO Nd% : StartingXiMin(i%) = XiMin(i%) : StartingXiMax(i%) = XiMax(i%) :
NEXT i%
        CASE "F15" '(4-D) Kowalik's Function
            Nd% = 4
            Np% = 16
            REDIM XiMin(1 TO Nd%), XiMax(1 TO Nd%) : FOR i% = 1 TO Nd% : XiMin(i%) = -5## : XiMax(i%) = 5## : NEXT i%
            REDIM StartingXiMin(1 TO Nd%), StartingXiMax(1 TO Nd%) : FOR i% = 1 TO Nd% : StartingXiMin(i%) = XiMin(i%) : StartingXiMax(i%) = XiMax(i%) :
NEXT i%
        CASE "F16" '(2-D) Camel Back
            Nd% = 2 : Np% = 8
            REDIM XiMin(1 TO Nd%), XiMax(1 TO Nd%) : FOR i% = 1 TO Nd% : XiMin(i%) = -5## : XiMax(i%) = 5## : NEXT i%
            REDIM StartingXiMin(1 TO Nd%), StartingXiMax(1 TO Nd%) : FOR i% = 1 TO Nd% : StartingXiMin(i%) = XiMin(i%) : StartingXiMax(i%) = XiMax(i%) :
NEXT i%
        CASE "F17" '(2-D) Branin
            Nd% = 2 : Np% = 8
            REDIM XiMin(1 TO Nd%), XiMax(1 TO Nd%) : XiMin(1) = -5## : XiMax(1) = 10## : XiMin(2) = 0## : XiMax(2) = 15##
            REDIM StartingXiMin(1 TO Nd%), StartingXiMax(1 TO Nd%) : FOR i% = 1 TO Nd% : StartingXiMin(i%) = XiMin(i%) : StartingXiMax(i%) = XiMax(i%) :
NEXT i%
        CASE "F18" '(2-D) Goldstein-Price
            Nd% = 2 : Np% = 8
            REDIM XiMin(1 TO Nd%), XiMax(1 TO Nd%) : XiMin(1) = -2## : XiMax(1) = 2## : XiMin(2) = -2## : XiMax(2) = 2##
            REDIM StartingXiMin(1 TO Nd%), StartingXiMax(1 TO Nd%) : FOR i% = 1 TO Nd% : StartingXiMin(i%) = XiMin(i%) : StartingXiMax(i%) = XiMax(i%) :
NEXT i%
        CASE "F19" '(3-D) Hartman's Family #1
            Nd% = 3 : Np% = 12
            REDIM XiMin(1 TO Nd%), XiMax(1 TO Nd%) : FOR i% = 1 TO Nd% : XiMin(i%) = 0## : XiMax(i%) = 1## : NEXT i%
            REDIM StartingXiMin(1 TO Nd%), StartingXiMax(1 TO Nd%) : FOR i% = 1 TO Nd% : StartingXiMin(i%) = XiMin(i%) : StartingXiMax(i%) = XiMax(i%) :
NEXT i%
        CASE "F20" '(6-D) Hartman's Family #2
```

```
        Nd% = 6 : Np% = 24

        REDIM XiMin(1 TO Nd%), XiMax(1 TO Nd%) : FOR i% = 1 TO Nd% : XiMin(i%) = 0## : XiMax(i%) = 1## : NEXT i%

        REDIM StartingXiMin(1 TO Nd%), StartingXiMax(1 TO Nd%) : FOR i% = 1 TO Nd% : StartingXiMin(i%) = XiMin(i%) : StartingXiMax(i%) = XiMax(i%) :
NEXT i%

    CASE "F21" '(4-D) Shekel's Family m=5

        Nd% = 4 : Np% = 16

        REDIM XiMin(1 TO Nd%), XiMax(1 TO Nd%) : FOR i% = 1 TO Nd% : XiMin(i%) = 0## : XiMax(i%) = 10## : NEXT i%

        REDIM StartingXiMin(1 TO Nd%), StartingXiMax(1 TO Nd%) : FOR i% = 1 TO Nd% : StartingXiMin(i%) = XiMin(i%) : StartingXiMax(i%) = XiMax(i%) :
NEXT i%

    CASE "F22" '(4-D) Shekel's Family m=7

        Nd% = 4 : Np% = 16

        REDIM XiMin(1 TO Nd%), XiMax(1 TO Nd%) : FOR i% = 1 TO Nd% : XiMin(i%) = 0## : XiMax(i%) = 10## : NEXT i%

        REDIM StartingXiMin(1 TO Nd%), StartingXiMax(1 TO Nd%) : FOR i% = 1 TO Nd% : StartingXiMin(i%) = XiMin(i%) : StartingXiMax(i%) = XiMax(i%) :
NEXT i%

    CASE "F23" '(4-D) Shekel's Family m=10

        Nd% = 4 : Np% = 16

        REDIM XiMin(1 TO Nd%), XiMax(1 TO Nd%) : FOR i% = 1 TO Nd% : XiMin(i%) = 0## : XiMax(i%) = 10## : NEXT i%

        REDIM StartingXiMin(1 TO Nd%), StartingXiMax(1 TO Nd%) : FOR i% = 1 TO Nd% : StartingXiMin(i%) = XiMin(i%) : StartingXiMax(i%) = XiMax(i%) :
NEXT i%

    CASE "PBM_1" '2-D

        Nd%                      = 2
        NumProbesPerDimension% = 2 '4 '20
        Np%                      = NumProbesPerDimension%*Nd%

        Nt&      = 100
        G        = 2##
        Alpha    = 2##
        Beta     = 2##
        DeltaT   = 1##
        Frep     = 0.5##

        PlaceInitialProbes$  = "UNIFORM ON-AXIS"
        InitialAcceleration$ = "ZERO"
        RepositionFactor$    = "VARIABLE" '"FIXED"

        Np% = NumProbesPerDimension%*Nd%

        REDIM XiMin(1 TO Nd%), XiMax(1 TO Nd%)

        XiMin(1) = 0.5## : XiMax(1) = 3## 'dipole length, L, in Wavelengths
        XiMin(2) = 0##   : XiMax(2) = Pi2 'polar angle, Theta, in Radians

        REDIM StartingXiMin(1 TO Nd%), StartingXiMax(1 TO Nd%) : FOR i% = 1 TO Nd% : StartingXiMin(i%) = XiMin(i%) : StartingXiMax(i%) = XiMax(i%) :
NEXT i%

        NN% = FREEFILE : OPEN "INFILE.DAT" FOR OUTPUT AS #NN% : PRINT #NN%,"PBM1.NEC" : PRINT #NN%,"PBM1.OUT" : CLOSE #NN% 'NEC Input/Output Files

    CASE "PBM_2" '2-D

        AddNoiseToPBM2$ = "NO" '"YES" '"NO" '"YES"

        Nd%                      = 2
        NumProbesPerDimension% = 4 '20
        Np%                      = NumProbesPerDimension%*Nd%

        Nt&      = 100
        G        = 2##
        Alpha    = 2##
        Beta     = 2##
        DeltaT   = 1##
        Frep     = 0.5##

        PlaceInitialProbes$  = "UNIFORM ON-AXIS"
        InitialAcceleration$ = "ZERO"
        RepositionFactor$    = "VARIABLE" '"FIXED"

        Np% = NumProbesPerDimension%*Nd%

        REDIM XiMin(1 TO Nd%), XiMax(1 TO Nd%)

        XiMin(1) = 5## : XiMax(1) = 15## 'dipole separation, D, in Wavelengths
        XiMin(2) = 0## : XiMax(2) = Pi   'polar angle, Theta, in Radians

        REDIM StartingXiMin(1 TO Nd%), StartingXiMax(1 TO Nd%) : FOR i% = 1 TO Nd% : StartingXiMin(i%) = XiMin(i%) : StartingXiMax(i%) = XiMax(i%) :
NEXT i%

        NN% = FREEFILE : OPEN "INFILE.DAT" FOR OUTPUT AS #NN% : PRINT #NN%,"PBM2.NEC" : PRINT #NN%,"PBM2.OUT" : CLOSE #NN%

    CASE "PBM_3" '2-D

        Nd%                      = 2
        NumProbesPerDimension% = 4 '20
        Np%                      = NumProbesPerDimension%*Nd%

        Nt&      = 100
        G        = 2##
        Alpha    = 2##
        Beta     = 2##
        DeltaT   = 1##
        Frep     = 0.5##

        PlaceInitialProbes$  = "UNIFORM ON-AXIS"
        InitialAcceleration$ = "ZERO"
        RepositionFactor$    = "VARIABLE" '"FIXED"

        Np% = NumProbesPerDimension%*Nd%

        REDIM XiMin(1 TO Nd%), XiMax(1 TO Nd%)

        XiMin(1) = 0## : XiMax(1) = 4## 'Phase Parameter, Beta (0-4)
        XiMin(2) = 0## : XiMax(2) = Pi  'polar angle, Theta, in Radians

        REDIM StartingXiMin(1 TO Nd%), StartingXiMax(1 TO Nd%) : FOR i% = 1 TO Nd% : StartingXiMin(i%) = XiMin(i%) : StartingXiMax(i%) = XiMax(i%) :
NEXT i%

        NN% = FREEFILE : OPEN "INFILE.DAT" FOR OUTPUT AS #NN% : PRINT #NN%,"PBM3.NEC" : PRINT #NN%,"PBM3.OUT" : CLOSE #NN%

    CASE "PBM_4" '2-D

        Nd%                      = 2
        NumProbesPerDimension% = 4 '6 '2 '4 '20
        Np%                      = NumProbesPerDimension%*Nd%
```

```
            Nt&      = 100
            G        = 2##
            Alpha    = 2##
            Beta     = 2##
            DeltaT   = 1##
            Frep     = 0.5##

            PlaceInitialProbes$  = "UNIFORM ON-AXIS"
            InitialAcceleration$ = "ZERO"
            RepositionFactor$    = "VARIABLE" '"FIXED"

            Np% = NumProbesPerDimension%*Nd%

            REDIM XiMin(1 TO Nd%), XiMax(1 TO Nd%)

            XiMin(1) = 0.5##   : XiMax(1) = 1.5##  'ARM LENGTH (NOT Total Length), wavelengths (0.5-1.5)
            XiMin(2) = Pi/18## : XiMax(2) = Pi/2## 'Inner angle, Alpha, in Radians (Pi/18-Pi/2)

            REDIM StartingXiMin(1 TO Nd%), StartingXiMax(1 TO Nd%) : FOR i% = 1 TO Nd% : StartingXiMin(i%) = XiMin(i%) : StartingXiMax(i%) = XiMax(i%) :
NEXT i%

            NN% = FREEFILE : OPEN "INFILE.DAT" FOR OUTPUT AS #NN% : PRINT #NN%,"PBM4.NEC" : PRINT #NN%,"PBM4.OUT" : CLOSE #NN%

       CASE "PBM_5"

            NumCollinearElements%  = 6 '30 'EVEN or ODD: 6,7,10,13,16,24 used by PBM

            Nd%                    = NumCollinearElements% - 1
            NumProbesPerDimension% = 4 '20
            Np%                    = NumProbesPerDimension%*Nd%

            Nt&      = 100
            G        = 2##
            Alpha    = 2##
            Beta     = 2##
            DeltaT   = 1##
            Frep     = 0.5##

            PlaceInitialProbes$  = "UNIFORM ON-AXIS"
            InitialAcceleration$ = "ZERO"
            RepositionFactor$    = "VARIABLE" '"FIXED"

            Nd% = NumCollinearElements% - 1

            Np% = NumProbesPerDimension%*Nd%

            REDIM XiMin(1 TO Nd%), XiMax(1 TO Nd%) : FOR i% = 1 TO Nd% : XiMin(i%) = 0.5## : XiMax(i%) = 1.5## : NEXT i%

            REDIM StartingXiMin(1 TO Nd%), StartingXiMax(1 TO Nd%) : FOR i% = 1 TO Nd% : StartingXiMin(i%) = XiMin(i%) : StartingXiMax(i%) = XiMax(i%) :
NEXT i%

            NN% = FREEFILE : OPEN "INFILE.DAT" FOR OUTPUT AS #NN% : PRINT #NN%,"PBM5.NEC" : PRINT #NN%,"PBM5.OUT" : CLOSE #NN%

'      ================================================= END PBM BENCHMARKS ================================================

       CASE "BOWTIE" 'FREE SPACE LOADED BOWTIE

'           --------- Check for NEC Executable ----------

            IF DIR$("NEC41D_4K_053011.EXE") = "" THEN
                MSGBOX("WARNING! NEC4.1D executable NEC41D_4K_053011.EXE not found!"+CHR$(13)+CHR$(13)+_
                       "           RUN TERMINATED.")+CHR$(13)+CHR$(13)
                EXIT SUB
            END IF

'           ------- Get Segmentation -------
'            IF DIR$("BowtieSeg.TXT") = "" THEN 'create BowtieSeg.TXT if it doesn't exist
'                    NN% = FREEFILE : OPEN "BowtieSeg.TXT" FOR OUTPUT AS NN% : PRINT #NN%,"Variable" : PRINT #NN%,"0.05" : CLOSE #NN%
'            END IF
'            NN% = FREEFILE : OPEN "BowtieSeg.TXT" FOR INPUT AS NN% : INPUT #NN%,BowtieSegmentLength$ : INPUT #NN%,BowtieSegmentLengthWvln : CLOSE #NN%
'            IF INSTR(UCASE$(BowtieSegmentLength$),"V") > 0 THEN BowtieSegmentLength$ = "VARIABLE"
'            IF INSTR(UCASE$(BowtieSegmentLength$),"F") > 0 THEN BowtieSegmentLength$ = "FIXED"
'            IF INSTR(UCASE$(BowtieSegmentLength$),"V") = 0 AND INSTR(UCASE$(BowtieSegmentLength$),"F") = 0 THEN BowtieSegmentLength$ = "VARIABLE"
'            IF BowtieSegmentLengthWvln < 0.02## OR BowtieSegmentLengthWvln > 0.2## THEN BowtieSegmentLengthWvln = 0.05## 'range 0.02-0.2 wavelengths

'           ----- Get Bowtie Fitness Function Coefficients -----  NOTE: OPTIONALLY USED IN FITNESS FUNCTION (SEE BOWTIE FUNCTION FOR ACTUAL CALCULATION)
            REDIM BowtieFitnessCoefficients(1 TO 3)
            IF DIR$("BowtieCoeff.TXT") = "" THEN 'create BowtieCoeff.TXT if it doesn't exist
                    NN% = FREEFILE : OPEN "BowtieCoeff.TXT" FOR OUTPUT AS NN% : PRINT #NN%,"40" : PRINT #NN%,"2" : PRINT #NN%,"3" : CLOSE #NN%
            END IF
            NN% = FREEFILE
            OPEN "BowtieCoeff.TXT" FOR INPUT AS NN%
                INPUT #NN%,BowtieFitnessCoefficients(1) : INPUT #NN%,BowtieFitnessCoefficients(2) : INPUT #NN%,BowtieFitnessCoefficients(3)
            CLOSE #NN%

            BowtieCoefficients$ = "A="+REMOVE$(STR$(BowtieFitnessCoefficients(1)),ANY" ")+_
                                  ", B="+REMOVE$(STR$(BowtieFitnessCoefficients(2)),ANY" ")+_
                                  ", C="+REMOVE$(STR$(BowtieFitnessCoefficients(3)),ANY" ")

'BOWTIE DECISION SPACE BOUNDARY ARRAY:

'    Array Element #                Design Variable
'    ---------------      ------------------------------------
'          1                Bowtie atm length (wavelengths)
'          2                Bowtie HALF angle (degrees)
'          3                Loading segment number #1
'          4                Loading resistance #1 (ohms)
'          5                Zo

            Nd% = 5 'FIXED DIMENSIONALITY BASED ON ABOVE DESIGN PARAMETERS

            REDIM XiMin(1 TO Nd%), XiMax(1 TO Nd%)

            XiMin(1) = 0.01## : XiMax(1) = 0.08## 'min/max arm length (METERS) (>~1/4-wave at 1 GHz)
            XiMin(2) = 10##   : XiMax(2) = 80##   'min/max HALF angle (degrees)
            XiMin(3) = 1      : XiMax(3) = 9      'min/max loading segment number #1 (NOTE FIXED SEGMENTATION AT 9 SEGS !!!)
            XiMin(4) = 1##    : XiMax(4) = 1000## 'min/max loading resistance, R1 (ohms)
            XiMin(5) = 50##   : XiMax(5) = 1000## 'Zo, ohms

            REDIM StartingXiMin(1 TO Nd%), StartingXiMax(1 TO Nd%)
            FOR i% = 1 TO Nd% : StartingXiMin(i%) = XiMin(i%) : StartingXiMax(i%) = XiMax(i%) : NEXT i%

            NN% = FREEFILE : OPEN "INFILE.DAT" FOR OUTPUT AS #NN% : PRINT #NN%,"BOWTIE.NEC" : PRINT #NN%,"BOWTIE.OUT" : CLOSE #NN%

'      ======================================================== END BOWTIE ================================================

       CASE "YAGI" 'FREE SPACE YAGI ARRAY

'           --------- Check for NEC Executable ----------

            IF DIR$("NEC41D_4K_053011.EXE") = "" THEN
                MSGBOX("WARNING! NEC4.1D executable NEC41D_4K_053011.EXE not found!"+CHR$(13)+CHR$(13)+_
                       "           RUN TERMINATED.")+CHR$(13)+CHR$(13)
                EXIT SUB
            END IF

'           ------- Get Segmentation -------
            IF DIR$("YagiSeg.TXT") = "" THEN 'create YagiSeg.TXT if it doesn't exist
                    NN% = FREEFILE : OPEN "YagiSeg.TXT" FOR OUTPUT AS NN% : PRINT #NN%,"Variable" : PRINT #NN%,"0.05" : CLOSE #NN%
```

```
            END IF
            NN% = FREEFILE : OPEN "YagiSeg.TXT" FOR INPUT AS NN% : INPUT #NN%,YagiSegmentLength$ : INPUT #NN%,YagiSegmentLengthWvln : CLOSE #NN%
            IF INSTR(UCASE$(YagiSegmentLength$),"V") > 0 THEN YagiSegmentLength$ = "VARIABLE"
            IF INSTR(UCASE$(YagiSegmentLength$),"F") > 0 THEN YagiSegmentLength$ = "FIXED"
            IF INSTR(UCASE$(YagiSegmentLength$),"V") = 0 AND INSTR(UCASE$(YagiSegmentLength$),"F") = 0 THEN YagiSegmentLength$ = "VARIABLE"
            IF YagiSegmentLengthWvln < 0.02## OR YagiSegmentLengthWvln > 0.2## THEN YagiSegmentLengthWvln = 0.05## 'range 0.02-0.2 wavelengths

'           ---------- Get Yagi Fitness Function Coefficients -----------
            REDIM YagiFitnessCoefficients(1 TO 4)
            IF DIR$("YagiCoeff.TXT") = "" THEN 'create YagiCoeff.TXT if it doesn't exist
'A=20, B=2, C=3, D=4.
                    NN% = FREEFILE : OPEN "YagiCoeff.TXT" FOR OUTPUT AS NN% : PRINT #NN%,"20" : PRINT #NN%,"2" : PRINT #NN%,"3" : PRINT #NN%,"4" : CLOSE
#NN%
            END IF
            NN% = FREEFILE
            OPEN "YagiCoeff.TXT" FOR INPUT AS NN%
                INPUT   #NN%,YagiFitnessCoefficients(1)   :   INPUT   #NN%,YagiFitnessCoefficients(2)   :   INPUT   #NN%,YagiFitnessCoefficients(3)   :   INPUT
#NN%,YagiFitnessCoefficients(4)
            CLOSE #NN%

            YagiCoefficients$ = "A="+REMOVE$(STR$(YagiFitnessCoefficients(1)),ANY" ")+_
                                ", B="+REMOVE$(STR$(YagiFitnessCoefficients(2)),ANY" ")+_
                                ", C="+REMOVE$(STR$(YagiFitnessCoefficients(3)),ANY" ")+_
                                ", D="+REMOVE$(STR$(YagiFitnessCoefficients(4)),ANY" ")

'           ----------------- # Array Elements -------------------
            A$ = INPUTBOX$("# Array Elements","YAGI ELEMENTS","5")

            NumYagiElements% = VAL(A$)

            Nd% =2*NumYagiElements% 'spacing and length are the two optimization parameters defining each element

            Nd% = Nd%+1 'INCREASE DECISION SPACE DIMENSION TO ADD Zo AS A DESIGN <<<VARIABLE>>>.
                                                                                        '========
'YAGI DECISION SPACE BOUNDARY ARRAY:

'    Array Element #                           Design Variable
'    ---------------      ---------------------------------------------------------------------
'           1
'           TO                 Yagi element spacing along boom from previous element, wavelengths
'     NumYagiElements

'   NumYagiElements+1
'           TO                                Yagi element length, wavelengths
'   2*NumYagiElements (Nd-1)

'           Nd                                           <<<< Zo >>>>

            REDIM XiMin(1 TO Nd%), XiMax(1 TO Nd%) '<<<<<<<<< FOLLOWING BOUNDS ARE FOR ZEPP ELEMENTS !!! >>>>>>>>>>>>>

            XiMin(1) = 0## : XiMax(1) = 0## : FOR i% = 2 TO NumYagiElements% : XiMin(i%) = 0.1## : XiMax(i%) = 0.5## : NEXT i% 'min/max SPACING, wvln
           '[NOTE: XiMin/max(1)=0 ALWAYS BECAUSE FIRST ELEMENT, REFLECTOR, IS AT X=0 (no spacing)]

            XiMin(NumYagiElements%+1) = 0.2## : XiMax(NumYagiElements%+1) = 0.6## 'min/max RREFLECTOR length, waves

            XiMin(NumYagiElements%+2) = 0.2## : XiMax(NumYagiElements%+2) = 0.6## 'min/max DRIVEN ELEMENT length, waves

            FOR i% = NumYagiElements% + 3 TO Nd%-1
                XiMin(i%) = 0.2## : XiMax(i%) = 0.6## 'min/max DIRECTOR lengths, waves
            NEXT i%

            XiMin(Nd%) = 5## : XiMax(i%) = 600##    'min/max values of Zo (OHMS)

            REDIM StartingXiMin(1 TO Nd%), StartingXiMax(1 TO Nd%)
            FOR i% = 1 TO Nd% : StartingXiMin(i%) = XiMin(i%) : StartingXiMax(i%) = XiMax(i%) : NEXT i%

            NN% = FREEFILE : OPEN "INFILE.DAT" FOR OUTPUT AS #NN% : PRINT #NN%,"YAGI.NEC" : PRINT #NN%,"YAGI.OUT" : CLOSE #NN%

'      =========================================================== END YAGI =====================================================

'   ==================================================================================
'   NOTE - DON'T FORGET TO ADD NEW TEST FUNCTIONS TO FUNCTION ObjectiveFunction() ABOVE !!
'   ==================================================================================

    END SELECT

    IF Nd% > 100 THEN Nt& = MIN(Nt&,200) 'to avoid array dimensioning problems

    DiagLength = 0## : FOR i% = 1 TO Nd% : DiagLength = DiagLength + (XiMax(i%)-XiMin(i%))^2 : NEXT i% : DiagLength = SQR(DiagLength) 'compute length of
decision space principal diagonal

END SUB 'GetFunctionRunParameters()

'--------------------------------

FUNCTION ParrottF4(R(),Nd%,p%,j&) 'Parrott F4 (1-D)

'MAXIMUM = 1 AT ~0.0796875... WITH ZERO OFFSET (SEEMS TO WORK BEST WITH JUST 3 PROBES, BUT NOT ALLOWED IN THIS VERSION...)

'References:

'Beasley, D., D. R. Bull, and R. R. Martin, "A Sequential Niche Technique for Multimodal
'Function Optimization," Evol. Comp. (MIT Press), vol. 1, no. 2, 1993, pp. 101-125
'(online at http://citeseer.ist.psu.edu/beasley93sequential.html).

'Parrott, D., and X. Li, "Locating and Tracking Multiple Dynamic Optima by a Particle Swarm
'Model Using Speciation," IEEE Trans. Evol. Computation, vol. 10, no. 4, Aug. 2006, pp. 440-458.


LOCAL Z, x, offset AS EXT

    offset = 0##

    x = R(p%,1,j&)

    Z = EXP(-2##*LOG(2##)*((x-0.08##-offset)/0.854##)^2)*(SIN(5##*Pi*((x-offset)^0.75##-0.05##)))^6 'WARNING! This is a NATURAL LOG, NOT Log10!!!

    ParrottF4 = Z

END FUNCTION 'ParrottF4()

'-----------------------------

FUNCTION SGO(R(),Nd%,p%,j&) 'SGO Function (2-D)

'MAXIMUM = ~130.8323226... @ ~(-2.8362075...,-2.8362075...) WITH ZERO OFFSET.

'Reference:

'Hsiao, Y., Chuang, C., Jiang, J., and Chien, C., "A Novel Optimization Algorithm: Space
'Gravitational Optimization," Proc. of 2005 IEEE International Conference on Systems, Man,
'and Cybernetics, 3, 2323-2328. (2005)

    LOCAL x1, x2, Z, t1, t2, SGOx1offset, SGOx2offset AS EXT

    SGOx1offset = 0## : SGOx2offset = 0##

'   SGOx1offset = 40## : SGOx2offset = 10##
```

```
    x1 = R(p%,1,j&) - SGOx1offset : x2 = R(p%,2,j&) - SGOx2offset

    t1 = x1^4 - 16##*x1^2 + 0.5##*x1 : t2 = x2^4 - 16##*x2^2 + 0.5##*x2

    Z = t1 + t2

    SGO = -Z

END FUNCTION 'SGO()

'------------------

FUNCTION GoldsteinPrice(R(),Nd%,p%,j&) 'Goldstein-Price Function (2-D)

'MAXIMUM = -3 @ (0,-1) WITH ZERO OFFSET.

'Reference:

'Cui, Z., Zeng, J., and Sun, G. (2006) 'A Fast Particle Swarm Optimization,' Int'l. J.
'Innovative Computing, Information and Control, vol. 2, no. 6, December, pp. 1365-1380.

    LOCAL Z, x1, x2, offset1, offset2, t1, t2 AS EXT

    offset1 = 0## : offset2 = 0##

'    offset1 = 20## : offset2 = -10##

    x1 = R(p%,1,j&)-offset1 : x2 = R(p%,2,j&)-offset2

    t1 = 1##+(x1+x2+1##)^2*(19##-14##*x1+3##*x1^2-14##*x2+6##*x1*x2+3##*x2^2)

    t2 = 30##+(2##*x1-3##*x2)^2*(18##-32##*x1+12##*x1^2+48##*x2-36##*x1*x2+27##*x2^2)

    Z  = t1*t2

    GoldsteinPrice = -Z

END FUNCTION 'GoldsteinPrice()

'-----------

FUNCTION StepFunction(R(),Nd%,p%,j&) 'Step Function (n-D)

'MAXIMUM VALUE = 0 @ [Offset]^n.

'Reference:

'Yao, X., Liu, Y., and Lin, G., "Evolutionary Programming Made Faster,"
'IEEE Trans. Evolutionary Computation, Vol. 3, No. 2, 82-102, Jul. 1999.


    LOCAL Offset, Z AS EXT

    LOCAL i%

    Z = 0## : Offset = 0## '75.123## '0##

    FOR i% = 1 TO Nd%

        IF Nd% = 2 AND i% = 1 THEN Offset = 75 '75##

        IF Nd% = 2 AND i% = 2 THEN Offset = 35 '30 '35##

        Z = Z + INT((R(p%,i%,j&)-Offset) + 0.5##)^2

    NEXT i%

    StepFunction = -Z

END FUNCTION 'StepFunction()

'-----------

FUNCTION Schwefel226(R(),Nd%,p%,j&) 'Schwefel Problem 2.26 (n-D)

'MAXIMUM = 12,569.5 @ [420.8687]^30 (30-D CASE).

'Reference:

'Yao, X., Liu, Y., and Lin, G., "Evolutionary Programming Made Faster,"
'IEEE Trans. Evolutionary Computation, Vol. 3, No. 2, 82-102, Jul. 1999.

    LOCAL Z, Xi AS EXT

    LOCAL i%

    Z = 0##

    FOR i% = 1 TO Nd%

        Xi = R(p%,i%,j&)

        Z = Z + Xi*SIN(SQR(ABS(Xi)))

    NEXT i%

    Schwefel226 = Z

END FUNCTION 'SCHWEFEL226()

'-----------

FUNCTION Colville(R(),Nd%,p%,j&) 'Colville Function (4-D)

'MAXIMUM = 0 @ (1,1,1,1) WITH ZERO OFFSET.

'Reference:

'Doo-Hyun, and Se-Young, O., "A New Mutation Rule for Evolutionary Programming Motivated from
'Backpropagation Learning," IEEE Trans. Evolutionary Computation, Vol. 4, No. 2, pp. 188-190,
'July 2000.

    LOCAL Z, x1, x2, x3, x4, offset AS EXT

    offset = 0## '7.123##

    x1 = R(p%,1,j&)-offset : x2 = R(p%,2,j&)-offset : x3 = R(p%,3,j&)-offset : x4 = R(p%,4,j&)-offset

    Z =  100##*(x2-x1^2)^2 + (1##-x1)^2  + _
          90##*(x4-x3^2)^2 + (1##-x3)^2  + _
        10.1##*((x2-1##)^2 + (x4-1##)^2) + _
        19.8##*(x2-1##)*(x4-1##)

    Colville = -Z

END FUNCTION 'Colville()

'-----------
```

```
FUNCTION Griewank(R(),Nd%,p%,j&) 'Griewank (n-D)

'Max of zero at (0,...,0)

'Reference: Yao, X., Liu, Y., and Lin, G., "Evolutionary Programming Made Faster,"
'IEEE Trans. Evolutionary Computation, Vol. 3, No. 2, 82-102, Jul. 1999.

    LOCAL Offset, Sum, Prod, Z, Xi AS EXT
    LOCAL i%
    Sum = 0## : Prod = 1##
    Offset = 75.123##
    FOR i% = 1 TO Nd%
        Xi = R(p%,i%,j&) - Offset
        Sum = Sum + Xi^2
        Prod = Prod*COS(Xi/SQR(i%))
    NEXT i%
    Z = Sum/4000## - Prod + 1##
    Griewank = -Z
END FUNCTION 'Griewank()
'-----------
FUNCTION Himmelblau(R(),Nd%,p%,j&) 'Himmelblau (2-D)
    LOCAL Z, x1, x2, offset AS EXT
    offset = 0##
    x1 = R(p%,1,j&)-offset : x2 = R(p%,2,j&)-offset
    Z = 200## - (x1^2 + x2 -11##)^2 - (x1+x2^2-7##)^2
    Himmelblau = Z
END FUNCTION 'Himmelblau()
'-----------
FUNCTION Rosenbrock(R(),Nd%,p%,j&) 'Rosenbrock (n-D)
'MAXIMUM = 0 @ [1,...,1]^n (n-D CASE).
'Reference: Yao, X., Liu, Y., and Lin, G., "Evolutionary Programming Made Faster,"
'IEEE Trans. Evolutionary Computation, Vol. 3, No. 2, 82-102, Jul. 1999.

    LOCAL Z, Xi, Xi1 AS EXT
    LOCAL i%
    Z = 0##
    FOR i% = 1 TO Nd%-1
        Xi  = R(p%,i%,j&) : Xi1 = R(p%,i%+1,j&)
        Z = Z + 100##*(Xi1-Xi^2)^2 + (Xi-1##)^2
    NEXT i%
    Rosenbrock = -Z
END FUNCTION 'ROSENBROCK()
'-----------
FUNCTION Sphere(R(),Nd%,p%,j&) 'Sphere (n-D)
'MAXIMUM = 0 @ [0,...,0]^n (n-D CASE).
'Reference: Yao, X., Liu, Y., and Lin, G., "Evolutionary Programming Made Faster,"
'IEEE Trans. Evolutionary Computation, Vol. 3, No. 2, 82-102, Jul. 1999.

    LOCAL Z, Xi, Xi1 AS EXT
    LOCAL i%
    Z = 0##
    FOR i% = 1 TO Nd%
        Xi  = R(p%,i%,j&)
        Z = Z + Xi^2
    NEXT i%
    Sphere = -Z
END FUNCTION 'SPHERE()
'-----------
FUNCTION HimmelblauNLO(R(),Nd%,p%,j&) 'Himmelblau non-linear optimization (5-D)
'MAXIMUM ~ 31025.5562644972 @ (78.0,33.0,27.0709971052,45.0,44.9692425501)
'Reference: "Constrained Optimization using CODEQ," Mahamed G.H. Omran & Ayed Salman,
'Chaos, Solitons and Tractals, 42(2009), 662-668
    LOCAL Z, x1, x2, x3, x4, x5, g1, g2, g3 AS EXT
    Z = 1E4200
    x1 = R(p%,1,j&) : x2 = R(p%,2,j&) : x3 = R(p%,3,j&) : x4 = R(p%,4,j&) : x5 = R(p%,5,j&)
    g1 = 85.334407## + 0.0056858##*x2*x5 + 0.00026##*x1*x4   - 0.0022053##*x3*x5
    g2 = 80.51249##  + 0.0071317##*x2*x5 + 0.0029955##*x1*x2 + 0.0021813##*x3*x3
    g3 = 9.300961##  + 0.0047026##*x3*x5 + 0.0012547##*x1*x3 + 0.0019085##*x3*x4
    IF g1## < 0 OR g1 > 92## OR g2 < 90## OR g2 > 110## OR g3 < 20## OR g3 > 25## THEN GOTO ExitHimmelblauNLO
    Z = 5.3578547##*x3*x3 + 0.8356891##*x1*x5 + 37.29329##*x1 - 40792.141##
```

```
ExitHimmelblauNLO:
    HimmelblauNLO = -Z
END FUNCTION 'HimmelblauNLO()
'----------
FUNCTION Tripod(R(),Nd%,p%,j&) 'Tripod (2-D)
'MAXIMUM = 0 at (0,-50)
'Reference: "Appendix: A mini-benchmark," Maurice Clerc
    LOCAL Z, x1, x2, s1, s2, t1, t2, t3 AS EXT
    x1 = R(p%,1,j&) : x2 = R(p%,2,j&)
    s1 = Sign(x1) : s2 = Sign(x2)
    t1 = (1##-s2)*(ABS(x1)+ABS(x2+50##))
    t2 = 0.5##*(1##+s2)*(1##-s1)*(1##+ABS(x1+50##)+ABS(x2-50##))
    t3 = (1##+s1)*(2##+ABS(x1-50##)+ABS(x2-50##))
    Z = 0.5##*(t1 + t2 + t3)
    Tripod = -Z
END FUNCTION 'Tripod()
'--------------------
FUNCTION Sign(X)
LOCAL Z AS EXT
    Z = 1## : IF X =< 0## THEN Z = -1##
    Sign = Z
END FUNCTION
'----------
FUNCTION RosenbrockF6(R(),Nd%,p%,j&) 'Rosenbrock F6 (10-D)
'WARNING !!  03-19-10  THIS FUNCTION CONTAINS ERRORS.  SEE CLERC's EMAIL!
'MAXIMUM = 394 at (0,-50)
'Reference: "Appendix: A mini-benchmark," Maurice Clerc (NOTE: Uses his notation...)
    LOCAL Z, Xi, Xi1, Zi, Zi1, Sum AS EXT
    LOCAL i%
    Sum = 0##
    FOR i% = 2 TO Nd%
        Xi = R(p%,i%,j&) : Xi1 = R(p%,i%-1,j&)
        Zi = Xi - XiOffset(i%) + 1## : Zi1 = Xi1 - XiOffset(i%-1) + 1##
        Sum = Sum + 100##*(Zi1^2-Zi)^2  + (Zi1-1##)^2
    NEXT i%
    Z = 390## + Sum
    RosenbrockF6 = -Z
END FUNCTION 'RosenbrockF6()
'--------------------
FUNCTION CompressionSpring(R(),Nd%,p%,j&) 'Compression Spring (3-D)
'MAXIMUM = 394 at (0,-50)
'Reference: "Appendix: A mini-benchmark," Maurice Clerc (NOTE: Uses his notation...)
LOCAL Z, x1, x2, x3, g1, g2, g3, g4, g5, Cf, Fmax, S, Lf, Lmax, SigP, SigPM, Fp, K, SigW AS EXT
    Z = 1E4200
    x1 = ROUND(R(p%,1,j&),0) : x2 = R(p%,2,j&)  : x3 = ROUND(R(p%,3,j&),3)
    Cf = 1## + 0.75##*x3/(x2-x3)+0.615##*x3/x2
    Fmax = 1000## : S = 189000## : Lmax = 14## : SigPM = 6## : Fp = 300## : SigW = 1.25##
    K = 11.5##*1E6*x3^4/(8##*x1*x2^3)
    Lf = Fmax/K + 1.05##*(x1+2##)*x3
    SigP = Fp/K
    g1 = 8##*Cf*Fmax*x2/(Pi*x3^3) - S
    g2 = Lf - Lmax
    g3 = SigP - SigPM
    g4 = SigP -Fp/K 'WARNING!  03-19-10.  THIS IS SATISFIED EXACTLY (SEE CLERC's EMAIL - TYPO IN HIS BENCHMARKS)
    g5 = SigW - (Fmax-Fp)/K
    IF g1 > 0## OR g2 > 0## OR g3 > 0## OR g4 > 0## OR g5 > 0## THEN GOTO ExitCompressionSpring
    Z = Pi^2*x2*x3^2*(x1+1##)/4##
ExitCompressionSpring:
    CompressionSpring = -Z
END FUNCTION 'CompressionSpring
'--------------------
FUNCTION GearTrain(R(),Nd%,p%,j&) 'GearTrain (4-D)
'MAXIMUM = 394 at (0,-50)
'Reference: "Appendix: A mini-benchmark," Maurice Clerc (NOTE: Uses his notation...)
LOCAL Z, x1, x2, x3, x4 AS EXT
```

```
    x1 = ROUND(R(p%,1,j&),0) : x2 = ROUND(R(p%,2,j&),0)
    x3 = ROUND(R(p%,3,j&),0) : x4 = ROUND(R(p%,4,j&),0)

    Z = (1##/6.931##-x1*x2/(x3*x4))^2

    GearTrain = -Z

END FUNCTION 'GearTrain

'---------------------

FUNCTION F1(R(),Nd%,p%,j&) 'F1 (n-D)

'MAXIMUM = ZERO (n-D CASE).

'Reference:

    LOCAL Z, Xi AS EXT

    LOCAL i%

    Z = 0##

    FOR i% = 1 TO Nd%

        Xi = R(p%,i%,j&)

        Z = Z + Xi^2

    NEXT i%

    F1 = -Z

END FUNCTION 'F1

'-----------

FUNCTION F2(R(),Nd%,p%,j&) 'F2 (n-D)

'MAXIMUM = ZERO (n-D CASE).

'Reference:

    LOCAL Sum, prod, Z, Xi AS EXT

    LOCAL i%

    Z = 0## : Sum = 0## : Prod = 1##

    FOR i% = 1 TO Nd%

        Xi = R(p%,i%,j&)

        Sum  = Sum+ ABS(Xi)

        Prod = Prod*ABS(Xi)

    NEXT i%

    Z = Sum + Prod

    F2 = -Z

END FUNCTION 'F2

'-----------

FUNCTION F3(R(),Nd%,p%,j&) 'F3 (n-D)

'MAXIMUM = ZERO (n-D CASE).

'Reference:

    LOCAL Z, Xk, Sum AS EXT

    LOCAL i%, k%

    Z = 0##

    FOR i% = 1 TO Nd%

        Sum = 0##

        FOR k% = 1 TO i%

            Xk = R(p%,k%,j&)

            Sum = Sum + Xk

        NEXT k%

        Z = Z + Sum^2

    NEXT i%

    F3 = -Z

END FUNCTION 'F3

'-----------

FUNCTION F4(R(),Nd%,p%,j&) 'F4 (n-D)

'MAXIMUM = ZERO (n-D CASE).

'Reference:

    LOCAL Z, Xi, MaxXi AS EXT

    LOCAL i%

    MaxXi = -1E4200

    FOR i% = 1 TO Nd%

        Xi = R(p%,i%,j&)

        IF ABS(Xi) >= MaxXi THEN MaxXi = ABS(Xi)

    NEXT i%

    F4 = -MaxXi

END FUNCTION 'F4

'-----------
```

```
FUNCTION F5(R(),Nd%,p%,j&) 'F5 (n-D)
'MAXIMUM = ZERO (n-D CASE).
'Reference:
    LOCAL Z, Xi, XiPlus1 AS EXT
    LOCAL i%
    Z = 0##
    FOR i% = 1 TO Nd%-1
        Xi      = R(p%,i%,j&)
        XiPlus1 = R(p%,i%+1,j&)
        Z = Z + (100##*(XiPlus1-Xi^2)^2+(Xi-1##))^2
    NEXT i%
    F5 = -Z
END FUNCTION 'F5
'-----------
FUNCTION F6(R(),Nd%,p%,j&) 'F6 (n-D STEP)
'MAXIMUM VALUE = 0 @ [Offset]^n.
'Reference:
'Yao, X., Liu, Y., and Lin, G., "Evolutionary Programming Made Faster,"
'IEEE Trans. Evolutionary Computation, Vol. 3, No. 2, 82-102, Jul. 1999.

    LOCAL Z AS EXT
    LOCAL i%
    Z = 0##
    FOR i% = 1 TO Nd%
        Z = Z + INT(R(p%,i%,j&) + 0.5##)^2
    NEXT i%
    F6 = -Z
END FUNCTION 'F6
'-----------
FUNCTION F7(R(),Nd%,p%,j&) 'F7
'MAXIMUM VALUE = 0 @ [Offset]^n.
'Reference:
'Yao, X., Liu, Y., and Lin, G., "Evolutionary Programming Made Faster,"
'IEEE Trans. Evolutionary Computation, Vol. 3, No. 2, 82-102, Jul. 1999.

    LOCAL Z, Xi AS EXT
    LOCAL i%
    Z = 0##
    FOR i% = 1 TO Nd%
        Xi = R(p%,i%,j&)
        Z = Z + i%*Xi^4
    NEXT i%
    F7 = -Z - RandomNum(0##,1##)
END FUNCTION 'F7
'-----------
FUNCTION F8(R(),Nd%,p%,j&) '(n-D) F8 [Schwefel Problem 2.26]
'MAXIMUM = 12,569.5 @ [420.8687]^30 (30-D CASE).
'Reference:
'Yao, X., Liu, Y., and Lin, G., "Evolutionary Programming Made Faster,"
'IEEE Trans. Evolutionary Computation, Vol. 3, No. 2, 82-102, Jul. 1999.
    LOCAL Z, Xi AS EXT
    LOCAL i%
    Z = 0##
    FOR i% = 1 TO Nd%
        Xi = R(p%,i%,j&)
        Z  = Z - Xi*SIN(SQR(ABS(Xi)))
    NEXT i%
    F8 = -Z
END FUNCTION 'F8
'-----------
FUNCTION F9(R(),Nd%,p%,j&) '(n-D) F9 [Rastrigin]
'MAXIMUM = ZERO (n-D CASE).
'Reference:
'Yao, X., Liu, Y., and Lin, G., "Evolutionary Programming Made Faster,"
'IEEE Trans. Evolutionary Computation, Vol. 3, No. 2, 82-102, Jul. 1999.
    LOCAL Z, Xi AS EXT
    LOCAL i%
    Z = 0##
```

```
    FOR i% = 1 TO Nd%
        Xi = R(p%,i%,j&)
        Z  = Z + (Xi^2 - 10##*COS(TwoPi*Xi) + 10##)^2
    NEXT i%
    F9 = -Z
END FUNCTION 'F9
'-----------
FUNCTION F10(R(),Nd%,p%,j&) '(n-D) F10 [Ackley's Function]
'MAXIMUM = ZERO (n-D CASE).
'Reference:
'Yao, X., Liu, Y., and Lin, G., "Evolutionary Programming Made Faster,"
'IEEE Trans. Evolutionary Computation, Vol. 3, No. 2, 82-102, Jul. 1999.
    LOCAL Z, Xi, Sum1, Sum2 AS EXT
    LOCAL i%
    Z = 0## : Sum1 = 0## : Sum2 = 0##
    FOR i% = 1 TO Nd%
        Xi   = R(p%,i%,j&)
        Sum1 = Sum1 + Xi^2
        Sum2 = Sum2 + COS(TwoPi*Xi)
    NEXT i%
    Z = -20##*EXP(-0.2##*SQR(Sum1/Nd%)) - EXP(Sum2/Nd%) + 20## + e
    F10 = -Z
END FUNCTION 'F10
'-----------
FUNCTION F11(R(),Nd%,p%,j&) '(n-D) F11
'MAXIMUM = ZERO (n-D CASE).
'Reference:
'Yao, X., Liu, Y., and Lin, G., "Evolutionary Programming Made Faster,"
'IEEE Trans. Evolutionary Computation, Vol. 3, No. 2, 82-102, Jul. 1999.
    LOCAL Z, Xi, Sum, Prod AS EXT
    LOCAL i%
    Z = 0## : Sum = 0## : Prod = 1##
    FOR i% = 1 TO Nd%
        Xi   = R(p%,i%,j&)
        Sum  = Sum + (Xi-100##)^2
        Prod = Prod*COS((Xi-100##)/SQR(i%))
    NEXT i%
    Z = Sum/4000## - Prod + 1##
    F11 = -Z
END FUNCTION 'F11
'-----
FUNCTION u(Xi,a,k,m)
LOCAL Z AS EXT
    Z = 0##
    SELECT CASE Xi
        CASE > a  : Z = k*(Xi-a)^m
        CASE < -a : Z = k*(-Xi-a)^m
    END SELECT
    u = Z
END FUNCTION
'-----------
FUNCTION F12(R(),Nd%,p%,j&) '(n-D) F12, Penalized #1
'Ref: Yao(1999).  Max=0 @ (-1,-1,...,-1), -50=<Xi=<50.
    LOCAL Offset, Sum1, Sum2, Z, X1, Y1, Xn, Yn, Xi, Yi, XiPlus1, YiPlus1 AS EXT
    LOCAL i%, m%, A$
    X1 = R(p%,1,j&)   : Y1 = 1## + (X1+1##)/4##
    Xn = R(p%,Nd%,j&) : Yn = 1## + (Xn+1##)/4##
    Sum1 = 0##
    FOR i% = 1 TO Nd%-1
        Xi      = R(p%,i%,j&)  : Yi      = 1## + (Xi+1##)/4##
        XiPlus1 = R(p%,i%+1,j&): YiPlus1 = 1## + (XiPlus1+1##)/4##
        Sum1 = Sum1 + (Yi-1##)^2*(1##+10##*(SIN(Pi*YiPlus1))^2)
    NEXT i%
    Sum1 = Sum1 + 10##*(SIN(Pi*Y1))^2 + (Yn-1##)^2
    Sum1 = Pi*Sum1/Nd%
```

```
    Sum2 = 0##
    FOR i% = 1 TO Nd%
        Xi = R(p%,i%,j&)
        Sum2 = Sum2 + u(Xi,10##,100##,4##)
    NEXT i%
    Z = Sum1 + Sum2
    F12 = -Z
END FUNCTION 'F12()
'------------------
FUNCTION F13(R(),Nd%,p%,j&) '(n-D) F13, Penalized #2
'Ref: Yao(1999).  Max=0 @ (1,1,...,1), -50=<Xi=<50.
    LOCAL Offset, Sum1, Sum2, Z, Xi, Xn, XiPlus1, X1 AS EXT
    LOCAL i%, m%, A$
    X1 = R(p%,1,j&) : Xn = R(p%,Nd%,j&)
    Sum1 = 0##
    FOR i% = 1 TO Nd%-1
        Xi  = R(p%,i%,j&) : XiPlus1 = R(p%,i%+1,j&)
        Sum1 = Sum1 + (Xi-1##)^2*(1##+(SIN(3##*Pi*XiPlus1))^2)
    NEXT i%
    Sum1 = Sum1 + (SIN(Pi*3##*X1))^2 +(Xn-1##)^2*(1##+(SIN(TwoPi*Xn))^2)
    Sum2 = 0##
    FOR i% = 1 TO Nd%
        Xi = R(p%,i%,j&)
        Sum2 = Sum2 + u(Xi,5##,100##,4##)
    NEXT i%
    Z = Sum1/10## + Sum2
    F13 = -Z
END FUNCTION 'F13()
'------------------
SUB FillArrayAij  'needed for function F14, Shekel's Foxholes

    Aij(1,1)=-32##  : Aij(1,2)=-16##  : Aij(1,3)=0##  : Aij(1,4)=16##  : Aij(1,5)=32##
    Aij(1,6)=-32##  : Aij(1,7)=-16##  : Aij(1,8)=0##  : Aij(1,9)=16##  : Aij(1,10)=32##
    Aij(1,11)=-32## : Aij(1,12)=-16## : Aij(1,13)=0## : Aij(1,14)=16## : Aij(1,15)=32##
    Aij(1,16)=-32## : Aij(1,17)=-16## : Aij(1,18)=0## : Aij(1,19)=16## : Aij(1,20)=32##
    Aij(1,21)=-32## : Aij(1,22)=-16## : Aij(1,23)=0## : Aij(1,24)=16## : Aij(1,25)=32##

    Aij(2,1)=-32##  : Aij(2,2)=-32## : Aij(2,3)=-32## : Aij(2,4)=-32## : Aij(2,5)=-32##
    Aij(2,6)=-16##  : Aij(2,7)=-16## : Aij(2,8)=-16## : Aij(2,9)=-16## : Aij(2,10)=-16##
    Aij(2,11)=0##   : Aij(2,12)=0##  : Aij(2,13)=0##  : Aij(2,14)=0##  : Aij(2,15)=0##
    Aij(2,16)=16##  : Aij(2,17)=16## : Aij(2,18)=16## : Aij(2,19)=16## : Aij(2,20)=16##
    Aij(2,21)=32##  : Aij(2,22)=32## : Aij(2,23)=32## : Aij(2,24)=32## : Aij(2,25)=32#

END SUB
'-----
FUNCTION F14(R(),Nd%,p%,j&) 'F14 (2-D) Shekel's Foxholes (INVERTED...)
    LOCAL Sum1, Sum2, Z, Xi AS EXT
    LOCAL i%, jj%
    Sum1 = 0##
    FOR jj% = 1 TO 25
        Sum2 = 0##
        FOR i% = 1 TO 2
            Xi = R(p%,i%,j&)
            Sum2 = Sum2 + (Xi-Aij(i%,jj%))^6
        NEXT i%
        Sum1 = Sum1 + 1##/(jj%+Sum2)
    NEXT j%
    Z = 1##/(0.002##+Sum1)
    F14 = -Z
END FUNCTION 'F14
'-----------
FUNCTION F16(R(),Nd%,p%,j&) 'F16 (2-D) 6-Hump Camel-Back
    LOCAL x1, x2, Z AS EXT
    x1 = R(p%,1,j&) : x2 = R(p%,2,j&)
    Z = 4##*x1^2 - 2.1##*x1^4 + x1^6/3## + x1*x2 - 4*x2^2 + 4*x2^4
    F16 = -Z
END FUNCTION 'F16
'-----------
FUNCTION F15(R(),Nd%,p%,j&) 'F15 (4-D) Kowalik's Function
'Global maximum = -0.0003075 @ (0.1928,0.1908,0.1231,0.1358)
    LOCAL x1, x2, x3, x4, Num, Denom, Z, Aj(), Bj() AS EXT
    LOCAL jj%
```

```
    REDIM Aj(1 TO 11), Bj(1 TO 11)

    Aj(1)  = 0.1957## : Bj(1)  = 1##/0.25##
    Aj(2)  = 0.1947## : Bj(2)  = 1##/0.50##
    Aj(3)  = 0.1735## : Bj(3)  = 1##/1.00##
    Aj(4)  = 0.1600## : Bj(4)  = 1##/2.00##
    Aj(5)  = 0.0844## : Bj(5)  = 1##/4.00##
    Aj(6)  = 0.0627## : Bj(6)  = 1##/6.00##
    Aj(7)  = 0.0456## : Bj(7)  = 1##/8.00##
    Aj(8)  = 0.0342## : Bj(8)  = 1##/10.0##
    Aj(9)  = 0.0323## : Bj(9)  = 1##/12.0##
    Aj(10) = 0.0235## : Bj(10) = 1##/14.0##
    Aj(11) = 0.0246## : Bj(11) = 1##/16.0##

    Z = 0##

    x1 = R(p%,1,j&) : x2 = R(p%,2,j&) : x3 = R(p%,3,j&) : x4 = R(p%,4,j&)

    FOR jj% = 1 TO 11

       Num   = x1*(Bj(jj%)^2+Bj(jj%)*x2)

       Denom = Bj(jj%)^2+Bj(jj%)*x3+x4

       Z = Z + (Aj(jj%)-Num/Denom)^2

    NEXT jj%

    F15 = -Z

END FUNCTION 'F15

'-----------

FUNCTION F17(R(),Nd%,p%,j&) 'F17, (2-D) Branin

'Global maximum = -0.398 @ (-3.142.12.275), (3.142,2.275), (9.425,2.425)

    LOCAL x1, x2, Z AS EXT

    x1 = R(p%,1,j&) : x2 = R(p%,2,j&)

    Z = (x2-5.1##*x1^2/(4##*Pi^2)+5##*x1/Pi-6##)^2 + 10##*(1##-1##/(8##*Pi))*COS(x1) + 10##

    F17 = -Z

END FUNCTION 'F17

'-----------

FUNCTION F18(R(),Nd%,p%,j&) 'Goldstein-Price 2-D Test Function

'Global maximum = -3 @ (0,-1)

    LOCAL Z, x1, x2, t1, t2 AS EXT

    x1 = R(p%,1,j&) : x2 = R(p%,2,j&)

    t1 = 1##+(x1+x2+1##)^2*(19##-14##*x1+3##*x1^2-14##*x2+6##*x1*x2+3##*x2^2)

    t2 = 30##+(2##*x1-3##*x2)^2*(18##-32##*x1+12##*x1^2+48##*x2-36##*x1*x2+27##*x2^2)

    Z  = t1*t2

    F18 = -Z

END FUNCTION 'F18()

'-----------

FUNCTION F19(R(),Nd%,p%,j&) 'F19 (3-D) Hartman's Family #1

'Global maximum = 3.86 @ (0.114,0.556,0.852)

    LOCAL Xi, Z, Sum, Aji(), Cj(), Pji() AS EXT

    LOCAL i%, jj%, m%

    REDIM Aji(1 TO 4, 1 TO 3), Cj(1 TO 4), Pji(1 TO 4, 1 TO 3)

    Aji(1,1) = 3.0## : Aji(1,2) = 10## : Aji(1,3) = 30## : Cj(1) = 1.0##
    Aji(2,1) = 0.1## : Aji(2,2) = 10## : Aji(2,3) = 35## : Cj(2) = 1.2##
    Aji(3,1) = 3.0## : Aji(3,2) = 10## : Aji(3,3) = 30## : Cj(3) = 3.0##
    Aji(4,1) = 0.1## : Aji(4,2) = 10## : Aji(4,3) = 35## : Cj(4) = 3.2##

    Pji(1,1) = 0.36890## : Pji(1,2) = 0.1170## : Pji(1,3) = 0.2673##
    Pji(2,1) = 0.46990## : Pji(2,2) = 0.4387## : Pji(2,3) = 0.7470##
    Pji(3,1) = 0.10910## : Pji(3,2) = 0.8732## : Pji(3,3) = 0.5547##
    Pji(4,1) = 0.03815## : Pji(4,2) = 0.5743## : Pji(4,3) = 0.8828##

    Z = 0##

    FOR jj% = 1 TO 4

        Sum = 0##

        FOR i% = 1 TO 3

            Xi = R(p%,i%,j&)

            Sum = Sum + Aji(jj%,i%)*(Xi-Pji(jj%,i%))^2

        NEXT i%

        Z = Z + Cj(jj%)*EXP(-Sum)

    NEXT jj%

    F19 = Z

END FUNCTION 'F19

'-----------

FUNCTION F20(R(),Nd%,p%,j&) 'F20 (6-D) Hartman's Family #2

'Global maximum = 3.32 @ (0.201,0.150,0.477,0.275,0.311,0.657)

    LOCAL Xi, Z, Sum, Aji(), Cj(), Pji() AS EXT

    LOCAL i%, jj%, m%

    REDIM Aji(1 TO 4, 1 TO 6), Cj(1 TO 4), Pji(1 TO 4, 1 TO 6)

    Aji(1,1) = 10.0## : Aji(1,2) = 3.00## : Aji(1,3) = 17.0## : Cj(1) = 1.0##
    Aji(2,1) = 0.05## : Aji(2,2) = 10.0## : Aji(2,3) = 17.0## : Cj(2) = 1.2##
    Aji(3,1) = 3.00## : Aji(3,2) = 3.50## : Aji(3,3) = 1.70## : Cj(3) = 3.0##
    Aji(4,1) = 17.0## : Aji(4,2) = 8.00## : Aji(4,3) = 0.05## : Cj(4) = 3.2##
```

```
    Aji(1,4) = 3.5## : Aji(1,5) = 1.7## : Aji(1,6) =  8##
    Aji(2,4) = 0.1## : Aji(2,5) =   8## : Aji(2,6) = 14##
    Aji(3,4) =  10## : Aji(3,5) =  17## : Aji(3,6) =  8##
    Aji(4,4) =  10## : Aji(4,5) = 0.1## : Aji(4,6) = 14##

    Pji(1,1) = 0.13120## : Pji(1,2) = 0.1696## : Pji(1,3) = 0.5569##
    Pji(2,1) = 0.23290## : Pji(2,2) = 0.4135## : Pji(2,3) = 0.8307##
    Pji(3,1) = 0.23480## : Pji(3,2) = 0.1415## : Pji(3,3) = 0.3522##
    Pji(4,1) = 0.40470## : Pji(4,2) = 0.8828## : Pji(4,3) = 0.8732##

    Pji(1,4) = 0.01240## : Pji(1,5) = 0.8283## : Pji(1,6) = 0.5886##
    Pji(2,4) = 0.37360## : Pji(2,5) = 0.1004## : Pji(2,6) = 0.9991##
    Pji(3,4) = 0.28830## : Pji(3,5) = 0.3047## : Pji(3,6) = 0.6650##
    Pji(4,4) = 0.57430## : Pji(4,5) = 0.1091## : Pji(4,6) = 0.0381##

    Z = 0##

    FOR jj% = 1 TO 4

        Sum = 0##

        FOR i% = 1 TO 6

            Xi = R(p%,i%,j&)

            Sum = Sum + Aji(jj%,i%)*(Xi-Pji(jj%,i%))^2

        NEXT i%

        Z = Z + Cj(jj%)*EXP(-Sum)

    NEXT jj%

    F20 = Z

END FUNCTION 'F20

'-----------

FUNCTION F21(R(),Nd%,p%,j&) 'F21 (4-D) Shekel's Family m=5

'Global maximum = 10

    LOCAL Xi, Z, Sum, Aji(), Cj() AS EXT

    LOCAL i%, jj%, m%

    m% = 5 : REDIM Aji(1 TO m%, 1 TO 4), Cj(1 TO m%)

    Aji(1,1)  = 4## : Aji(1,2)  =   4## : Aji(1,3)  = 4## : Aji(1,4)  =   4## : Cj(1)  = 0.1##
    Aji(2,1)  = 1## : Aji(2,2)  =   1## : Aji(2,3)  = 1## : Aji(2,4)  =   1## : Cj(2)  = 0.2##
    Aji(3,1)  = 8## : Aji(3,2)  =   8## : Aji(3,3)  = 8## : Aji(3,4)  =   8## : Cj(3)  = 0.2##
    Aji(4,1)  = 6## : Aji(4,2)  =   6## : Aji(4,3)  = 6## : Aji(4,4)  =   6## : Cj(4)  = 0.4##
    Aji(5,1)  = 3## : Aji(5,2)  =   7## : Aji(5,3)  = 3## : Aji(5,4)  =   7## : Cj(5)  = 0.4##

    Z = 0##

    FOR jj% = 1 TO m%  'NOTE:  Index jj% is used to avoid same variable name as j&

        Sum = 0##

        FOR i% = 1 TO 4 'Shekel's family is 4-D only

            Xi = R(p%,i%,j&)

            Sum = Sum + (Xi-Aji(jj%,i%))^2

        NEXT i%

        Z = Z + 1##/(Sum + Cj(jj%))

    NEXT jj%

    F21 = Z

END FUNCTION 'F21

'-----------

FUNCTION F22(R(),Nd%,p%,j&) 'F22 (4-D) Shekel's Family m=7

'Global maximum = 10

    LOCAL Xi, Z, Sum, Aji(), Cj() AS EXT

    LOCAL i%, jj%, m%

    m% = 7 : REDIM Aji(1 TO m%, 1 TO 4), Cj(1 TO m%)

    Aji(1,1)  = 4## : Aji(1,2)  =   4## : Aji(1,3)  = 4## : Aji(1,4)  =   4## : Cj(1)  = 0.1##
    Aji(2,1)  = 1## : Aji(2,2)  =   1## : Aji(2,3)  = 1## : Aji(2,4)  =   1## : Cj(2)  = 0.2##
    Aji(3,1)  = 8## : Aji(3,2)  =   8## : Aji(3,3)  = 8## : Aji(3,4)  =   8## : Cj(3)  = 0.2##
    Aji(4,1)  = 6## : Aji(4,2)  =   6## : Aji(4,3)  = 6## : Aji(4,4)  =   6## : Cj(4)  = 0.4##
    Aji(5,1)  = 3## : Aji(5,2)  =   7## : Aji(5,3)  = 3## : Aji(5,4)  =   7## : Cj(5)  = 0.4##
    Aji(6,1)  = 2## : Aji(6,2)  =   9## : Aji(6,3)  = 2## : Aji(6,4)  =   9## : Cj(6)  = 0.6##
    Aji(7,1)  = 5## : Aji(7,2)  =   5## : Aji(7,3)  = 3## : Aji(7,4)  =   3## : Cj(7)  = 0.3##

    Z = 0##

    FOR jj% = 1 TO m%  'NOTE:  Index jj% is used to avoid same variable name as j&

        Sum = 0##

        FOR i% = 1 TO 4 'Shekel's family is 4-D only

            Xi = R(p%,i%,j&)

            Sum = Sum + (Xi-Aji(jj%,i%))^2

        NEXT i%

        Z = Z + 1##/(Sum + Cj(jj%))

    NEXT jj%

    F22 = Z

END FUNCTION 'F22

'-----------

FUNCTION F23(R(),Nd%,p%,j&) 'F23 (4-D) Shekel's Family m=10

'Global maximum = 10

    LOCAL Xi, Z, Sum, Aji(), Cj() AS EXT

    LOCAL i%, jj%, m%
```

```
m% = 10 : REDIM Aji(1 TO m%, 1 TO 4), Cj(1 TO m%)

Aji(1,1)  = 4## : Aji(1,2)  =   4## : Aji(1,3)  = 4## : Aji(1,4)  =   4## : Cj(1)  = 0.1##
Aji(2,1)  = 1## : Aji(2,2)  =   1## : Aji(2,3)  = 1## : Aji(2,4)  =   1## : Cj(2)  = 0.2##
Aji(3,1)  = 8## : Aji(3,2)  =   8## : Aji(3,3)  = 8## : Aji(3,4)  =   8## : Cj(3)  = 0.2##
Aji(4,1)  = 6## : Aji(4,2)  =   6## : Aji(4,3)  = 6## : Aji(4,4)  =   6## : Cj(4)  = 0.4##
Aji(5,1)  = 3## : Aji(5,2)  =   7## : Aji(5,3)  = 3## : Aji(5,4)  =   7## : Cj(5)  = 0.4##
Aji(6,1)  = 2## : Aji(6,2)  =   9## : Aji(6,3)  = 2## : Aji(6,4)  =   9## : Cj(6)  = 0.6##
Aji(7,1)  = 5## : Aji(7,2)  =   5## : Aji(7,3)  = 3## : Aji(7,4)  =   3## : Cj(7)  = 0.3##
Aji(8,1)  = 8## : Aji(8,2)  =   1## : Aji(8,3)  = 8## : Aji(8,4)  =   1## : Cj(8)  = 0.7##
Aji(9,1)  = 6## : Aji(9,2)  =   2## : Aji(9,3)  = 6## : Aji(9,4)  =   2## : Cj(9)  = 0.5##
Aji(10,1) = 7## : Aji(10,2) = 3.6## : Aji(10,3) = 7## : Aji(10,4) = 3.6## : Cj(10) = 0.5##

Z = 0##

FOR jj% = 1 TO m%  'NOTE:  Index jj% is used to avoid same variable name as j&

    Sum = 0##

    FOR i% = 1 TO 4 'Shekel's family is 4-D only

        Xi = R(p%,i%,j&)

        Sum = Sum + (Xi-Aji(jj%,i%))^2

    NEXT i%

    Z = Z + 1##/(Sum + Cj(jj%))

NEXT jj%

F23 = Z

END FUNCTION 'F23

'==================================================== END FUNCTION DEFINITIONS ===============================================================

SUB Plot2DbestProbeTrajectories(NumTrajectories%,M(),R(),Np%,Nd%,LastStep&,FunctionName$)

LOCAL TrajectoryNumber%, ProbeNumber%, StepNumber&, N%, M%, ProcID???

LOCAL MaximumFitness, MinimumFitness AS EXT

LOCAL BestProbeThisStep%()

LOCAL BestFitnessThisStep(), TempFitness() AS EXT

LOCAL Annotation$, xCoord$, yCoord$, GnuPlotEXE$, PlotWithLines$

    Annotation$    = ""

    PlotWithLines$ = "YES" '"NO"

    NumTrajectories% = MIN(Np%,NumTrajectories%)

    GnuPlotEXE$ = "wgnuplot.exe"

'   ---------------- Get Min/Max Fitnesses -----------------

    MaximumFitness = M(1,0) : MinimumFitness = M(1,0)  'Note:  M(p%,j&)

    FOR StepNumber& = 0 TO LastStep&

        FOR ProbeNumber% = 1 TO Np%

            IF M(ProbeNumber%,StepNumber&) >= MaximumFitness THEN MaximumFitness = M(ProbeNumber%,StepNumber&)

            IF M(ProbeNumber%,StepNumber&) =< MinimumFitness THEN MinimumFitness = M(ProbeNumber%,StepNumber&)

        NEXT ProbeNumber%

    NEXT StepNumber%

'   ------------- Copy Fitness Array M() into TempFitness to Preserve M() ----------------

    REDIM TempFitness(1 TO Np%, 0 TO LastStep&)

    FOR StepNumber& = 0 TO LastStep&

        FOR ProbeNumber% = 1 TO Np%

            TempFitness(ProbeNumber%,StepNumber&) = M(ProbeNumber%,StepNumber&)

        NEXT ProbeNumber%

    NEXT StepNumber%

'   ------------ LOOP ON TRAJECTORIES -----------

    FOR TrajectoryNumber% = 1 TO NumTrajectories%

'       --------------- Get Trajectory Coordinate Data -----------------

        REDIM BestFitnessThisStep(0 TO LastStep&), BestProbeThisStep%(0 TO LastStep&)

        FOR StepNumber& = 0 TO LastStep&

            BestFitnessThisStep(StepNumber&) = TempFitness(1,StepNumber&)

            FOR ProbeNumber% = 1 TO Np%

                IF TempFitness(ProbeNumber%,StepNumber&) >= BestFitnessThisStep(StepNumber&) THEN

                    BestFitnessThisStep(StepNumber&) = TempFitness(ProbeNumber%,StepNumber&)

                    BestProbeThisStep%(StepNumber&)  = ProbeNumber%

                END IF

            NEXT ProbeNumber%

        NEXT StepNumber&

'   ----- Create Plot Data File -----

    N% = FREEFILE

    SELECT CASE TrajectoryNumber%

        CASE 1  : OPEN "t1"  FOR OUTPUT AS #N%
        CASE 2  : OPEN "t2"  FOR OUTPUT AS #N%
        CASE 3  : OPEN "t3"  FOR OUTPUT AS #N%
        CASE 4  : OPEN "t4"  FOR OUTPUT AS #N%
        CASE 5  : OPEN "t5"  FOR OUTPUT AS #N%
        CASE 6  : OPEN "t6"  FOR OUTPUT AS #N%
        CASE 7  : OPEN "t7"  FOR OUTPUT AS #N%
        CASE 8  : OPEN "t8"  FOR OUTPUT AS #N%
        CASE 9  : OPEN "t9"  FOR OUTPUT AS #N%
```

```
        CASE 10 : OPEN "t10" FOR OUTPUT AS #N%

    END SELECT

'   ------------ Write Plot File Data ------------

    FOR StepNumber& = 0 TO LastStep&

        PRINT                                                #N%,                                                   USING$("######.########
######.########",R(BestProbeThisStep%(StepNumber&),1,StepNumber&),R(BestProbeThisStep%(StepNumber&),2,StepNumber&))

        TempFitness(BestProbeThisStep%(StepNumber&),StepNumber&) = MinimumFitness 'so that same max will not be found for next trajectory

    NEXT StepNumber%

    CLOSE #N%

    NEXT TrajectoryNumber%

'   ------------------------- Plot Trajectories --------------------------

    CALL CreateGNUplotINIfile(0.13##*ScreenWidth&,0.18##*ScreenHeight&,0.7##*ScreenHeight&,0.7##*ScreenHeight&)

    Annotation$ = ""

    N% = FREEFILE

    OPEN "cmd2d.gp" FOR OUTPUT AS #N%

        PRINT #N%, "set xrange ["+REMOVE$(STR$(XiMin(1)),ANY"" )+":"+REMOVE$(STR$(XiMax(1)),ANY" ")+"]"
        PRINT #N%, "set yrange ["+REMOVE$(STR$(XiMin(2)),ANY" ")+":"+REMOVE$(STR$(XiMax(2)),ANY" ")+"]"

        'PRINT #N%, "set label "      + Quote$ + Annotation$ + Quote$ + " at graph " + xCoord$ + "," + yCoord$
        PRINT #N%, "set grid xtics " + "10"
        PRINT #N%, "set grid ytics " + "10"
        PRINT #N%, "set grid mxtics"
        PRINT #N%, "set grid mytics"
        PRINT #N%, "show grid"
        PRINT #N%, "set title "   + Quote$ + "2D "+ FunctionName$+" TRAJECTORIES OF PROBES WITH BEST\nFITNESSES (ORDERED BY FITNESS)" + "\n" + RunID$ +
Quote$
        PRINT #N%, "set xlabel " + Quote$ + "x1\n\n"                                   + Quote$
        PRINT #N%, "set ylabel " + Quote$ + "\nx2"                                     + Quote$

        IF PlotWithLines$ = "YES" THEN

            SELECT CASE NumTrajectories%

                CASE 1  : PRINT #N%, "plot "+Quote$+"t1"+Quote$+" w l lw 3"
                CASE 2  : PRINT #N%, "plot "+Quote$+"t1"+Quote$+" w l lw 3,"+Quote$+"t2"+Quote$+" w l"
                CASE 3  : PRINT #N%, "plot "+Quote$+"t1"+Quote$+" w l lw 3,"+Quote$+"t2"+Quote$+" w l,"+Quote$+"t3"+Quote$+" w l"
                CASE 4   : PRINT #N%, "plot "+Quote$+"t1"+Quote$+" w l lw 3,"+Quote$+"t2"+Quote$+" w l,"+Quote$+"t3"+Quote$+" w l,"+Quote$+"t4"+Quote$+" w
l"
                CASE 5   : PRINT #N%, "plot "+Quote$+"t1"+Quote$+" w l lw 3,"+Quote$+"t2"+Quote$+" w l,"+Quote$+"t3"+Quote$+" w l,"+Quote$+"t4"+Quote$+" w
l,"+Quote$+"t5"+Quote$+" w l"
                CASE 6   : PRINT #N%, "plot "+Quote$+"t1"+Quote$+" w l lw 3,"+Quote$+"t2"+Quote$+" w l,"+Quote$+"t3"+Quote$+" w l,"+Quote$+"t4"+Quote$+" w
l,"+Quote$+"t5"+Quote$+" w l,"+Quote$+"t6"+Quote$+" w l"
                CASE 7   : PRINT #N%, "plot "+Quote$+"t1"+Quote$+" w l lw 3,"+Quote$+"t2"+Quote$+" w l,"+Quote$+"t3"+Quote$+" w l,"+Quote$+"t4"+Quote$+" w
l,"+Quote$+"t5"+Quote$+" w l,"+Quote$+"t6"+Quote$+" w l,"+_
                                             Quote$+"t7"+Quote$+" w l"
                CASE 8   : PRINT #N%, "plot "+Quote$+"t1"+Quote$+" w l lw 3,"+Quote$+"t2"+Quote$+" w l,"+Quote$+"t3"+Quote$+" w l,"+Quote$+"t4"+Quote$+" w
l,"+Quote$+"t5"+Quote$+" w l,"+Quote$+"t6"+Quote$+" w l,"+_
                                             Quote$+"t7"+Quote$+" w l,"     +Quote$+"t8"+Quote$+" w l"
                CASE 9   : PRINT #N%, "plot "+Quote$+"t1"+Quote$+" w l lw 3,"+Quote$+"t2"+Quote$+" w l,"+Quote$+"t3"+Quote$+" w l,"+Quote$+"t4"+Quote$+" w
l,"+Quote$+"t5"+Quote$+" w l,"+Quote$+"t6"+Quote$+" w l,"+_
                                             Quote$+"t7"+Quote$+" w l,"     +Quote$+"t8"+Quote$+" w l,"+Quote$+"t9"+Quote$+" w l"
                CASE 10  : PRINT #N%, "plot "+Quote$+"t1"+Quote$+" w l lw 3,"+Quote$+"t2"+Quote$+" w l,"+Quote$+"t3"+Quote$+" w l,"+Quote$+"t4"+Quote$+" w
l,"+Quote$+"t5"+Quote$+" w l,"+Quote$+"t6"+Quote$+" w l,"+_
                                             Quote$+"t7"+Quote$+" w l,"      +Quote$+"t8"+Quote$+" w l,"+Quote$+"t9"+Quote$+" w l,"+Quote$+"t10"+Quote$+" w
l"
            END SELECT

        ELSE

            SELECT CASE NumTrajectories%

                CASE 1  : PRINT #N%, "plot "+Quote$+"t1"+Quote$+" lw 2"
                CASE 2  : PRINT #N%, "plot "+Quote$+"t1"+Quote$+" lw 2,"+Quote$+"t2"+Quote$
                CASE 3  : PRINT #N%, "plot "+Quote$+"t1"+Quote$+" lw 2,"+Quote$+"t2"+Quote$+" ,"+Quote$+"t3"+Quote$
                CASE 4  : PRINT #N%, "plot "+Quote$+"t1"+Quote$+" lw 2,"+Quote$+"t2"+Quote$+" ,"+Quote$+"t3"+Quote$+" ,"+Quote$+"t4"+Quote$
                CASE  5     :  PRINT  #N%,   "plot   "+Quote$+"t1"+Quote$+"   lw  2,"+Quote$+"t2"+Quote$+"   ,"+Quote$+"t3"+Quote$+"   ,"+Quote$+"t4"+Quote$+"
,"+Quote$+"t5"+Quote$
                CASE  6     :  PRINT  #N%,   "plot  "+Quote$+"t1"+Quote$+"   lw  2,"+Quote$+"t2"+Quote$+"   ,"+Quote$+"t3"+Quote$+"   ,"+Quote$+"t4"+Quote$+"
,"+Quote$+"t5"+Quote$+" ,"+Quote$+"t6"+Quote$
                CASE  7     :  PRINT  #N%,   "plot  "+Quote$+"t1"+Quote$+"   lw  2,"+Quote$+"t2"+Quote$+"   ,"+Quote$+"t3"+Quote$+"   ,"+Quote$+"t4"+Quote$+"
,"+Quote$+"t5"+Quote$+" ,"+Quote$+"t6"+Quote$+" ,"+_
                                             Quote$+"t7"+Quote$
                CASE  8     :  PRINT  #N%,   "plot   "+Quote$+"t1"+Quote$+"   lw  2,"+Quote$+"t2"+Quote$+"   ,"+Quote$+"t3"+Quote$+"   ,"+Quote$+"t4"+Quote$+"
,"+Quote$+"t5"+Quote$+" ,"+Quote$+"t6"+Quote$+" ,"+_
                                             Quote$+"t7"+Quote$+" ,"    +Quote$+"t8"+Quote$
                CASE  9     :  PRINT  #N%,   "plot  "+Quote$+"t1"+Quote$+"   lw  2,"+Quote$+"t2"+Quote$+"   ,"+Quote$+"t3"+Quote$+"   ,"+Quote$+"t4"+Quote$+"
,"+Quote$+"t5"+Quote$+" ,"+Quote$+"t6"+Quote$+" ,"+_
                                             Quote$+"t7"+Quote$+" ,"    +Quote$+"t8"+Quote$+" ,"+Quote$+"t9"+Quote$
                CASE  10   :   PRINT   #N%,   "plot   "+Quote$+"t1"+Quote$+"   lw   2,"+Quote$+"t2"+Quote$+"   ,"+Quote$+"t3"+Quote$+"   ,"+Quote$+"t4"+Quote$+"
,"+Quote$+"t5"+Quote$+" ,"+Quote$+"t6"+Quote$+" ,"+_
                                             Quote$+"t7"+Quote$+" ,"    +Quote$+"t8"+Quote$+" ,"+Quote$+"t9"+Quote$+" ,"+Quote$+"t10"+Quote$
            END SELECT

        END IF

    CLOSE #N%

    ProcID??? = SHELL(GnuPlotEXE$+" cmd2d.gp -") : CALL Delay(1##)

END SUB 'Plot2DbestProbeTrajectories()

'----

SUB Plot2DindividualProbeTrajectories(NumTrajectories%,M(),R(),Np%,Nd%,LastStep&,FunctionName$)

LOCAL ProbeNumber%, StepNumber&, N%, ProcID???

LOCAL Annotation$, xCoord$, yCoord$, GnuPlotEXE$, PlotWithLines$

    NumTrajectories% = MIN(Np%,NumTrajectories%)

    Annotation$    = ""

    PlotWithLines$ = "YES" '"NO"

    GnuPlotEXE$ = "wgnuplot.exe"

'   -------------- LOOP ON PROBES ---------------

    FOR ProbeNumber% = 1 TO MIN(NumTrajectories%,Np%)

'   ----- Create Plot Data File -----

    N% = FREEFILE
```

```
    SELECT CASE ProbeNumber%

        CASE 1  : OPEN "p1"  FOR OUTPUT AS #N%
        CASE 2  : OPEN "p2"  FOR OUTPUT AS #N%
        CASE 3  : OPEN "p3"  FOR OUTPUT AS #N%
        CASE 4  : OPEN "p4"  FOR OUTPUT AS #N%
        CASE 5  : OPEN "p5"  FOR OUTPUT AS #N%
        CASE 6  : OPEN "p6"  FOR OUTPUT AS #N%
        CASE 7  : OPEN "p7"  FOR OUTPUT AS #N%
        CASE 8  : OPEN "p8"  FOR OUTPUT AS #N%
        CASE 9  : OPEN "p9"  FOR OUTPUT AS #N%
        CASE 10 : OPEN "p10" FOR OUTPUT AS #N%
        CASE 11 : OPEN "p11" FOR OUTPUT AS #N%
        CASE 12 : OPEN "p12" FOR OUTPUT AS #N%
        CASE 13 : OPEN "p13" FOR OUTPUT AS #N%
        CASE 14 : OPEN "p14" FOR OUTPUT AS #N%
        CASE 15 : OPEN "p15" FOR OUTPUT AS #N%
        CASE 16 : OPEN "p16" FOR OUTPUT AS #N%

    END SELECT

'   ------------ Write Plot File Data ------------

    FOR StepNumber& = 0 TO LastStep&

        PRINT #N%, USING$("######.######## ######.########",R(ProbeNumber%,1,StepNumber&),R(ProbeNumber%,2,StepNumber&))

    NEXT StepNumber%

    CLOSE #N%

    NEXT ProbeNumber%

'   ------------------------------------------- Plot Trajectories --------------------------------------------

'usage:  CALL CreateGNUplotINIfile(PlotWindowULC_X%,PlotWindowULC_Y%,PlotWindowWidth%,PlotWindowHeight%)

    CALL CreateGNUplotINIfile(0.17##*ScreenWidth&,0.22##*ScreenHeight&,0.7##*ScreenHeight&,0.7##*ScreenHeight&)

    Annotation$ = ""

    N% = FREEFILE

    OPEN "cmd2d.gp" FOR OUTPUT AS #N%

        PRINT #N%, "set xrange ["+REMOVE$(STR$(XiMin(1)),ANY"" )+":"+REMOVE$(STR$(XiMax(1)),ANY" ")+"]"
        PRINT #N%, "set yrange ["+REMOVE$(STR$(XiMin(2)),ANY" ")+":"+REMOVE$(STR$(XiMax(2)),ANY" ")+"]"

        PRINT #N%, "set grid xtics " + "10"
        PRINT #N%, "set grid ytics " + "10"
        PRINT #N%, "set grid mxtics"
        PRINT #N%, "set grid mytics"
        PRINT #N%, "show grid"
        PRINT #N%, "set title "  + Quote$ + "2D "+ FunctionName$+" INDIVIDUAL PROBE TRAJECTORIES\n(ORDERED BY PROBE #)" + "\n" + RunID$ + Quote$
        PRINT #N%, "set xlabel " + Quote$ + "x1\n\n"                                     + Quote$
        PRINT #N%, "set ylabel " + Quote$ + "\nx2"                                       + Quote$

        IF PlotWithLines$ = "YES" THEN

            SELECT CASE NumTrajectories%

                CASE 1  : PRINT #N%, "plot "+Quote$+"p1"  +Quote$+" w l lw 1"
                CASE 2  : PRINT #N%, "plot "+Quote$+"p1"  +Quote$+" w l lw 1,"+Quote$+"p2"+Quote$+" w l"
                CASE 3  : PRINT #N%, "plot "+Quote$+"p1"  +Quote$+" w l lw 1,"+Quote$+"p2"+Quote$+" w l,"+Quote$+"p3"+Quote$+" w l"
                CASE 4  : PRINT #N%, "plot "+Quote$+"p1"  +Quote$+" w l lw 1,"+Quote$+"p2"+Quote$+" w l,"+Quote$+"p3"+Quote$+" w l,"+Quote$+"p4"+Quote$+" w
l"
                CASE 5  : PRINT #N%, "plot "+Quote$+"p1"  +Quote$+" w l lw 1,"+Quote$+"p2"+Quote$+" w l,"+Quote$+"p3"+Quote$+" w l,"+Quote$+"p4"+Quote$+" w
l,"+Quote$+"p5"+Quote$+" w l"
                CASE 6  : PRINT #N%, "plot "+Quote$+"p1"  +Quote$+" w l lw 1,"+Quote$+"p2"+Quote$+" w l,"+Quote$+"p3"+Quote$+" w l,"+Quote$+"p4"+Quote$+" w
l,"+Quote$+"p5"+Quote$+" w l,"+Quote$+"p6"+Quote$+" w l"
                CASE 7  : PRINT #N%, "plot "+Quote$+"p1"  +Quote$+" w l lw 1,"+Quote$+"p2"+Quote$+" w l,"+Quote$+"p3"+Quote$+" w l,"+Quote$+"p4"+Quote$+" w
l,"+Quote$+"p5"+Quote$+" w l,"+Quote$+"p6"+Quote$+" w l,"+_
                                      Quote$+"p7"  +Quote$+" w l"
                CASE 8  : PRINT #N%, "plot "+Quote$+"p1"  +Quote$+" w l lw 1,"+Quote$+"p2"+Quote$+" w l,"+Quote$+"p3"+Quote$+" w l,"+Quote$+"p4"+Quote$+" w
l,"+Quote$+"p5"+Quote$+" w l,"+Quote$+"p6"+Quote$+" w l,"+_
                                      Quote$+"p7"  +Quote$+" w l,"     +Quote$+"p8"+Quote$+" w l"
                CASE 9  : PRINT #N%, "plot "+Quote$+"p1"  +Quote$+" w l lw 1,"+Quote$+"p2"+Quote$+" w l,"+Quote$+"p3"+Quote$+" w l,"+Quote$+"p4"+Quote$+" w
l,"+Quote$+"p5"+Quote$+" w l,"+Quote$+"p6"+Quote$+" w l,"+_
                                      Quote$+"p7"  +Quote$+" w l,"     +Quote$+"p8"+Quote$+" w l,"+Quote$+"p9"+Quote$+" w l"

                CASE  10  :  PRINT  #N%,  "plot  "+Quote$+"p1"   +Quote$+"  w  l  lw  1,"+Quote$+"p2"  +Quote$+"  w  l,"+Quote$+"p3"  +Quote$+"  w  l,"+Quote$+"p4"
+Quote$+" w l,"+Quote$+"p5"+Quote$+" w l,"+Quote$+"p6"+Quote$+" w l,"+_
                                      Quote$+"p7"      +Quote$+"   w   l,"              +Quote$+"p8"   +Quote$+"   w   l,"+Quote$+"p9"   +Quote$+"   w
l,"+Quote$+"p10"+Quote$+" w l"

                CASE  11  :  PRINT  #N%,  "plot  "+Quote$+"p1"   +Quote$+"  w  l  lw  1,"+Quote$+"p2"  +Quote$+"  w  l,"+Quote$+"p3"  +Quote$+"  w  l,"+Quote$+"p4"
+Quote$+" w l,"+Quote$+"p5" +Quote$+" w l,"+Quote$+"p6" +Quote$+" w l,"+_
                                      Quote$+"p7"      +Quote$+"   w   l,"              +Quote$+"p8"   +Quote$+"   w   l,"+Quote$+"p9"   +Quote$+"   w
l,"+Quote$+"p10"+Quote$+" w l,"+Quote$+"p11"+Quote$+" w l"

                CASE  12  :  PRINT  #N%,  "plot  "+Quote$+"p1"   +Quote$+"  w  l  lw  1,"+Quote$+"p2"  +Quote$+"  w  l,"+Quote$+"p3"  +Quote$+"  w  l,"+Quote$+"p4"
+Quote$+" w l,"+Quote$+"p5" +Quote$+" w l,"+Quote$+"p6" +Quote$+" w l,"+_
                                      Quote$+"p7"      +Quote$+"   w   l,"              +Quote$+"p8"   +Quote$+"   w   l,"+Quote$+"p9"   +Quote$+"   w
l,"+Quote$+"p10"+Quote$+" w l,"+Quote$+"p11"+Quote$+" w l,"+Quote$+"p12"+Quote$+" w l"

                CASE  13  :  PRINT  #N%,  "plot  "+Quote$+"p1"   +Quote$+"  w  l  lw  1,"+Quote$+"p2"  +Quote$+"  w  l,"+Quote$+"p3"  +Quote$+"  w  l,"+Quote$+"p4"
+Quote$+" w l,"+Quote$+"p5" +Quote$+" w l,"+Quote$+"p6" +Quote$+" w l,"+_
                                      Quote$+"p7"      +Quote$+"   w   l,"              +Quote$+"p8"   +Quote$+"   w   l,"+Quote$+"p9"   +Quote$+"   w
l,"+Quote$+"p10"+Quote$+" w l,"+Quote$+"p11"+Quote$+" w l,"+Quote$+"p12"+Quote$+" w l,"+_
                                      Quote$+"p13 "+Quote$+" w l"

                CASE  14  :  PRINT  #N%,  "plot  "+Quote$+"p1"   +Quote$+"  w  l  lw  1,"+Quote$+"p2"  +Quote$+"  w  l,"+Quote$+"p3"  +Quote$+"  w  l,"+Quote$+"p4"
+Quote$+" w l,"+Quote$+"p5" +Quote$+" w l,"+Quote$+"p6" +Quote$+" w l,"+_
                                      Quote$+"p7"      +Quote$+"   w   l,"              +Quote$+"p8"   +Quote$+"   w   l,"+Quote$+"p9"   +Quote$+"   w
l,"+Quote$+"p10"+Quote$+" w l,"+Quote$+"p11"+Quote$+" w l,"+Quote$+"p12"+Quote$+" w l,"+_
                                      Quote$+"p13" +Quote$+" w l,"     +Quote$+"p14"+Quote$+" w l"

                CASE  15  :  PRINT  #N%,  "plot  "+Quote$+"p1"   +Quote$+"  w  l  lw  1,"+Quote$+"p2"  +Quote$+"  w  l,"+Quote$+"p3"  +Quote$+"  w  l,"+Quote$+"p4"
+Quote$+" w l,"+Quote$+"p5" +Quote$+" w l,"+Quote$+"p6" +Quote$+" w l,"+_
                                      Quote$+"p7"      +Quote$+"   w   l,"              +Quote$+"p8"   +Quote$+"   w   l,"+Quote$+"p9"   +Quote$+"   w
l,"+Quote$+"p10"+Quote$+" w l,"+Quote$+"p11"+Quote$+" w l,"+Quote$+"p12"+Quote$+" w l,"+_
                                      Quote$+"p13" +Quote$+" w l,"     +Quote$+"p14"+Quote$+" w l,"+Quote$+"p15"+Quote$+" w l"

                CASE  16  :  PRINT  #N%,  "plot  "+Quote$+"p1"   +Quote$+"  w  l  lw  1,"+Quote$+"p2"  +Quote$+"  w  l,"+Quote$+"p3"  +Quote$+"  w  l,"+Quote$+"p4"
+Quote$+" w l,"+Quote$+"p5" +Quote$+" w l,"+Quote$+"p6" +Quote$+" w l,"+_
                                      Quote$+"p7"      +Quote$+"   w   l,"              +Quote$+"p8"   +Quote$+"   w   l,"+Quote$+"p9"   +Quote$+"   w
l,"+Quote$+"p10"+Quote$+" w l,"+Quote$+"p11"+Quote$+" w l,"+Quote$+"p12"+Quote$+" w l,"+_
                                      Quote$+"p13"    +Quote$+"    w    l,"              +Quote$+"p14"+Quote$+"    w    l,"+Quote$+"p15"+Quote$+"    w
l,"+Quote$+"p16"+Quote$+" w l"
            END SELECT

        ELSE

            SELECT CASE NumTrajectories%

                CASE 1  : PRINT #N%, "plot "+Quote$+"p1"+Quote$+" lw 1"
                CASE 2  : PRINT #N%, "plot "+Quote$+"p1"+Quote$+" lw 1,"+Quote$+"p2"+Quote$
                CASE 3  : PRINT #N%, "plot "+Quote$+"p1"+Quote$+" lw 1,"+Quote$+"p2"+Quote$+" ,"+Quote$+"p3"+Quote$
                CASE 4  : PRINT #N%, "plot "+Quote$+"p1"+Quote$+" lw 1,"+Quote$+"p2"+Quote$+" ,"+Quote$+"p3"+Quote$+" ,"+Quote$+"p4"+Quote$
```

```
                CASE  5    :  PRINT  #N%,  "plot  "+Quote$+"p1"+Quote$+"  lw  1,"+Quote$+"p2"+Quote$+"  ,"+Quote$+"p3"+Quote$+"  ,"+Quote$+"p4"+Quote$+"
,"+Quote$+"p5"+Quote$
                CASE  6    :  PRINT  #N%,  "plot  "+Quote$+"p1"+Quote$+"  lw  1,"+Quote$+"p2"+Quote$+"  ,"+Quote$+"p3"+Quote$+"  ,"+Quote$+"p4"+Quote$+"
,"+Quote$+"p5"+Quote$+" ,"+Quote$+"p6"+Quote$
                CASE  7    :  PRINT  #N%,  "plot  "+Quote$+"p1"+Quote$+"  lw  1,"+Quote$+"p2"+Quote$+"  ,"+Quote$+"p3"+Quote$+"  ,"+Quote$+"p4"+Quote$+"
,"+Quote$+"p5"+Quote$+" ,"+Quote$+"p6"+Quote$+" ,"+_
                                             Quote$+"p7"+Quote$
                CASE  8    :  PRINT  #N%,  "plot  "+Quote$+"p1"+Quote$+"  lw  1,"+Quote$+"p2"+Quote$+"  ,"+Quote$+"p3"+Quote$+"  ,"+Quote$+"p4"+Quote$+"
,"+Quote$+"p5"+Quote$+" ,"+Quote$+"p6"+Quote$+" ,"+_
                                             Quote$+"p7"+Quote$+" ,"    +Quote$+"p8"+Quote$
                CASE  9    :  PRINT  #N%,  "plot  "+Quote$+"p1"+Quote$+"  lw  1,"+Quote$+"p2"+Quote$+"  ,"+Quote$+"p3"+Quote$+"  ,"+Quote$+"p4"+Quote$+"
,"+Quote$+"p5"+Quote$+" ,"+Quote$+"p6"+Quote$+" ,"+_
                                             Quote$+"p7"+Quote$+" ,"    +Quote$+"p8"+Quote$+" ,"+Quote$+"p9"+Quote$

                CASE 10  :  PRINT #N%,  "plot "+Quote$+"p1"+Quote$+"  lw  1,"  +Quote$+"p2"  +Quote$+"  ,"+Quote$+"p3"+Quote$+"  ,"+Quote$+"p4"   +Quote$+"
,"+Quote$+"p5"+Quote$+" ,"+Quote$+"p6"+Quote$+" ,"+_
                                             Quote$+"p7"+Quote$+" ,"     +Quote$+"p8" +Quote$+" ,"+Quote$+"p9"+Quote$+" ,"+Quote$+"p10" +Quote$

                CASE 11  :  PRINT #N%,  "plot "+Quote$+"p1"+Quote$+"  lw  1,"  +Quote$+"p2"  +Quote$+"  ,"+Quote$+"p3"+Quote$+"  ,"+Quote$+"p4"   +Quote$+"
,"+Quote$+"p5" +Quote$+" ,"+Quote$+"p6"+Quote$+" ,"+_
                                             Quote$+"p7"+Quote$+"  ,"         +Quote$+"p8"  +Quote$+"  ,"+Quote$+"p9"+Quote$+"  ,"+Quote$+"p10"  +Quote$+"
,"+Quote$+"p11"+Quote$

                CASE 12  :  PRINT #N%,  "plot "+Quote$+"p1"+Quote$+"  lw  1,"  +Quote$+"p2"  +Quote$+"  ,"+Quote$+"p3"+Quote$+"  ,"+Quote$+"p4"   +Quote$+"
,"+Quote$+"p5" +Quote$+" ,"+Quote$+"p6" +Quote$+" ,"+_
                                             Quote$+"p7"+Quote$+"  ,"         +Quote$+"p8"  +Quote$+"  ,"+Quote$+"p9"+Quote$+"  ,"+Quote$+"p10"  +Quote$+"
,"+Quote$+"p11"+Quote$+" ,"+Quote$+"p12"+Quote$

                CASE 13  :  PRINT #N%,  "plot "+Quote$+"p1"  +Quote$+"  lw  1,"+Quote$+"p2"  +Quote$+"  ,"+Quote$+"p3"+Quote$+"  ,"+Quote$+"p4"   +Quote$+"
,"+Quote$+"p5" +Quote$+" ," +Quote$+"p6" +Quote$+" ,"+_
                                             Quote$+"p7"  +Quote$+"  ,"        +Quote$+"p8"  +Quote$+"  ,"+Quote$+"p9"+Quote$+"  ,"+Quote$+"p10"  +Quote$+"
,"+Quote$+"p11"+Quote$+" ," +Quote$+"p12"+Quote$+" ,"+_
                                             Quote$+"p13"+Quote$

                CASE 14  :  PRINT #N%,  "plot "+Quote$+"p1"  +Quote$+"  lw  1,"+Quote$+"p2"  +Quote$+"  ,"+Quote$+"p3"+Quote$+"  ,"+Quote$+"p4"   +Quote$+"
,"+Quote$+"p5" +Quote$+" ," +Quote$+"p6" +Quote$+" ,"+_
                                             Quote$+"p7"  +Quote$+"  ,"        +Quote$+"p8"  +Quote$+"  ,"+Quote$+"p9"+Quote$+"  ,"+Quote$+"p10"  +Quote$+"
,"+Quote$+"p11"+Quote$+" ," +Quote$+"p12"+Quote$+" ,"+_
                                             Quote$+"p13"+Quote$+" ,"    +Quote$+"p14"+Quote$

                CASE 15  :  PRINT #N%,  "plot "+Quote$+"p1"  +Quote$+"  lw  1,"+Quote$+"p2"  +Quote$+"  ,"+Quote$+"p3"  +Quote$+"  ,"+Quote$+"p4"  +Quote$+"
,"+Quote$+"p5" +Quote$+" ," +Quote$+"p6" +Quote$+" ,"+_
                                             Quote$+"p7"  +Quote$+"  ,"        +Quote$+"p8"  +Quote$+"  ,"+Quote$+"p9"  +Quote$+"  ,"+Quote$+"p10"+Quote$+"
,"+Quote$+"p11"+Quote$+" ," +Quote$+"p12"+Quote$+" ,"+_
                                             Quote$+"p13"+Quote$+" ,"    +Quote$+"p14"+Quote$+" ,"+Quote$+"p15"+Quote$

                CASE 16  :  PRINT #N%,  "plot "+Quote$+"p1"  +Quote$+"  lw  1,"+Quote$+"p2"  +Quote$+"  ,"+Quote$+"p3"  +Quote$+"  ,"+Quote$+"p4"  +Quote$+"
,"+Quote$+"p5" +Quote$+" ," +Quote$+"p6" +Quote$+" ,"+_
                                             Quote$+"p7"  +Quote$+"  ,"        +Quote$+"p8"  +Quote$+"  ,"+Quote$+"p9"  +Quote$+"  ,"+Quote$+"p10"+Quote$+"
,"+Quote$+"p11"+Quote$+" ," +Quote$+"p12"+Quote$+" ,"+_
                                             Quote$+"p13"+Quote$+" ,"    +Quote$+"p14"+Quote$+" ,"+Quote$+"p15"+Quote$+" ,"+Quote$+"p16"+Quote$
            END SELECT

        END IF

    CLOSE #N%

    ProcID??? = SHELL(GnuPlotEXE$+" cmd2d.gp -") : CALL Delay(1##)

END SUB 'Plot2DindividualProbeTrajectories()

'----

SUB Plot3DbestProbeTrajectories(NumTrajectories%,M(),R(),Np%,Nd%,LastStep&,FunctionName$) 'XYZZY

LOCAL TrajectoryNumber%, ProbeNumber%, StepNumber&, N%, M%, ProcID???

LOCAL MaximumFitness, MinimumFitness AS EXT

LOCAL BestProbeThisStep%()

LOCAL BestFitnessThisStep(), TempFitness() AS EXT

LOCAL Annotation$, xCoord$, yCoord$, zCoord$, GnuPlotEXE$, PlotWithLines$

    Annotation$      = ""

    PlotWithLines$   = "NO" '"YES" '"NO"

    NumTrajectories% = MIN(Np%,NumTrajectories%)

    GnuPlotEXE$ = "wgnuplot.exe"

'   ---------------- Get Min/Max Fitnesses -----------------

    MaximumFitness = M(1,0) : MinimumFitness = M(1,0)  'Note:  M(p%,j&)

    FOR StepNumber& = 0 TO LastStep&

        FOR ProbeNumber% = 1 TO Np%

            IF M(ProbeNumber%,StepNumber&) >= MaximumFitness THEN MaximumFitness = M(ProbeNumber%,StepNumber&)

            IF M(ProbeNumber%,StepNumber&) =< MinimumFitness THEN MinimumFitness = M(ProbeNumber%,StepNumber&)

        NEXT ProbeNumber%

    NEXT StepNumber%

'   ------------- Copy Fitness Array M() into TempFitness to Preserve M() ----------------

    REDIM TempFitness(1 TO Np%, 0 TO LastStep&)

    FOR StepNumber& = 0 TO LastStep&

        FOR ProbeNumber% = 1 TO Np%

            TempFitness(ProbeNumber%,StepNumber&) = M(ProbeNumber%,StepNumber&)

        NEXT ProbeNumber%

    NEXT StepNumber%

'   ------------ LOOP ON TRAJECTORIES -----------

    FOR TrajectoryNumber% = 1 TO NumTrajectories%

'       --------------- Get Trajectory Coordinate Data -----------------

        REDIM BestFitnessThisStep(0 TO LastStep&), BestProbeThisStep%(0 TO LastStep&)

        FOR StepNumber& = 0 TO LastStep&

            BestFitnessThisStep(StepNumber&) = TempFitness(1,StepNumber&)

            FOR ProbeNumber% = 1 TO Np%

                IF TempFitness(ProbeNumber%,StepNumber&) >= BestFitnessThisStep(StepNumber&) THEN
```

```
                BestFitnessThisStep(StepNumber&) = TempFitness(ProbeNumber%,StepNumber&)

                BestProbeThisStep%(StepNumber&)  = ProbeNumber%

            END IF

        NEXT ProbeNumber%

    NEXT StepNumber&

'   ----- Create Plot Data File -----

    N% = FREEFILE

    SELECT CASE TrajectoryNumber%

        CASE 1  : OPEN "t1"  FOR OUTPUT AS #N%
        CASE 2  : OPEN "t2"  FOR OUTPUT AS #N%
        CASE 3  : OPEN "t3"  FOR OUTPUT AS #N%
        CASE 4  : OPEN "t4"  FOR OUTPUT AS #N%
        CASE 5  : OPEN "t5"  FOR OUTPUT AS #N%
        CASE 6  : OPEN "t6"  FOR OUTPUT AS #N%
        CASE 7  : OPEN "t7"  FOR OUTPUT AS #N%
        CASE 8  : OPEN "t8"  FOR OUTPUT AS #N%
        CASE 9  : OPEN "t9"  FOR OUTPUT AS #N%
        CASE 10 : OPEN "t10" FOR OUTPUT AS #N%

    END SELECT

'   ------------ Write Plot File Data ------------

    FOR StepNumber& = 0 TO LastStep&

        PRINT                      #N%,                      USING$("######.########                  ######.########
######.########",R(BestProbeThisStep%(StepNumber&),1,StepNumber&),R(BestProbeThisStep%(StepNumber&),2,StepNumber&),R(BestProbeThisStep%(StepNumber&),3,Step
Number&))+CHR$(13)

        TempFitness(BestProbeThisStep%(StepNumber&),StepNumber&) = MinimumFitness 'so that same max will not be found for next trajectory

    NEXT StepNumber%

    CLOSE #N%

    NEXT TrajectoryNumber%

'   ------------------------- Plot Trajectories --------------------------

   'CALL CreateGNUplotINIfile(0.1##*ScreenWidth&,0.25##*ScreenHeight&,0.6##*ScreenHeight&,0.6##*ScreenHeight&)

    Annotation$ = ""

    N% = FREEFILE

    OPEN "cmd3d.gp" FOR OUTPUT AS #N%

    PRINT #N%, "set pm3d"
    PRINT #N%, "show pm3d"
    PRINT #N%, "set hidden3d"
    PRINT #N%, "set view 45, 45, 1, 1"

    PRINT #N%, "unset colorbox"

    PRINT #N%, "set xrange [" + REMOVE$(STR$(XiMin(1)),ANY"" ) + ":" + REMOVE$(STR$(XiMax(1)),ANY"" ) + "]"
    PRINT #N%, "set yrange [" + REMOVE$(STR$(XiMin(2)),ANY"" ) + ":" + REMOVE$(STR$(XiMax(2)),ANY"" ) + "]"
    PRINT #N%, "set zrange [" + REMOVE$(STR$(XiMin(3)),ANY"" ) + ":" + REMOVE$(STR$(XiMax(3)),ANY"" ) + "]"

    PRINT #N%, "set grid xtics ytics ztics"
    PRINT #N%, "show grid"
    PRINT #N%, "set title "  + Quote$ + "3D " + FunctionName$ + " PROBE TRAJECTORIES" + "\n" + RunID$ + Quote$
    PRINT #N%, "set xlabel " + Quote$ + "x1"                                        + Quote$
    PRINT #N%, "set ylabel " + Quote$ + "x2"                                        + Quote$
    PRINT #N%, "set zlabel " + Quote$ + "x3"                                        + Quote$

    IF PlotWithLines$ = "YES" THEN

        SELECT CASE NumTrajectories%

            CASE 1  : PRINT #N%, "splot "+Quote$+"t1"+Quote$+" w l lw 3"
            CASE 2  : PRINT #N%, "splot "+Quote$+"t1"+Quote$+" w l lw 3,"+Quote$+"t2"+Quote$+" w l"
            CASE 3  : PRINT #N%, "splot "+Quote$+"t1"+Quote$+" w l lw 3,"+Quote$+"t2"+Quote$+" w l,"+Quote$+"t3"+Quote$+" w l"
            CASE 4  : PRINT #N%, "splot "+Quote$+"t1"+Quote$+" w l lw 3,"+Quote$+"t2"+Quote$+" w l,"+Quote$+"t3"+Quote$+" w l,"+Quote$+"t4"+Quote$+" w l"
            CASE 5   : PRINT #N%, "splot "+Quote$+"t1"+Quote$+" w l lw 3,"+Quote$+"t2"+Quote$+" w l,"+Quote$+"t3"+Quote$+" w l,"+Quote$+"t4"+Quote$+" w
l,"+Quote$+"t5"+Quote$+" w l"
            CASE 6   : PRINT #N%, "splot "+Quote$+"t1"+Quote$+" w l lw 3,"+Quote$+"t2"+Quote$+" w l,"+Quote$+"t3"+Quote$+" w l,"+Quote$+"t4"+Quote$+" w
l,"+Quote$+"t5"+Quote$+" w l,"+Quote$+"t6"+Quote$+" w l"
            CASE 7   : PRINT #N%, "splot "+Quote$+"t1"+Quote$+" w l lw 3,"+Quote$+"t2"+Quote$+" w l,"+Quote$+"t3"+Quote$+" w l,"+Quote$+"t4"+Quote$+" w
l,"+Quote$+"t5"+Quote$+" w l,"+Quote$+"t6"+Quote$+" w l,"+_
                                      Quote$+"t7"+Quote$+" w l"
            CASE 8   : PRINT #N%, "splot "+Quote$+"t1"+Quote$+" w l lw 3,"+Quote$+"t2"+Quote$+" w l,"+Quote$+"t3"+Quote$+" w l,"+Quote$+"t4"+Quote$+" w
l,"+Quote$+"t5"+Quote$+" w l,"+Quote$+"t6"+Quote$+" w l,"+_
                                      Quote$+"t7"+Quote$+" w l,"     +Quote$+"t8"+Quote$+" w l"
            CASE 9   : PRINT #N%, "splot "+Quote$+"t1"+Quote$+" w l lw 3,"+Quote$+"t2"+Quote$+" w l,"+Quote$+"t3"+Quote$+" w l,"+Quote$+"t4"+Quote$+" w
l,"+Quote$+"t5"+Quote$+" w l,"+Quote$+"t6"+Quote$+" w l,"+_
                                      Quote$+"t7"+Quote$+" w l,"     +Quote$+"t8"+Quote$+" w l,"+Quote$+"t9"+Quote$+" w l"
            CASE 10 : PRINT #N%, "splot "+Quote$+"t1"+Quote$+" w l lw 3,"+Quote$+"t2"+Quote$+" w l,"+Quote$+"t3"+Quote$+" w l,"+Quote$+"t4"+Quote$+" w
l,"+Quote$+"t5"+Quote$+" w l,"+Quote$+"t6"+Quote$+" w l,"+_
                                      Quote$+"t7"+Quote$+" w l,"     +Quote$+"t8"+Quote$+" w l,"+Quote$+"t9"+Quote$+" w l,"+Quote$+"t10"+Quote$+" w l"
        END SELECT

    ELSE

        SELECT CASE NumTrajectories%

            CASE 1  : PRINT #N%, "splot "+Quote$+"t1"+Quote$+" lw 2"
            CASE 2  : PRINT #N%, "splot "+Quote$+"t1"+Quote$+" lw 2,"+Quote$+"t2"+Quote$
            CASE 3  : PRINT #N%, "splot "+Quote$+"t1"+Quote$+" lw 2,"+Quote$+"t2"+Quote$+" ,"+Quote$+"t3"+Quote$
            CASE 4  : PRINT #N%, "splot "+Quote$+"t1"+Quote$+" lw 2,"+Quote$+"t2"+Quote$+" ,"+Quote$+"t3"+Quote$+" ,"+Quote$+"t4"+Quote$
            CASE 5    :  PRINT  #N%,  "splot  "+Quote$+"t1"+Quote$+"  lw  2,"+Quote$+"t2"+Quote$+"  ,"+Quote$+"t3"+Quote$+"  ,"+Quote$+"t4"+Quote$+"
,"+Quote$+"t5"+Quote$
            CASE 6    :  PRINT  #N%,  "splot  "+Quote$+"t1"+Quote$+"  lw  2,"+Quote$+"t2"+Quote$+"  ,"+Quote$+"t3"+Quote$+"  ,"+Quote$+"t4"+Quote$+"
,"+Quote$+"t5"+Quote$+" ,"+Quote$+"t6"+Quote$
            CASE 7    :  PRINT  #N%,  "splot  "+Quote$+"t1"+Quote$+"  lw  2,"+Quote$+"t2"+Quote$+"  ,"+Quote$+"t3"+Quote$+"  ,"+Quote$+"t4"+Quote$+"
,"+Quote$+"t5"+Quote$+" ,"+Quote$+"t6"+Quote$+" ,"+_
                                      Quote$+"t7"+Quote$
            CASE 8    :  PRINT  #N%,  "splot  "+Quote$+"t1"+Quote$+"  lw  2,"+Quote$+"t2"+Quote$+"  ,"+Quote$+"t3"+Quote$+"  ,"+Quote$+"t4"+Quote$+"
,"+Quote$+"t5"+Quote$+" ,"+Quote$+"t6"+Quote$+" ,"+_
                                      Quote$+"t7"+Quote$+" ,"    +Quote$+"t8"+Quote$
            CASE 9    :  PRINT  #N%,  "splot  "+Quote$+"t1"+Quote$+"  lw  2,"+Quote$+"t2"+Quote$+"  ,"+Quote$+"t3"+Quote$+"  ,"+Quote$+"t4"+Quote$+"
,"+Quote$+"t5"+Quote$+" ,"+Quote$+"t6"+Quote$+" ,"+_
                                      Quote$+"t7"+Quote$+" ,"    +Quote$+"t8"+Quote$+" ,"+Quote$+"t9"+Quote$
            CASE 10   :  PRINT  #N%,  "splot  "+Quote$+"t1"+Quote$+"  lw  2,"+Quote$+"t2"+Quote$+"  ,"+Quote$+"t3"+Quote$+"  ,"+Quote$+"t4"+Quote$+"
,"+Quote$+"t5"+Quote$+" ,"+Quote$+"t6"+Quote$+" ,"+_
                                      Quote$+"t7"+Quote$+" ,"    +Quote$+"t8"+Quote$+" ,"+Quote$+"t9"+Quote$+" ,"+Quote$+"t10"+Quote$
        END SELECT

    END IF

    CLOSE #N%
```

```
    ProcID??? = SHELL(GnuPlotEXE$+" cmd3d.gp -") : CALL Delay(1##)
END SUB 'Plot3DbestProbeTrajectories()
'-----------
FUNCTION HasDAVGsaturated$(Nsteps&,j&,Np%,Nd%,M(),R(),DiagLength)
LOCAL A$
LOCAL k&
LOCAL SumOfDavg, DavgStepJ AS EXT
LOCAL DavgSatTOL AS EXT
    A$ = "NO"
    DavgSatTOL = 0.0005## 'tolerance for DAVG saturation
    IF j& < Nsteps& + 10 THEN GOTO ExitHasDAVGsaturated 'execute at least 10 steps after averaging interval before performing this check
    DavgStepJ = DavgThisStep(j&,Np%,Nd%,M(),R(),DiagLength)
    SumOfDavg = 0##
    FOR k& = j&-Nsteps&+1 TO j& 'check this step and previous (Nsteps&-1) steps
        SumOfDavg = SumOfDavg + DavgThisStep(k&,Np%,Nd%,M(),R(),DiagLength)
    NEXT k&
    IF ABS(SumOfDavg/Nsteps&-DavgStepJ) =< DavgSatTOL THEN A$ = "YES" 'saturation if (avg value - last value) are within TOL
ExitHasDAVGsaturated:
    HasDAVGsaturated$ = A$
END FUNCTION 'HasDAVGsaturated$()
'-----------
FUNCTION OscillationInDavg$(j&,Np%,Nd%,M(),R(),DiagLength)
LOCAL A$
LOCAL k&, NumSlopeChanges%
    A$ = "NO"
    NumSlopeChanges% = 0
    IF j& < 15 THEN GOTO ExitDavgOscillation 'wait at least 15 steps
    FOR k& = j&-10 TO j&-1 'check previous ten steps
        IF (DavgThisStep(k&,Np%,Nd%,M(),R(),DiagLength)-DavgThisStep(k&-1,Np%,Nd%,M(),R(),DiagLength))* _
           (DavgThisStep(k&+1,Np%,Nd%,M(),R(),DiagLength)-DavgThisStep(k&,Np%,Nd%,M(),R(),DiagLength)) < 0## THEN INCR NumSlopeChanges%
    NEXT j&
    IF NumSlopeChanges% >= 3 THEN A$ = "YES"
ExitDavgOscillation:
    OscillationInDavg$ = A$
END FUNCTION 'OscillationInDavg()
'------
FUNCTION DavgThisStep(j&,Np%,Nd%,M(),R(),DiagLength)
LOCAL BestFitness, TotalDistanceAllProbes, SumSQ AS EXT
LOCAL p%, k&, N%, i%, BestProbeNumber%, BestTimeStep&
'   ----------- Best Probe #, etc. -----------
    FOR k& = 0 TO j&
        BestFitness = M(1,k&)
        FOR p% = 1 TO Np%
            IF M(p%,k&) >= BestFitness THEN
                BestFitness = M(p%,k&) : BestProbeNumber% = p% : BestTimeStep& = k&
            END IF
        NEXT p% 'probe #
    NEXT k& 'time step
'   --------- Average Distance to Best Probe -----------
    TotalDistanceAllProbes = 0##
    FOR p% = 1 TO Np%
        SumSQ = 0##
        FOR i% = 1 TO Nd%
            SumSQ = SumSQ + (R(BestProbeNumber%,i%,BestTimeStep&)-R(p%,i%,j&))^2 'do not exclude p%=BestProbeNumber%(j&) from sum because it adds zero
         NEXT i%
        TotalDistanceAllProbes = TotalDistanceAllProbes + SQR(SumSQ)
    NEXT p%
    DavgThisStep = TotalDistanceAllProbes/(DiagLength*(Np%-1)) 'but exclude best prove from average
END FUNCTION 'DavgThisStep()
'-----------
SUB
PlotBestFitnessEvolution(Nd%,Np%,LastStep&,G,DeltaT,Alpha,Beta,Frep,Mbest(),PlaceInitialProbes$,InitialAcceleration$,RepositionFactor$,FunctionName$,Gamma)
LOCAL BestFitness(), GlobalBestFitness AS EXT
LOCAL PlotAnnotation$, PlotTitle$
LOCAL p%, j&, N%
    REDIM BestFitness(0 TO LastStep&)
```

```
    CALL
GetPlotAnnotation(PlotAnnotation$,Nd%,Np%,LastStep&,G,DeltaT,Alpha,Beta,Frep,Mbest(),PlaceInitialProbes$,InitialAcceleration$,RepositionFactor$,FunctionNam
e$,Gamma)

    GlobalBestFitness = Mbest(1,0)

    FOR j& = 0 TO LastStep&

'       BestFitness(j&) = Mbest(1,j&) 'orig code 03-23-2010

        BestFitness(j&) = -1E4200 'added 03-23-2010

        FOR p% = 1 TO Np%

            IF Mbest(p%,j&) >= BestFitness(j&)   THEN BestFitness(j&)   = Mbest(p%,j&)

            IF Mbest(p%,j&) >= GlobalBestFitness THEN GlobalBestFitness = Mbest(p%,j&)

        NEXT p% 'probe #

    NEXT j& 'time step

    N% = FREEFILE

    OPEN "Fitness" FOR OUTPUT AS #N%

'    print #N%,"From Sub PlotBestFitness..."

        FOR j& = 0 TO LastStep&

'            PRINT #N%, USING$("###### ##.######^^^^^^",j&,BestFitness(j&))
            PRINT #N%, USING$("###### #######.#########",j&,BestFitness(j&))

        NEXT j&

    CLOSE #N%

    PlotAnnotation$ = PlotAnnotation$ + "Best Fitness = " + REMOVE$(STR$(ROUND(GlobalBestFitness,8)),ANY" ")

    PlotTitle$ = "Best Fitness vs Time Step\n" + "[" + REMOVE$(STR$(Np%),ANY" ") + " probes, "+REMOVE$(STR$(LastStep&),ANY" ")+" time steps]"

    CALL CreateGNUplotINIfile(0.1##*ScreenWidth&,0.1##*ScreenHeight&,0.6##*ScreenWidth&,0.6##*ScreenHeight&)

    CALL TwoDplot("Fitness","Best Fitness","0.7","0.7","Time Step\n\n.",".\n\nBest Fitness(X)", _
                "","","","","","","","","wgnuplot.exe"," with lines linewidth 2",PlotAnnotation$)

END SUB 'PlotBestFitnessEvolution()

'------

SUB
PlotAverageDistance(Nd%,Np%,LastStep&,G,DeltaT,Alpha,Beta,Frep,Mbest(),PlaceInitialProbes$,InitialAcceleration$,RepositionFactor$,FunctionName$,R(),DiagLen
gth,Gamma)

LOCAL Davg(), BestFitness(), TotalDistanceAllProbes, SumSQ AS EXT

LOCAL PlotAnnotation$, PlotTitle$

LOCAL p%, j&, N%, i%, BestProbeNumber%(), BestTimeStep&()

    REDIM Davg(0 TO LastStep&), BestFitness(0 TO LastStep&), BestProbeNumber%(0 TO LastStep&), BestTimeStep&(0 TO LastStep&)

    CALL
GetPlotAnnotation(PlotAnnotation$,Nd%,Np%,LastStep&,G,DeltaT,Alpha,Beta,Frep,Mbest(),PlaceInitialProbes$,InitialAcceleration$,RepositionFactor$,FunctionNam
e$,Gamma)

'   ----------- Best Probe #, etc. -----------

    FOR j& = 0 TO LastStep&

        BestFitness(j&) = Mbest(1,j&)

        FOR p% = 1 TO Np%

            IF Mbest(p%,j&) >= BestFitness(j&) THEN

                BestFitness(j&) = Mbest(p%,j&) : BestProbeNumber%(j&) = p% : BestTimeStep&(j&) = j& 'only probe number is used at this time, but other data
are computed for possible future use.

            END IF

        NEXT p% 'probe #

    NEXT j& 'time step

    N% = FREEFILE

'   --------- Average Distance to Best Probe -----------

    FOR j& = 0 TO LastStep&

        TotalDistanceAllProbes = 0##

        FOR p% = 1 TO Np%

            SumSQ = 0##

            FOR i% = 1 TO Nd%

                SumSQ = SumSQ + (R(BestProbeNumber%(j&),i%,j&)-R(p%,i%,j&))^2 'do not exclude p%=BestProbeNumber%(j&) from sum because it adds zero

            NEXT i%

            TotalDistanceAllProbes = TotalDistanceAllProbes + SQR(SumSQ)

        NEXT p%

        Davg(j&) = TotalDistanceAllProbes/(DiagLength*(Np%-1)) 'but exclude best prove from average

    NEXT j&

'   ------------ Create Plot Data File -----------

    OPEN "Davg" FOR OUTPUT AS #N%

        FOR j& = 0 TO LastStep&

            PRINT #N%, USING$("###### #######.#######",j&,Davg(j&))

        NEXT j&

    CLOSE #N%

    PlotTitle$ = "Average Distance of " + REMOVE$(STR$(Np%-1),ANY" ") + " Probes to Best Probe\nNormalized to Size of Decision Space\n" + _
               "[" + REMOVE$(STR$(Np%),ANY" ") + " probes, " + REMOVE$(STR$(LastStep&),ANY" ") + " time steps]"

    CALL CreateGNUplotINIfile(0.2##*ScreenWidth&,0.2##*ScreenHeight&,0.6##*ScreenWidth&,0.6##*ScreenHeight&)
```

```
    CALL TwoDplot("Davg",PlotTitle$,"0.7","0.9","Time Step\n\n.",".\n\n<D>/Ldiag", _
              "","","","","","","","","wgnuplot.exe"," with lines linewidth 2",PlotAnnotation$)

END SUB 'PlotAverageDistance()

'------

SUB
GetPlotAnnotation(PlotAnnotation$,Nd%,Np%,LastStep&,G,DeltaT,Alpha,Beta,Frep,Mbest(),PlaceInitialProbes$,InitialAcceleration$,RepositionFactor$,FunctionNam
e$,Gamma)

LOCAL A$

    A$ = "" : IF PlaceInitialProbes$ = "UNIFORM ON-AXIS" AND Nd% > 1 THEN A$ = " ("+REMOVE$(STR$(Np%/Nd%),ANY" ") + "/axis)"

    PlotAnnotation$ = RunID$ + "\n" + _
                      FunctionName$ + " Function" + " ("+ FormatInteger$(Nd%) + "-D) \n"    +_
                      FormatInteger$(Np%) + " probes"       + A$ + "\n" +_
                      "G = " + FormatFP$(G,2)               + "\n" +_
                      "Alpha = "     + FormatFP$(Alpha,1)   + "\n" +_
                      "Beta = "      + FormatFP$(Beta,1)    + "\n" +_
                      "DelT = "      + FormatFP$(DeltaT,1)  + "\n" +_
                      "Gamma = "     + FormatFP$(Gamma,3)   + "\n" +_
                      "Init Probes " + PlaceInitialProbes$  + "\n" +_
                      "Init Accel "  + InitialAcceleration$ + "\n" +_
                      "Frep "        + RepositionFactor$ + "\n"

END SUB

'------

SUB
PlotBestProbeVsTimeStep(Nd%,Np%,LastStep&,G,DeltaT,Alpha,Beta,Frep,Mbest(),PlaceInitialProbes$,InitialAcceleration$,RepositionFactor$,FunctionName$,Gamma)

LOCAL BestFitness AS EXT

LOCAL PlotAnnotation$, PlotTitle$

LOCAL p%, j&, N%, BestProbeNumber%()

    REDIM BestProbeNumber%(0 TO LastStep&)

    CALL
GetPlotAnnotation(PlotAnnotation$,Nd%,Np%,LastStep&,G,DeltaT,Alpha,Beta,Frep,Mbest(),PlaceInitialProbes$,InitialAcceleration$,RepositionFactor$,FunctionNam
e$,Gamma)

    FOR j& = 0 TO LastStep&

        Bestfitness = Mbest(1,j&)

        FOR p% = 1 TO Np%

            IF Mbest(p%,j&) >= BestFitness THEN

                BestFitness = Mbest(p%,j&) : BestProbeNumber%(j&) = p%

            END IF

        NEXT p% 'probe #

    NEXT j& 'time step

    N% = FREEFILE

    OPEN "Best Probe" FOR OUTPUT AS #N%

        FOR j& = 0 TO LastStep&

            PRINT #N%, USING$("###### #####",j&,BestProbeNumber%(j&))

        NEXT j&

    CLOSE #N%

    PlotTitle$ = "Best Probe Number vs Time Step\n" + "[" +REMOVE$(STR$(Np%),ANY" ") + " probes, " + REMOVE$(STR$(LastStep&),ANY" ") + " time steps]"

    CALL CreateGNUplotINIfile(0.15#*ScreenWidth&,0.15##*ScreenHeight&,0.6##*ScreenWidth&,0.6##*ScreenHeight&)

'USAGE:                                                                                                                                               CALL
TwoDplot(PlotFileName$,PlotTitle$,xCoord$,yCoord$,XaxisLabel$,YaxisLabel$,LogXaxis$,LogYaxis$,xMin$,xMax$,yMin$,yMax$,xTics$,yTics$,GnuPlotEXE$,LineType$,A
nnotation$)

    CALL TwoDplot("Best Probe",PlotTitle$,"0.7","0.7","Time Step\n\n.",".\n\nBest Probe #","","","","","","0",NoSpaces$(Np%+1,0),"","","wgnuplot.exe"," pt 8
ps .5 lw 1",PlotAnnotation$) 'pt, pointtype; ps, pointsize; lw, linewidth

END SUB 'PlotBestProbeVsTimeStep()

'------

FUNCTION FormatInteger$(M%) : FormatInteger$ = REMOVE$(STR$(M%),ANY" ") : END FUNCTION

'------

FUNCTION FormatFP$(X,Ndigits%)

LOCAL A$

    IF X = 0## THEN

        A$ = "0." : GOTO ExitFormatFP

    END IF

    A$ = REMOVE$(STR$(ROUND(ABS(X),Ndigits%)),ANY" ")

    IF ABS(X) < 1## THEN

        IF X > 0## THEN

            A$ = "0" + A$

        ELSE

            A$ = "-0" + A$

        END IF

    ELSE

        IF X < 0## THEN A$ = "-" + A$

    END IF

ExitFormatFP:

    FormatFP$ = A$

END FUNCTION
```

```
'-----------
SUB InitialProbeDistribution(Np%,Nd%,Nt&,R(),PlaceInitialProbes$,Gamma)
LOCAL DeltaXi, DelX1, DelX2, Di AS EXT
LOCAL NumProbesPerDimension%, p%, i%, k%, NumX1points%, NumX2points%, x1pointNum%, x2pointNum%, A$
    SELECT CASE PlaceInitialProbes$
        CASE "UNIFORM ON-AXIS"
            IF Nd% > 1 THEN
                NumProbesPerDimension% = Np%\Nd% 'even #
            ELSE
                NumProbesPerDimension% = Np%
            END IF
            FOR i% = 1 TO Nd%
                FOR p% = 1 TO Np%
                    R(p%,i%,0) = XiMin(i%) + Gamma*(XiMax(i%)-XiMin(i%))
                NEXT Np%
            NEXT i%
            FOR i% = 1 TO Nd% 'place probes probe line-by-probe line (i% is dimension number)
                DeltaXi = (XiMax(i%)-XiMin(i%))/(NumProbesPerDimension%-1)
                FOR k% = 1 TO NumProbesPerDimension%
                    p% = k% + NumProbesPerDimension%*(i%-1) 'probe #
                    R(p%,i%,0) = XiMin(i%) + (k%-1)*DeltaXi
                NEXT k%
            NEXT i%
        CASE "UNIFORM ON-DIAGONAL"
            FOR p% = 1 TO Np%
                FOR i% = 1 TO Nd%
                    DeltaXi = (XiMax(i%)-XiMin(i%))/(Np%-1)
                    R(p%,i%,0) = XiMin(i%) + (p%-1)*DeltaXi
                NEXT i%
            NEXT p%
        CASE "2D GRID"
            NumProbesPerDimension% = SQR(Np%) : NumX1points% = NumProbesPerDimension% : NumX2points% = NumX1points% 'broken down for possible future use
            DelX1 = (XiMax(1)-XiMin(1))/(NumX1points%-1)
            DelX2 = (XiMax(2)-XiMin(2))/(NumX2points%-1)
            FOR x1pointNum% = 1 TO NumX1points%
                FOR x2pointNum% = 1 TO NumX2points%
                    p% = NumX1points%*(x1pointNum%-1)+x2pointNum% 'probe #
                    R(p%,1,0) = XiMin(1) + DelX1*(x1pointNum%-1) 'x1 coord
                    R(p%,2,0) = XiMin(2) + DelX2*(x2pointNum%-1) 'x2 coord
                NEXT x2pointNum%
            NEXT x1pointNum%
        CASE "RANDOM"
            FOR p% = 1 TO Np%
                FOR i% = 1 TO Nd%
                    R(p%,i%,0) = XiMin(i%) + RandomNum(0##,1##)*(XiMax(i%)-XiMin(i%))
                NEXT i%
            NEXT p%
    END SELECT
END SUB 'InitialProbeDistribution()
'------
SUB
ChangeRunParameters(NumProbesPerDimension%,Np%,Nd%,Nt&,G,Alpha,Beta,DeltaT,Frep,PlaceInitialProbes$,InitialAcceleration$,RepositionFactor$,FunctionName$)
'THIS PROCEDURE NOT USED
LOCAL A$, DefaultValue$
    A$ = INPUTBOX$("# dimensions?","Change # Dimensions ("+FunctionName$+")",NoSpaces$(Nd%+0,0)) : Nd%    = VAL(A$) : IF Nd% < 1 OR Nd% > 500 THEN Nd% = 2
    IF Nd% > 1 THEN NumProbesPerDimension% = 2*((NumProbesPerDimension%+1)\2) 'require an even # probes on each probe line to avoid overlapping at origin
(in symmetrical spaces at least...)
    IF Nd% = 1 THEN NumProbesPerDimension% = MAX(NumProbesPerDimension%,3)    'at least 3 probes on x-axis for 1-D functions
    Np% = NumProbesPerDimension%*Nd%
    A$ = INPUTBOX$("# time steps?","Change # Steps ("+FunctionName$+")",NoSpaces$(Nt&+0,0)) : Nt&    = VAL(A$) : IF Nt& < 3                        THEN Nt&
= 50
    A$ = INPUTBOX$("Grav Const G?","Change G ("+FunctionName$+")",NoSpaces$(G,2))            : G       = VAL(A$) : IF G < -100##    OR G > 100##    THEN G
= 2##
    A$ = INPUTBOX$("Alpha?","Change Alpha ("+FunctionName$+")",NoSpaces$(Alpha,2))            : Alpha   = VAL(A$) : IF Alpha < -50## OR Alpha > 50## THEN
Alpha = 2##
    A$ = INPUTBOX$("Beta?","Change Beta ("+FunctionName$+")",NoSpaces$(Beta,2))              : Beta    = VAL(A$) : IF Beta  < -50## OR Beta  > 50## THEN Beta
= 2##'
    A$ = INPUTBOX$("Delta T","Change Delta-T ("+FunctionName$+")",NoSpaces$(DeltaT,2))       : DeltaT = VAL(A$) : IF DeltaT =< 0##                     THEN
DeltaT = 1##
```

```
    A$ = INPUTBOX$("Frep [0-1]?","Change Frep ("+FunctionName$+")",NoSpaces$(Frep,3))       : Frep   = VAL(A$) : IF Frep < 0##    OR Frep > 1##   THEN Frep
= 0.5##

'   ------------ Initial Probe Distribution ------------

    SELECT CASE PlaceInitialProbes$
        CASE "UNIFORM ON-AXIS"     : DefaultValue$ = "1"
        CASE "UNIFORM ON-DIAGONAL" : DefaultValue$ = "2"
        CASE "2D GRID"             : DefaultValue$ = "3"
        CASE "RANDOM"              : DefaultValue$ = "4"
    END SELECT

    A$  =  INPUTBOX$("Initial  Probes?"+CHR$(13)+"1  -  UNIFORM  ON-AXIS"+CHR$(13)+"2  -  UNIFORM  ON-DIAGONAL"+CHR$(13)+"3  -  2D  GRID"+CHR$(13)+"4  -
RANDOM","Initial Probe Distribution ("+FunctionName$+")",DefaultValue$)

    IF VAL(A$) < 1 OR VAL(A$) > 4 THEN A$ = "1"

    SELECT CASE VAL(A$)
        CASE 1 : PlaceInitialProbes$ = "UNIFORM ON-AXIS"
        CASE 2 : PlaceInitialProbes$ = "UNIFORM ON-DIAGONAL"
        CASE 3 : PlaceInitialProbes$ = "2D GRID"
        CASE 4 : PlaceInitialProbes$ = "RANDOM"
    END SELECT

    IF Nd% = 1  AND PlaceInitialProbes$ = "UNIFORM ON-DIAGONAL" THEN PlaceInitialProbes$ = "UNIFORM ON-AXIS" 'cannot do diagonal in 1-D space

    IF Nd% <> 2 AND PlaceInitialProbes$ = "2D GRID" THEN PlaceInitialProbes$ = "UNIFORM ON-AXIS" '2D grid is available only in 2 dimensions!

'   ------------- Initial Acceleration -----------------

    SELECT CASE InitialAcceleration$
        CASE "ZERO"   : DefaultValue$ = "1"
        CASE "FIXED"  : DefaultValue$ = "2"
        CASE "RANDOM" : DefaultValue$ = "3"
    END SELECT

    A$    =    INPUTBOX$("Initial    Acceleration?"+CHR$(13)+"1    -    ZERO"+CHR$(13)+"2    -    FIXED"+CHR$(13)+"3    -    RANDOM","Initial    Acceleration
("+FunctionName$+")",DefaultValue$)

   IF VAL(A$) < 1 OR VAL(A$) > 3 THEN A$ = "1"

    SELECT CASE VAL(A$)
        CASE 1 : InitialAcceleration$ = "ZERO"
        CASE 2 : InitialAcceleration$ = "FIXED"
        CASE 3 : InitialAcceleration$ = "RANDOM"
    END SELECT

'   ----------- Reposition Factor ---------------

    SELECT CASE RepositionFactor$
        CASE "FIXED"    : DefaultValue$ = "1"
        CASE "VARIABLE" : DefaultValue$ = "2"
        CASE "RANDOM"   : DefaultValue$ = "3"
    END SELECT

    A$    =    INPUTBOX$("Reposition    Factor?"+CHR$(13)+"1    -    FIXED"+CHR$(13)+"2    -    VARIABLE"+CHR$(13)+"3    -    RANDOM","Retrieve    Probes
("+FunctionName$+")",DefaultValue$)

    IF VAL(A$) < 1 OR VAL(A$) > 3 THEN A$ = "1"

    SELECT CASE VAL(A$)
        CASE 1 : RepositionFactor$ = "FIXED"
        CASE 2 : RepositionFactor$ = "VARIABLE"
        CASE 3 : RepositionFactor$ = "RANDOM"
    END SELECT

END SUB 'ChangeRunParameters()

'------

FUNCTION NoSpaces$(X,NumDigits%) :  NoSpaces$ = REMOVE$(STR$(X,NumDigits%),ANY" ") : END FUNCTION

'-----------

FUNCTION TerminateNowForSaturation$(j&,Nd%,Np%,Nt&,G,DeltaT,Alpha,Beta,R(),A(),M())

LOCAL A$, i&, p%, NumStepsForAveraging&

LOCAL BestFitness, AvgFitness, FitnessTOL AS EXT 'terminate if avg fitness does not change over NumStepsForAveraging& time steps

    FitnessTOL = 0.00001## : NumStepsForAveraging& = 10

    A$ = "NO"

    IF j& >= NumStepsForAveraging+10 THEN 'wait until step 10 to start checking for fitness saturation

        AvgFitness = 0##

        FOR i& = j&-NumStepsForAveraging&+1 TO j& 'avg fitness over current step & previous NumStepsForAveraging&-1 steps

            BestFitness = M(1,i&)

            FOR p% = 1 TO Np%

                IF M(p%,i&) >= BestFitness THEN BestFitness = M(p%,i&)

            NEXT p%

            AvgFitness = AvgFitness + BestFitness

        NEXT i&

        AvgFitness = AvgFitness/NumStepsForAveraging&

        IF ABS(AvgFitness-BestFitness) < FitnessTOL THEN A$ = "YES" 'compare avg fitness to best fitness at this step

    END IF

    TerminateNowForSaturation$ = A$

END FUNCTION 'TerminateNowForSaturation$()

'-----------

FUNCTION MagVector(V(),N%) 'returns magnitude of Nx1 column vector V

LOCAL SumSq AS EXT

LOCAL i%

    SumSQ = 0## : FOR i% = 1 TO N% : SumSQ = SumSQ + V(i%)^2 : NEXT i% : MagVector = SQR(SumSQ)

END FUNCTION 'MagVector()

'---

FUNCTION UnitStep(X)
```

```
LOCAL Z AS EXT
    IF X < 0## THEN
        Z = 0##
    ELSE
        Z = 1##
    END IF
    UnitStep = Z
END FUNCTION 'UnitStep()
'---
SUB Plot1Dfunction(FunctionName$,R()) 'plots 1D function on-screen
LOCAL NumPoints%, i%, N%
LOCAL DeltaX, X AS EXT
    NumPoints% = 32001
    DeltaX = (XiMax(1)-XiMin(1))/(NumPoints%-1)
    N% = FREEFILE
    SELECT CASE FunctionName$
        CASE "ParrottF4" 'PARROTT F4 FUNCTION
            OPEN "ParrottF4" FOR OUTPUT AS #N%
                FOR i% = 1 TO NumPoints%
                    R(1,1,0) = XiMin(1) + (i%-1)*DeltaX
                    PRINT #N%, USING$("#.######  #.######",R(1,1,0),ParrottF4(R(),1,1,0))
                NEXT i%
            CLOSE #N%
            CALL CreateGNUplotINIfile(0.2##*ScreenWidth&,0.2##*ScreenHeight&,0.6##*ScreenWidth&,0.6##*ScreenHeight&)
            CALL  TwoDplot("ParrottF4","Parrott  F4  Function","0.7","0.7","X\n\n.",".\n\nParrott  F4(X)","","","0","1","0","1","","","wgnuplot.exe","  with
lines linewidth 2","")
    END SELECT
END SUB
'------
SUB CLEANUP 'probe coordinate plot files
    IF DIR$("P1")  <> "" THEN KILL "P1"
    IF DIR$("P2")  <> "" THEN KILL "P2"
    IF DIR$("P3")  <> "" THEN KILL "P3"
    IF DIR$("P4")  <> "" THEN KILL "P4"
    IF DIR$("P5")  <> "" THEN KILL "P5"
    IF DIR$("P6")  <> "" THEN KILL "P6"
    IF DIR$("P7")  <> "" THEN KILL "P7"
    IF DIR$("P8")  <> "" THEN KILL "P8"
    IF DIR$("P9")  <> "" THEN KILL "P9"
    IF DIR$("P10") <> "" THEN KILL "P10"
    IF DIR$("P11") <> "" THEN KILL "P11"
    IF DIR$("P12") <> "" THEN KILL "P12"
    IF DIR$("P13") <> "" THEN KILL "P13"
    IF DIR$("P14") <> "" THEN KILL "P14"
    IF DIR$("P15") <> "" THEN KILL "P15"
END SUB
'------
SUB Plot2Dfunction(FunctionName$,R())
LOCAL A$
LOCAL NumPoints%, i%, k%, N%
LOCAL Delx1, Delx2, Z AS EXT
    SELECT CASE FunctionName$
        CASE "BOWTIE" : Numpoints% = 5
        CASE "YAGI"   : Numpoints% = 5
        CASE "PBM_1","PBM_2","PBM_3","PBM_4","PBM_5" : NumPoints% = 25
        CASE ELSE : NumPoints% = 100
    END SELECT
    N% = FREEFILE : OPEN "TwoDplot.DAT" FOR OUTPUT AS #N%
    Delx1 = (XiMax(1)-XiMin(1))/(NumPoints%-1) : Delx2 = (XiMax(2)-XiMin(2))/(NumPoints%-1)
    FOR i% = 1 TO NumPoints%
        R(1,1,0) = XiMin(1) + (i%-1)*Delx1 'x1 value
        FOR k% = 1 TO NumPoints%
            R(1,2,0) = XiMin(2) + (k%-1)*Delx2 'x2 value
            Z = ObjectiveFunction(R(),2,1,0,FunctionName$)
            PRINT #N%, USING$("######.###### ######.###### #######.######^^^^",R(1,1,0),R(1,2,0),Z)
        NEXT k%
        PRINT #N%, ""
    NEXT i%
    CLOSE #N%
    CALL CreateGNUplotINIfile(0.1##*ScreenWidth&,0.1##*ScreenHeight&,0.6##*ScreenWidth&,0.6##*ScreenHeight&)
    A$ = "" : IF INSTR(FunctionName$,"PBM_") > 0 THEN A$ = "Coarse "
    CALL ThreeDplot2("TwoDplot.DAT",A$+"Plot of "+FunctionName$+" Function","","0.6","0.6","1.2", _
                  "x1","x2","z=F(x1,x2)","","","wgnuplot.exe","","","","","")
```

```
END SUB
'------

    SUB TwoDplot3curves(NumCurves%,PlotFileName1$,PlotFileName2$,PlotFileName3$,PlotTitle$,Annotation$,xCoord$,yCoord$,XaxisLabel$,YaxisLabel$, _
                    LogXaxis$,LogYaxis$,xMin$,xMax$,yMin$,yMax$,xTics$,yTics$,GnuPlotEXE$)

        LOCAL N%

        LOCAL LineSize$

        LineSize$ = "2"

        N% = FREEFILE

        OPEN "cmd2d.gp" FOR OUTPUT AS #N%

            IF LogXaxis$ = "YES" AND LogYaxis$ = "NO"  THEN PRINT #N%, "set logscale x"
            IF LogXaxis$ = "NO"  AND LogYaxis$ = "YES" THEN PRINT #N%, "set logscale y"
            IF LogXaxis$ = "YES" AND LogYaxis$ = "YES" THEN PRINT #N%, "set logscale xy"

            IF xMin$ <> "" AND xMax$ <> "" THEN  PRINT #N%, "set xrange ["+xMin$+":"+xMax$+"]"

            IF yMin$ <> "" AND yMax$ <> "" THEN  PRINT #N%, "set yrange ["+yMin$+":"+yMax$+"]"

            PRINT #N%, "set label "+Quote$+AnnoTation$+Quote$+" at graph "+xCoord$+","+yCoord$
            PRINT #N%, "set grid xtics"
            PRINT #N%, "set grid ytics"
            PRINT #N%, "set xtics "+xTics$
            PRINT #N%, "set ytics "+yTics$
            PRINT #N%, "set grid mxtics"
            PRINT #N%, "set grid mytics"
            PRINT #N%, "set title " +Quote$+PlotTitle$+Quote$
            PRINT #N%, "set xlabel "+Quote$+XaxisLabel$+Quote$
            PRINT #N%, "set ylabel "+Quote$+YaxisLabel$+Quote$

            SELECT CASE NumCurves%

            CASE 1
            PRINT #N%, "plot " + Quote$ + PlotFileName1$ + Quote$ + " with lines linewidth " + LineSize$

            CASE 2
            PRINT #N%, "plot " + Quote$ + PlotFileName1$ + Quote$ + " with lines linewidth " + LineSize$+", " + _
                                 Quote$ + PlotFileName2$ + Quote$ + " with lines linewidth " + LineSize$
            CASE 3
            PRINT #N%, "plot " + Quote$ + PlotFileName1$ + Quote$ + " with lines linewidth " + LineSize$+", " + _
                                 Quote$ + PlotFileName2$ + Quote$ + " with lines linewidth " + LineSize$+", " + _
                                 Quote$ + PlotFileName3$ + Quote$ + " with lines linewidth " + LineSize$
            END SELECT

        CLOSE #N%

        SHELL(GnuPlotEXE$+" cmd2d.gp -")

        CALL Delay(1##)

    END SUB 'TwoDplot3Curves()

'---

FUNCTION Fibonacci&&(N%) 'RETURNS Nth FIBONACCI NUMBER

LOCAL i%, Fn&&, Fn1&&, Fn2&&

LOCAL A$

    IF N% > 91 OR N% < 0 THEN

        MSGBOX("ERROR!  Fibonacci argument"+STR$(N%)+" > 91.  Out of range or < 0...") : EXIT FUNCTION

    END IF

    SELECT CASE N%

        CASE 0: Fn&& = 1

        CASE ELSE

            Fn&& = 0 : Fn2&& = 1 : i% = 0

            FOR i% = 1 TO N%

                Fn&& = Fn1&& + Fn2&&

                Fn1&& = Fn2&&

                Fn2&& = Fn&&

            NEXT i% 'LOOP

    END SELECT

    Fibonacci&& = Fn&&

END FUNCTION 'Fibonacci&&()

'-----------

FUNCTION RandomNum(a,b) 'Returns random number X, a=< X < b.

    RandomNum = a + (b-a)*RND

END FUNCTION 'RandomNum()

'-----------

FUNCTION GaussianDeviate(Mu,Sigma) 'returns NORMAL (Gaussian) random deviate with mean Mu and standard deviation Sigma (variance = Sigma^2)

'Refs: (1) Press, W.H., Flannery, B.P., Teukolsky, S.A., and Vetterling, W.T., "Numerical Recipes: The Art of Scientific Computing,"
'          §7.2, Cambridge University Press, Cambridge, UK, 1986.
'      (2) Shinzato, T., "Box Muller Method," 2007, http://www.sp.dis.titech.ac.jp/~shinzato/boxmuller.pdf

LOCAL s, t, Z AS EXT

    s = RND : t = RND

    Z = Mu + Sigma*SQR(-2##*LOG(s))*COS(TwoPi*t)

    GaussianDeviate = Z

END FUNCTION 'GaussianDeviate()

'-----------

    SUB ContourPlot(PlotFileName$,PlotTitle$,Annotation$,xCoord$,yCoord$,zCoord$, _
                XaxisLabel$,YaxisLabel$,ZaxisLabel$,zMin$,zMax$,GnuPlotEXE$,A$)
        LOCAL N%
```

```
        N% = FREEFILE

        OPEN "cmd3d.gp" FOR OUTPUT AS #N%

            PRINT #N%, "show surface"
            PRINT #N%, "set hidden3d"
            IF zMin$ <> "" AND zMax$ <> "" THEN  PRINT #N%, "set zrange ["+zMin$+":"+zMax$+"]"
            PRINT #N%, "set label "+Quote$+AnnoTation$+Quote$+" at graph "+xCoord$+","+yCoord$+","+zCoord$
            PRINT #N%, "show label"
            PRINT #N%, "set grid xtics ytics ztics"
            PRINT #N%, "show grid"
            PRINT #N%, "set title "+Quote$+PlotTitle$+Quote$
            PRINT #N%, "set xlabel "+Quote$+XaxisLabel$+Quote$
            PRINT #N%, "set ylabel "+Quote$+YaxisLabel$+Quote$
            PRINT #N%, "set zlabel "+Quote$+ZaxisLabel$+Quote$
            PRINT #N%, "splot "+Quote$+PlotFileName$+Quote$+A$  '" notitle with linespoints" 'A$'" notitle with lines"
        CLOSE #N%

        SHELL(GnuPlotEXE$+" cmd3d.gp -")

    END SUB 'ContourPlot()

'---

    SUB ThreeDplot(PlotFileName$,PlotTitle$,Annotation$,xCoord$,yCoord$,zCoord$, _
                XaxisLabel$,YaxisLabel$,ZaxisLabel$,zMin$,zMax$,GnuPlotEXE$,A$)

        LOCAL N%, ProcessID???

        N% = FREEFILE

        OPEN "cmd3d.gp" FOR OUTPUT AS #N%

            PRINT #N%, "set pm3d"
            PRINT #N%, "show pm3d"
            IF zMin$ <> "" AND zMax$ <> "" THEN  PRINT #N%, "set zrange ["+zMin$+":"+zMax$+"]"
            PRINT #N%, "set label "+Quote$+AnnoTation$+Quote$+" at graph "+xCoord$+","+yCoord$+","+zCoord$
            PRINT #N%, "show label"
            PRINT #N%, "set grid xtics ytics ztics"
            PRINT #N%, "show grid"
            PRINT #N%, "set title "+Quote$+PlotTitle$+Quote$
            PRINT #N%, "set xlabel "+Quote$+XaxisLabel$+Quote$
            PRINT #N%, "set ylabel "+Quote$+YaxisLabel$+Quote$
            PRINT #N%, "set zlabel "+Quote$+ZaxisLabel$+Quote$
            PRINT #N%, "splot "+Quote$+PlotFileName$+Quote$+A$+" notitle"' with lines"
        CLOSE #N%

        SHELL(GnuPlotEXE$+" cmd3d.gp -") : CALL Delay(1##)

    END SUB 'ThreeDplot()

'---

    SUB ThreeDplot2(PlotFileName$,PlotTitle$,Annotation$,xCoord$,yCoord$,zCoord$, _
                 XaxisLabel$,YaxisLabel$,ZaxisLabel$,zMin$,zMax$,GnuPlotEXE$,A$,xStart$,xStop$,yStart$,yStop$)

        LOCAL N%

        N% = FREEFILE

        OPEN "cmd3d.gp" FOR OUTPUT AS #N%

            PRINT #N%, "set pm3d"
            PRINT #N%, "show pm3d"
            PRINT #N%, "set hidden3d"
            PRINT #N%, "set view 45, 45, 1, 1"

            IF zMin$ <> "" AND zMax$ <> "" THEN  PRINT #N%, "set zrange ["+zMin$+":"+zMax$+"]"

            PRINT #N%, "set xrange [" + xStart$ + ":" + xStop$ + "]"
            PRINT #N%, "set yrange [" + yStart$ + ":" + yStop$ + "]"

            PRINT #N%, "set label "   + Quote$  + AnnoTation$ + Quote$+" at graph "+xCoord$+","+yCoord$+","+zCoord$
            PRINT #N%, "show label"
            PRINT #N%, "set grid xtics ytics ztics"
            PRINT #N%, "show grid"
            PRINT #N%, "set title "  + Quote$+PlotTitle$    + Quote$
            PRINT #N%, "set xlabel " + Quote$+XaxisLabel$   + Quote$
            PRINT #N%, "set ylabel " + Quote$+YaxisLabel$   + Quote$
            PRINT #N%, "set zlabel " + Quote$+ZaxisLabel$   + Quote$
            PRINT #N%, "splot "      + Quote$+PlotFileName$ + Quote$ + A$ + " notitle with lines"
        CLOSE #N%

        SHELL(GnuPlotEXE$+" cmd3d.gp -")

    END SUB 'ThreeDplot2()

'---

     SUB TwoDplot2Curves(PlotFileName1$,PlotFileName2$,PlotTitle$,Annotation$,xCoord$,yCoord$,XaxisLabel$,YaxisLabel$, _
                    LogXaxis$,LogYaxis$,xMin$,xMax$,yMin$,yMax$,xTics$,yTics$,GnuPlotEXE$,LineSize)

        LOCAL N%, ProcessID???

        N% = FREEFILE

        OPEN "cmd2d.gp" FOR OUTPUT AS #N%
            'print #N%, "set output "+Quote$+"test.plt"+Quote$ 'tried this 3/11/06, didn't work...

            IF LogXaxis$ = "YES" AND LogYaxis$ = "NO"  THEN PRINT #N%, "set logscale x"
            IF LogXaxis$ = "NO"  AND LogYaxis$ = "YES" THEN PRINT #N%, "set logscale y"
            IF LogXaxis$ = "YES" AND LogYaxis$ = "YES" THEN PRINT #N%, "set logscale xy"

            IF xMin$ <> "" AND xMax$ <> "" THEN  PRINT #N%, "set xrange ["+xMin$+":"+xMax$+"]"

            IF yMin$ <> "" AND yMax$ <> "" THEN  PRINT #N%, "set yrange ["+yMin$+":"+yMax$+"]"

            PRINT #N%, "set label "+Quote$+AnnoTation$+Quote$+" at graph "+xCoord$+","+yCoord$
            PRINT #N%, "set grid xtics"
            PRINT #N%, "set grid ytics"
            PRINT #N%, "set xtics "+xTics$
            PRINT #N%, "set ytics "+yTics$
            PRINT #N%, "set grid mxtics"
            PRINT #N%, "set grid mytics"
            PRINT #N%, "set title "+Quote$+PlotTitle$+Quote$
            PRINT #N%, "set xlabel "+Quote$+XaxisLabel$+Quote$
            PRINT #N%, "set ylabel "+Quote$+YaxisLabel$+Quote$

            PRINT #N%, "plot "+Quote$+PlotFileName1$+Quote$+" with lines linewidth "+REMOVE$(STR$(LineSize),ANY" ")+","+ _
                               Quote$+PlotFileName2$+Quote$+" with points pointsize 0.05"'+REMOVE$(STR$(LineSize),ANY" ")

        CLOSE #N%

        ProcessID??? = SHELL(GnuPlotEXE$+" cmd2d.gp -") : CALL Delay(1##)

    END SUB 'TwoDplot2Curves()

'---
```

```
    SUB Probe2Dplots(ProbePlotsFileList$,PlotTitle$,Annotation$,xCoord$,yCoord$,XaxisLabel$,YaxisLabel$, _
                     LogXaxis$,LogYaxis$,xMin$,xMax$,yMin$,yMax$,xTics$,yTics$,GnuPlotEXE$)

        LOCAL N%, ProcessID???

        N% = FREEFILE

        OPEN "cmd2d.gp" FOR OUTPUT AS #N%

            IF LogXaxis$ = "YES" AND LogYaxis$ = "NO"  THEN PRINT #N%, "set logscale x"
            IF LogXaxis$ = "NO"  AND LogYaxis$ = "YES" THEN PRINT #N%, "set logscale y"
            IF LogXaxis$ = "YES" AND LogYaxis$ = "YES" THEN PRINT #N%, "set logscale xy"

            IF xMin$ <> "" AND xMax$ <> "" THEN  PRINT #N%, "set xrange ["+xMin$+":"+xMax$+"]"

            IF yMin$ <> "" AND yMax$ <> "" THEN  PRINT #N%, "set yrange ["+yMin$+":"+yMax$+"]"

            PRINT #N%, "set label "+Quote$+AnnoTation$+Quote$+" at graph "+xCoord$+","+yCoord$
            PRINT #N%, "set grid xtics"
            PRINT #N%, "set grid ytics"
            PRINT #N%, "set xtics "+xTics$
            PRINT #N%, "set ytics "+yTics$
            PRINT #N%, "set grid mxtics"
            PRINT #N%, "set grid mytics"
            PRINT #N%, "set title "+Quote$+PlotTitle$+Quote$
            PRINT #N%, "set xlabel "+Quote$+XaxisLabel$+Quote$
            PRINT #N%, "set ylabel "+Quote$+YaxisLabel$+Quote$

            PRINT #N%, ProbePlotsFileList$

        CLOSE #N%

        ProcessID??? = SHELL(GnuPlotEXE$+" cmd2d.gp -") : CALL Delay(1##)

    END SUB 'Probe2Dplots()
'---
SUB Show2Dprobes(R(),Np%,Nt&,j&,Frep,BestFitness,BestProbeNumber%,BestTimeStep&,FunctionName$,RepositionFactor$,Gamma)

    LOCAL N%, p%

    LOCAL A$, PlotFileName$, PlotTitle$, Symbols$

    LOCAL xMin$, xMax$, yMin$, yMax$

    LOCAL s1, s2, s3, s4 AS EXT

    PlotFileName$ = "Probes("+REMOVE$(STR$(j&),ANY" ")+")"

    IF j& > 0 THEN 'PLOT PROBES AT THIS TIME STEP

        PlotTitle$ = "\nLOCATIONS OF "+REMOVE$(STR$(Np%),ANY" ") + " PROBES AT TIME STEP" + STR$(j&) + " / " + REMOVE$(STR$(Nt&),ANY" ") + "\n" + _
                     "Fitness = "+REMOVE$(STR$(ROUND(BestFitness,3)),ANY" ") + ", Probe #" + REMOVE$(STR$(BestProbeNumber%),ANY" ") + " at Step #" +
REMOVE$(STR$(BestTimeStep&),ANY" ") + _
                     "  [Frep = "+REMOVE$(STR$(Frep,4),ANY" ") + " " + RepositionFactor$ + "]\n"

    ELSE 'PLOT INITIAL PROBE DISTRIBUTION

        PlotTitle$ = "\nLOCATIONS OF "+REMOVE$(STR$(Np%),ANY" ") + " INITIAL PROBES FOR " + FunctionName$ + " FUNCTION\n[gamma =
"+STR$(ROUND(Gamma,3))+"]\n"

    END IF

    N% = FREEFILE : OPEN PlotFileName$ FOR OUTPUT AS #N%

        FOR p% = 1 TO Np% : PRINT #N%, USING$("######.#####    ######.#####",R(p%,1,j&),R(p%,2,j&)) : NEXT p%

    CLOSE #N%

    s1 = 1.1## : s2 = 1.1## : s3 = 1.1## : s4 = 1.1## 'expand plots axes by 10%

    IF XiMin(1) > 0## THEN s1 = 0.9##
    IF XiMax(1) < 0## THEN s2 = 0.9##
    IF XiMin(2) > 0## THEN s3 = 0.9##
    IF XiMax(2) < 0## THEN s4 = 0.9##

    xMin$ = REMOVE$(STR$(s1*XiMin(1),2),ANY" ")
    xMax$ = REMOVE$(STR$(s2*XiMax(1),2),ANY" ")
    yMin$ = REMOVE$(STR$(s3*XiMin(2),2),ANY" ")
    yMax$ = REMOVE$(STR$(s4*XiMax(2),2),ANY" ")

    CALL TwoDplot(PlotFileName$,PlotTitle$,"0.6","0.7","x1\n\n","\nx2","NO","NO",xMin$,xMax$,yMin$,yMax$,"5","5","wgnuplot.exe"," pointsize 1 linewidth
2","")

    KILL PlotFileName$ 'erase plot data file after probes have been displayed

END SUB 'Show2Dprobes()
'---
SUB Show3Dprobes(R(),Np%,Nd%,Nt&,j&,Frep,BestFitness,BestProbeNumber%,BestTimeStep&,FunctionName$,RepositionFactor$,Gamma)

    LOCAL N%, p%, PlotWindowULC_X%, PlotWindowULC_Y%, PlotWindowWidth%, PlotWindowHeight%, PlotWindowOffset%

    LOCAL A$, PlotFileName$, PlotTitle$, Symbols$, Annotation$

    LOCAL xMin$, xMax$, yMin$, yMax$, zMin$, zMax$

    LOCAL s1, s2, s3, s4, s5, s6 AS EXT

    PlotFileName$ = "Probes("+REMOVE$(STR$(j&),ANY" ")+")"

    IF j& > 0 THEN 'PLOT PROBES AT THIS TIME STEP

        PlotTitle$ = "\nLOCATIONS OF "+REMOVE$(STR$(Np%),ANY" ") + " PROBES AT TIME STEP" + STR$(j&) + " / " + REMOVE$(STR$(Nt&),ANY" ") + "\n" + _
                     "Fitness = "+REMOVE$(STR$(ROUND(BestFitness,3)),ANY" ") + ", Probe #" + REMOVE$(STR$(BestProbeNumber%),ANY" ") + " at Step #" +
REMOVE$(STR$(BestTimeStep&),ANY" ") + _
                     "  [Frep = "+REMOVE$(STR$(Frep,4),ANY" ") + " " + RepositionFactor$ + "]\n"

    ELSE 'PLOT INITIAL PROBE DISTRIBUTION

        A$ = "" : IF Gamma > 0## AND Gamma < 1## THEN A$ = "0"

        PlotTitle$ = "\n"+REMOVE$(STR$(Np%),ANY" ") + "-PROBE IPD FOR FUNCTION " + FunctionName$ + ", GAMMA = "+A$+REMOVE$(STR$(ROUND(Gamma,3)),ANY" ")

    END IF

'   ---------------- Probe Coordinates -----------------

    N% = FREEFILE : OPEN PlotFileName$ FOR OUTPUT AS #N%

        PRINT #N%, USING$("######.#####     ######.#####     ######.#####",R(1,1,j&),R(1,2,j&),R(1,3,j&)) 'This line repeats Probe #1's coordinates.  It's
necessary
                                                                                                                'to deal with a plotting artifact in Gnuplot!
        FOR p% = 1 TO Np% : PRINT #N%, USING$("######.#####    ######.#####    ######.#####",R(p%,1,j&),R(p%,2,j&),R(p%,3,j&)) : NEXT p%

    CLOSE #N%
```

```
'   ------------ Principal Diagonal -------------

    N% = FREEFILE : OPEN "diag" FOR OUTPUT AS #N%

        PRINT #N%, USING$("######.#####    ######.#####    ######.#####",XiMin(1),XiMin(2),XiMin(3))
        PRINT #N%, ""
        PRINT #N%, USING$("######.#####    ######.#####    ######.#####",XiMax(1),XiMax(2),XiMax(3))

    CLOSE #N%

'   ------------------- Probe Line #1 ------------------

    N% = FREEFILE : OPEN "probeline1" FOR OUTPUT AS #N%

        PRINT #N%, USING$("######.#####    ######.#####    ######.#####",R(1,1,j&),R(1,2,j&),R(1,3,j&))
        PRINT #N%, ""
        PRINT #N%, USING$("######.#####    ######.#####    ######.#####",R(Np%/Nd%,1,j&),R(Np%/Nd%,2,j&),R(Np%/Nd%,3,j&))

    CLOSE #N%

'   ------------------- Probe Line #2 ------------------

    N% = FREEFILE : OPEN "probeline2" FOR OUTPUT AS #N%

        PRINT #N%, USING$("######.#####    ######.#####    ######.#####",R(1+Np%/Nd%,1,j&),R(1+Np%/Nd%,2,j&),R(1+Np%/Nd%,3,j&))
        PRINT #N%, ""
        PRINT #N%, USING$("######.#####    ######.#####    ######.#####",R(2*Np%/Nd%,1,j&),R(2*Np%/Nd%,2,j&),R(2*Np%/Nd%,3,j&))

    CLOSE #N%

'   ------------------- Probe Line #3 ------------------

    N% = FREEFILE : OPEN "probeline3" FOR OUTPUT AS #N%

        PRINT #N%, USING$("######.#####    ######.#####    ######.#####",R(1+2*Np%/Nd%,1,j&),R(1+2*Np%/Nd%,2,j&),R(1+2*Np%/Nd%,3,j&))
        PRINT #N%, ""
        PRINT #N%, USING$("######.#####    ######.#####    ######.#####",R(3*Np%/Nd%,1,j&),R(3*Np%/Nd%,2,j&),R(3*Np%/Nd%,3,j&))

    CLOSE #N%

'   -------- RE-PLOT PROBE #1 BECAUSE OF SOME ARTIFACT THAT DROPS IT FROM PROBE LINE #1 ?????? ------------------

    N% = FREEFILE : OPEN "probe1" FOR OUTPUT AS #N%

        PRINT #N%, USING$("######.#####    ######.#####    ######.#####",R(1,1,j&),R(1,2,j&),R(1,3,j&))
        PRINT #N%, USING$("######.#####    ######.#####    ######.#####",R(1,1,j&),R(1,2,j&),R(1,3,j&))

    CLOSE #N%

    s1 = 1.1## : s2 = s1 : s3 = s1 : s4 = s1 : s5 = s1 : s6 = s1 'expand plots axes by 10%

    IF XiMin(1) > 0## THEN s1 = 0.9##
    IF XiMax(1) < 0## THEN s2 = 0.9##
    IF XiMin(2) > 0## THEN s3 = 0.9##
    IF XiMax(2) < 0## THEN s4 = 0.9##
    IF XiMin(3) > 0## THEN s5 = 0.9##
    IF XiMax(3) < 0## THEN s6 = 0.9##

    xMin$ = REMOVE$(STR$(s1*XiMin(1),2),ANY" ")
    xMax$ = REMOVE$(STR$(s2*XiMax(1),2),ANY" ")
    yMin$ = REMOVE$(STR$(s3*XiMin(2),2),ANY" ")
    yMax$ = REMOVE$(STR$(s4*XiMax(2),2),ANY" ")
    zMin$ = REMOVE$(STR$(s5*XiMin(2),2),ANY" ")
    zMax$ = REMOVE$(STR$(s6*XiMax(2),2),ANY" ")

'USAGE: CALL ThreeDplot3(PlotFileName$,PlotTitle$,Annotation$,xCoord$,yCoord$,zCoord$, _
'                        XaxisLabel$,YaxisLabel$,ZaxisLabel$,zMin$,zMax$,GnuPlotEXE$,xStart$,xStop$,yStart$,yStop$)

    PlotWindowULC_X% = 50 : PlotWindowULC_Y% = 50 : PlotWindowWidth% = 1000 : PlotWindowHeight% = 800

    PlotWindowOffset% = 100*Gamma

    CALL CreateGNUplotINIfile(PlotWindowULC_X%+PlotWindowOffset%,PlotWindowULC_Y%+PlotWindowOffset%,PlotWindowWidth%,PlotWindowHeight%)

    CALL ThreeDplot3(PlotFileName$,PlotTitle$,Annotation$,"0.6","0.7","0.8", _
                     "x1","x2","x3",zMin$,zMax$,"wgnuplot.exe",xMin$,xMax$,yMin$,yMax$)

    'KILL PlotFileName$ 'erase plot data file after probes have been displayed

END SUB 'Show3Dprobes()

'---

    SUB ThreeDplot3(PlotFileName$,PlotTitle$,Annotation$,xCoord$,yCoord$,zCoord$, _
                    XaxisLabel$,YaxisLabel$,ZaxisLabel$,zMin$,zMax$,GnuPlotEXE$,xStart$,xStop$,yStart$,yStop$)

        LOCAL N%, ProcID???

        N% = FREEFILE

        OPEN "cmd3d.gp" FOR OUTPUT AS #N%

            PRINT #N%, "set pm3d"
            PRINT #N%, "show pm3d"
            PRINT #N%, "set hidden3d"
'           PRINT #N%, "set view 45, 45, 1, 1"

            PRINT #N%, "set view 45, 60, 1, 1"

            IF zMin$ <> "" AND zMax$ <> "" THEN  PRINT #N%, "set zrange ["+zMin$+":"+zMax$+"]"

            PRINT #N%, "set xrange [" + xStart$ + ":" + xStop$ + "]"
            PRINT #N%, "set yrange [" + yStart$ + ":" + yStop$ + "]"

            PRINT #N%, "set label "   + Quote$  + AnnoTation$ + Quote$+" at graph "+xCoord$+","+yCoord$+","+zCoord$
            PRINT #N%, "show label"
            PRINT #N%, "set grid xtics ytics ztics"
            PRINT #N%, "show grid"
            PRINT #N%, "set title "  + Quote$+PlotTitle$     + Quote$
            PRINT #N%, "set xlabel " + Quote$+XaxisLabel$    + Quote$
            PRINT #N%, "set ylabel " + Quote$+YaxisLabel$    + Quote$
            PRINT #N%, "set zlabel " + Quote$+ZaxisLabel$    + Quote$
            PRINT #N%, "unset colorbox"
'           print #N%, "set style fill"

            PRINT #N%, "splot "      + Quote$+PlotFileName$  + Quote$ + " notitle lw 1 pt 8," _
                                     + Quote$ + "diag"       + Quote$ + " notitle w l," _
                                     + Quote$ + "probeline1" + Quote$ + " notitle w l," _
                                     + Quote$ + "probeline2" + Quote$ + " notitle w l," _
                                     + Quote$ + "probeline3" + Quote$ + " notitle w l"
        CLOSE #N%

        ProcID??? = SHELL(GnuPlotEXE$+" cmd3d.gp -")

        CALL Delay(1##)

    END SUB 'ThreeDplot3()

'----
```

```
    SUB TwoDplot(PlotFileName$,PlotTitle$,xCoord$,yCoord$,XaxisLabel$,YaxisLabel$, _
             LogXaxis$,LogYaxis$,xMin$,xMax$,yMin$,yMax$,xTics$,yTics$,GnuPlotEXE$,LineType$,Annotation$)

        LOCAL N%, ProcessID???

        N% = FREEFILE

        OPEN "cmd2d.gp" FOR OUTPUT AS #N%

            IF LogXaxis$ = "YES" AND LogYaxis$ = "NO"  THEN PRINT #N%, "set logscale x"
            IF LogXaxis$ = "NO"  AND LogYaxis$ = "YES" THEN PRINT #N%, "set logscale y"
            IF LogXaxis$ = "YES" AND LogYaxis$ = "YES" THEN PRINT #N%, "set logscale xy"

            IF xMin$ <> "" AND xMax$ <> "" THEN  PRINT #N%, "set xrange ["+xMin$+":"+xMax$+"]"
            IF yMin$ <> "" AND yMax$ <> "" THEN  PRINT #N%, "set yrange ["+yMin$+":"+yMax$+"]"

            PRINT #N%, "set label "      + Quote$ + Annotation$ + Quote$ + " at graph " + xCoord$ + "," + yCoord$
            PRINT #N%, "set grid xtics " + XTics$
            PRINT #N%, "set grid ytics " + yTics$
            PRINT #N%, "set grid mxtics"
            PRINT #N%, "set grid mytics"
            PRINT #N%, "show grid"
            PRINT #N%, "set title "  + Quote$+PlotTitle$+Quote$
            PRINT #N%, "set xlabel " + Quote$+XaxisLabel$+Quote$
            PRINT #N%, "set ylabel " + Quote$+YaxisLabel$+Quote$

            PRINT #N%, "plot "+Quote$+PlotFileName$+Quote$+" notitle"+LineType$

        CLOSE #N%

        ProcessID??? = SHELL(GnuPlotEXE$+" cmd2d.gp -") : CALL Delay(1##)

    END SUB 'TwoDplot()

'-----

    SUB CreateGNUplotINIfile(PlotWindowULC_X%,PlotWindowULC_Y%,PlotWindowWidth%,PlotWindowHeight%)

    LOCAL N%, WinPath$, A$, B$, WindowsDirectory$

    WinPath$ = UCASE$(ENVIRON$("Path"))'DIR$("C:\WINDOWS",23)

    DO
        B$ = A$

        A$ = EXTRACT$(WinPath$,";")

        WinPath$ = REMOVE$(WinPath$,A$+";")

        IF RIGHT$(A$,7) = "WINDOWS" OR A$ = B$ THEN EXIT LOOP

        IF RIGHT$(A$,5) = "WINNT"   OR A$ = B$ THEN EXIT LOOP

    LOOP

    WindowsDirectory$ = A$

    N% = FREEFILE

'   ----------- WGNUPLOT.INPUT FILE -----------
    OPEN WindowsDirectory$+"\wgnuplot.ini" FOR OUTPUT AS #N%

        PRINT #N%,"[WGNUPLOT]"
        PRINT #N%,"TextOrigin=0 0"
        PRINT #N%,"TextSize=640 150"
        PRINT #N%,"TextFont=Terminal,9"
        PRINT #N%,"GraphOrigin="+REMOVE$(STR$(PlotWindowULC_X%),ANY" ")+" "+REMOVE$(STR$(PlotWindowULC_Y%),ANY" ")
        PRINT #N%,"GraphSize="  +REMOVE$(STR$(PlotWindowWidth%),ANY" ")+" "+REMOVE$(STR$(PlotWindowHeight%),ANY" ")
        PRINT #N%,"GraphFont=Arial,10"
        PRINT #N%,"GraphColor=1"
        PRINT #N%,"GraphToTop=1"
        PRINT #N%,"GraphBackground=255 255 255"
        PRINT #N%,"Border=0 0 0 0 0"
        PRINT #N%,"Axis=192 192 192 2 2"
        PRINT #N%,"Line1=0 0 255 0 0"
        PRINT #N%,"Line2=0 255 0 0 1"
        PRINT #N%,"Line3=255 0 0 0 2"
        PRINT #N%,"Line4=255 0 255 0 3"
        PRINT #N%,"Line5=0 0 128 0 4"

    CLOSE #N%

    END SUB 'CreateGNUplotINIfile()

'------

    SUB Delay(NumSecs)

        LOCAL StartTime, StopTime AS EXT

        StartTime = TIMER

        DO UNTIL (StopTime-StartTime) >= NumSecs

            StopTime = TIMER

        LOOP

    END SUB 'Delay()

'-----

SUB MathematicalConstants
    EulerConst    = 0.577215664901532860606512##
    Pi            = 3.141592653589793238462643##
    Pi2           = Pi/2##
    Pi4           = Pi/4##
    TwoPi         = 2##*Pi
    FourPi        = 4##*Pi
    e             = 2.718281828459045235360287##
    Root2         = 1.4142135623730950488##
END SUB

'-----

SUB AlphabetAndDigits
    Alphabet$   = "ABCDEFGHIJKLMNOPQRSTUVWXYZabcdefghijklmnopqrstuvwxyz"
    Digits$     = "0123456789"
    RunID$      = REMOVE$(DATE$+"_"+TIME$,ANY Alphabet$+" :-/")
END SUB

'------

SUB SpecialSymbols
    Quote$             = CHR$(34) 'Quotation mark "
    SpecialCharacters$ = "'(),#:;/_"
END SUB
```

```
'-----

SUB EMconstants
    Mu0  = 4E-7##*Pi      'hy/meter
    Eps0 = 8.854##*1E-12 'fd/meter
    c    = 2.998E8##      'velocity of light, 1##/SQR(Mu0*Eps0) 'meters/sec
    eta0 = SQR(Mu0/Eps0) 'impedance of free space, ohms
END SUB

'------

SUB ConversionFactors
    Rad2Deg       = 180##/Pi
    Deg2Rad       = 1##/Rad2Deg
    Feet2Meters   = 0.3048##
    Meters2Feet   = 1##/Feet2Meters
    Inches2Meters = 0.0254##
    Meters2Inches = 1##/Inches2Meters
    Miles2Meters  = 1609.344##
    Meters2Miles  = 1##/Miles2Meters
    NautMi2Meters = 1852##
    Meters2NautMi = 1##/NautMi2Meters
END SUB

'------

SUB ShowConstants 'puts up msgbox showing all constants

LOCAL A$

A$ = _
"Mathematical Constants:"+CHR$(13)+_
"Euler const="+STR$(EulerConst)+CHR$(13)+_
"Pi="+STR$(Pi)+CHR$(13)+_
"Pi/2="+STR$(Pi2)+CHR$(13)+_
"Pi/4="+STR$(Pi4)+CHR$(13)+_
"2Pi="+STR$(TwoPi)+CHR$(13)+_
"4Pi="+STR$(FourPi)+CHR$(13)+_
"e="+STR$(e)+CHR$(13)+_
"Sqr2="+STR$(Root2)+CHR$(13)+CHR$(13)+_
"Alphabet, Digits & Special Characters:"+CHR$(13)+_
"Alphabet="+Alphabet$+CHR$(13)+_
"Digits="+Digits$+CHR$(13)+_
"quote="+Quote$+CHR$(13)+_
"Spec chars="+SpecialCharacters$+CHR$(13)+CHR$(13)+_
"E&M Constants:"+CHR$(13)+_
"Mu0="+STR$(Mu0)+CHR$(13)+_
"Eps0="+STR$(Eps0)+CHR$(13)+_
"c="+STR$(c)+CHR$(13)+_
"Eta0="+STR$(eta0)+CHR$(13)+CHR$(13)+_
"Conversion Factors:"+CHR$(13)+_
"Rad2Deg="+STR$(Rad2Deg)+CHR$(13)+_
"Deg2Rad="+STR$(Deg2Rad)+CHR$(13)+_
"Ft2meters="+STR$(Feet2Meters)+CHR$(13)+_
"Meters2Ft="+STR$(Meters2Feet)+CHR$(13)+_
"Inches2Meters="+STR$(Inches2Meters)+CHR$(13)+_
"Meters2Inches="+STR$(Meters2Inches)+CHR$(13)+_
"Miles2Meters="+STR$(Miles2Meters)+CHR$(13)+_
"Meters2Miles="+STR$(Meters2Miles)+CHR$(13)+_
"NautMi2Meters="+STR$(NautMi2Meters)+CHR$(13)+_
"Meters2NautMi="+STR$(Meters2NautMi)+CHR$(13)+CHR$(13)

MSGBOX(A$)

END SUB

'------

SUB DisplayRmatrix(Np%,Nd%,Nt&,R())

LOCAL p%, i%, j&, A$

    A$ = "Position Vector Matrix R()"+CHR$(13)

    FOR p% = 1 TO Np%

        FOR i% = 1 TO Nd%

            FOR j& = 0 TO Nt&

                A$ = A$ + "R("+STR$(p%)+", "+STR$(i%)+", "+STR$(j&)+ ") ="+STR$(R(p%,i%,j&)) + CHR$(13)

            NEXT j&

        NEXT i%

    NEXT p%

    MSGBOX(A$)

END SUB

'------

SUB DisplayRmatrixThisTimeStep(Np%,Nd%,j&,R(),Gamma)

LOCAL p%, i%, A$, B$

    A$ = "Position Vector Matrix R() at step "+STR$(j&)+", Gamma ="+STR$(Gamma)+":"+CHR$(13)+CHR$(13)

    FOR p% = 1 TO Np%

        A$ = A$ + "Probe#"+REMOVE$(STR$(p%),ANY" ")+": "

        B$ = ""

        FOR i% = 1 TO Nd%

             B$ = B$ + "  " + USING$("####.##",R(p%,i%,j&))

        NEXT i%

        A$ = A$ + B$ + CHR$(13)

    NEXT p%

    MSGBOX(A$)

END SUB

'------

SUB DisplayAmatrix(Np%,Nd%,Nt&,A())

LOCAL p%, i%, j&, A$

    A$ = "Acceleration Vector Matrix A()"+CHR$(13)

    FOR p% = 1 TO Np%
```

```
        FOR i% = 1 TO Nd%
            FOR j& = 0 TO Nt&
                A$ = A$ + "A("+STR$(p%)+", "+STR$(i%)+", "+STR$(j&)+ ") ="+STR$(A(p%,i%,j&)) + CHR$(13)
            NEXT j&
        NEXT i%
    NEXT p%
    MSGBOX(A$)
END SUB
'------
SUB DisplayAmatrixThisTimeStep(Np%,Nd%,j&,A())
LOCAL p%, i%, A$
    A$ = "Acceleration matrix A() at step "+STR$(j&)+":"+CHR$(13)
    FOR p% = 1 TO Np%
        FOR i% = 1 TO Nd%
            A$ = A$ + "A("+STR$(p%)+", "+STR$(i%)+", "+STR$(j&)+ ") ="+STR$(A(p%,i%,j&)) + CHR$(13)
        NEXT i%
    NEXT p%
    MSGBOX(A$)
END SUB
'------
SUB DisplayMmatrix(Np%,Nt&,M())
LOCAL p%, j&, A$
    A$ = "Fitness Matrix M()"+CHR$(13)
    FOR p% = 1 TO Np%
       FOR j& = 0 TO Nt&
          A$ = A$ + "M("+STR$(p%)+", "+STR$(j&)+ ") ="+STR$(M(p%,j&)) + CHR$(13)
       NEXT j&
    NEXT p%
    MSGBOX(A$)
END SUB
'------
SUB DisplayMbestMatrix(Np%,Nt&,Mbest())
LOCAL p%, j&, A$
    A$ = "Np= "+STR$(Np%)+"  Nt="+STR$(Nt&)+CHR$(13)+"Fitness Matrix Mbest()"+CHR$(13)
    FOR p% = 1 TO Np%
       FOR j& = 0 TO Nt&
          A$ = A$ + "Mbest("+STR$(p%)+", "+STR$(j&)+ ") ="+STR$(Mbest(p%,j&)) + CHR$(13)
       NEXT j&
    NEXT p%
    MSGBOX(A$)
END SUB
'------
SUB DisplayMmatrixThisTimeStep(Np%,j&,M())
LOCAL p%, A$
    A$ = "Fitness matrix M() at step "+STR$(j&)+":"+CHR$(13)
    FOR p% = 1 TO Np%
          A$ = A$ + "M("+STR$(p%)+", "+STR$(j&)+ ") ="+STR$(M(p%,j&)) + CHR$(13)
    NEXT p%
    MSGBOX(A$)
END SUB
'------
SUB DisplayXiMinMax(Nd%,XiMin(),XiMax())
LOCAL i%, A$
    A$ = ""
    FOR i% = 1 TO Nd%
        A$ = A$ + "XiMin("+STR$(i%)+" ) = "+STR$(XiMin(i%))+"   XiMax("+STR$(i%)+" ) = "+STR$(XiMax(i%)) + CHR$(13)
    NEXT i%
    MSGBOX(A$)
END SUB
'------
SUB DisplayRunParameters2(FunctionName$,Nd%,Np%,Nt&,G,DeltaT,Alpha,Beta,Frep,PlaceInitialProbes$,InitialAcceleration$,RepositionFactor$)
LOCAL A$
    A$ = "Function = "+ FunctionName$+CHR$(13)+_
         "Nd = "+STR$(Nd%)+CHR$(13)+_
         "Np = "+STR$(Np%)+CHR$(13)+_
         "Nt = "+STR$(Nt&)+CHR$(13)+_
```

```
          "G  = "+STR$(G)+CHR$(13)+_
          "DeltaT = "+STR$(DeltaT)+CHR$(13)+_
          "Alpha = "+STR$(Alpha)+CHR$(13)+_
          "Beta  = "+STR$(Beta)+CHR$(13)+_
          "Frep  = "+STR$(Frep)+CHR$(13)+_
          "Init Probes: "+PlaceInitialProbes$+CHR$(13)+_
          "Init Accel:  "+InitialAcceleration$+CHR$(13)+_
          "Retrive Method: "+RepositionFactor$+CHR$(13)

    MSGBOX(A$)

END SUB

'------

SUB
Tabulate1DprobeCoordinates(Max1DprobesPlotted%,Nd%,Np%,LastStep&,G,DeltaT,Alpha,Beta,Frep,R(),M(),PlaceInitialProbes$,InitialAcceleration$,RepositionFactor
$,FunctionName$,Gamma)

LOCAL N%, ProbeNum%, FileHeader$, A$, B$, C$, D$, E$, F$, H$, StepNum&, FieldNumber% 'kludgy, yes, but it accomplishes its purpose...

        CALL
GetPlotAnnotation(FileHeader$,Nd%,Np%,LastStep&,G,DeltaT,Alpha,Beta,Frep,M(),PlaceInitialProbes$,InitialAcceleration$,RepositionFactor$,FunctionName$,Gamma
)

        REPLACE "\n" WITH ", " IN FileHeader$

        FileHeader$ = LEFT$(FileHeader$,LEN(FileHeader$)-2)

        FileHeader$ = "PROBE COORDINATES" + CHR$(13) +_
                      "-----------------" + CHR$(13) + FileHeader$

        N% = FREEFILE : OPEN "ProbeCoordinates.DAT" FOR OUTPUT AS #N%

            A$ = "   Step #     " : B$ = "   ------     " : C$ = ""

            FOR ProbeNum% = 1 TO Np% 'create out data file header

                SELECT CASE ProbeNum%
                    CASE   1 TO   9 : E$ = ""   : F$ = "            " : H$ = "            "
                    CASE  10 TO  99 : E$ = "-"  : F$ = "           "  : H$ = "           "
                    CASE 100 TO 999 : E$ = "--" : F$ = "          "   : H$ = "          "
                END SELECT

                A$ = A$ + "P" + NoSpaces$(ProbeNum%+0,0) + F$ 'note: adding zero to ProbeNum% necessary to convert to floating point...

                B$ = B$ + E$ + "--" + H$

                C$ = C$ + "######.###    "

'                C$ = C$ + "##.#######"

            NEXT ProbeNum%

            PRINT #N%, FileHeader$ + CHR$(13) : PRINT #N%, A$ : PRINT #N%, B$

            FOR StepNum& = 0 TO LastStep&

                D$ = USING$("######   ",StepNum&)

                FOR ProbeNum% = 1 TO Np% : D$ = D$ + USING$(C$,R(ProbeNum%,1,StepNum&)) : NEXT ProbeNum%

                PRINT #N%, D$

            NEXT StepNum&

        CLOSE #N%

END SUB 'Tabulate1DprobeCoordinates()

'------

SUB
Plot1DprobePositions(Max1DprobesPlotted%,Nd%,Np%,LastStep&,G,DeltaT,Alpha,Beta,Frep,R(),M(),PlaceInitialProbes$,InitialAcceleration$,RepositionFactor$,Func
tionName$,Gamma)
   'plots on-screen 1D function probe positions vs time step if Np =< 10

LOCAL ProcessID???, N%, n1%, n2%, n3%, n4%, n5%, n6%, n7%, n8%, n9%, n10%, n11%, n12%, n13%, n14%, n15%, ProbeNum%, StepNum&, A$

LOCAL PlotAnnotation$

    IF Np% > Max1DprobesPlotted% THEN EXIT SUB

    CALL CLEANUP 'delete old "Px" plot files, if any

    ProbeNum% = 0

    DO 'create output data files, probe-by-probe
        INCR ProbeNum% : n1%  = FREEFILE : OPEN "P"+REMOVE$(STR$(ProbeNum%),ANY" ") FOR OUTPUT AS #n1%  : IF ProbeNum% = Np% THEN EXIT LOOP
        INCR ProbeNum% : n2%  = FREEFILE : OPEN "P"+REMOVE$(STR$(ProbeNum%),ANY" ") FOR OUTPUT AS #n2%  : IF ProbeNum% = Np% THEN EXIT LOOP
        INCR ProbeNum% : n3%  = FREEFILE : OPEN "P"+REMOVE$(STR$(ProbeNum%),ANY" ") FOR OUTPUT AS #n3%  : IF ProbeNum% = Np% THEN EXIT LOOP
        INCR ProbeNum% : n4%  = FREEFILE : OPEN "P"+REMOVE$(STR$(ProbeNum%),ANY" ") FOR OUTPUT AS #n4%  : IF ProbeNum% = Np% THEN EXIT LOOP
        INCR ProbeNum% : n5%  = FREEFILE : OPEN "P"+REMOVE$(STR$(ProbeNum%),ANY" ") FOR OUTPUT AS #n5%  : IF ProbeNum% = Np% THEN EXIT LOOP
        INCR ProbeNum% : n6%  = FREEFILE : OPEN "P"+REMOVE$(STR$(ProbeNum%),ANY" ") FOR OUTPUT AS #n6%  : IF ProbeNum% = Np% THEN EXIT LOOP
        INCR ProbeNum% : n7%  = FREEFILE : OPEN "P"+REMOVE$(STR$(ProbeNum%),ANY" ") FOR OUTPUT AS #n7%  : IF ProbeNum% = Np% THEN EXIT LOOP
        INCR ProbeNum% : n8%  = FREEFILE : OPEN "P"+REMOVE$(STR$(ProbeNum%),ANY" ") FOR OUTPUT AS #n8%  : IF ProbeNum% = Np% THEN EXIT LOOP
        INCR ProbeNum% : n9%  = FREEFILE : OPEN "P"+REMOVE$(STR$(ProbeNum%),ANY" ") FOR OUTPUT AS #n9%  : IF ProbeNum% = Np% THEN EXIT LOOP
        INCR ProbeNum% : n10% = FREEFILE : OPEN "P"+REMOVE$(STR$(ProbeNum%),ANY" ") FOR OUTPUT AS #n10% : IF ProbeNum% = Np% THEN EXIT LOOP
        INCR ProbeNum% : n11% = FREEFILE : OPEN "P"+REMOVE$(STR$(ProbeNum%),ANY" ") FOR OUTPUT AS #n11% : IF ProbeNum% = Np% THEN EXIT LOOP
        INCR ProbeNum% : n12% = FREEFILE : OPEN "P"+REMOVE$(STR$(ProbeNum%),ANY" ") FOR OUTPUT AS #n12% : IF ProbeNum% = Np% THEN EXIT LOOP
        INCR ProbeNum% : n13% = FREEFILE : OPEN "P"+REMOVE$(STR$(ProbeNum%),ANY" ") FOR OUTPUT AS #n13% : IF ProbeNum% = Np% THEN EXIT LOOP
        INCR ProbeNum% : n14% = FREEFILE : OPEN "P"+REMOVE$(STR$(ProbeNum%),ANY" ") FOR OUTPUT AS #n14  : IF ProbeNum% = Np% THEN EXIT LOOP
        INCR ProbeNum% : n15% = FREEFILE : OPEN "P"+REMOVE$(STR$(ProbeNum%),ANY" ") FOR OUTPUT AS #n15% : IF ProbeNum% = Np% THEN EXIT LOOP
    LOOP

    ProbeNum% = 0

    DO 'output probe positions as a function of time step
        INCR ProbeNum% : FOR StepNum& = 0 TO LastStep& : PRINT #n1%,  USING$("######  ######.########",StepNum&,R(ProbeNum%,1,StepNum&)) : NEXT StepNum& :
IF ProbeNum% = Np% THEN EXIT LOOP
        INCR ProbeNum% : FOR StepNum& = 0 TO LastStep& : PRINT #n2%,  USING$("######  ######.########",StepNum&,R(ProbeNum%,1,StepNum&)) : NEXT StepNum& :
IF ProbeNum% = Np% THEN EXIT LOOP
        INCR ProbeNum% : FOR StepNum& = 0 TO LastStep& : PRINT #n3%,  USING$("######  ######.########",StepNum&,R(ProbeNum%,1,StepNum&)) : NEXT StepNum& :
IF ProbeNum% = Np% THEN EXIT LOOP
        INCR ProbeNum% : FOR StepNum& = 0 TO LastStep& : PRINT #n4%,  USING$("######  ######.########",StepNum&,R(ProbeNum%,1,StepNum&)) : NEXT StepNum& :
IF ProbeNum% = Np% THEN EXIT LOOP
        INCR ProbeNum% : FOR StepNum& = 0 TO LastStep& : PRINT #n5%,  USING$("######  ######.########",StepNum&,R(ProbeNum%,1,StepNum&)) : NEXT StepNum& :
IF ProbeNum% = Np% THEN EXIT LOOP
        INCR ProbeNum% : FOR StepNum& = 0 TO LastStep& : PRINT #n6%,  USING$("######  ######.########",StepNum&,R(ProbeNum%,1,StepNum&)) : NEXT StepNum& :
IF ProbeNum% = Np% THEN EXIT LOOP
        INCR ProbeNum% : FOR StepNum& = 0 TO LastStep& : PRINT #n7%,  USING$("######  ######.########",StepNum&,R(ProbeNum%,1,StepNum&)) : NEXT StepNum& :
IF ProbeNum% = Np% THEN EXIT LOOP
        INCR ProbeNum% : FOR StepNum& = 0 TO LastStep& : PRINT #n8%,  USING$("######  ######.########",StepNum&,R(ProbeNum%,1,StepNum&)) : NEXT StepNum& :
IF ProbeNum% = Np% THEN EXIT LOOP
        INCR ProbeNum% : FOR StepNum& = 0 TO LastStep& : PRINT #n9%,  USING$("######  ######.########",StepNum&,R(ProbeNum%,1,StepNum&)) : NEXT StepNum& :
IF ProbeNum% = Np% THEN EXIT LOOP
        INCR ProbeNum% : FOR StepNum& = 0 TO LastStep& : PRINT #n10%, USING$("######  ######.########",StepNum&,R(ProbeNum%,1,StepNum&)) : NEXT StepNum& :
IF ProbeNum% = Np% THEN EXIT LOOP
```

```
        INCR ProbeNum% : FOR StepNum& = 0 TO LastStep& : PRINT #n11,  USING$("######  ######.########",StepNum&,R(ProbeNum%,1,StepNum&)) : NEXT StepNum& :
IF ProbeNum% = Np% THEN EXIT LOOP
        INCR ProbeNum% : FOR StepNum& = 0 TO LastStep& : PRINT #n12%, USING$("######  ######.########",StepNum&,R(ProbeNum%,1,StepNum&)) : NEXT StepNum& :
IF ProbeNum% = Np% THEN EXIT LOOP
        INCR ProbeNum% : FOR StepNum& = 0 TO LastStep& : PRINT #n13%, USING$("######  ######.########",StepNum&,R(ProbeNum%,1,StepNum&)) : NEXT StepNum& :
IF ProbeNum% = Np% THEN EXIT LOOP
        INCR ProbeNum% : FOR StepNum& = 0 TO LastStep& : PRINT #n14%, USING$("######  ######.########",StepNum&,R(ProbeNum%,1,StepNum&)) : NEXT StepNum& :
IF ProbeNum% = Np% THEN EXIT LOOP
        INCR ProbeNum% : FOR StepNum& = 0 TO LastStep& : PRINT #n15%, USING$("######  ######.########",StepNum&,R(ProbeNum%,1,StepNum&)) : NEXT StepNum& :
IF ProbeNum% = Np% THEN EXIT LOOP
    LOOP

    ProbeNum% = 0

    DO 'close output data files
        INCR ProbeNum% : CLOSE #n1%  : IF ProbeNum% = Np% THEN EXIT LOOP
        INCR ProbeNum% : CLOSE #n2%  : IF ProbeNum% = Np% THEN EXIT LOOP
        INCR ProbeNum% : CLOSE #n3%  : IF ProbeNum% = Np% THEN EXIT LOOP
        INCR ProbeNum% : CLOSE #n4%  : IF ProbeNum% = Np% THEN EXIT LOOP
        INCR ProbeNum% : CLOSE #n5%  : IF ProbeNum% = Np% THEN EXIT LOOP
        INCR ProbeNum% : CLOSE #n6%  : IF ProbeNum% = Np% THEN EXIT LOOP
        INCR ProbeNum% : CLOSE #n7%  : IF ProbeNum% = Np% THEN EXIT LOOP
        INCR ProbeNum% : CLOSE #n8%  : IF ProbeNum% = Np% THEN EXIT LOOP
        INCR ProbeNum% : CLOSE #n9%  : IF ProbeNum% = Np% THEN EXIT LOOP
        INCR ProbeNum% : CLOSE #n10% : IF ProbeNum% = Np% THEN EXIT LOOP
        INCR ProbeNum% : CLOSE #n11% : IF ProbeNum% = Np% THEN EXIT LOOP
        INCR ProbeNum% : CLOSE #n12% : IF ProbeNum% = Np% THEN EXIT LOOP
        INCR ProbeNum% : CLOSE #n13% : IF ProbeNum% = Np% THEN EXIT LOOP
        INCR ProbeNum% : CLOSE #n14% : IF ProbeNum% = Np% THEN EXIT LOOP
        INCR ProbeNum% : CLOSE #n15% : IF ProbeNum% = Np% THEN EXIT LOOP
    LOOP

    ProbeNum% = 0 : A$ = ""

    DO 'create file string for plot command file
        INCR ProbeNum% : A$ = A$ + Quote$ + "P"+REMOVE$(STR$(ProbeNum%),ANY" ") + Quote$ + " w l lw 2, " : IF ProbeNum% = Np% THEN EXIT LOOP
        INCR ProbeNum% : A$ = A$ + Quote$ + "P"+REMOVE$(STR$(ProbeNum%),ANY" ") + Quote$ + " w l lw 2, " : IF ProbeNum% = Np% THEN EXIT LOOP
        INCR ProbeNum% : A$ = A$ + Quote$ + "P"+REMOVE$(STR$(ProbeNum%),ANY" ") + Quote$ + " w l lw 2, " : IF ProbeNum% = Np% THEN EXIT LOOP
        INCR ProbeNum% : A$ = A$ + Quote$ + "P"+REMOVE$(STR$(ProbeNum%),ANY" ") + Quote$ + " w l lw 2, " : IF ProbeNum% = Np% THEN EXIT LOOP
        INCR ProbeNum% : A$ = A$ + Quote$ + "P"+REMOVE$(STR$(ProbeNum%),ANY" ") + Quote$ + " w l lw 2, " : IF ProbeNum% = Np% THEN EXIT LOOP
        INCR ProbeNum% : A$ = A$ + Quote$ + "P"+REMOVE$(STR$(ProbeNum%),ANY" ") + Quote$ + " w l lw 2, " : IF ProbeNum% = Np% THEN EXIT LOOP
        INCR ProbeNum% : A$ = A$ + Quote$ + "P"+REMOVE$(STR$(ProbeNum%),ANY" ") + Quote$ + " w l lw 2, " : IF ProbeNum% = Np% THEN EXIT LOOP
        INCR ProbeNum% : A$ = A$ + Quote$ + "P"+REMOVE$(STR$(ProbeNum%),ANY" ") + Quote$ + " w l lw 2, " : IF ProbeNum% = Np% THEN EXIT LOOP
        INCR ProbeNum% : A$ = A$ + Quote$ + "P"+REMOVE$(STR$(ProbeNum%),ANY" ") + Quote$ + " w l lw 2, " : IF ProbeNum% = Np% THEN EXIT LOOP
        INCR ProbeNum% : A$ = A$ + Quote$ + "P"+REMOVE$(STR$(ProbeNum%),ANY" ") + Quote$ + " w l lw 2, " : IF ProbeNum% = Np% THEN EXIT LOOP
        INCR ProbeNum% : A$ = A$ + Quote$ + "P"+REMOVE$(STR$(ProbeNum%),ANY" ") + Quote$ + " w l lw 2, " : IF ProbeNum% = Np% THEN EXIT LOOP
        INCR ProbeNum% : A$ = A$ + Quote$ + "P"+REMOVE$(STR$(ProbeNum%),ANY" ") + Quote$ + " w l lw 2, " : IF ProbeNum% = Np% THEN EXIT LOOP
        INCR ProbeNum% : A$ = A$ + Quote$ + "P"+REMOVE$(STR$(ProbeNum%),ANY" ") + Quote$ + " w l lw 2, " : IF ProbeNum% = Np% THEN EXIT LOOP
        INCR ProbeNum% : A$ = A$ + Quote$ + "P"+REMOVE$(STR$(ProbeNum%),ANY" ") + Quote$ + " w l lw 2, " : IF ProbeNum% = Np% THEN EXIT LOOP
        INCR ProbeNum% : A$ = A$ + Quote$ + "P"+REMOVE$(STR$(ProbeNum%),ANY" ") + Quote$ + " w l lw 2, " : IF ProbeNum% = Np% THEN EXIT LOOP
    LOOP

    A$ = LEFT$(A$,LEN(A$)-2)

    CALL
GetPlotAnnotation(PlotAnnotation$,Nd%,Np%,LastStep&,G,DeltaT,Alpha,Beta,Frep,M(),PlaceInitialProbes$,InitialAcceleration$,RepositionFactor$,FunctionName$,G
amma)

    N% = FREEFILE

    OPEN "cmd2d.gp" FOR OUTPUT AS #N%

        PRINT #N%, "set label "      + Quote$ + PlotAnnotation$ + Quote$ + " at graph 0.5,0.95"
        PRINT #N%, "set grid xtics"
        PRINT #N%, "set grid ytics"
        PRINT #N%, "set title "  + Quote$ + "Evolution of "    + FunctionName$ + " Probe Positions"+ "\n" + RunID$ + Quote$
        PRINT #N%, "set xlabel " + Quote$ + "Time Step"        + Quote$
        PRINT #N%, "set ylabel " + Quote$ + "Probe Coordinate" + Quote$
        PRINT #N%, "plot "       + A$

    CLOSE #N%

    CALL            CreateGNUplotINIfile(0.2##*ScreenWidth&,0.2##*ScreenHeight&,0.6##*ScreenWidth&,0.6##*ScreenHeight&)            'USAGE:            CALL
CreateGNUplotINIfile(PlotWindowULC_X%,PlotWindowULC_Y%,PlotWindowWidth%,PlotWindowHeight%)

    ProcessID??? = SHELL("wgnuplot.exe"+" cmd2d.gp -") : CALL Delay(5##) 'before SUB Cleanup is called

END SUB

'------

SUB
DisplayRunParameters(FunctionName$,Nd%,Np%,Nt&,G,DeltaT,Alpha,Beta,Frep,R(),A(),M(),PlaceInitialProbes$,InitialAcceleration$,RepositionFactor$,RunCFO$,Shri
nkDS$,CheckForEarlyTermination$)

LOCAL A$, B$, YN&

    B$ = "" : IF PlaceInitialProbes$ = "UNIFORM ON-AXIS" AND Nd% > 1 THEN B$ = "  ["+REMOVE$(STR$(Np%/Nd%),ANY" ") + "/axis]"

    RunCFO$ = "NO"

 A$ = "RUN CFO WITH THE" + CHR$(13) +_
        "FOLLOWING PARAMETERS?"                        + CHR$(13) + CHR$(13) +_
        "Function "          + FunctionName$           + " (" + REMOVE$(STR$(Nd%),ANY" ") + "-D)" + CHR$(13) +_
        "# time steps = "  + REMOVE$(STR$(Nt&),ANY" ")   + CHR$(13) + _
        "Grav Const G = "  + REMOVE$(STR$(G,2),ANY" ")   + CHR$(13) + _
        "Delta-T = "       + REMOVE$(STR$(DeltaT,3),ANY" ") + CHR$(13) + _
        "Exp Alpha = "     + REMOVE$(STR$(Alpha,3),ANY" ")  + CHR$(13) + _
        "Exp Beta = "      + REMOVE$(STR$(Beta,3),ANY" ")   + CHR$(13) + _
        "Frep = "          + REMOVE$(STR$(Frep,4),ANY" ")   + " ("+RepositionFactor$ + ")" + CHR$(13) + _
        "Initial Probes: " + PlaceInitialProbes$            + CHR$(13) + _
        "Initial Accel: "  + InitialAcceleration$           + CHR$(13) + _
        "Check for Early Termination? " + CheckForEarlyTermination$ + CHR$(13) + _
        "Shrink Decision Space? "       + ShrinkDS$ + CHR$(13) +CHR$(13)

'   lResult& = MSGBOX(txt$ [, [style&], title$])

    A$ = "RUN CFO ON FUNCTION " + FunctionName$ + "?"

    YN& = MSGBOX(A$,%MB_YESNO,"CONFIRM RUN")

    IF YN& = %IDYES THEN RunCFO$ = "YES"

END SUB

'------

SUB StatusWindow(FunctionName$,StatusWindowHandle???)

    GRAPHIC     WINDOW     "Run     Progress,     "+FunctionName$,0.08##*ScreenWidth&,0.08##*ScreenHeight&,0.25##*ScreenWidth&,0.17##*ScreenHeight&     TO
StatusWindowHandle???

    GRAPHIC ATTACH StatusWindowHandle???,0,REDRAW

    GRAPHIC FONT "Lucida Console",8,0 '"Courier New",8,0 'Fixed width fonts

    GRAPHIC SET PIXEL (35,15) : GRAPHIC PRINT "  Initializing...      " : GRAPHIC REDRAW

END SUB
```

```
'------
SUB GetTestFunctionNumber(FunctionName$)
  LOCAL hDlg AS DWORD
  LOCAL N%, M%
  LOCAL FrameWidth&, FrameHeight&, BoxWidth&, BoxHeight&
' BoxWidth& = 276 : BoxHeight& = 300 : FrameWidth& = 82 : FrameHeight& = BoxHeight&-5
  BoxWidth& = 276 : BoxHeight& = 300 : FrameWidth& = 90 : FrameHeight& = BoxHeight&-5
  DIALOG NEW 0, "CENTRAL FORCE OPTIMIZATION TEST FUNCTIONS",,, BoxWidth&, BoxHeight&, %WS_CAPTION OR %WS_SYSMENU, 0 TO hDlg
'-------------------------------------------------------------------

  CONTROL ADD FRAME,  hDlg, %IDC_FRAME1,  "Test Functions",     5,   2, FrameWidth&, FrameHeight&
  CONTROL ADD FRAME,  hDlg, %IDC_FRAME2,  "GSO Test Functions",  105, 2, FrameWidth&, 255

  CONTROL ADD OPTION, hDlg, %IDC_Function_Number1,  "Parrott F4",     10,  14, 70, 10, %WS_GROUP OR %WS_TABSTOP
  CONTROL ADD OPTION, hDlg, %IDC_Function_Number2,  "SGO",            10,  24, 70, 10
  CONTROL ADD OPTION, hDlg, %IDC_Function_Number3,  "Goldstein-Price",10,  34, 70, 10
  CONTROL ADD OPTION, hDlg, %IDC_Function_Number4,  "Step",           10,  44, 70, 10
  CONTROL ADD OPTION, hDlg, %IDC_Function_Number5,  "Schwefel 2.26",  10,  54, 70, 10
  CONTROL ADD OPTION, hDlg, %IDC_Function_Number6,  "Colville",       10,  64, 70, 10
  CONTROL ADD OPTION, hDlg, %IDC_Function_Number7,  "Griewank",       10,  74, 70, 10

  CONTROL ADD OPTION, hDlg, %IDC_Function_Number31, "PBM #1",         10,  84, 70, 10
  CONTROL ADD OPTION, hDlg, %IDC_Function_Number32, "PBM #2",         10,  94, 70, 10
  CONTROL ADD OPTION, hDlg, %IDC_Function_Number33, "PBM #3",         10, 104, 70, 10
  CONTROL ADD OPTION, hDlg, %IDC_Function_Number34, "PBM #4",         10, 114, 70, 10
  CONTROL ADD OPTION, hDlg, %IDC_Function_Number35, "PBM #5",         10, 124, 70, 10
  CONTROL ADD OPTION, hDlg, %IDC_Function_Number36, "Himmelblau",     10, 134, 70, 10
  CONTROL ADD OPTION, hDlg, %IDC_Function_Number37, "Rosenbrock",     10, 144, 70, 10
  CONTROL ADD OPTION, hDlg, %IDC_Function_Number38, "Sphere",         10, 154, 70, 10
  CONTROL ADD OPTION, hDlg, %IDC_Function_Number39, "HimmelblauNLO",  10, 164, 70, 10
  CONTROL ADD OPTION, hDlg, %IDC_Function_Number40, "Tripod",         10, 174, 70, 10
  CONTROL ADD OPTION, hDlg, %IDC_Function_Number41, "Rosenbrock F6",  10, 184, 70, 10
  CONTROL ADD OPTION, hDlg, %IDC_Function_Number42, "Comp Spring",    10, 194, 70, 10
  CONTROL ADD OPTION, hDlg, %IDC_Function_Number43, "Gear Train",     10, 204, 70, 10
  CONTROL ADD OPTION, hDlg, %IDC_Function_Number44, "Loaded Bowtie",  10, 214, 70, 10
  CONTROL ADD OPTION, hDlg, %IDC_Function_Number45, "Yagi Array",     10, 224, 70, 10
  CONTROL ADD OPTION, hDlg, %IDC_Function_Number46, "Reserved",       10, 234, 70, 10
  CONTROL ADD OPTION, hDlg, %IDC_Function_Number47, "Reserved",       10, 244, 70, 10
  CONTROL ADD OPTION, hDlg, %IDC_Function_Number48, "Reserved",       10, 254, 70, 10
  CONTROL ADD OPTION, hDlg, %IDC_Function_Number49, "Reserved",       10, 264, 70, 10
  CONTROL ADD OPTION, hDlg, %IDC_Function_Number50, "Reserved",       10, 274, 70, 10

' -------------------- Test Functions from GSO Paper --------------------
  CONTROL ADD OPTION, hDlg, %IDC_Function_Number8,  "f1" , 120,  14, 40, 10
  CONTROL ADD OPTION, hDlg, %IDC_Function_Number9,  "f2" , 120,  24, 40, 10
  CONTROL ADD OPTION, hDlg, %IDC_Function_Number10, "f3" , 120,  34, 40, 10
  CONTROL ADD OPTION, hDlg, %IDC_Function_Number11, "f4" , 120,  44, 40, 10
  CONTROL ADD OPTION, hDlg, %IDC_Function_Number12, "f5" , 120,  54, 40, 10
  CONTROL ADD OPTION, hDlg, %IDC_Function_Number13, "f6" , 120,  64, 40, 10
  CONTROL ADD OPTION, hDlg, %IDC_Function_Number14, "f7" , 120,  74, 40, 10
  CONTROL ADD OPTION, hDlg, %IDC_Function_Number15, "f8" , 120,  84, 40, 10
  CONTROL ADD OPTION, hDlg, %IDC_Function_Number16, "f9" , 120,  94, 40, 10
  CONTROL ADD OPTION, hDlg, %IDC_Function_Number17, "f10", 120, 104, 40, 10
  CONTROL ADD OPTION, hDlg, %IDC_Function_Number18, "f11", 120, 114, 40, 10
  CONTROL ADD OPTION, hDlg, %IDC_Function_Number19, "f12", 120, 124, 40, 10
  CONTROL ADD OPTION, hDlg, %IDC_Function_Number20, "f13", 120, 134, 40, 10
  CONTROL ADD OPTION, hDlg, %IDC_Function_Number21, "f14", 120, 144, 40, 10
  CONTROL ADD OPTION, hDlg, %IDC_Function_Number22, "f15", 120, 154, 40, 10
  CONTROL ADD OPTION, hDlg, %IDC_Function_Number23, "f16", 120, 164, 40, 10
  CONTROL ADD OPTION, hDlg, %IDC_Function_Number24, "f17", 120, 174, 40, 10
  CONTROL ADD OPTION, hDlg, %IDC_Function_Number25, "f18", 120, 184, 40, 10
  CONTROL ADD OPTION, hDlg, %IDC_Function_Number26, "f19", 120, 194, 40, 10
  CONTROL ADD OPTION, hDlg, %IDC_Function_Number27, "f20", 120, 204, 40, 10
  CONTROL ADD OPTION, hDlg, %IDC_Function_Number28, "f21", 120, 214, 40, 10
  CONTROL ADD OPTION, hDlg, %IDC_Function_Number29, "f22", 120, 224, 40, 10
  CONTROL ADD OPTION, hDlg, %IDC_Function_Number30, "f23", 120, 234, 40, 10

  CONTROL SET OPTION  hDlg, %IDC_Function_Number1, %IDC_Function_Number1, %IDC_Function_Number3 'default to Parrott F4
'------------------------------------------------------------------------

  CONTROL ADD BUTTON, hDlg, %IDOK, "&OK", 200, 0.45##*BoxHeight&, 50, 14
'------------------------------------------------------------------------

  DIALOG SHOW MODAL hDlg CALL DlgProc
  CALL Delay(1##)
  IF FunctionNumber% < 1 OR FunctionNumber% > 45 THEN
    FunctionNumber% = 1 : MSGBOX("Error in function number...")
  END IF
    SELECT CASE FunctionNumber%
        CASE 1 : FunctionName$ = "ParrottF4"
        CASE 2 : FunctionName$ = "SGO"
        CASE 3 : FunctionName$ = "GP"
        CASE 4 : FunctionName$ = "STEP"
        CASE 5 : FunctionName$ = "SCHWEFEL_226"
        CASE 6 : FunctionName$ = "COLVILLE"
        CASE 7 : FunctionName$ = "GRIEWANK"
        CASE 8 : FunctionName$ = "F1"
        CASE 9 : FunctionName$ = "F2"
        CASE 10: FunctionName$ = "F3"
        CASE 11: FunctionName$ = "F4"
        CASE 12: FunctionName$ = "F5"
        CASE 13: FunctionName$ = "F6"
        CASE 14: FunctionName$ = "F7"
        CASE 15: FunctionName$ = "F8"
        CASE 16: FunctionName$ = "F9"
        CASE 17: FunctionName$ = "F10"
        CASE 18: FunctionName$ = "F11"
        CASE 19: FunctionName$ = "F12"
        CASE 20: FunctionName$ = "F13"
        CASE 21: FunctionName$ = "F14"
        CASE 22: FunctionName$ = "F15"
        CASE 23: FunctionName$ = "F16"
        CASE 24: FunctionName$ = "F17"
        CASE 25: FunctionName$ = "F18"
        CASE 26: FunctionName$ = "F19"
        CASE 27: FunctionName$ = "F20"
        CASE 28: FunctionName$ = "F21"
        CASE 29: FunctionName$ = "F22"
        CASE 30: FunctionName$ = "F23"
        CASE 31: FunctionName$ = "PBM_1"
        CASE 32: FunctionName$ = "PBM_2"
        CASE 33: FunctionName$ = "PBM_3"
```

```
        CASE 34: FunctionName$ = "PBM_4"
        CASE 35: FunctionName$ = "PBM_5"
        CASE 36: FunctionName$ = "HIMMELBLAU"
        CASE 37: FunctionName$ = "ROSENBROCK"
        CASE 38: FunctionName$ = "SPHERE"
        CASE 39: FunctionName$ = "HIMMELBLAUNLO"
        CASE 40: FunctionName$ = "TRIPOD"
        CASE 41: FunctionName$ = "ROSENBROCKF6"
        CASE 42: FunctionName$ = "COMPRESSIONSPRING"
        CASE 43: FunctionName$ = "GEARTRAIN"
        CASE 44: FunctionName$ = "BOWTIE"
        CASE 45: FunctionName$ = "YAGI"

    END SELECT

END SUB

'-----------

CALLBACK FUNCTION DlgProc() AS LONG

  '--------------------------------------------------------------------
  ' Callback procedure for the main dialog
  '------------------------------------------------------------------
  LOCAL c, lRes AS LONG, sText AS STRING

  SELECT CASE AS LONG CBMSG

  CASE %WM_INITDIALOG' %WM_INITDIALOG is sent right before the dialog is shown.

  CASE %WM_COMMAND              ' <- a control is calling

     SELECT CASE AS LONG CBCTL  ' <- look at control's id

     CASE %IDOK                 ' <- OK button or Enter key was pressed

         IF CBCTLMSG = %BN_CLICKED THEN
             '----------------------------------------
             ' Loop through the Function_Number controls
             ' to see which one is selected
             '----------------------------------------
             FOR c = %IDC_Function_Number1 TO %IDC_Function_Number50

                 CONTROL GET CHECK CBHNDL, c TO lRes

                 IF lRes THEN EXIT FOR

             NEXT 'c holds the id for selected test function.

             FunctionNumber% = c-120

             DIALOG END CBHNDL

         END IF

     END SELECT

  END SELECT

END FUNCTION

'----------------------------- PBM ANTENNA BENCHMARK FUNCTIONS ---------------------------

'Reference for benchmarks PBM_1 through PBM_5:

'Pantoja, M F., Bretones, A. R., Martin, R. G., "Benchmark Antenna Problems for Evolutionary
'Optimization Algorithms," IEEE Trans. Antennas & Propagation, vol. 55, no. 4, April 2007,
'pp. 1111-1121

FUNCTION PBM_1(R(),Nd%,p%,j&) 'PBM Benchmark #1: Max D for Variable-Length CF Dipole

    LOCAL Z, LengthWaves, ThetaRadians AS EXT

    LOCAL N%, Nsegs%, FeedSegNum%

    LOCAL NumSegs$, FeedSeg$, HalfLength$, Radius$, ThetaDeg$, Lyne$, GainDB$

    LengthWaves  = R(p%,1,j&)

    ThetaRadians = R(p%,2,j&)

    ThetaDeg$ = REMOVE$(STR$(ROUND(ThetaRadians*Rad2Deg,2)),ANY" ")

    IF TALLY(ThetaDeg$,".") = 0 THEN ThetaDeg$ = ThetaDeg$+"."

    Nsegs% = 2*(INT(100*LengthWaves)\2)+1 '100 segs per wavelength, must be an odd #, VOLTAGE SOURCE

    FeedSegNum% = Nsegs%\2 + 1 'center segment number, VOLTAGE SOURCE

    NumSegs$    = REMOVE$(STR$(Nsegs%),ANY" ")

    FeedSeg$    = REMOVE$(STR$(FeedSegNum%),ANY" ")

    HalfLength$ = REMOVE$(STR$(ROUND(LengthWaves/2##,6)),ANY" ")

    IF TALLY(HalfLength$,".") = 0 THEN HalfLength$ = HalfLength$+"."

    Radius$     = "0.00001" 'REMOVE$(STR$(ROUND(LengthWaves/1000##,6)),ANY" ")

    N% = FREEFILE

    OPEN "PBM1.NEC" FOR OUTPUT AS #N%

        PRINT #N%,"CM File: PBM1.NEC"
        PRINT #N%,"CM Run ID "+DATE$+" "+TIME$
        PRINT #N%,"CM Nd="+STR$(Nd%)+", p="STR$(p%)+", j="+STR$(j&)
        PRINT #N%,"CM R(p,1,j)="+STR$(R(p%,1,j&))+", R(p,2,j)="+STR$(R(p%,2,j&))
        PRINT #N%,"CE"
        PRINT #N%,"GW 1,"+NumSegs$+",0.,0.,-"+HalfLength$+",0.,0.,"+HalfLength$+","+Radius$
        PRINT #N%,"GE"
        PRINT #N%,"EX 0,1,"+FeedSeg$+",0,1.,0." 'VOLTAGE SOURCE
        PRINT #N%,"FR 0,1,0,0,299.79564,0."
        PRINT #N%,"RP 0,1,1,1001,"+ThetaDeg$+",0.,0.,0.,1000." 'gain at 1000 wavelengths range
        PRINT #N%,"XQ"
        PRINT #N%,"EN"

    CLOSE #N%

'      - - ANGLES - -           - POWER GAINS -        - - - POLARIZATION - - -    - - - E(THETA) - - -    - - - E(PHI) - - -
'  THETA     PHI        VERT.   HOR.    TOTAL      AXIAL     TILT   SENSE     MAGNITUDE    PHASE      MAGNITUDE    PHASE
' DEGREES  DEGREES       DB      DB      DB        RATIO     DEG.              VOLTS/M    DEGREES      VOLTS/M    DEGREES
'   90.00     0.00       3.91 -999.99    3.91    0.00000     0.00  LINEAR    1.29504E-04     5.37    0.00000E+00    -5.24
'123456789x123456789x123456789x123456789x123456789x123456789x123456789x123456789x123456789x123456789x123456789x123456789x
'        10        20        30        40        50        60        70        80        90       100       110       120

    SHELL "n41_2k1.exe",0

    N% = FREEFILE
```

```
    OPEN "PBM1.OUT" FOR INPUT AS #N%

        WHILE NOT EOF(N%)

            LINE INPUT #N%, Lyne$

            IF INSTR(Lyne$,"DEGREES  DEGREES") > 0 THEN EXIT LOOP

        WEND 'position at next data line

        LINE INPUT #N%, Lyne$

    CLOSE #N%

    GainDB$ = REMOVE$(MID$(Lyne$,37,8),ANY" ")

    PBM_1 = 10^(VAL(GainDB$)/10##) 'Directivity

END FUNCTION 'PBM_1()

'----

FUNCTION PBM_2(R(),Nd%,p%,j&) 'PBM Benchmark #2: Max D for Variable-Separation Array of CF Dipoles

    LOCAL Z, DipoleSeparationWaves, ThetaRadians AS EXT

    LOCAL N%, i%

    LOCAL NumSegs$, FeedSeg$, Radius$, ThetaDeg$, Lyne$, GainDB$, Xcoord$, WireNum$

    DipoleSeparationWaves = R(p%,1,j&)

    ThetaRadians          = R(p%,2,j&)

    ThetaDeg$ = REMOVE$(STR$(ROUND(ThetaRadians*Rad2Deg,2)),ANY" ")

    IF TALLY(ThetaDeg$,".") = 0 THEN ThetaDeg$ = ThetaDeg$+"."

    NumSegs$ = "49"

    FeedSeg$ = "25"

    Radius$  = "0.00001"

    N% = FREEFILE

    OPEN "PBM2.NEC" FOR OUTPUT AS #N%

        PRINT #N%,"CM File: PBM2.NEC"
        PRINT #N%,"CM Run ID "+DATE$+" "+TIME$
        PRINT #N%,"CM Nd="+STR$(Nd%)+", p="STR$(p%)+", j="+STR$(j&)
        PRINT #N%,"CM R(p,1,j)="+STR$(R(p%,1,j&))+", R(p,2,j)="+STR$(R(p%,2,j&))
        PRINT #N%,"CE"

        FOR i% = -9 TO 9 STEP 2
            WireNum$ = REMOVE$(STR$((i%+11)\2),ANY" ")
            Xcoord$  = REMOVE$(STR$(i%*DipoleSeparationWaves/2##),ANY" ")
            PRINT #N%,"GW "+WireNum$+","+NumSegs$+","+Xcoord$+",0.,-0.25,"+Xcoord$+",0.,0.25,"+Radius$
        NEXT i%

        PRINT #N%,"GE"

        FOR i% = 1 TO 10
            PRINT #N%,"EX 0,"+REMOVE$(STR$(i%),ANY" ")+","+FeedSeg$+",0,1.,0." 'VOLTAGE SOURCE
        NEXT i%
        PRINT #N%,"FR 0,1,0,0,299.79564,0."
        PRINT #N%,"RP 0,1,1,1001,"+ThetaDeg$+",90.,0.,0.,1000." 'gain at 1000 wavelengths range
        PRINT #N%,"XQ"
        PRINT #N%,"EN"

    CLOSE #N%

'      - - ANGLES - -          - POWER GAINS -        - - - POLARIZATION - - -    - - - E(THETA) - - -    - - - E(PHI) - - -
'  THETA     PHI        VERT.   HOR.    TOTAL      AXIAL     TILT   SENSE     MAGNITUDE    PHASE      MAGNITUDE    PHASE
' DEGREES  DEGREES       DB      DB      DB        RATIO     DEG.              VOLTS/M    DEGREES      VOLTS/M    DEGREES
'   90.00     0.00       3.91 -999.99    3.91    0.00000     0.00  LINEAR    1.29504E-04     5.37    0.00000E+00    -5.24
'123456789x123456789x123456789x123456789x123456789x123456789x123456789x123456789x123456789x123456789x123456789x123456789x
'        10        20        30        40        50        60        70        80        90       100       110       120

    SHELL "n41_2k1.exe",0

    N% = FREEFILE

    OPEN "PBM2.OUT" FOR INPUT AS #N%

        WHILE NOT EOF(N%)

            LINE INPUT #N%, Lyne$

            IF INSTR(Lyne$,"DEGREES  DEGREES") > 0 THEN EXIT LOOP

        WEND 'position at next data line

        LINE INPUT #N%, Lyne$

    CLOSE #N%

    GainDB$ = REMOVE$(MID$(Lyne$,37,8),ANY" ")

    IF AddNoiseToPBM2$ = "YES" THEN

        Z = 10^(VAL(GainDB$)/10##) + GaussianDeviate(0##,0.4472##) 'Directivity with Gaussian noise (zero mean, 0.2 variance)

    ELSE

        Z = 10^(VAL(GainDB$)/10##) 'Directivity without noise

    END IF

    PBM_2 = Z

END FUNCTION 'PBM_2()

'----

FUNCTION PBM_3(R(),Nd%,p%,j&) 'PBM Benchmark #3: Max D for Circular Dipole Array

    LOCAL Beta, ThetaRadians, Alpha, ReV, ImV AS EXT

    LOCAL N%, i%

    LOCAL NumSegs$, FeedSeg$, Radius$, ThetaDeg$, Lyne$, GainDB$, Xcoord$, Ycoord$, WireNum$, ReEX$, ImEX$

    Beta         = R(p%,1,j&)

    ThetaRadians = R(p%,2,j&)

    ThetaDeg$ = REMOVE$(STR$(ROUND(ThetaRadians*Rad2Deg,2)),ANY" ")
```

```
    IF TALLY(ThetaDeg$,".") = 0 THEN ThetaDeg$ = ThetaDeg$+"."

    NumSegs$ = "49"

    FeedSeg$ = "25"

    Radius$  = "0.00001"

    N% = FREEFILE

    OPEN "PBM3.NEC" FOR OUTPUT AS #N%

        PRINT #N%,"CM File: PBM3.NEC"
        PRINT #N%,"CM Run ID "+DATE$+" "+TIME$
        PRINT #N%,"CM Nd="+STR$(Nd%)+", p="STR$(p%)+", j="+STR$(j&)
        PRINT #N%,"CM R(p,1,j)="+STR$(R(p%,1,j&))+", R(p,2,j)="+STR$(R(p%,2,j&))
        PRINT #N%,"CE"

        FOR i% = 1 TO 8
            WireNum$ = REMOVE$(STR$(i%),ANY" ")

            SELECT CASE i%
                CASE 1 : Xcoord$ = "1"        : Ycoord$ = "0"
                CASE 2 : Xcoord$ = "0.70711"  : Ycoord$ = "0.70711"
                CASE 3 : Xcoord$ = "0"        : Ycoord$ = "1"
                CASE 4 : Xcoord$ = "-0.70711" : Ycoord$ = "0.70711"
                CASE 5 : Xcoord$ = "-1"       : Ycoord$ = "0"
                CASE 6 : Xcoord$ = "-0.70711" : Ycoord$ = "-0.70711"
                CASE 7 : Xcoord$ = "0"        : Ycoord$ = "-1"
                CASE 8 : Xcoord$ = "0.70711"  : Ycoord$ = "-0.70711"
            END SELECT

            PRINT #N%,"GW "+WireNum$+","+NumSegs$+","+Xcoord$+","+Ycoord$+",-0.25,"+Xcoord$+","Ycoord$+",0.25,"+Radius$
        NEXT i%

        PRINT #N%,"GE"

        FOR i% = 1 TO 8
            Alpha = -COS(TwoPi*Beta*(i%-1))

            ReV   = COS(Alpha)
            ImV   = SIN(Alpha)

            ReEX$ = REMOVE$(STR$(ROUND(ReV,6)),ANY" ")
            ImEX$ = REMOVE$(STR$(ROUND(ImV,6)),ANY" ")

            IF TALLY(ReEX$,".") = 0 THEN ReEX$ = ReEX$+"."
            IF TALLY(ImEX$,".") = 0 THEN ImEX$ = ImEX$+"."

            PRINT #N%,"EX 0,"+REMOVE$(STR$(i%),ANY" ")+","+FeedSeg$+",0,"+ReEX$+","+ImEX$ 'VOLTAGE SOURCE
        NEXT i%

        PRINT #N%,"FR 0,1,0,0,299.79564,0."
        PRINT #N%,"RP 0,1,1,1001,"+ThetaDeg$+",0.,0.,0.,1000." 'gain at 1000 wavelengths range
        PRINT #N%,"XQ"
        PRINT #N%,"EN"

    CLOSE #N%

'       - - ANGLES - -           - POWER GAINS -        - - - POLARIZATION - - -    - - - E(THETA) - - -    - - - E(PHI) - - -
'  THETA     PHI        VERT.   HOR.    TOTAL      AXIAL     TILT   SENSE     MAGNITUDE    PHASE      MAGNITUDE    PHASE
' DEGREES  DEGREES       DB      DB      DB        RATIO     DEG.              VOLTS/M    DEGREES      VOLTS/M    DEGREES
'   90.00     0.00       3.91 -999.99    3.91    0.00000     0.00  LINEAR    1.29504E-04     5.37    0.00000E+00    -5.24
'123456789x123456789x123456789x123456789x123456789x123456789x123456789x123456789x123456789x123456789x123456789x123456789x
'        10        20        30        40        50        60        70        80        90        100       110       120

    SHELL "n41_2k1.exe",0

    N% = FREEFILE

    OPEN "PBM3.OUT" FOR INPUT AS #N%

        WHILE NOT EOF(N%)

            LINE INPUT #N%, Lyne$

            IF INSTR(Lyne$,"DEGREES  DEGREES") > 0 THEN EXIT LOOP

        WEND 'position at next data line

        LINE INPUT #N%, Lyne$

    CLOSE #N%

    GainDB$ = REMOVE$(MID$(Lyne$,37,8),ANY" ")

    PBM_3 = 10^(VAL(GainDB$)/10##) 'Directivity

END FUNCTION 'PBM_3()

'----

FUNCTION PBM_4(R(),Nd%,p%,j&) 'PBM Benchmark #4: Max D for Vee Dipole

    LOCAL TotalLengthWaves, AlphaRadians, ArmLength, Xlength, Zlength, Lfeed AS EXT

    LOCAL N%, i%, Nsegs%, FeedZcoord$

    LOCAL NumSegs$, Lyne$, GainDB$, Xcoord$, Zcoord$

    TotalLengthWaves = 2##*R(p%,1,j&)

    AlphaRadians     = R(p%,2,j&)

    Lfeed            = 0.01##

    FeedZcoord$      = REMOVE$(STR$(Lfeed),ANY" ")

    ArmLength = (TotalLengthWaves-2##*Lfeed)/2##

    Xlength  = ROUND(ArmLength*COS(AlphaRadians),6)

    Xcoord$  = REMOVE$(STR$(Xlength),ANY" ") : IF TALLY(Xcoord$,".") = 0 THEN Xcoord$ = Xcoord$+"."

    Zlength  = ROUND(ArmLength*SIN(AlphaRadians),6)

    Zcoord$  = REMOVE$(STR$(Zlength+Lfeed),ANY" ") : IF TALLY(Zcoord$,".") = 0 THEN Zcoord$ = Zcoord$+"."

    Nsegs%   = 2*(INT(TotalLengthWaves*100)\2) 'even number, total # segs

    NumSegs$ = REMOVE$(STR$(Nsegs%\2),ANY" ") '# segs per arm

    N% = FREEFILE

    OPEN "PBM4.NEC" FOR OUTPUT AS #N%

        PRINT #N%,"CM File: PBM4.NEC"
        PRINT #N%,"CM Run ID "+DATE$+" "+TIME$
        PRINT #N%,"CM Nd="+STR$(Nd%)+", p="STR$(p%)+", j="+STR$(j&)
```

```
        PRINT #N%,"CM R(p,1,j)="+STR$(R(p%,1,j&))+", R(p,2,j)="+STR$(R(p%,2,j&))
        PRINT #N%,"CE"

        PRINT #N%,"GW 1,5,0.,0.,-"+FeedZcoord$+",0.,0.,"+FeedZcoord$+",0.00001" 'feed wire, 1 segment, 0.01 wvln

        PRINT #N%,"GW 2,"+NumSegs$+",0.,0.,"+FeedZcoord$+","+Xcoord$+",0.,"+Zcoord$+",0.00001" 'upper arm

        PRINT #N%,"GW 3,"+NumSegs$+",0.,0.,-"+FeedZcoord$+","+Xcoord$+",0.,-"+Zcoord$+",0.00001" 'lower arm

        PRINT #N%,"GE"

        PRINT #N%,"EX 0,1,3,0,1.,0." 'VOLTAGE SOURCE

        PRINT #N%,"FR 0,1,0,0,299.79564,0."
        PRINT #N%,"RP 0,1,1,1001,90.,0.,0.,0.,1000." 'ENDFIRE gain at 1000 wavelengths range
        PRINT #N%,"XQ"
        PRINT #N%,"EN"

    CLOSE #N%

'       - - ANGLES - -           - POWER GAINS -         - - - POLARIZATION - - -    - - - E(THETA) - - -    - - - E(PHI) - - -
'  THETA     PHI        VERT.   HOR.    TOTAL       AXIAL     TILT   SENSE      MAGNITUDE    PHASE      MAGNITUDE    PHASE
' DEGREES  DEGREES       DB      DB      DB         RATIO      DEG.              VOLTS/M     DEGREES      VOLTS/M     DEGREES
'   90.00     0.00       3.91 -999.99     3.91     0.00000     0.00  LINEAR    1.29504E-04     5.37    0.00000E+00    -5.24
'123456789x123456789x123456789x123456789x123456789x123456789x123456789x123456789x123456789x123456789x123456789x123456789x
'        10        20        30        40        50        60        70        80        90        100       110       120

    SHELL "n41_2k1.exe",0

    N% = FREEFILE

    OPEN "PBM4.OUT" FOR INPUT AS #N%

        WHILE NOT EOF(N%)

            LINE INPUT #N%, Lyne$

            IF INSTR(Lyne$,"DEGREES  DEGREES") > 0 THEN EXIT LOOP

        WEND 'position at next data line

        LINE INPUT #N%, Lyne$

    CLOSE #N%

    GainDB$ = REMOVE$(MID$(Lyne$,37,8),ANY" ")

    PBM_4 = 10^(VAL(GainDB$)/10##) 'Directivity

END FUNCTION 'PBM_4()

'----

FUNCTION PBM_5(R(),Nd%,p%,j&) 'PBM Benchmark #5: N-element collinear array (Nd=N-1)

    LOCAL TotalLengthWaves, Di(), Ystart, Y1, Y2, SumDi AS EXT

    LOCAL N%, i%, q%

    LOCAL Lyne$, GainDB$

    REDIM Di(1 TO Nd%)

    FOR i% = 1 TO Nd%

        Di(i%) = R(p%,i%,j&) 'dipole separation, wavelengths

    NEXT i%

    TotalLengthWaves = 0##

    FOR i% = 1 TO Nd%

        TotalLengthwaves = TotalLengthWaves + Di(i%)

    NEXT i%

    TotalLengthWaves = TotalLengthWaves + 0.5## 'add half-wavelength of 1 meter at 299.8 MHz

    Ystart = -TotalLengthWaves/2##

    N% = FREEFILE

    OPEN "PBM5.NEC" FOR OUTPUT AS #N%

        PRINT #N%,"CM File: PBM5.NEC"
        PRINT #N%,"CM Run ID "+DATE$+" "+TIME$
        PRINT #N%,"CM Nd="+STR$(Nd%)+", p="STR$(p%)+", j="+STR$(j&)
        PRINT #N%,"CM R(p,1,j)="+STR$(R(p%,1,j&))+", R(p,2,j)="+STR$(R(p%,2,j&))
        PRINT #N%,"CE"

        FOR i% = 1 TO Nd%+1

            SumDi = 0##

            FOR q% = 1 TO i%-1

                SumDi = SumDi + Di(q%)

            NEXT q%

            Y1 = ROUND(Ystart + SumDi,6)

            Y2 = ROUND(Y1+0.5##,6) 'add one-half wavelength for other end of dipole

            PRINT #N%,"GW "+REMOVE$(STR$(i%),ANY" ")+",49,0.,"+REMOVE$(STR$(Y1),ANY" ")+",0.,0.,"+REMOVE$(STR$(Y2),ANY" ")+",0.,0.00001"

        NEXT i%

        PRINT #N%,"GE"

        FOR i% = 1 TO Nd%+1
            PRINT #N%,"EX 0,"+REMOVE$(STR$(i%),ANY" ")+",25,0,1.,0." 'VOLTAGE SOURCES
        NEXT i%

        PRINT #N%,"FR 0,1,0,0,299.79564,0."
        PRINT #N%,"RP 0,1,1,1001,90.,0.,0.,0.,1000." 'gain at 1000 wavelengths range
        PRINT #N%,"XQ"
        PRINT #N%,"EN"

    CLOSE #N%

'       - - ANGLES - -           - POWER GAINS -         - - - POLARIZATION - - -    - - - E(THETA) - - -    - - - E(PHI) - - -
'  THETA     PHI        VERT.   HOR.    TOTAL       AXIAL     TILT   SENSE      MAGNITUDE    PHASE      MAGNITUDE    PHASE
' DEGREES  DEGREES       DB      DB      DB         RATIO      DEG.              VOLTS/M     DEGREES      VOLTS/M     DEGREES
'   90.00     0.00       3.91 -999.99     3.91     0.00000     0.00  LINEAR    1.29504E-04     5.37    0.00000E+00    -5.24
'123456789x123456789x123456789x123456789x123456789x123456789x123456789x123456789x123456789x123456789x123456789x123456789x
'        10        20        30        40        50        60        70        80        90        100       110       120
```

```
    SHELL "n41_2k1.exe",0

    N% = FREEFILE

    OPEN "PBM5.OUT" FOR INPUT AS #N%

        WHILE NOT EOF(N%)

            LINE INPUT #N%, Lyne$

            IF INSTR(Lyne$,"DEGREES  DEGREES") > 0 THEN EXIT LOOP

        WEND 'position at next data line

        LINE INPUT #N%, Lyne$

    CLOSE #N%

    GainDB$ = REMOVE$(MID$(Lyne$,37,8),ANY" ")

    PBM_5 = 10^(VAL(GainDB$)/10##) 'Directivity

END FUNCTION 'PBM_5()

'----

FUNCTION BOWTIE(R(),Nd%,p%,j&) 'FREE SPACE BOWTIE

    LOCAL N%, i%, Nsegs%, NumFreqs%, ElemNum%, NumRadPattAngles%, ExcitedSegment%

    LOCAL Lyne$, FR1$, FileStatus$, FileID$

    LOCAL A, B, C AS EXT 'coefficients for Bowtie Fitness function

    LOCAL a_meters, SegLength, x1, x2, y1, y2, z1, z2 AS EXT

    LOCAL Zo, FrequencyMHZ(), RadEfficiencyPCT(), MaxGainDBI(), MinGainDBI(), RinOhms(), XinOhms(), VSWR(), ForwardGainDBi() AS EXT

    LOCAL  Fitness,  MinimumRadiationEfficiency,  MaximumRadiationEfficiency,  MinimumMaxGain,  MaximumMaxGain,  MinVSWR,  MaxVSWR,  MinRin,  MaxRin,  MinXin,
MaxXin, MaxFwdGain, MinFwdGain AS EXT

    LOCAL StartFreqMHZ!, StopFreqMHZ!, FreqStepMHZ!

    LOCAL ArmLengthMeters, AngleRadians, LoadResistance1Ohms, LoadResistance2Ohms AS EXT

    LOCAL LoadedSegNum1%, LoadedSegNum2%

    REDIM  FrequencyMHZ(1  TO  1),  RadEfficiencyPCT(1  TO  1),  MaxGainDBI(1  TO  1),  MinGainDBI(1  TO  1),  RinOhms(1  TO  1),  XinOhms(1  TO  1),  VSWR(1  TO  1),
ForwardGainDBi(1 TO 1)

'DECISION SPACE BOUNDARY ARRAY:

'DECISION SPACE BOUNDARY ARRAY:

'    Array Element #                Design Variable
'    ---------------      ------------------------------------
'          1                Bowtie atm length (meters)
'          2                Bowtie HALF angle (degrees)
'          3                Loading segment number #1
'          4                Loading resistance #1 (ohms)
'          5                Zo

'   IMPORTANT NOTE: SEGMENTATION IS FIXED AT 9 !!!!  BE SURE Nsegs% = 9 BELOW, OR CHANGE IT HERE AND IN SUB GetFunctionRunParameters().
'   -------------------------------------------------------------------------------------------***-----------------------------------
    Nsegs% = 9

    ArmLengthMeters     = ROUND(R(p%,1,j&),3)
    AngleRadians        = ROUND(R(p%,2,j&)*Deg2Rad,3) 'remember, this is the HALF angle
    LoadedSegNum1%      = MAX(INT(R(p%,3,j&)),1)
    LoadResistance1Ohms = ROUND(R(p%,4,j&),2)
    Zo                  = ROUND(R(p%,5,j&),1)

    a_meters = 0.0005## 'meters

'    Zo = 300## 'feed system characteristic impedance, OHMS, for computing VSWR

    NumRadPattAngles% = 19

'    IF BowtieSegmentLength$ = "FIXED" THEN SegLength = BowtieSegmentLengthWvln 'wavelengths, FIXED [SEE HEADER NOTE ON SEGMENTATION]

    StartFreqMHZ! = 800! : StopFreqMHZ! = 12000! : FreqStepMHZ! = 100! : NumFreqs% = 1 + (StopFreqMHZ!-StartFreqMHZ!)/FreqStepMHZ!

    FR1$ = "FR 0,"+Int2String$(NumFreqs%)+",0,0,"+FP2String2$(StartFreqMHZ!)+","+FP2String2$(FreqStepMHZ!)

'    PROCESS SET PRIORITY %NORMAL_PRIORITY_CLASS 'NORMAL PRIORITY TO AVOID PROBLEMS WRITING/READING NEC FILES

'    PROCESS SET PRIORITY %REALTIME_PRIORITY_CLASS 'PREEMPTS ALL OTHER PROCESSES -> CAN CAUSE PROBLEMS...

    N% = FREEFILE

    OPEN "BOWTIE.NEC" FOR OUTPUT AS #N%

        FileID$  = REMOVE$(DATE$+TIME$,ANY Alphabet$+" -:/")

        PRINT #N%,"CM File: BOWTIE.NEC"
        PRINT #N%,"CM R-LOADED BOWTIE IN FREE SPACE WITH"
        PRINT #N%,"CM Zo AS AN OPTIMIZATION PARAMETER."
        PRINT #N%,"CM Antenna in Y-Z plane."
        PRINT #N%,"CM Run ID: "+RunID$
        PRINT #N%,"CM Fitness function:
        PRINT #N%,"CM [Min(Eff)+5*Min(Gmax)]/[|Zo-MaxRin|*(MaxVSWR-MinVSWR)*(MaxXin-MinXin)]
        PRINT #N%,"CM Arm Length = "+REMOVE$(STR$(ArmLengthMeters),ANY" ")+" meters"
        PRINT #N%,"CM Bowtie HALF Angle = "+REMOVE$(STR$(ROUND(R(p%,2,j&),2)),ANY" ")+" degrees"
        PRINT #N%,"CM Zo = "+REMOVE$(STR$(Zo),ANY" ")+" ohms"
        PRINT #N%,"CM Rload = "+REMOVE$(STR$(ROUND(LoadResistance1Ohms,2)),ANY" ")+ " ohms"
        PRINT #N%,"CM Loaded Seg # = "+REMOVE$(STR$(LoadedSegNum1%),ANY" ")+"/"+REMOVE$(STR$(Nsegs%),ANY" ")
        PRINT #N%,"CM File ID "+FileID$
        PRINT #N%,"CM Nd ="+STR$(Nd%)+", p ="STR$(p%)+", j ="+STR$(j&)
        PRINT #N%,"CE"

        x1 = 0## : x2 = x1 : y1 = -0.01## : y2 = -y1 : z1 = 0## : z2 = z1 'feed wire coords (wvln)
        PRINT
#N%,"GW1,3,"+FP2String$(x1)+","+FP2String$(y1)+","+FP2String$(z1)+","+FP2String$(x2)+","FP2String$(y2)+","+FP2String$(z2)+","+FP2String$(a_meters)    'feed
wire, 3 segs

        y1 = 0.01## : y2 = ROUND(y1+ArmLengthMeters*COS(AngleRadians),3) : z1 = 0## : z2 = ROUND(ArmLengthMeters*SIN(AngleRadians),3) 'upper right arm
        PRINT
#N%,"GW2,"+Int2String$(Nsegs%)+","+FP2String$(x1)+","+FP2String$(y1)+","+FP2String$(z1)+","+FP2String$(x2)+","FP2String$(y2)+","+FP2String$(z2)+","+FP2Stri
ng$(a_meters) 'feed wire, 3 segs

        y1 = 0.01## : y2 = ROUND(y1+ArmLengthMeters*COS(AngleRadians),3) : z1 = 0## : z2 = -ROUND(ArmLengthMeters*SIN(AngleRadians),3) 'lower right arm
        PRINT
#N%,"GW3,"+Int2String$(Nsegs%)+","+FP2String$(x1)+","+FP2String$(y1)+","+FP2String$(z1)+","+FP2String$(x2)+","FP2String$(y2)+","+FP2String$(z2)+","+FP2Stri
ng$(a_meters) 'feed wire, 3 segs

        y1 = ROUND(0.01##+ArmLengthMeters*COS(AngleRadians),3) : y2 = y1 : z1 = ROUND(ArmLengthMeters*SIN(AngleRadians),3) : z2 = -z1 'right arm vertical
connecting wire
```

```
        PRINT
#N%,"GW4,"+Int2String$(Nsegs%)+","+FP2String$(x1)+","+FP2String$(y1)+","+FP2String$(z1)+","+FP2String$(x2)+","FP2String$(y2)+","+FP2String$(z2)+","+FP2String$(a_meters) 'feed wire, 3 segs

        y1 = -0.01## : y2 = ROUND(y1-ArmLengthMeters*COS(AngleRadians),3) : z1 = 0## : z2 = ROUND(ArmLengthMeters*SIN(AngleRadians),3) 'upper left arm
        PRINT
#N%,"GW5,"+Int2String$(Nsegs%)+","+FP2String$(x1)+","+FP2String$(y1)+","+FP2String$(z1)+","+FP2String$(x2)+","FP2String$(y2)+","+FP2String$(z2)+","+FP2String$(a_meters) 'feed wire, 3 segs

        y1 = -0.01## : y2 = ROUND(y1-ArmLengthMeters*COS(AngleRadians),3) : z1 = 0## : z2 = -ROUND(ArmLengthMeters*SIN(AngleRadians),3) 'lower left arm
        PRINT
#N%,"GW6,"+Int2String$(Nsegs%)+","+FP2String$(x1)+","+FP2String$(y1)+","+FP2String$(z1)+","+FP2String$(x2)+","FP2String$(y2)+","+FP2String$(z2)+","+FP2String$(a_meters) 'feed wire, 3 segs

        y1 = -ROUND(0.01##+ArmLengthMeters*COS(AngleRadians),3) : y2 = y1 : z1 = ROUND(ArmLengthMeters*SIN(AngleRadians),3) : z2 = -z1 'left arm vertical connecting wire
        PRINT
#N%,"GW7,"+Int2String$(Nsegs%)+","+FP2String$(x1)+","+FP2String$(y1)+","+FP2String$(z1)+","+FP2String$(x2)+","FP2String$(y2)+","+FP2String$(z2)+","+FP2String$(a_meters) 'feed wire, 3 segs

        PRINT #N%,"GE"

'       ----------------------------------- LOAD RESISTOR #1 --------------------------------------
        IF LoadResistance1Ohms <> 0## THEN 'load bowtie only if loading resistance is not zero
            PRINT #N%,"LD0,2,"+Int2String$(LoadedSegNum1%)+","+Int2String$(LoadedSegNum1%)+","_
                              +FP2String2$(ROUND(LoadResistance1Ohms,2))+",0.,0."'load upper right arm
            PRINT #N%,"LD0,3,"+Int2String$(LoadedSegNum1%)+","+Int2String$(LoadedSegNum1%)+","_
                              +FP2String2$(ROUND(LoadResistance1Ohms,2))+",0.,0." 'load lower right arm
            PRINT #N%,"LD0,5,"+Int2String$(LoadedSegNum1%)+","+Int2String$(LoadedSegNum1%)+","_
                              +FP2String2$(ROUND(LoadResistance1Ohms,2))+",0.,0." 'load upper left arm
            PRINT #N%,"LD0,6,"+Int2String$(LoadedSegNum1%)+","+Int2String$(LoadedSegNum1%)+","_
                              +FP2String2$(ROUND(LoadResistance1Ohms,2))+",0.,0." 'load lower left arm
        END IF

'       ----------------------------------- LOAD RESISTOR #2 --------------------------------------
'        IF LoadResistance2Ohms <> 0## THEN 'load bowtie only if loading resistance is not zero
'        PRINT #N%,"LD0,2,"+Int2String$(LoadedSegNum2%)+","+Int2String$(LoadedSegNum2%)+","_
 '                        +FP2String2$(ROUND(LoadResistance2Ohms,2))+",0.,0."'load upper right arm
  '      PRINT #N%,"LD0,3,"+Int2String$(LoadedSegNum2%)+","+Int2String$(LoadedSegNum2%)+","_
   '                      +FP2String2$(ROUND(LoadResistance2Ohms,2))+",0.,0." 'load lower right arm
    '    PRINT #N%,"LD0,5,"+Int2String$(LoadedSegNum2%)+","+Int2String$(LoadedSegNum2%)+","_
     '                   +FP2String2$(ROUND(LoadResistance2Ohms,2))+",0.,0." 'load upper left arm
      '  PRINT #N%,"LD0,6,"+Int2String$(LoadedSegNum2%)+","+Int2String$(LoadedSegNum2%)+","_
       '                 +FP2String2$(ROUND(LoadResistance2Ohms,2))+",0.,0." 'load lower left arm
        'END IF

        PRINT #N%, FR1$ ' frequency card

        PRINT #N%, "EX 0,1,2,1,1.,0." 'excite center segment in DRIVEN ELEMENT #2 with 1+j0 volts

        PRINT   #N%,   "RP   0,"+Int2String$(NumRadPattAngles%)+",1,1001,0.,0.,"+FP2String2$(ROUND(90##/(NumRadPattAngles%-1),2))+",0.,100000."   'vertical radiation pattern at Phi=0 (+X-axis)

        PRINT #N%, "EN"

    CLOSE #N%

'    SHELL "NEC2D_200_02-22-2011.EXE",0

    SHELL "NEC41D_4K_053011.EXE",0

    CALL
GetNECdata("BOWTIE.OUT",NumFreqs%,NumRadPattAngles%,Zo,FrequencyMHZ(),RadEfficiencyPCT(),MaxGainDBI(),MinGainDBI(),RinOhms(),XinOhms(),VSWR(),ForwardGainDBi(),FileStatus$,FileID$)

    Fitness = -98765## 'default value

    IF FileStatus$ = "OK" THEN

        MinimumRadiationEfficiency = RadEfficiencyPCT(1) : MinimumMaxGain = MaxGainDBI(1) : MinVSWR = VSWR(1) : MaxVSWR = VSWR(1)

        MinRin = RinOhms(1) : MaxRin = RinOhms(1) :  MinXin = XinOhms(1) : MaxXin = XinOhms(1)

        MinFwdGain = ForwardGainDBi(1) : MaxFwdGain = ForwardGainDBi(1)

        FOR i% = 1 TO NumFreqs%
            IF RadEfficiencyPCT(i%) =< MinimumRadiationEfficiency THEN MinimumRadiationEfficiency = RadEfficiencyPCT(i%)
            IF MinimumRadiationEfficiency < 0## THEN EXIT FOR 'bad run -> use default fitness & exit
            IF RadEfficiencyPCT(i%) >= MaximumRadiationEfficiency THEN MaximumRadiationEfficiency = RadEfficiencyPCT(i%)
            IF MaxGainDBI(i%)       =< MinimumMaxGain             THEN MinimumMaxGain             = MaxGainDBI(i%)
            IF MaxGainDBI(i%)       >= MaximumMaxGain             THEN MaximumMaxGain             = MaxGainDBI(i%)
            IF VSWR(i%)             =< MinVSWR                    THEN MinVSWR                    = VSWR(i%)
            IF VSWR(i%)             >= MaxVSWR                    THEN MaxVSWR                    = VSWR(i%)
            IF RinOhms(i%)          =< MinRin                     THEN MinRin                     = RinOhms(i%)
            IF RinOhms(i%)          >= MaxRin                     THEN MaxRin                     = RinOhms(i%)
            IF XinOhms(i%)          =< MinXin                     THEN MinXin                     = XinOhms(i%)
            IF XinOhms(i%)          >= MaxXin                     THEN MaxXin                     = XinOhms(i%)
            IF ForwardGainDBi(i%)   >= MaxFwdGain                 THEN MaxFwdGain                 = ForwardGainDBi(i%)
            IF ForwardGainDBi(i%)   =< MinFwdGain                 THEN MinFwdGain                 = ForwardGainDBi(i%)
        NEXT i%

'        A = BowtieFitnessCoefficients(1) : B = BowtieFitnessCoefficients(2) : C = BowtieFitnessCoefficients(3) 'use this notation for consistency

'NOTE: Forward gain is in direction on +X-axis.

'        Fitness = (A*ForwardGainDBi-B*ABS(Zo-MaxRin)-C*(MAX(ABS(MaxXin),ABS(MinXin))))/(A+B+C) 'NOTE: THIS IS ONE OF AN INFINITY OF FITNESS FUNCTIONS.
                                                                                               'CHANGING THE FUNCTION CHANGES THE DECISION SPACE LANDSCAPE
                                                                                               'RESULTING IN COMPLETELY DIFFERENT ANTENNA DESIGNS.

        IF  MinimumRadiationEfficiency  >=  0##  THEN  Fitness  =  (MinimumRadiationEfficiency+5##*MinimumMaxGain)/(ABS(Zo-MaxRin)*(MaxVSWR-MinVSWR)*(MaxXin-MinXin)) 'compute only if run is OK and min eff >= 0

    END IF

    N% = FREEFILE
    OPEN "BOWTIE.NEC" FOR APPEND AS #N%
        PRINT #N%,""
        PRINT #N%,"CM R-LOADED BOWTIE IN FREE SPACE WITH"
        PRINT #N%,"CM Zo AS AN OPTIMIZATION PARAMETER."
        PRINT #N%,"Run ID: "+RunID$
        PRINT #N%,USING$ ("Fitness           =#.####^^^^^^",Fitness)
        PRINT #N%,"Fwd Gain (dBi) Min/Max    ="+STR$(MinFwdGain)+"/"+STR$(MinFwdGain)
        PRINT #N%,"VSWR Min/Max      ="+STR$(ROUND(MinVSWR,2))+"/"+STR$(ROUND(MaxVSWR,2))+"//"+REMOVE$(STR$(Zo),ANY" ")
        PRINT #N%,"Rin Min/Max       ="+STR$(ROUND(MinRin,2))+"/"+STR$(ROUND(MaxRin,2))
        PRINT #N%,"Xin Min/Max       ="+STR$(ROUND(MinXin,2))+"/"+STR$(ROUND(MaxXin,2))
        PRINT #N%,"Gmax(dBi) Min/Max ="+STR$(ROUND(MinimumMaxGain,2))+"/"+STR$(ROUND(MaximumMaxGain,2))
        PRINT #N%,"Eff(%) Min/Max    ="+STR$(ROUND(MinimumRadiationEfficiency,2))+"/"+STR$(ROUND(MaximumRadiationEfficiency,2))
        PRINT #N%,""
    CLOSE #N%

    N% = FREEFILE
    OPEN "BOWTIE.OUT" FOR APPEND AS #N%
        PRINT #N%,""
        PRINT #N%,"CM R-LOADED BOWTIE IN FREE SPACE WITH"
        PRINT #N%,"CM Zo AS AN OPTIMIZATION PARAMETER."
        PRINT #N%,"Run ID: "+RunID$
        PRINT #N%,USING$ ("Fitness           =#.####^^^^^^",Fitness)
        PRINT #N%,"Fwd Gain (dBi) Min/Max    ="+STR$(MinFwdGain)+"/"+STR$(MinFwdGain)
        PRINT #N%,"VSWR Min/Max      ="+STR$(ROUND(MinVSWR,2))+"/"+STR$(ROUND(MaxVSWR,2))+"//"+REMOVE$(STR$(Zo),ANY" ")
```

```
        PRINT #N%,"Rin Min/Max       ="+STR$(ROUND(MinRin,2))+"/"+STR$(ROUND(MaxRin,2))
        PRINT #N%,"Xin Min/Max       ="+STR$(ROUND(MinXin,2))+"/"+STR$(ROUND(MaxXin,2))
        PRINT #N%,"Gmax(dBi) Min/Max ="+STR$(ROUND(MinimumMaxGain,2))+"/"+STR$(ROUND(MaximumMaxGain,2))
        PRINT #N%,"Eff(%) Min/Max    ="+STR$(ROUND(MinimumRadiationEfficiency,2))+"/"+STR$(ROUND(MaximumRadiationEfficiency,2))
        PRINT #N%,""
    CLOSE #N%

'       - - ANGLES - -           - POWER GAINS -        - - - POLARIZATION - - -    - - - E(THETA) - - -    - - - E(PHI) - - -
'  THETA     PHI        VERT.   HOR.    TOTAL      AXIAL     TILT   SENSE     MAGNITUDE    PHASE      MAGNITUDE    PHASE
' DEGREES  DEGREES       DB      DB      DB        RATIO     DEG.              VOLTS/M    DEGREES      VOLTS/M    DEGREES
'   90.00     0.00      3.91 -999.99    3.91    0.00000     0.00  LINEAR    1.29504E-04     5.37    0.00000E+00    -5.24
'123456789x123456789x123456789x123456789x123456789x123456789x123456789x123456789x123456789x123456789x123456789x123456789x
'        10        20        30        40        50        60        70        80        90       100       110       120

    BOWTIE = Fitness

END FUNCTION 'BOWTIE()

'--------------===-------

SUB ReplaceCommentCard(NECfile$)

LOCAL N%, M%

LOCAL Lyne$

    N% = FREEFILE : OPEN NECfile$ FOR INPUT AS #N%

    M% = FREEFILE : OPEN "NECtemp" FOR OUTPUT AS #M%

    WHILE NOT EOF(N%)
        LINE INPUT #N%, Lyne$
        IF INSTR(Lyne$,"BOWTIE.NEC") > 0 THEN REPLACE "BOWTIE.NEC" WITH "BESTBOWTIE.NEC" IN Lyne$
        PRINT #M%, Lyne$
    WEND

    CLOSE #M% : CLOSE #N%

    KILL NecFile$ : NAME "NECtemp" AS NECfile$

END SUB

'------

    FUNCTION FP2String2$(X!)
    LOCAL A$
        A$=LTRIM$(RTRIM$(STR$(X!)))
        IF TALLY(A$,".") = 0! THEN A$ = A$ + "."
        FP2String2$ = A$
    END FUNCTION

'---

    FUNCTION FP2String$(X)
    LOCAL A$
        A$=LTRIM$(RTRIM$(STR$(X)))
        IF TALLY(A$,".") = 0## THEN A$ = A$ + "."
        FP2String$ = A$
    END FUNCTION

'---

    FUNCTION Int2String$(X%)
    LOCAL A$
       A$=LTRIM$(RTRIM$(STR$(X%)))
       Int2String$ = A$
    END FUNCTION

'---

    SUB
GetNECdata(NECoutputFile$,NumFreqs%,NumRadPattAngles%,Zo,FrequencyMHZ(),RadEfficiencyPCT(),MaxGainDBI(),MinGainDBI(),RinOhms(),XinOhms(),VSWR(),ForwardGain
DBi(),FileStatus$,FileID$)

    LOCAL N%, idx%, AngleNum%

    LOCAL Lyne$, Dum$

    LOCAL GmaxDBI, GminDBI, FwdGainDBi AS EXT

    REDIM FrequencyMHZ(1 TO NumFreqs%),RadEfficiencyPCT(1 TO NumFreqs%),MaxGainDBI(1 TO NumFreqs%),MinGainDBI(1 TO NumFreqs%), _
          RinOhms(1 TO NumFreqs%),XinOhms(1 TO NumFreqs%),VSWR(1 TO NumFreqs%), ForwardGainDBi(1 TO NumFreqs%)

    FileStatus$ = "NOK"

    OPEN NECoutputFile$ FOR INPUT AS #N%

        WHILE NOT EOF(N%)

            LINE INPUT #N%, Lyne$

            IF INSTR(Lyne$,"RUN TIME") > 0 THEN FileStatus$ = "OK"

        WEND

    CLOSE #N%

    IF FileStatus$ <> "OK" THEN EXIT SUB

    OPEN NECoutputFile$ FOR INPUT AS #N%

        idx% = 1

        WHILE NOT EOF(N%)

            LINE INPUT #N%, Lyne$

            IF INSTR(Lyne$,"File ID") > 0 THEN 'CHECK THAT NEC OUTPUT FILE WAS COMPUTED FROM CURRENT NEC INPUT FILE BY COMPARIN FILE ID's (TO AVOID
CACHE/BUFFER PROBLEMS CREATD BY OS)
                IF FileID$ <> REMOVE$(Lyne$,ANY Alphabet$+" ") THEN MSGBOX("WARNING! NEC I/O File ID's Don't Match!"+CHR$(13)+"Lyne$ =
"+Lyne$+CHR$(13)+"FileID$ = "+FileID$)
            END IF

            IF INSTR(Lyne$,"FREQUENCY=") > 0 THEN
                Lyne$ = REMOVE$(Lyne$,"MHZ") : Lyne$ = REMOVE$(Lyne$,"FREQUENCY= ") : FrequencyMHZ(idx%) = VAL(Lyne$)
'MSGBOX("idx="+STR$(idx%)+"  F="+STR$(FrequencyMHZ(idx%)))
            END IF

            IF INSTR(Lyne$,"INPUT PARAMETERS") > 0 THEN
                LINE INPUT #N%, Dum$ : LINE INPUT #N%, Dum$ : LINE INPUT #N%, Dum$ 'skip three lines
                LINE INPUT #N%, Lyne$ 'input next line with impedance data
                RinOhms(idx%) = VAL(MID$(Lyne$,61,12)) : XinOhms(idx%) = VAL(MID$(Lyne$,73,12)) : VSWR(idx%) =
StandingWaveRatio(Zo,RinOhms(idx%),XinOhms(idx%))
            END IF
            IF INSTR(Lyne$,"EFFICIENCY") > 0 THEN RadEfficiencyPCT(idx%) = VAL(REMOVE$(Lyne$,ANY Alphabet$+" ="))
            IF INSTR(Lyne$,"E(THETA)") > 0 THEN
                LINE INPUT #N%, Dum$ : LINE INPUT #N%, Dum$ 'skip two lines
                GmaxDBI = -9999## : GminDBI = -GmaxDBI
                FOR AngleNum% = 1 TO NumRadPattAngles%
```

```
                LINE INPUT #N%, Lyne$ 'input next TEN lines with pattern data
                IF VAL(MID$(Lyne$,38,7)) >= GmaxDBI THEN GmaxDBI = VAL(MID$(Lyne$,38,7)) 'get max gain
                IF (VAL(MID$(Lyne$,38,7)) =< GminDBI AND VAL(MID$(Lyne$,38,7)) >= -999.99##) THEN GminDBI = VAL(MID$(Lyne$,38,7)) 'get min gain
                IF AngleNum% = NumRadPattAngles% THEN FwdGainDBi = VAL(MID$(Lyne$,38,7)) 'forward gain
            NEXT AngleNum%
            MaxGainDBI(idx%) = GmaxDBI
            MinGainDBI(idx%) = GminDBI
            ForwardGainDBi(idx%) = FwdGainDBi
            INCR idx%
        END IF

'msgbox("idx="+STR$(idx%))

       WEND

    CLOSE #N%

'   TAG   SEG.    VOLTAGE (VOLTS)         CURRENT (AMPS)         IMPEDANCE (OHMS)        ADMITTANCE (MHOS)      POWER
'   NO.   NO.     REAL        IMAG.       REAL        IMAG.       REAL        IMAG.       REAL        IMAG.     (WATTS)
'     1     1 1.00000E+00 0.00000E+00 7.17910E-06 8.93193E-04 8.99811E+00-1.11951E+03 7.17910E-06 8.93193E-04 3.58955E-06
'123456789x123456789x123456789x123456789x123456789x123456789x123456789x123456789x123456789x123456789x123456789x123456789x123456789x
'         10        20        30        40        50        60        70        80        90       100       110       120       130

'  - - ANGLES - -           - POWER GAINS -       - - - POLARIZATION - - -    - - - E(THETA) - - -    - - - E(PHI) - - -
'  THETA     PHI        VERT.    HOR.    TOTAL       AXIAL     TILT   SENSE     MAGNITUDE    PHASE      MAGNITUDE    PHASE
' DEGREES  DEGREES       DB       DB      DB         RATIO     DEG.             VOLTS/M    DEGREES      VOLTS/M    DEGREES
'    0.00     0.00    -999.99 -999.99 -999.99     0.00000     0.00             0.00000E+00  -240.17    0.00000E+00  -240.17
'   10.00     0.00     -18.97 -999.99  -18.97     0.00000     0.00  LINEAR     2.69380E-08   -61.44    0.00000E+00  -240.17
'123456789x123456789x123456789x123456789x123456789x123456789x123456789x123456789x123456789x123456789x123456789x123456789x123456789x
'         10        20        30        40        50        60        70        80        90       100       110       120       130

    END SUB 'GetNECdata()

'------------------------

    FUNCTION StandingWaveRatio(Zo,ReZ,ImZ)

    LOCAL ReRho, ImRho, MagRho, SWR AS EXT

        SWR = 9999##

        CALL ComplexDivide(ReZ-Zo,ImZ,ReZ+Zo,ImZ,ReRho,ImRho)

        MagRho = SQR(ReRho*ReRho+ImRho*ImRho)  'reflection coefficient

        IF MagRho <> 1## THEN SWR=(1##+MagRho)/(1##-MagRho)

        StandingWaveRatio = SWR

    END FUNCTION 'StandingWaveRatio()

'-----

    SUB ComplexMultiply(ReA,ImA,ReB,ImB,ReC,ImC)

'   Returns real and imaginary parts of product C=A*B

        ReC = ReA*ReB-ImA*ImB
        ImC = ImA*ReB+ReA*ImB
    END SUB

'-----

    SUB ComplexDivide(ReA,ImA,ReB,ImB,ReC,ImC)

'   Returns real and imaginary parts of quotient C=A/B

        LOCAL Denom AS EXT

        Denom = ReB*ReB+ImB*ImB
        ReC = (ReA*ReB+ImA*ImB)/Denom
        ImC = (ImA*ReB-ReA*ImB)/Denom
    END SUB

'----

FUNCTION YAGI_ARRAY(R(),Nd%,p%,j&) 'FREE SPACE YAGI

    LOCAL N%, i%, Nsegs%, NumFreqs%, ElemNum%, NumRadPattAngles%, ExcitedSegment%

    LOCAL Lyne$, FR1$, FileStatus$, FileID$

    LOCAL A, B, C, D AS EXT 'coefficients for Yagi Fitness function

    LOCAL a_wvln, SegLength, x1, x2, y1, y2, BoomDistance, Spacing, Length AS EXT

    LOCAL Zo, FrequencyMHZ(), RadEfficiencyPCT(), MaxGainDBI(), MinGainDBI(), RinOhms(), XinOhms(), VSWR(), ForwardGainDBi() AS EXT

    LOCAL  Fitness,  MinimumRadiationEfficiency,  MaximumRadiationEfficiency,  MinimumMaxGain,  MaximumMaxGain,  MinVSWR,  MaxVSWR,  MinRin,  MaxRin,  MinXin,
MaxXin, MinFwdGain, MaxFwdGain AS EXT

    LOCAL StartFreqMHZ!, StopFreqMHZ!, FreqStepMHZ!

    REDIM  FrequencyMHZ(1  TO  1),  RadEfficiencyPCT(1  TO  1),  MaxGainDBI(1  TO  1),  MinGainDBI(1  TO  1),  RinOhms(1  TO  1),  XinOhms(1  TO  1),  VSWR(1  TO  1),
ForwardGainDBi(1 TO 1)

'INFO: YAGI DECISION SPACE BOUNDARY ARRAY:

'    Array Element #                            Design Variable
'    ---------------      ----------------------------------------------------------------------
'          1
'          TO               Yagi element spacing along boom from previous element, wavelengths
'     NumYagiElements

'   NumYagiElements+1
'          TO                               Yagi element length, wavelengths
'   2*NumYagiElements (Nd-1)

'          Nd                                         <<<< Zo >>>>

'   ==========================================================================================

    a_wvln = 0.00635## 'wire radius WAVELENGTHS (corresponds to 1/2" DIAMETER elements at 299.8 MHz)

    Zo = ROUND(R(p%,Nd%,j&),2) 'feed system characteristic impedance, OHMS, as a design VARIABLE for computing VSWR

    NumRadPattAngles% = 19

    IF YagiSegmentLength$ = "FIXED" THEN SegLength = YagiSegmentLengthWvln 'wavelengths @ 299.8 MHz, FIXED [SEE HEADER NOTE ON SEGMENTATION]

    StartFreqMHZ! = 275! : StopFreqMHZ! = 325! : FreqStepMHZ! = 10!

    NumFreqs% = 1 + (StopFreqMHZ!-StartFreqMHZ!)/FreqStepMHZ! 'Center Frequency, Fc = 299.8 MHz (wvln = 1.000 meters)

    NumFreqs% = 2*(NumFreqs%\2)+1 'make it an odd # so three-point averaging can be done

    FR1$ = "FR 0,"+Int2String$(NumFreqs%)+",0,0,"+FP2String2$(StartFreqMHZ!)+","+FP2String2$(FreqStepMHZ!)

'    PROCESS SET PRIORITY %NORMAL_PRIORITY_CLASS 'NORMAL PRIORITY TO AVOID PROBLEMS WRITING/READING NEC FILES
```

```
'    PROCESS SET PRIORITY %REALTIME_PRIORITY_CLASS 'PREEMPTS ALL OTHER PROCESSES -> CAN CAUSE PROBLEMS...

    N% = FREEFILE

    OPEN "YAGI.NEC" FOR OUTPUT AS #N%

        FileID$  = REMOVE$(DATE$+TIME$,ANY Alphabet$+" -:/")

        PRINT #N%,"CM File: YAGI.NEC"
        PRINT #N%,"CM YAGI ARRAY IN FREE SPACE"
        PRINT #N%,"CM Band center frequency, Fc = 299.8 MHz"
        PRINT #N%,"CM Run ID: "+RunID$
        PRINT #N%,"CM Fitness function:
        PRINT #N%,"CM Gfwd(L)/5-2*VSWR(L)+Gfwd(M)-4*VSWR(M)+Gfwd(U)-2*VSWR(U)/5"
        PRINT #N%,"CM where L,M,U are lower/mid/upper frequencies"
'         PRINT #N%,"CM MaxGfwd/[|Zo-MaxRin|*(MaxVSWR-MinVSWR)*(MaxXin-MinXin)]
'         PRINT #N%,"CM (A*MaxFwdGain-B*|Zo-MaxRin|-C*(MAX(|MaxXin|,|MinXin|))"
'         PRINT #N%,"CM               -D*(MaxVSWR-MinVSWR))/(A+B+C+D)"
'         PrINT #N%,"CM where "+YagiCoefficients$
        PRINT #N%,"CM Zo="+REMOVE$(STR$(Zo),ANY" ")+" ohms"
        PRINT #N%,"CM File ID "+FileID$
        PRINT #N%,"CM Nd="+STR$(Nd%)+", p="STR$(p%)+", j="+STR$(j&)
        PRINT #N%,"CE"

        FOR ElemNum% = 1 TO NumYagiElements%

            BoomDistance = 0##

            FOR i% = 1 TO ElemNum% 'add up element spacings to get position along boom
                Spacing = ROUND(R(p%,i%,j&),3) 'element spacing from previous element, WAVELENGTHS (note Reflector spacing = 0 always)
                BoomDistance = BoomDistance + Spacing
            NEXT i%

            Length  = R(p%,ElemNum%+NumYagiElements%,j&) 'element length, WAVELENGTHS

            Nsegs% = 9 'use same # segs each element unless seg LENGTH is FIXED, in which case use a different # segs in each element as follows

            IF YagiSegmentLength$ = "FIXED" THEN 'adjust # segments in each element and change element length to be an integer # of segments
                NSegs% = INT(Length/SegLength)
                NSegs% = 2*(Nsegs%\2)+1    'must be odd # to preserve symmetry about excitation at center
                Length = NSegs%*SegLength 'round element length to an integer multiple of segment length to guaranty segment alignment as recommended in
NEC Manual
            END IF

            IF ElemNum% = 2 THEN ExcitedSegment% = Nsegs%\2+1

            x1 = ROUND(BoomDistance,3) : x2 =  x1 'Yagi elements are parallel to Y-axis arrayed along +X-axis.  Round spacings & length to nearest 0.001
wavelength.

            y1 = ROUND(-Length/2##,3)  : y2 = -y1

            PRINT
#N%,"GW"+Int2String$(ElemNum%)+","+Int2String$(NSegs%)+","+FP2String$(x1)+","+FP2String$(y1)+",0.,"+FP2String$(x2)+","FP2String$(y2)+",0.,"+FP2String$(a_wv
ln)

        NEXT ElemNum%

        PRINT #N%,"GE"

        PRINT #N%, FR1$ ' frequency card

        PRINT #N%, "EX 0,2,"+Int2String$(ExcitedSegment%)+",1,1.,0." 'excite center segment in DRIVEN ELEMENT #2 with 1+j0 volts

        PRINT   #N%,   "RP   0,"+Int2String$(NumRadPattAngles%)+",1,1001,0.,0.,"+FP2String2$(ROUND(90##/(NumRadPattAngles%-1),2))+",0.,100000."   'vertical
radiation pattern at Phi=0 (+X-axis)

        PRINT #N%, "EN"

    CLOSE #N%

'    SHELL "NEC2D_200_02-22-2011.EXE",0

    SHELL "NEC41D_4K_053011.EXE",0

'                                                                                                                                                    CALL
GetNECdata(NumFreqs%,NumRadPattAngles%,Zo,FrequencyMHZ(),RadEfficiencyPCT(),MaxGainDBI(),MinGainDBI(),RinOhms(),XinOhms(),VSWR(),ForwardGainDBi,FileStatus$
,FileID$)
    CALL
GetNECdata("YAGI.OUT",NumFreqs%,NumRadPattAngles%,Zo,FrequencyMHZ(),RadEfficiencyPCT(),MaxGainDBI(),MinGainDBI(),RinOhms(),XinOhms(),VSWR(),ForwardGainDBi(
),FileStatus$,FileID$)

    Fitness = -98765## 'default value

    IF FileStatus$ = "OK" THEN

        MinimumRadiationEfficiency = RadEfficiencyPCT(1) : MinimumMaxGain = MaxGainDBI(1) : MinVSWR = VSWR(1) : MaxVSWR = VSWR(1)

        MinRin = RinOhms(1) : MaxRin = RinOhms(1) :  MinXin = XinOhms(1) : MaxXin = XinOhms(1)

        MinFwdGain = ForwardGainDBi(1) : MaxFwdGain = ForwardGainDBi(1)

        FOR i% = 1 TO NumFreqs%
            IF RadEfficiencyPCT(i%) =< MinimumRadiationEfficiency THEN MinimumRadiationEfficiency = RadEfficiencyPCT(i%)
            IF RadEfficiencyPCT(i%) >= MaximumRadiationEfficiency THEN MaximumRadiationEfficiency = RadEfficiencyPCT(i%)
            IF MaxGainDBI(i%)       =< MinimumMaxGain             THEN MinimumMaxGain             = MaxGainDBI(i%)
            IF MaxGainDBI(i%)       >= MaximumMaxGain             THEN MaximumMaxGain             = MaxGainDBI(i%)
            IF VSWR(i%)             =< MinVSWR                    THEN MinVSWR                    = VSWR(i%)
            IF VSWR(i%)             >= MaxVSWR                    THEN MaxVSWR                    = VSWR(i%)
            IF RinOhms(i%)          =< MinRin                     THEN MinRin                     = RinOhms(i%)
            IF RinOhms(i%)          >= MaxRin                     THEN MaxRin                     = RinOhms(i%)
            IF XinOhms(i%)          =< MinXin                     THEN MinXin                     = XinOhms(i%)
            IF XinOhms(i%)          >= MaxXin                     THEN MaxXin                     = XinOhms(i%)
            IF ForwardGainDBi(i%)   >= MaxFwdGain                 THEN MaxFwdGain                 = ForwardGainDBi(i%)
            IF ForwardGainDBi(i%)   =< MinFwdGain                 THEN MinFwdGain                 = ForwardGainDBi(i%)
        NEXT i%

        A  =  YagiFitnessCoefficients(1)  :  B  =  YagiFitnessCoefficients(2)  :  C  =  YagiFitnessCoefficients(3)  :  D  =  YagiFitnessCoefficients(4)  'use  this
notation for consistency

'          Fitness  =  (A*MaxFwdGain-B*ABS(Zo-MaxRin)-C*(MAX(ABS(MaxXin),ABS(MinXin)))-D*(MaxVSWR-MinVSWR))/(A+B+C+D)   'Yagi  radiation  efficiency  =  100%
because it isn't loaded.
        'NOTE:  THIS  IS  ONE  OF  AN  INFINITY  OF  FITNESS  FUNCTIONS.  CHANGING  THE  FUNCTION  CHANGES  THE  DECISION  SPACE  LANDSCAPE  RESULTING  IN  COMPLETELY
DIFFERENT ANTENNA DESIGNS.

'         Fitness = MaxFwdGain/(ABS(Zo-MaxRin)*(MaxVSWR-MinVSWR)*(MaxXin-MinXin)) 'compute only if run is OK

        Fitness = ForwardGainDBi(1)/5##-2##*VSWR(1)+ForwardGainDBi(NumFreqs%\2+1)-4##*VSWR(NumFreqs%\2+1)+ForwardGainDBi(NumFreqs%)-2##*VSWR(NumFreqs%)/5##

    END IF

    N% = FREEFILE
    OPEN "YAGI.NEC" FOR APPEND AS #N%
        PRINT #N%,""
        PRINT #N%,"YAGI ARRAY IN FREE SPACE"
        PRINT #N%,"CM Band center frequency, Fc = 299.8 MHz"
        PRINT #N%,"Run ID: "+RunID$
        PRINT #N%,USING$ ("Fitness           =#.####^^^^^^",Fitness)
        PRINT #N%,"Max Fwd Gain (dBi) ="+STR$(MaxFwdGain)
```

```
        PRINT #N%,"VSWR Min/Max       ="+STR$(ROUND(MinVSWR,2))+"/"+STR$(ROUND(MaxVSWR,2))+"//"+REMOVE$(STR$(Zo),ANY" ")
        PRINT #N%,"Rin Min/Max        ="+STR$(ROUND(MinRin,2))+"/"+STR$(ROUND(MaxRin,2))
        PRINT #N%,"Xin Min/Max        ="+STR$(ROUND(MinXin,2))+"/"+STR$(ROUND(MaxXin,2))
        PRINT #N%,"Gmax(dBi) Min/Max  ="+STR$(ROUND(MinimumMaxGain,2))+"/"+STR$(ROUND(MaximumMaxGain,2))
        PRINT #N%,"Eff(%) Min/Max     ="+STR$(ROUND(MinimumRadiationEfficiency,2))+"/"+STR$(ROUND(MaximumRadiationEfficiency,2))
        PRINT #N%,""
    CLOSE #N%

    N% = FREEFILE
    OPEN "YAGI.OUT" FOR APPEND AS #N%
        PRINT #N%,""
        PRINT #N%,"YAGI ARRAY IN FREE SPACE"
        PRINT #N%,"CM Band center frequency, Fc = 299.8 MHz"
        PRINT #N%,"Run ID: "+RunID$
        PRINT #N%,USING$ ("Fitness            =#.####^^^^^^",Fitness)
        PRINT #N%,"Max Fwd Gain (dBi) ="+STR$(MaxFwdGain)
        PRINT #N%,"VSWR Min/Max       ="+STR$(ROUND(MinVSWR,2))+"/"+STR$(ROUND(MaxVSWR,2))+"//"+REMOVE$(STR$(Zo),ANY" ")
        PRINT #N%,"Rin Min/Max        ="+STR$(ROUND(MinRin,2))+"/"+STR$(ROUND(MaxRin,2))
        PRINT #N%,"Xin Min/Max        ="+STR$(ROUND(MinXin,2))+"/"+STR$(ROUND(MaxXin,2))
        PRINT #N%,"Gmax(dBi) Min/Max  ="+STR$(ROUND(MinimumMaxGain,2))+"/"+STR$(ROUND(MaximumMaxGain,2))
        PRINT #N%,"Eff(%) Min/Max     ="+STR$(ROUND(MinimumRadiationEfficiency,2))+"/"+STR$(ROUND(MaximumRadiationEfficiency,2))
        PRINT #N%,""
    CLOSE #N%

'       - - ANGLES - -           - POWER GAINS -         - - - POLARIZATION - - -    - - - E(THETA) - - -    - - - E(PHI) - - -
'  THETA      PHI        VERT.   HOR.    TOTAL       AXIAL     TILT   SENSE     MAGNITUDE    PHASE      MAGNITUDE    PHASE
' DEGREES  DEGREES       DB      DB      DB         RATIO     DEG.              VOLTS/M    DEGREES      VOLTS/M    DEGREES
'   90.00     0.00       3.91 -999.99    3.91    0.00000     0.00  LINEAR    1.29504E-04     5.37    0.00000E+00    -5.24
'123456789x123456789x123456789x123456789x123456789x123456789x123456789x123456789x123456789x123456789x123456789x123456789x
'        10        20        30        40        50        60        70        80        90       100       110       120

    YAGI_ARRAY = Fitness

END FUNCTION 'YAGI_ARRAY()

'------------------------

**************************************  END PROGRAM 'CFO_R_ONLY_LOADED_BOWTIE_YAGI_07-10-2011.BAS'  ***********************************
```

# Appendix II

```
CM File: BESTBOWTIE.NEC
CM R-LOADED BOWTIE IN FREE SPACE WITH
CM Zo AS AN OPTIMIZATION PARAMETER.
CM Antenna in Y-Z plane.
CM Run ID: 07022011_221747
CM Fitness function:
CM [Min(Eff)+5*Min(Gmax)]/[|Zo-MaxRin|*(MaxVSWR-MinVSWR)*(MaxXin-MinXin)]
CM Arm Length = .051 meters
CM Bowtie HALF Angle = 39.4 degrees
CM Zo = 715 ohms
CM Rload = 166.93 ohms
CM Loaded Seg # = 6/9
CM File ID 07032011225840
CM Nd = 5, p = 1, j = 35
R-LOADED BOWTIE ANTENNA IN FREE SPACE: SUMMARY NEC DATA
-------------------------------------------------------
```

| F(MHz) | Rad Eff (%) | Max Gain (dBi) | Min Gain (dBi) | Fwd Gain (dBi) | Rin (ohms) | Xin (ohms) | VSWR//715 |
|---|---|---|---|---|---|---|---|
| 200.00 | 0.000 | -13.22 | -13.23 | -13.22 | 52.28 | -808.22 | 31.19 |
| 215.00 | 0.000 | -12.62 | -12.64 | -12.62 | 52.79 | -745.39 | 28.30 |
| 230.00 | 0.000 | -12.07 | -12.08 | -12.07 | 53.33 | -690.31 | 25.94 |
| 245.00 | 0.000 | -11.55 | -11.57 | -11.55 | 53.92 | -641.56 | 23.97 |
| 260.00 | 0.000 | -11.07 | -11.09 | -11.07 | 54.54 | -598.04 | 22.31 |
| 275.00 | 0.370 | -10.62 | -10.65 | -10.62 | 55.20 | -558.90 | 20.89 |
| 290.00 | 0.960 | -10.20 | -10.23 | -10.20 | 55.91 | -523.46 | 19.67 |
| 305.00 | 1.570 | -9.80 | -9.83 | -9.80 | 56.66 | -491.17 | 18.60 |
| 320.00 | 2.200 | -9.42 | -9.46 | -9.42 | 57.45 | -461.59 | 17.66 |
| 335.00 | 2.840 | -9.07 | -9.11 | -9.07 | 58.28 | -434.35 | 16.82 |
| 350.00 | 3.500 | -8.73 | -8.78 | -8.73 | 59.16 | -409.16 | 16.06 |
| 365.00 | 4.180 | -8.42 | -8.46 | -8.42 | 60.08 | -385.75 | 15.38 |
| 380.00 | 4.860 | -8.11 | -8.17 | -8.11 | 61.05 | -363.93 | 14.76 |
| 395.00 | 5.560 | -7.83 | -7.88 | -7.83 | 62.06 | -343.50 | 14.20 |
| 410.00 | 6.260 | -7.55 | -7.61 | -7.55 | 63.13 | -324.33 | 13.67 |
| 425.00 | 6.980 | -7.29 | -7.36 | -7.29 | 64.24 | -306.26 | 13.19 |
| 440.00 | 7.690 | -7.04 | -7.12 | -7.04 | 65.40 | -289.19 | 12.73 |
| 455.00 | 8.420 | -6.81 | -6.88 | -6.81 | 66.61 | -273.02 | 12.31 |
| 470.00 | 9.140 | -6.58 | -6.66 | -6.58 | 67.88 | -257.66 | 11.91 |
| 485.00 | 9.870 | -6.36 | -6.45 | -6.36 | 69.19 | -243.04 | 11.54 |
| 500.00 | 10.610 | -6.15 | -6.25 | -6.15 | 70.57 | -229.10 | 11.18 |
| 515.00 | 11.340 | -5.96 | -6.06 | -5.96 | 71.99 | -215.77 | 10.84 |
| 530.00 | 12.070 | -5.77 | -5.87 | -5.77 | 73.47 | -203.00 | 10.52 |
| 545.00 | 12.800 | -5.58 | -5.69 | -5.58 | 75.02 | -190.74 | 10.22 |
| 560.00 | 13.530 | -5.41 | -5.53 | -5.41 | 76.62 | -178.97 | 9.92 |
| 575.00 | 14.250 | -5.24 | -5.36 | -5.24 | 78.28 | -167.63 | 9.64 |
| 590.00 | 14.970 | -5.08 | -5.21 | -5.08 | 80.00 | -156.71 | 9.37 |
| 605.00 | 15.680 | -4.92 | -5.06 | -4.92 | 81.79 | -146.16 | 9.11 |
| 620.00 | 16.390 | -4.77 | -4.92 | -4.77 | 83.64 | -135.96 | 8.86 |
| 635.00 | 17.100 | -4.63 | -4.78 | -4.63 | 85.56 | -126.10 | 8.62 |
| 650.00 | 17.790 | -4.49 | -4.65 | -4.49 | 87.54 | -116.53 | 8.39 |
| 665.00 | 18.480 | -4.35 | -4.52 | -4.35 | 89.60 | -107.26 | 8.16 |
| 680.00 | 19.160 | -4.22 | -4.40 | -4.22 | 91.73 | -98.26 | 7.94 |
| 695.00 | 19.830 | -4.10 | -4.29 | -4.10 | 93.93 | -89.51 | 7.73 |
| 710.00 | 20.490 | -3.98 | -4.18 | -3.98 | 96.20 | -81.01 | 7.53 |
| 725.00 | 21.140 | -3.86 | -4.07 | -3.86 | 98.55 | -72.73 | 7.33 |
| 740.00 | 21.790 | -3.75 | -3.97 | -3.75 | 100.98 | -64.68 | 7.14 |
| 755.00 | 22.420 | -3.64 | -3.87 | -3.64 | 103.49 | -56.83 | 6.95 |
| 770.00 | 23.040 | -3.54 | -3.77 | -3.54 | 106.08 | -49.18 | 6.77 |
| 785.00 | 23.650 | -3.44 | -3.68 | -3.44 | 108.76 | -41.72 | 6.60 |
| 800.00 | 24.250 | -3.34 | -3.59 | -3.34 | 111.51 | -34.45 | 6.43 |
| 815.00 | 24.820 | -3.24 | -3.51 | -3.24 | 114.34 | -27.34 | 6.26 |
| 830.00 | 25.400 | -3.15 | -3.43 | -3.15 | 117.27 | -20.42 | 6.10 |
| 845.00 | 25.960 | -3.06 | -3.35 | -3.06 | 120.29 | -13.67 | 5.95 |
| 860.00 | 26.520 | -2.97 | -3.28 | -2.97 | 123.40 | -7.09 | 5.79 |
| 875.00 | 27.060 | -2.89 | -3.20 | -2.89 | 126.60 | -0.66 | 5.65 |
| 890.00 | 27.590 | -2.81 | -3.13 | -2.81 | 129.90 | 5.60 | 5.50 |
| 905.00 | 28.100 | -2.73 | -3.07 | -2.73 | 133.29 | 11.71 | 5.37 |
| 920.00 | 28.610 | -2.65 | -3.01 | -2.65 | 136.78 | 17.66 | 5.23 |
| 935.00 | 29.100 | -2.58 | -2.94 | -2.58 | 140.37 | 23.45 | 5.10 |
| 950.00 | 29.580 | -2.51 | -2.89 | -2.51 | 144.05 | 29.09 | 4.97 |
| 965.00 | 30.050 | -2.44 | -2.83 | -2.44 | 147.84 | 34.58 | 4.85 |
| 980.00 | 30.510 | -2.37 | -2.78 | -2.37 | 151.72 | 39.90 | 4.73 |
| 995.00 | 30.950 | -2.30 | -2.72 | -2.30 | 155.70 | 45.08 | 4.61 |
| 1010.00 | 31.380 | -2.24 | -2.67 | -2.24 | 159.79 | 50.09 | 4.50 |
| 1025.00 | 31.800 | -2.18 | -2.63 | -2.18 | 163.97 | 54.94 | 4.39 |
| 1040.00 | 32.210 | -2.12 | -2.58 | -2.12 | 168.26 | 59.62 | 4.28 |
| 1055.00 | 32.610 | -2.06 | -2.54 | -2.06 | 172.64 | 64.14 | 4.18 |
| 1070.00 | 32.990 | -2.00 | -2.50 | -2.00 | 177.13 | 68.49 | 4.08 |
| 1085.00 | 33.360 | -1.94 | -2.46 | -1.94 | 181.71 | 72.67 | 3.98 |
| 1100.00 | 33.720 | -1.89 | -2.42 | -1.89 | 186.38 | 76.66 | 3.88 |
| 1115.00 | 34.070 | -1.84 | -2.38 | -1.84 | 191.16 | 80.48 | 3.79 |
| 1130.00 | 34.410 | -1.78 | -2.35 | -1.78 | 196.02 | 84.10 | 3.70 |
| 1145.00 | 34.730 | -1.73 | -2.31 | -1.73 | 200.97 | 87.54 | 3.62 |
| 1160.00 | 35.050 | -1.68 | -2.28 | -1.68 | 206.00 | 90.77 | 3.53 |
| 1175.00 | 35.350 | -1.63 | -2.25 | -1.63 | 211.12 | 93.81 | 3.45 |
| 1190.00 | 35.640 | -1.59 | -2.22 | -1.59 | 216.31 | 96.63 | 3.37 |
| 1205.00 | 35.920 | -1.54 | -2.20 | -1.54 | 221.58 | 99.24 | 3.30 |
| 1220.00 | 36.190 | -1.50 | -2.17 | -1.50 | 226.91 | 101.63 | 3.22 |
| 1235.00 | 36.450 | -1.45 | -2.15 | -1.45 | 232.32 | 103.80 | 3.15 |
| 1250.00 | 36.690 | -1.41 | -2.13 | -1.41 | 237.76 | 105.73 | 3.08 |
| 1265.00 | 36.930 | -1.37 | -2.10 | -1.37 | 243.25 | 107.43 | 3.01 |
| 1280.00 | 37.150 | -1.33 | -2.08 | -1.33 | 248.77 | 108.88 | 2.95 |
| 1295.00 | 37.360 | -1.29 | -2.07 | -1.29 | 254.32 | 110.09 | 2.89 |
| 1310.00 | 37.570 | -1.25 | -2.05 | -1.25 | 259.89 | 111.04 | 2.83 |
| 1325.00 | 37.760 | -1.21 | -2.03 | -1.21 | 265.46 | 111.74 | 2.77 |
| 1340.00 | 37.940 | -1.17 | -2.02 | -1.17 | 271.03 | 112.18 | 2.71 |
| 1355.00 | 38.110 | -1.13 | -2.00 | -1.13 | 276.58 | 112.35 | 2.66 |
| 1370.00 | 38.270 | -1.10 | -1.99 | -1.10 | 282.11 | 112.26 | 2.61 |
| 1385.00 | 38.410 | -1.06 | -1.98 | -1.06 | 287.59 | 111.89 | 2.56 |
| 1400.00 | 38.550 | -1.03 | -1.97 | -1.03 | 293.01 | 111.26 | 2.51 |
| 1415.00 | 38.680 | -0.99 | -1.96 | -0.99 | 298.36 | 110.35 | 2.47 |
| 1430.00 | 38.800 | -0.96 | -1.95 | -0.96 | 303.63 | 109.18 | 2.42 |
| 1445.00 | 38.910 | -0.93 | -1.95 | -0.93 | 308.79 | 107.73 | 2.38 |
| 1460.00 | 39.000 | -0.90 | -1.94 | -0.90 | 313.84 | 106.02 | 2.34 |
| 1475.00 | 39.090 | -0.86 | -1.94 | -0.86 | 318.76 | 104.04 | 2.30 |
| 1490.00 | 39.170 | -0.83 | -1.94 | -0.83 | 323.52 | 101.81 | 2.27 |
| 1505.00 | 39.230 | -0.80 | -1.93 | -0.80 | 328.12 | 99.32 | 2.23 |
| 1520.00 | 39.290 | -0.77 | -1.93 | -0.77 | 332.54 | 96.59 | 2.20 |
| 1535.00 | 39.340 | -0.75 | -1.93 | -0.75 | 336.77 | 93.63 | 2.17 |
| 1550.00 | 39.380 | -0.72 | -1.94 | -0.72 | 340.78 | 90.44 | 2.14 |
| 1565.00 | 39.400 | -0.69 | -1.94 | -0.69 | 344.57 | 87.03 | 2.11 |
| 1580.00 | 39.420 | -0.66 | -1.94 | -0.66 | 348.12 | 83.42 | 2.09 |
| 1595.00 | 39.430 | -0.64 | -1.95 | -0.64 | 351.41 | 79.62 | 2.07 |
| 1610.00 | 39.420 | -0.61 | -1.95 | -0.61 | 354.44 | 75.65 | 2.05 |
| 1625.00 | 39.410 | -0.59 | -1.96 | -0.59 | 357.18 | 71.52 | 2.03 |
| 1640.00 | 39.390 | -0.56 | -1.97 | -0.56 | 359.64 | 67.26 | 2.01 |
| 1655.00 | 39.360 | -0.54 | -1.98 | -0.54 | 361.81 | 62.86 | 2.00 |
| 1670.00 | 39.320 | -0.52 | -1.99 | -0.52 | 363.66 | 58.37 | 1.98 |
| 1685.00 | 39.270 | -0.49 | -2.00 | -0.49 | 365.21 | 53.79 | 1.97 |
| 1700.00 | 39.210 | -0.47 | -2.01 | -0.47 | 366.44 | 49.14 | 1.96 |
| 1715.00 | 39.140 | -0.45 | -2.03 | -0.45 | 367.34 | 44.45 | 1.96 |
| 1730.00 | 39.060 | -0.43 | -2.04 | -0.43 | 367.93 | 39.73 | 1.95 |
| 1745.00 | 38.970 | -0.41 | -2.06 | -0.41 | 368.19 | 35.01 | 1.95 |
| 1760.00 | 38.870 | -0.39 | -2.08 | -0.39 | 368.13 | 30.31 | 1.95 |
| 1775.00 | 38.760 | -0.37 | -2.09 | -0.37 | 367.75 | 25.64 | 1.95 |
| 1790.00 | 38.640 | -0.36 | -2.11 | -0.36 | 367.05 | 21.02 | 1.95 |
| 1805.00 | 38.510 | -0.34 | -2.14 | -0.34 | 366.04 | 16.49 | 1.95 |
| 1820.00 | 38.380 | -0.32 | -2.16 | -0.32 | 364.73 | 12.04 | 1.96 |
| 1835.00 | 38.230 | -0.31 | -2.18 | -0.31 | 363.12 | 7.71 | 1.97 |
| 1850.00 | 38.080 | -0.29 | -2.21 | -0.29 | 361.22 | 3.51 | 1.98 |
| 1865.00 | 37.910 | -0.28 | -2.23 | -0.28 | 359.04 | -0.54 | 1.99 |
| 1880.00 | 37.740 | -0.26 | -2.26 | -0.26 | 356.60 | -4.44 | 2.01 |
| 1895.00 | 37.560 | -0.25 | -2.29 | -0.25 | 353.90 | -8.16 | 2.02 |
| 1910.00 | 37.370 | -0.24 | -2.31 | -0.24 | 350.97 | -11.68 | 2.04 |
| 1925.00 | 37.170 | -0.23 | -2.34 | -0.23 | 347.80 | -15.01 | 2.06 |
| 1940.00 | 36.960 | -0.22 | -2.38 | -0.22 | 344.42 | -18.13 | 2.08 |
| 1955.00 | 36.740 | -0.21 | -2.41 | -0.21 | 340.83 | -21.02 | 2.10 |
| 1970.00 | 36.520 | -0.21 | -2.44 | -0.21 | 337.06 | -23.68 | 2.12 |
| 1985.00 | 36.290 | -0.20 | -2.48 | -0.20 | 333.12 | -26.11 | 2.15 |
| 2000.00 | 36.040 | -0.20 | -2.51 | -0.20 | 329.02 | -28.28 | 2.18 |
| 2015.00 | 35.790 | -0.19 | -2.55 | -0.19 | 324.75 | -30.17 | 2.21 |
| 2030.00 | 35.530 | -0.19 | -2.59 | -0.19 | 320.35 | -31.80 | 2.24 |
| 2045.00 | 35.260 | -0.19 | -2.62 | -0.19 | 315.84 | -33.18 | 2.27 |
| 2060.00 | 34.980 | -0.19 | -2.66 | -0.19 | 311.24 | -34.29 | 2.30 |
| 2075.00 | 34.700 | -0.19 | -2.70 | -0.19 | 306.54 | -35.14 | 2.34 |
| 2090.00 | 34.420 | -0.20 | -2.74 | -0.20 | 301.78 | -35.72 | 2.38 |
| 2105.00 | 34.130 | -0.20 | -2.78 | -0.20 | 296.97 | -36.03 | 2.42 |
| 2120.00 | 33.830 | -0.21 | -2.83 | -0.21 | 292.11 | -36.08 | 2.46 |
| 2135.00 | 33.540 | -0.22 | -2.87 | -0.22 | 287.22 | -35.86 | 2.50 |
| 2150.00 | 33.240 | -0.23 | -2.91 | -0.23 | 282.31 | -35.38 | 2.54 |
| 2165.00 | 32.950 | -0.24 | -2.95 | -0.24 | 277.39 | -34.64 | 2.58 |
| 2180.00 | 32.650 | -0.26 | -2.99 | -0.26 | 272.48 | -33.63 | 2.63 |
| 2195.00 | 32.360 | -0.28 | -3.03 | -0.28 | 267.58 | -32.36 | 2.68 |
| 2210.00 | 32.070 | -0.30 | -3.07 | -0.30 | 262.70 | -30.84 | 2.73 |
| 2225.00 | 31.780 | -0.33 | -3.11 | -0.33 | 257.87 | -29.06 | 2.78 |
| 2240.00 | 31.510 | -0.35 | -3.15 | -0.35 | 253.07 | -27.03 | 2.83 |
| 2255.00 | 31.240 | -0.39 | -3.19 | -0.39 | 248.33 | -24.75 | 2.88 |
| 2270.00 | 30.980 | -0.42 | -3.22 | -0.42 | 243.65 | -22.23 | 2.94 |
| 2285.00 | 30.740 | -0.46 | -3.25 | -0.46 | 239.04 | -19.46 | 2.99 |
| 2300.00 | 30.520 | -0.50 | -3.27 | -0.50 | 234.51 | -16.45 | 3.05 |
| 2315.00 | 30.310 | -0.55 | -3.30 | -0.55 | 230.07 | -13.21 | 3.11 |
| 2330.00 | 30.120 | -0.61 | -3.31 | -0.61 | 225.72 | -9.73 | 3.17 |
| 2345.00 | 29.970 | -0.66 | -3.33 | -0.66 | 221.48 | -6.02 | 3.23 |
| 2360.00 | 29.840 | -0.73 | -3.33 | -0.73 | 217.35 | -2.08 | 3.29 |
| 2375.00 | 29.750 | -0.80 | -3.33 | -0.80 | 213.33 | 2.09 | 3.35 |
| 2390.00 | 29.690 | -0.87 | -3.32 | -0.87 | 209.44 | 6.47 | 3.41 |
| 2405.00 | 29.680 | -0.96 | -3.31 | -0.96 | 205.68 | 11.08 | 3.48 |
| 2420.00 | 29.720 | -1.05 | -3.29 | -1.05 | 202.07 | 15.91 | 3.54 |
| 2435.00 | 29.810 | -1.15 | -3.28 | -1.15 | 198.60 | 20.96 | 3.60 |
| 2450.00 | 29.960 | -1.25 | -3.27 | -1.25 | 195.29 | 26.22 | 3.67 |
| 2465.00 | 30.170 | -1.37 | -3.27 | -1.37 | 192.14 | 31.69 | 3.73 |
| 2480.00 | 30.460 | -1.50 | -3.29 | -1.50 | 189.17 | 37.37 | 3.79 |
| 2495.00 | 30.820 | -1.63 | -3.31 | -1.63 | 186.38 | 43.26 | 3.85 |
| 2510.00 | 31.270 | -1.78 | -3.34 | -1.78 | 183.77 | 49.36 | 3.91 |
| 2525.00 | 31.810 | -1.94 | -3.39 | -1.94 | 181.37 | 55.66 | 3.97 |
| 2540.00 | 32.440 | -2.11 | -3.44 | -2.11 | 179.18 | 62.16 | 4.02 |
| 2555.00 | 33.180 | -2.30 | -3.51 | -2.30 | 177.20 | 68.86 | 4.07 |
| 2570.00 | 34.030 | -2.41 | -3.59 | -2.50 | 175.46 | 75.76 | 4.12 |
| 2585.00 | 34.980 | -2.26 | -3.69 | -2.72 | 173.96 | 82.85 | 4.17 |
| 2600.00 | 36.060 | -2.10 | -3.80 | -2.95 | 172.72 | 90.14 | 4.21 |
| 2615.00 | 37.260 | -1.93 | -3.93 | -3.20 | 171.74 | 97.61 | 4.25 |
| 2630.00 | 38.570 | -1.75 | -4.08 | -3.46 | 171.04 | 105.27 | 4.28 |
| 2645.00 | 40.020 | -1.57 | -4.25 | -3.74 | 170.64 | 113.11 | 4.30 |
| 2660.00 | 41.590 | -1.38 | -4.44 | -4.04 | 170.55 | 121.12 | 4.32 |
| 2675.00 | 43.280 | -1.19 | -4.65 | -4.36 | 170.79 | 129.30 | 4.33 |
| 2690.00 | 45.080 | -1.00 | -4.89 | -4.69 | 171.37 | 137.64 | 4.34 |
| 2705.00 | 47.010 | -0.80 | -5.15 | -5.03 | 172.31 | 146.14 | 4.33 |
| 2720.00 | 49.030 | -0.61 | -5.45 | -5.39 | 173.64 | 154.78 | 4.32 |
| 2735.00 | 51.150 | -0.42 | -5.77 | -5.76 | 175.36 | 163.56 | 4.30 |
| 2750.00 | 53.360 | -0.23 | -6.13 | -6.13 | 177.51 | 172.46 | 4.28 |
| 2765.00 | 55.640 | -0.04 | -6.49 | -6.49 | 180.10 | 181.46 | 4.24 |
| 2780.00 | 57.980 | 0.14 | -6.84 | -6.84 | 183.16 | 190.55 | 4.20 |
| 2795.00 | 60.360 | 0.31 | -7.17 | -7.17 | 186.70 | 199.71 | 4.15 |
| 2810.00 | 62.770 | 0.48 | -7.47 | -7.47 | 190.76 | 208.92 | 4.09 |
| 2825.00 | 65.190 | 0.64 | -7.72 | -7.72 | 195.35 | 218.15 | 4.03 |
| 2840.00 | 67.600 | 0.79 | -7.90 | -7.90 | 200.50 | 227.38 | 3.95 |
| 2855.00 | 69.990 | 0.93 | -8.02 | -8.02 | 206.22 | 236.56 | 3.88 |
| 2870.00 | 72.330 | 1.07 | -8.06 | -8.06 | 212.55 | 245.67 | 3.79 |
| 2885.00 | 74.630 | 1.19 | -8.02 | -8.02 | 219.50 | 254.67 | 3.71 |
| 2900.00 | 76.850 | 1.31 | -7.91 | -7.91 | 227.10 | 263.50 | 3.62 |
| 2915.00 | 79.000 | 1.42 | -7.73 | -7.73 | 235.34 | 272.12 | 3.52 |
| 2930.00 | 81.050 | 1.52 | -7.49 | -7.49 | 244.26 | 280.48 | 3.43 |
| 2945.00 | 83.000 | 1.61 | -7.21 | -7.21 | 253.85 | 288.52 | 3.33 |
| 2960.00 | 84.840 | 1.69 | -6.90 | -6.90 | 264.13 | 296.17 | 3.23 |
| 2975.00 | 86.570 | 1.76 | -6.57 | -6.57 | 275.08 | 303.37 | 3.13 |
| 2990.00 | 88.180 | 1.82 | -6.22 | -6.22 | 286.70 | 310.04 | 3.03 |
| 3005.00 | 89.660 | 1.88 | -5.87 | -5.87 | 298.97 | 316.12 | 2.94 |
| 3020.00 | 91.030 | 1.93 | -5.53 | -5.53 | 311.86 | 321.51 | 2.84 |
| 3035.00 | 92.280 | 1.97 | -5.19 | -5.19 | 325.34 | 326.16 | 2.75 |
| 3050.00 | 93.400 | 2.01 | -4.86 | -4.86 | 339.36 | 329.97 | 2.65 |
| 3065.00 | 94.410 | 2.04 | -4.54 | -4.54 | 353.85 | 332.87 | 2.56 |
| 3080.00 | 95.300 | 2.06 | -4.23 | -4.23 | 368.74 | 334.80 | 2.48 |
| 3095.00 | 96.080 | 2.08 | -3.94 | -3.94 | 383.96 | 335.68 | 2.39 |
| 3110.00 | 96.760 | 2.09 | -3.66 | -3.66 | 399.40 | 335.46 | 2.31 |
| 3125.00 | 97.340 | 2.10 | -3.39 | -3.39 | 414.96 | 334.10 | 2.23 |
| 3140.00 | 97.810 | 2.10 | -3.14 | -3.14 | 430.52 | 331.55 | 2.16 |
| 3155.00 | 98.200 | 2.10 | -2.90 | -2.90 | 445.96 | 327.80 | 2.08 |
| 3170.00 | 98.510 | 2.09 | -2.67 | -2.67 | 461.15 | 322.85 | 2.02 |
| 3185.00 | 98.730 | 2.08 | -2.46 | -2.46 | 475.96 | 316.72 | 1.95 |
| 3200.00 | 98.880 | 2.07 | -2.26 | -2.26 | 490.26 | 309.43 | 1.89 |
| 3215.00 | 98.950 | 2.05 | -2.06 | -2.06 | 503.92 | 301.04 | 1.83 |
| 3230.00 | 98.970 | 2.03 | -1.88 | -1.88 | 516.82 | 291.62 | 1.77 |
| 3245.00 | 98.920 | 2.01 | -1.71 | -1.71 | 528.85 | 281.25 | 1.72 |
| 3260.00 | 98.810 | 1.99 | -1.55 | -1.55 | 539.91 | 270.05 | 1.67 |
| 3275.00 | 98.660 | 1.96 | -1.39 | -1.39 | 549.93 | 258.12 | 1.62 |
| 3290.00 | 98.450 | 1.93 | -1.25 | -1.25 | 558.84 | 245.59 | 1.58 |
| 3305.00 | 98.200 | 1.90 | -1.11 | -1.11 | 566.59 | 232.59 | 1.54 |
| 3320.00 | 97.910 | 1.87 | -0.98 | -0.98 | 573.15 | 219.26 | 1.50 |
| 3335.00 | 97.580 | 1.84 | -0.86 | -0.86 | 578.53 | 205.72 | 1.46 |
| 3350.00 | 97.220 | 1.80 | -0.74 | -0.74 | 582.72 | 192.12 | 1.43 |
| 3365.00 | 96.820 | 1.76 | -0.63 | -0.63 | 585.77 | 178.56 | 1.40 |
| 3380.00 | 96.400 | 1.73 | -0.53 | -0.53 | 587.70 | 165.17 | 1.38 |
| 3395.00 | 95.950 | 1.69 | -0.43 | -0.43 | 588.57 | 152.05 | 1.35 |
| 3410.00 | 95.470 | 1.65 | -0.33 | -0.33 | 588.45 | 139.30 | 1.34 |
| 3425.00 | 94.980 | 1.60 | -0.25 | -0.25 | 587.41 | 126.99 | 1.32 |
| 3440.00 | 94.460 | 1.56 | -0.16 | -0.16 | 585.52 | 115.20 | 1.31 |
| 3455.00 | 93.920 | 1.52 | -0.08 | -0.08 | 582.88 | 103.98 | 1.30 |
| 3470.00 | 93.370 | 1.48 | -0.01 | -0.01 | 579.56 | 93.38 | 1.29 |
| 3485.00 | 92.800 | 1.43 | 0.06 | 0.06 | 575.65 | 83.42 | 1.29 |
| 3500.00 | 92.220 | 1.39 | 0.12 | 0.12 | 571.23 | 74.13 | 1.29 |
| 3515.00 | 91.630 | 1.35 | 0.19 | 0.19 | 566.37 | 65.53 | 1.29 |
| 3530.00 | 91.020 | 1.30 | 0.24 | 0.24 | 561.15 | 57.61 | 1.30 |
| 3545.00 | 90.410 | 1.26 | 0.30 | 0.30 | 555.64 | 50.37 | 1.30 |
| 3560.00 | 89.790 | 1.22 | 0.35 | 0.35 | 549.92 | 43.80 | 1.31 |
| 3575.00 | 89.160 | 1.18 | 0.39 | 0.39 | 544.03 | 37.90 | 1.32 |
| 3590.00 | 88.530 | 1.15 | 0.43 | 0.43 | 538.04 | 32.64 | 1.34 |
| 3605.00 | 87.890 | 1.11 | 0.47 | 0.47 | 531.99 | 27.99 | 1.35 |
| 3620.00 | 87.250 | 1.08 | 0.51 | 0.51 | 525.93 | 23.94 | 1.36 |
| 3635.00 | 86.610 | 1.06 | 0.54 | 0.54 | 519.90 | 20.46 | 1.38 |
| 3650.00 | 85.970 | 1.03 | 0.57 | 0.57 | 513.95 | 17.52 | 1.39 |
| 3665.00 | 85.330 | 1.01 | 0.60 | 0.60 | 508.09 | 15.09 | 1.41 |
| 3680.00 | 84.690 | 0.99 | 0.62 | 0.62 | 502.36 | 13.14 | 1.42 |
| 3695.00 | 84.050 | 0.96 | 0.64 | 0.64 | 496.79 | 11.66 | 1.44 |
| 3710.00 | 83.410 | 0.95 | 0.65 | 0.65 | 491.39 | 10.60 | 1.46 |
| 3725.00 | 82.780 | 0.93 | 0.66 | 0.66 | 486.19 | 9.94 | 1.47 |
| 3740.00 | 82.150 | 0.91 | 0.67 | 0.67 | 481.20 | 9.65 | 1.49 |
| 3755.00 | 81.540 | 0.89 | 0.68 | 0.68 | 476.44 | 9.72 | 1.50 |
| 3770.00 | 80.920 | 0.87 | 0.68 | 0.68 | 471.91 | 10.10 | 1.52 |
| 3785.00 | 80.320 | 0.85 | 0.68 | 0.68 | 467.63 | 10.78 | 1.53 |
| 3800.00 | 79.730 | 0.84 | 0.68 | 0.68 | 463.60 | 11.73 | 1.54 |
| 3815.00 | 79.150 | 0.82 | 0.68 | 0.68 | 459.84 | 12.93 | 1.56 |
| 3830.00 | 78.580 | 0.80 | 0.67 | 0.67 | 456.34 | 14.36 | 1.57 |
| 3845.00 | 78.020 | 0.78 | 0.65 | 0.65 | 453.12 | 15.99 | 1.58 |
| 3860.00 | 77.480 | 0.76 | 0.64 | 0.64 | 450.18 | 17.81 | 1.59 |
| 3875.00 | 76.950 | 0.74 | 0.62 | 0.62 | 447.51 | 19.80 | 1.60 |
| 3890.00 | 76.440 | 0.73 | 0.60 | 0.60 | 445.13 | 21.93 | 1.61 |
| 3905.00 | 75.950 | 0.71 | 0.58 | 0.58 | 443.03 | 24.19 | 1.62 |
| 3920.00 | 75.470 | 0.70 | 0.55 | 0.55 | 441.21 | 26.56 | 1.62 |
| 3935.00 | 75.020 | 0.71 | 0.52 | 0.52 | 439.68 | 29.02 | 1.63 |
| 3950.00 | 74.580 | 0.72 | 0.49 | 0.49 | 438.43 | 31.55 | 1.64 |
| 3965.00 | 74.160 | 0.75 | 0.45 | 0.45 | 437.47 | 34.15 | 1.64 |
| 3980.00 | 73.760 | 0.77 | 0.41 | 0.41 | 436.79 | 36.78 | 1.64 |
| 3995.00 | 73.390 | 0.81 | 0.37 | 0.37 | 436.39 | 39.44 | 1.65 |
| 4010.00 | 73.040 | 0.84 | 0.33 | 0.33 | 436.27 | 42.11 | 1.65 |
| 4025.00 | 72.700 | 0.88 | 0.28 | 0.28 | 436.43 | 44.78 | 1.65 |
| 4040.00 | 72.400 | 0.92 | 0.23 | 0.23 | 436.87 | 47.42 | 1.65 |
| 4055.00 | 72.110 | 0.96 | 0.18 | 0.18 | 437.57 | 50.03 | 1.65 |
| 4070.00 | 71.850 | 1.01 | 0.13 | 0.13 | 438.54 | 52.59 | 1.64 |
| 4085.00 | 71.620 | 1.06 | 0.07 | 0.07 | 439.78 | 55.07 | 1.64 |
| 4100.00 | 71.410 | 1.12 | 0.01 | 0.01 | 441.28 | 57.48 | 1.64 |
| 4115.00 | 71.220 | 1.17 | -0.05 | -0.05 | 443.03 | 59.79 | 1.63 |
| 4130.00 | 71.050 | 1.23 | -0.11 | -0.11 | 445.04 | 61.99 | 1.63 |
| 4145.00 | 70.910 | 1.29 | -0.18 | -0.18 | 447.28 | 64.07 | 1.62 |
| 4160.00 | 70.790 | 1.36 | -0.24 | -0.24 | 449.76 | 66.00 | 1.61 |
| 4175.00 | 70.700 | 1.42 | -0.31 | -0.31 | 452.48 | 67.77 | 1.60 |
| 4190.00 | 70.630 | 1.49 | -0.38 | -0.38 | 455.41 | 69.38 | 1.59 |
| 4205.00 | 70.580 | 1.55 | -0.46 | -0.46 | 458.55 | 70.80 | 1.58 |
| 4220.00 | 70.550 | 1.62 | -0.53 | -0.53 | 461.89 | 72.02 | 1.57 |
| 4235.00 | 70.550 | 1.69 | -0.60 | -0.60 | 465.43 | 73.03 | 1.56 |
| 4250.00 | 70.560 | 1.76 | -0.68 | -0.68 | 469.14 | 73.81 | 1.55 |
| 4265.00 | 70.600 | 1.83 | -0.76 | -0.76 | 473.02 | 74.35 | 1.54 |
| 4280.00 | 70.650 | 1.90 | -0.84 | -0.84 | 477.06 | 74.64 | 1.53 |
| 4295.00 | 70.720 | 1.97 | -0.91 | -0.91 | 481.23 | 74.66 | 1.51 |
| 4310.00 | 70.810 | 2.04 | -0.99 | -0.99 | 485.52 | 74.41 | 1.50 |
| 4325.00 | 70.920 | 2.11 | -1.07 | -1.07 | 489.92 | 73.87 | 1.49 |
| 4340.00 | 71.040 | 2.19 | -1.15 | -1.15 | 494.41 | 73.02 | 1.47 |
| 4355.00 | 71.180 | 2.26 | -1.24 | -1.24 | 498.96 | 71.87 | 1.46 |
| 4370.00 | 71.340 | 2.33 | -1.32 | -1.32 | 503.57 | 70.40 | 1.45 |
| 4385.00 | 71.500 | 2.40 | -1.40 | -1.40 | 508.20 | 68.61 | 1.43 |
| 4400.00 | 71.680 | 2.46 | -1.48 | -1.48 | 512.85 | 66.49 | 1.42 |
| 4415.00 | 71.870 | 2.53 | -1.56 | -1.56 | 517.48 | 64.04 | 1.40 |
| 4430.00 | 72.070 | 2.60 | -1.64 | -1.64 | 522.07 | 61.24 | 1.39 |
| 4445.00 | 72.280 | 2.67 | -1.72 | -1.72 | 526.60 | 58.11 | 1.38 |
| 4460.00 | 72.500 | 2.73 | -1.80 | -1.80 | 531.04 | 54.64 | 1.36 |
| 4475.00 | 72.730 | 2.80 | -1.88 | -1.88 | 535.38 | 50.83 | 1.35 |
| 4490.00 | 72.960 | 2.86 | -1.96 | -1.96 | 539.58 | 46.69 | 1.34 |
| 4505.00 | 73.210 | 2.92 | -2.03 | -2.03 | 543.62 | 42.22 | 1.33 |
| 4520.00 | 73.450 | 2.98 | -2.11 | -2.11 | 547.48 | 37.44 | 1.31 |
| 4535.00 | 73.710 | 3.04 | -2.18 | -2.18 | 551.14 | 32.34 | 1.30 |
| 4550.00 | 73.960 | 3.10 | -2.26 | -2.26 | 554.56 | 26.95 | 1.29 |
| 4565.00 | 74.220 | 3.16 | -2.33 | -2.33 | 557.73 | 21.27 | 1.28 |
| 4580.00 | 74.490 | 3.22 | -2.40 | -2.40 | 560.62 | 15.33 | 1.28 |
| 4595.00 | 74.750 | 3.27 | -2.47 | -2.47 | 563.22 | 9.13 | 1.27 |
| 4610.00 | 75.020 | 3.33 | -2.54 | -2.54 | 565.50 | 2.70 | 1.26 |
| 4625.00 | 75.290 | 3.38 | -2.61 | -2.61 | 567.44 | -3.93 | 1.26 |
| 4640.00 | 75.560 | 3.43 | -2.68 | -2.68 | 569.04 | -10.76 | 1.26 |
| 4655.00 | 75.830 | 3.48 | -2.74 | -2.74 | 570.27 | -17.75 | 1.26 |
| 4670.00 | 76.100 | 3.53 | -2.81 | -2.81 | 571.12 | -24.88 | 1.26 |
| 4685.00 | 76.370 | 3.58 | -2.87 | -2.87 | 571.59 | -32.12 | 1.26 |
| 4700.00 | 76.640 | 3.63 | -2.93 | -2.93 | 571.66 | -39.45 | 1.26 |
| 4715.00 | 76.910 | 3.68 | -2.99 | -2.99 | 571.33 | -46.84 | 1.27 |
| 4730.00 | 77.170 | 3.72 | -3.05 | -3.05 | 570.60 | -54.26 | 1.27 |
| 4745.00 | 77.440 | 3.77 | -3.10 | -3.10 | 569.45 | -61.69 | 1.28 |
| 4760.00 | 77.700 | 3.81 | -3.16 | -3.16 | 567.90 | -69.09 | 1.29 |
| 4775.00 | 77.960 | 3.85 | -3.22 | -3.22 | 565.95 | -76.44 | 1.30 |
| 4790.00 | 78.220 | 3.89 | -3.27 | -3.27 | 563.60 | -83.72 | 1.31 |
| 4805.00 | 78.470 | 3.93 | -3.32 | -3.32 | 560.85 | -90.89 | 1.33 |
| 4820.00 | 78.720 | 3.97 | -3.37 | -3.37 | 557.72 | -97.94 | 1.34 |
| 4835.00 | 78.960 | 4.01 | -3.43 | -3.43 | 554.22 | -104.83 | 1.35 |
| 4850.00 | 79.210 | 4.04 | -3.48 | -3.48 | 550.37 | -111.55 | 1.37 |
| 4865.00 | 79.450 | 4.08 | -3.53 | -3.53 | 546.16 | -118.08 | 1.39 |
| 4880.00 | 79.680 | 4.12 | -3.57 | -3.57 | 541.63 | -124.38 | 1.41 |
| 4895.00 | 79.910 | 4.15 | -3.62 | -3.62 | 536.78 | -130.46 | 1.43 |
| 4910.00 | 80.140 | 4.18 | -3.67 | -3.67 | 531.63 | -136.29 | 1.45 |
| 4925.00 | 80.360 | 4.21 | -3.72 | -3.72 | 526.21 | -141.85 | 1.47 |
| 4940.00 | 80.580 | 4.24 | -3.77 | -3.77 | 520.52 | -147.14 | 1.49 |
| 4955.00 | 80.800 | 4.27 | -3.81 | -3.81 | 514.60 | -152.13 | 1.51 |
| 4970.00 | 81.010 | 4.30 | -3.86 | -3.86 | 508.45 | -156.83 | 1.53 |
| 4985.00 | 81.210 | 4.33 | -3.91 | -3.91 | 502.10 | -161.22 | 1.56 |
| 5000.00 | 81.410 | 4.36 | -3.95 | -3.95 | 495.57 | -165.29 | 1.58 |
| 5015.00 | 81.610 | 4.38 | -4.00 | -4.00 | 488.88 | -169.05 | 1.60 |
| 5030.00 | 81.800 | 4.41 | -4.05 | -4.05 | 482.04 | -172.48 | 1.63 |
| 5045.00 | 81.990 | 4.43 | -4.10 | -4.10 | 475.08 | -175.58 | 1.66 |
| 5060.00 | 82.170 | 4.46 | -4.15 | -4.15 | 468.01 | -178.36 | 1.68 |
| 5075.00 | 82.350 | 4.48 | -4.20 | -4.20 | 460.85 | -180.81 | 1.71 |
| 5090.00 | 82.520 | 4.50 | -4.25 | -4.25 | 453.62 | -182.93 | 1.74 |
| 5105.00 | 82.690 | 4.52 | -4.30 | -4.30 | 446.34 | -184.72 | 1.77 |
| 5120.00 | 82.850 | 4.54 | -4.36 | -4.36 | 439.02 | -186.20 | 1.80 |
| 5135.00 | 83.010 | 4.56 | -4.41 | -4.41 | 431.68 | -187.35 | 1.83 |
| 5150.00 | 83.160 | 4.57 | -4.47 | -4.47 | 424.32 | -188.19 | 1.86 |
| 5165.00 | 83.310 | 4.59 | -4.53 | -4.53 | 416.98 | -188.72 | 1.89 |
| 5180.00 | 83.460 | 4.60 | -4.59 | -4.59 | 409.64 | -188.94 | 1.92 |
| 5195.00 | 83.600 | 4.61 | -4.65 | -4.65 | 402.34 | -188.86 | 1.95 |
| 5210.00 | 83.730 | 4.63 | -4.72 | -4.72 | 395.09 | -188.49 | 1.98 |
| 5225.00 | 83.860 | 4.64 | -4.79 | -4.79 | 387.88 | -187.83 | 2.02 |
| 5240.00 | 83.990 | 4.65 | -4.86 | -4.86 | 380.74 | -186.89 | 2.05 |
| 5255.00 | 84.110 | 4.66 | -4.93 | -4.93 | 373.67 | -185.68 | 2.09 |
| 5270.00 | 84.230 | 4.67 | -5.01 | -5.01 | 366.68 | -184.19 | 2.12 |
| 5285.00 | 84.340 | 4.68 | -5.09 | -5.09 | 359.78 | -182.45 | 2.16 |
| 5300.00 | 84.440 | 4.68 | -5.17 | -5.17 | 352.97 | -180.45 | 2.19 |
| 5315.00 | 84.550 | 4.69 | -5.26 | -5.26 | 346.27 | -178.21 | 2.23 |
| 5330.00 | 84.640 | 4.70 | -5.35 | -5.35 | 339.68 | -175.72 | 2.27 |
| 5345.00 | 84.740 | 4.70 | -5.44 | -5.44 | 333.21 | -173.00 | 2.30 |
| 5360.00 | 84.820 | 4.71 | -5.54 | -5.54 | 326.86 | -170.05 | 2.34 |
| 5375.00 | 84.910 | 4.71 | -5.65 | -5.65 | 320.63 | -166.88 | 2.38 |
| 5390.00 | 84.980 | 4.71 | -5.76 | -5.76 | 314.54 | -163.50 | 2.42 |
| 5405.00 | 85.060 | 4.71 | -5.87 | -5.87 | 308.58 | -159.90 | 2.46 |
| 5420.00 | 85.130 | 4.71 | -5.99 | -5.99 | 302.77 | -156.10 | 2.50 |
| 5435.00 | 85.190 | 4.71 | -6.11 | -6.11 | 297.10 | -152.11 | 2.54 |
| 5450.00 | 85.250 | 4.70 | -6.24 | -6.24 | 291.58 | -147.92 | 2.58 |
| 5465.00 | 85.300 | 4.70 | -6.38 | -6.38 | 286.21 | -143.54 | 2.62 |
| 5480.00 | 85.350 | 4.69 | -6.52 | -6.52 | 281.00 | -138.98 | 2.66 |
| 5495.00 | 85.400 | 4.68 | -6.67 | -6.67 | 275.94 | -134.25 | 2.70 |
| 5510.00 | 85.440 | 4.67 | -6.82 | -6.82 | 271.05 | -129.34 | 2.74 |
| 5525.00 | 85.470 | 4.66 | -6.98 | -6.98 | 266.32 | -124.26 | 2.78 |
| 5540.00 | 85.500 | 4.64 | -7.15 | -7.15 | 261.76 | -119.02 | 2.82 |
| 5555.00 | 85.520 | 4.62 | -7.32 | -7.32 | 257.37 | -113.61 | 2.86 |
| 5570.00 | 85.540 | 4.59 | -7.50 | -7.50 | 253.16 | -108.05 | 2.90 |
| 5585.00 | 85.560 | 4.57 | -7.68 | -7.68 | 249.12 | -102.33 | 2.94 |
| 5600.00 | 85.570 | 4.55 | -7.86 | -7.86 | 245.26 | -96.47 | 2.98 |
| 5615.00 | 85.570 | 4.52 | -8.05 | -8.05 | 241.59 | -90.45 | 3.01 |
| 5630.00 | 85.570 | 4.48 | -8.25 | -8.25 | 238.10 | -84.29 | 3.05 |
| 5645.00 | 85.560 | 4.44 | -8.44 | -8.44 | 234.80 | -77.99 | 3.09 |
| 5660.00 | 85.550 | 4.40 | -8.63 | -8.63 | 231.70 | -71.55 | 3.12 |
| 5675.00 | 85.530 | 4.35 | -8.81 | -8.81 | 228.79 | -64.98 | 3.15 |
| 5690.00 | 85.510 | 4.30 | -8.99 | -8.99 | 226.09 | -58.27 | 3.19 |
| 5705.00 | 85.490 | 4.24 | -9.16 | -9.16 | 223.59 | -51.43 | 3.22 |
| 5720.00 | 85.460 | 4.17 | -9.31 | -9.31 | 221.31 | -44.46 | 3.24 |
| 5735.00 | 85.420 | 4.10 | -9.45 | -9.45 | 219.24 | -37.37 | 3.27 |
| 5750.00 | 85.380 | 4.03 | -9.56 | -9.56 | 217.38 | -30.15 | 3.30 |
| 5765.00 | 85.330 | 3.95 | -9.64 | -9.64 | 215.76 | -22.81 | 3.32 |
| 5780.00 | 85.290 | 3.87 | -9.69 | -9.69 | 214.37 | -15.35 | 3.34 |
| 5795.00 | 85.230 | 3.78 | -9.70 | -9.70 | 213.22 | -7.78 | 3.35 |
| 5810.00 | 85.170 | 3.68 | -9.67 | -9.67 | 212.31 | -0.09 | 3.37 |
| 5825.00 | 85.110 | 3.57 | -9.59 | -9.59 | 211.65 | 7.70 | 3.38 |
| 5840.00 | 85.050 | 3.46 | -9.53 | -9.48 | 211.25 | 15.61 | 3.39 |
| 5855.00 | 84.980 | 3.33 | -9.46 | -9.32 | 211.12 | 23.62 | 3.39 |
| 5870.00 | 84.910 | 3.20 | -9.34 | -9.13 | 211.26 | 31.72 | 3.39 |
| 5885.00 | 84.830 | 3.05 | -9.35 | -8.90 | 211.69 | 39.93 | 3.39 |
| 5900.00 | 84.760 | 2.90 | -9.40 | -8.64 | 212.41 | 48.22 | 3.38 |
| 5915.00 | 84.680 | 2.74 | -9.41 | -8.35 | 213.43 | 56.60 | 3.37 |
| 5930.00 | 84.600 | 2.56 | -9.37 | -8.04 | 214.77 | 65.06 | 3.36 |
| 5945.00 | 84.510 | 2.37 | -9.30 | -7.72 | 216.43 | 73.60 | 3.34 |
| 5960.00 | 84.430 | 2.19 | -9.19 | -7.38 | 218.43 | 82.20 | 3.32 |
| 5975.00 | 84.340 | 1.99 | -9.41 | -7.04 | 220.77 | 90.86 | 3.30 |
| 5990.00 | 84.250 | 1.78 | -9.61 | -6.69 | 223.48 | 99.57 | 3.27 |
| 6005.00 | 84.170 | 1.56 | -9.78 | -6.35 | 226.55 | 108.32 | 3.24 |
| 6020.00 | 84.080 | 1.32 | -9.90 | -6.01 | 230.01 | 117.11 | 3.20 |
| 6035.00 | 83.990 | 1.08 | -9.97 | -5.67 | 233.88 | 125.91 | 3.16 |
| 6050.00 | 83.900 | 0.81 | -10.00 | -5.34 | 238.15 | 134.71 | 3.12 |
| 6065.00 | 83.810 | 0.54 | -9.97 | -5.02 | 242.85 | 143.50 | 3.08 |
| 6080.00 | 83.730 | 0.25 | -9.97 | -4.70 | 247.99 | 152.27 | 3.03 |
| 6095.00 | 83.640 | -0.06 | -10.41 | -4.40 | 253.59 | 160.99 | 2.98 |
| 6110.00 | 83.550 | -0.38 | -10.83 | -4.11 | 259.66 | 169.64 | 2.93 |
| 6125.00 | 83.470 | -0.72 | -11.23 | -3.82 | 266.21 | 178.21 | 2.88 |
| 6140.00 | 83.380 | -1.08 | -11.59 | -3.55 | 273.26 | 186.66 | 2.82 |
| 6155.00 | 83.300 | -1.46 | -11.90 | -3.29 | 280.83 | 194.97 | 2.77 |
| 6170.00 | 83.210 | -1.85 | -12.14 | -3.04 | 288.92 | 203.11 | 2.71 |
| 6185.00 | 83.130 | -2.27 | -12.30 | -2.80 | 297.54 | 211.05 | 2.65 |
| 6200.00 | 83.040 | -2.57 | -12.37 | -2.57 | 306.70 | 218.76 | 2.59 |
| 6215.00 | 82.960 | -2.35 | -12.37 | -2.35 | 316.42 | 226.18 | 2.53 |
| 6230.00 | 82.880 | -2.14 | -12.96 | -2.14 | 326.70 | 233.30 | 2.47 |
| 6245.00 | 82.790 | -1.95 | -13.80 | -1.95 | 337.54 | 240.06 | 2.42 |
| 6260.00 | 82.710 | -1.76 | -14.66 | -1.76 | 348.94 | 246.41 | 2.36 |
| 6275.00 | 82.620 | -1.58 | -15.52 | -1.58 | 360.90 | 252.31 | 2.30 |
| 6290.00 | 82.540 | -1.40 | -16.35 | -1.40 | 373.40 | 257.70 | 2.24 |
| 6305.00 | 82.450 | -1.24 | -17.08 | -1.24 | 386.45 | 262.54 | 2.18 |
| 6320.00 | 82.370 | -1.09 | -17.65 | -1.09 | 400.01 | 266.77 | 2.13 |
| 6335.00 | 82.280 | -0.94 | -17.98 | -0.94 | 414.07 | 270.33 | 2.07 |
| 6350.00 | 82.190 | -0.80 | -18.03 | -0.80 | 428.59 | 273.16 | 2.01 |
| 6365.00 | 82.100 | -0.67 | -17.80 | -0.67 | 443.54 | 275.21 | 1.96 |
| 6380.00 | 82.000 | -0.54 | -19.38 | -0.54 | 458.88 | 276.42 | 1.91 |
| 6395.00 | 81.910 | -0.42 | -21.62 | -0.42 | 474.55 | 276.74 | 1.86 |
| 6410.00 | 81.810 | -0.31 | -24.54 | -0.31 | 490.49 | 276.11 | 1.81 |
| 6425.00 | 81.710 | -0.20 | -28.76 | -0.20 | 506.64 | 274.49 | 1.76 |
| 6440.00 | 81.600 | -0.10 | -35.79 | -0.10 | 522.92 | 271.83 | 1.71 |
| 6455.00 | 81.500 | 0.00 | -35.99 | 0.00 | 539.25 | 268.10 | 1.67 |
| 6470.00 | 81.390 | 0.09 | -29.10 | 0.09 | 555.53 | 263.27 | 1.62 |
| 6485.00 | 81.270 | 0.17 | -25.04 | 0.17 | 571.68 | 257.32 | 1.58 |
| 6500.00 | 81.160 | 0.25 | -27.53 | 0.25 | 587.58 | 250.24 | 1.54 |
| 6515.00 | 81.040 | 0.33 | -34.00 | 0.33 | 603.15 | 242.03 | 1.50 |
| 6530.00 | 80.910 | 0.40 | -31.28 | 0.40 | 618.27 | 232.71 | 1.46 |
| 6545.00 | 80.780 | 0.47 | -25.68 | 0.47 | 632.83 | 222.31 | 1.42 |
| 6560.00 | 80.650 | 0.54 | -22.08 | 0.54 | 646.73 | 210.86 | 1.38 |
| 6575.00 | 80.510 | 0.60 | -19.97 | 0.60 | 659.87 | 198.43 | 1.35 |
| 6590.00 | 80.370 | 0.65 | -18.58 | 0.65 | 672.16 | 185.07 | 1.31 |
| 6605.00 | 80.230 | 0.71 | -17.25 | 0.71 | 683.50 | 170.87 | 1.28 |
| 6620.00 | 80.080 | 0.76 | -16.03 | 0.76 | 693.81 | 155.91 | 1.25 |
| 6635.00 | 79.920 | 0.81 | -14.95 | 0.81 | 703.04 | 140.30 | 1.22 |
| 6650.00 | 79.760 | 0.85 | -14.07 | 0.85 | 711.11 | 124.13 | 1.19 |
| 6665.00 | 79.590 | 0.89 | -13.29 | 0.89 | 718.00 | 107.53 | 1.16 |
| 6680.00 | 79.420 | 0.93 | -12.57 | 0.93 | 723.68 | 90.60 | 1.13 |
| 6695.00 | 79.240 | 0.96 | -11.91 | 0.96 | 728.12 | 73.47 | 1.11 |
| 6710.00 | 79.060 | 1.00 | -11.30 | 1.00 | 731.33 | 56.25 | 1.08 |
| 6725.00 | 78.870 | 1.03 | -10.73 | 1.03 | 733.33 | 39.06 | 1.06 |
| 6740.00 | 78.680 | 1.05 | -10.21 | 1.05 | 734.13 | 21.99 | 1.04 |
| 6755.00 | 78.480 | 1.08 | -9.72 | 1.08 | 733.79 | 5.17 | 1.03 |
| 6770.00 | 78.280 | 1.10 | -9.26 | 1.10 | 732.34 | -11.32 | 1.03 |
| 6785.00 | 78.060 | 1.16 | -8.85 | 1.12 | 729.85 | -27.40 | 1.04 |
| 6800.00 | 77.850 | 1.29 | -8.49 | 1.14 | 726.38 | -42.98 | 1.06 |
| 6815.00 | 77.630 | 1.41 | -8.14 | 1.15 | 722.00 | -58.00 | 1.08 |
| 6830.00 | 77.400 | 1.53 | -7.82 | 1.16 | 716.78 | -72.39 | 1.11 |
| 6845.00 | 77.160 | 1.64 | -7.52 | 1.17 | 710.81 | -86.12 | 1.13 |
| 6860.00 | 76.920 | 1.76 | -7.23 | 1.18 | 704.15 | -99.15 | 1.15 |
| 6875.00 | 76.680 | 1.86 | -6.96 | 1.19 | 696.89 | -111.44 | 1.17 |
| 6890.00 | 76.420 | 1.97 | -6.70 | 1.19 | 689.11 | -122.98 | 1.20 |
| 6905.00 | 76.170 | 2.07 | -6.45 | 1.19 | 680.87 | -133.75 | 1.22 |
| 6920.00 | 75.900 | 2.16 | -6.21 | 1.19 | 672.25 | -143.76 | 1.24 |
| 6935.00 | 75.630 | 2.26 | -5.98 | 1.19 | 663.32 | -153.00 | 1.26 |
| 6950.00 | 75.360 | 2.35 | -5.77 | 1.18 | 654.15 | -161.49 | 1.29 |
| 6965.00 | 75.080 | 2.43 | -5.56 | 1.17 | 644.79 | -169.22 | 1.31 |
| 6980.00 | 74.790 | 2.52 | -5.36 | 1.16 | 635.30 | -176.24 | 1.33 |
| 6995.00 | 74.500 | 2.60 | -5.17 | 1.15 | 625.72 | -182.54 | 1.35 |
| 7010.00 | 74.200 | 2.67 | -5.03 | 1.13 | 616.12 | -188.15 | 1.38 |
| 7025.00 | 73.900 | 2.75 | -4.89 | 1.11 | 606.53 | -193.10 | 1.40 |
| 7040.00 | 73.590 | 2.82 | -4.76 | 1.09 | 596.98 | -197.42 | 1.42 |
| 7055.00 | 73.280 | 2.89 | -4.63 | 1.07 | 587.52 | -201.13 | 1.44 |
| 7070.00 | 72.960 | 2.95 | -4.51 | 1.04 | 578.18 | -204.26 | 1.46 |
| 7085.00 | 72.640 | 3.02 | -4.39 | 1.01 | 568.98 | -206.83 | 1.48 |
| 7100.00 | 72.310 | 3.08 | -4.28 | 0.98 | 559.95 | -208.88 | 1.50 |
| 7115.00 | 71.980 | 3.13 | -4.17 | 0.94 | 551.11 | -210.43 | 1.52 |
| 7130.00 | 71.650 | 3.19 | -4.06 | 0.90 | 542.47 | -211.52 | 1.54 |
| 7145.00 | 71.320 | 3.24 | -3.96 | 0.86 | 534.06 | -212.16 | 1.56 |
| 7160.00 | 70.980 | 3.29 | -3.87 | 0.81 | 525.89 | -212.39 | 1.58 |
| 7175.00 | 70.640 | 3.33 | -3.77 | 0.76 | 517.97 | -212.22 | 1.60 |
| 7190.00 | 70.300 | 3.37 | -3.69 | 0.71 | 510.32 | -211.69 | 1.62 |
| 7205.00 | 69.960 | 3.42 | -3.62 | 0.65 | 502.93 | -210.81 | 1.64 |
| 7220.00 | 69.620 | 3.45 | -3.58 | 0.59 | 495.82 | -209.62 | 1.66 |
| 7235.00 | 69.280 | 3.49 | -3.55 | 0.53 | 489.00 | -208.13 | 1.67 |
| 7250.00 | 68.940 | 3.52 | -3.52 | 0.46 | 482.46 | -206.36 | 1.69 |
| 7265.00 | 68.600 | 3.55 | -3.49 | 0.38 | 476.22 | -204.33 | 1.70 |
| 7280.00 | 68.260 | 3.57 | -3.47 | 0.31 | 470.29 | -202.06 | 1.72 |
| 7295.00 | 67.930 | 3.60 | -3.45 | 0.22 | 464.65 | -199.58 | 1.73 |
| 7310.00 | 67.610 | 3.62 | -3.43 | 0.13 | 459.32 | -196.90 | 1.74 |
| 7325.00 | 67.280 | 3.63 | -3.42 | 0.04 | 454.29 | -194.03 | 1.76 |
| 7340.00 | 66.970 | 3.65 | -3.40 | -0.06 | 449.58 | -190.99 | 1.77 |
| 7355.00 | 66.660 | 3.66 | -3.39 | -0.16 | 445.17 | -187.80 | 1.78 |
| 7370.00 | 66.360 | 3.66 | -3.38 | -0.27 | 441.08 | -184.48 | 1.79 |
| 7385.00 | 66.080 | 3.67 | -3.38 | -0.39 | 437.30 | -181.04 | 1.79 |
| 7400.00 | 65.800 | 3.67 | -3.40 | -0.51 | 433.84 | -177.50 | 1.80 |
| 7415.00 | 65.530 | 3.67 | -3.46 | -0.64 | 430.69 | -173.86 | 1.81 |
| 7430.00 | 65.280 | 3.66 | -3.53 | -0.78 | 427.86 | -170.15 | 1.81 |
| 7445.00 | 65.040 | 3.66 | -3.59 | -0.92 | 425.34 | -166.38 | 1.82 |
| 7460.00 | 64.820 | 3.64 | -3.66 | -1.07 | 423.14 | -162.57 | 1.82 |
| 7475.00 | 64.610 | 3.63 | -3.74 | -1.23 | 421.27 | -158.72 | 1.82 |
| 7490.00 | 64.420 | 3.61 | -3.82 | -1.40 | 419.71 | -154.87 | 1.82 |
| 7505.00 | 64.250 | 3.59 | -3.89 | -1.57 | 418.47 | -151.01 | 1.82 |
| 7520.00 | 64.100 | 3.57 | -3.98 | -1.76 | 417.55 | -147.17 | 1.82 |
| 7535.00 | 63.970 | 3.54 | -4.06 | -1.95 | 416.95 | -143.36 | 1.82 |
| 7550.00 | 63.870 | 3.51 | -4.15 | -2.16 | 416.67 | -139.60 | 1.81 |
| 7565.00 | 63.790 | 3.48 | -4.23 | -2.37 | 416.70 | -135.90 | 1.81 |
| 7580.00 | 63.730 | 3.44 | -4.36 | -2.60 | 417.05 | -132.29 | 1.80 |
| 7595.00 | 63.690 | 3.40 | -4.54 | -2.84 | 417.72 | -128.78 | 1.79 |
| 7610.00 | 63.690 | 3.36 | -4.72 | -3.08 | 418.69 | -125.39 | 1.79 |
| 7625.00 | 63.700 | 3.32 | -4.91 | -3.34 | 419.97 | -122.14 | 1.78 |
| 7640.00 | 63.750 | 3.27 | -5.10 | -3.62 | 421.55 | -119.05 | 1.77 |
| 7655.00 | 63.820 | 3.22 | -5.29 | -3.90 | 423.43 | -116.14 | 1.76 |
| 7670.00 | 63.920 | 3.17 | -5.49 | -4.20 | 425.59 | -113.42 | 1.74 |
| 7685.00 | 64.050 | 3.11 | -5.69 | -4.51 | 428.04 | -110.93 | 1.73 |
| 7700.00 | 64.210 | 3.06 | -5.89 | -4.84 | 430.75 | -108.68 | 1.72 |
| 7715.00 | 64.390 | 3.00 | -6.08 | -5.18 | 433.72 | -106.70 | 1.71 |
| 7730.00 | 64.600 | 2.95 | -6.33 | -5.54 | 436.93 | -105.00 | 1.69 |
| 7745.00 | 64.840 | 2.89 | -6.65 | -5.91 | 440.37 | -103.62 | 1.68 |
| 7760.00 | 65.100 | 2.83 | -6.96 | -6.30 | 444.01 | -102.56 | 1.66 |
| 7775.00 | 65.390 | 2.77 | -7.29 | -6.70 | 447.84 | -101.86 | 1.65 |
| 7790.00 | 65.700 | 2.71 | -7.61 | -7.12 | 451.84 | -101.53 | 1.63 |
| 7805.00 | 66.030 | 2.65 | -7.93 | -7.56 | 455.96 | -101.59 | 1.62 |
| 7820.00 | 66.390 | 2.59 | -8.25 | -8.00 | 460.20 | -102.06 | 1.61 |
| 7835.00 | 66.770 | 2.54 | -8.61 | -8.47 | 464.51 | -102.96 | 1.59 |
| 7850.00 | 67.170 | 2.48 | -9.04 | -8.94 | 468.86 | -104.29 | 1.58 |
| 7865.00 | 67.590 | 2.43 | -9.47 | -9.41 | 473.22 | -106.08 | 1.57 |
| 7880.00 | 68.020 | 2.38 | -9.90 | -9.89 | 477.54 | -108.32 | 1.56 |
| 7895.00 | 68.470 | 2.33 | -10.37 | -10.37 | 481.79 | -111.02 | 1.55 |
| 7910.00 | 68.930 | 2.28 | -10.83 | -10.83 | 485.92 | -114.18 | 1.54 |
| 7925.00 | 69.410 | 2.24 | -11.26 | -11.26 | 489.89 | -117.80 | 1.53 |
| 7940.00 | 69.890 | 2.20 | -11.66 | -11.66 | 493.66 | -121.87 | 1.53 |
| 7955.00 | 70.390 | 2.17 | -12.01 | -12.01 | 497.19 | -126.37 | 1.52 |
| 7970.00 | 70.890 | 2.14 | -12.29 | -12.29 | 500.43 | -131.28 | 1.52 |
| 7985.00 | 71.400 | 2.11 | -12.50 | -12.50 | 503.34 | -136.58 | 1.52 |
| 8000.00 | 71.910 | 2.09 | -12.62 | -12.62 | 505.88 | -142.24 | 1.52 |
| 8015.00 | 72.420 | 2.07 | -12.64 | -12.64 | 508.01 | -148.24 | 1.52 |
| 8030.00 | 72.940 | 2.06 | -12.58 | -12.58 | 509.72 | -154.52 | 1.53 |
| 8045.00 | 73.450 | 2.06 | -12.44 | -12.44 | 510.95 | -161.05 | 1.53 |
| 8060.00 | 73.970 | 2.05 | -12.22 | -12.22 | 511.70 | -167.79 | 1.54 |
| 8075.00 | 74.480 | 2.06 | -11.95 | -11.95 | 511.94 | -174.68 | 1.55 |
| 8090.00 | 74.980 | 2.07 | -11.63 | -11.63 | 511.67 | -181.68 | 1.56 |
| 8105.00 | 75.480 | 2.08 | -11.29 | -11.29 | 510.87 | -188.74 | 1.58 |
| 8120.00 | 75.970 | 2.10 | -10.93 | -10.93 | 509.54 | -195.82 | 1.59 |
| 8135.00 | 76.460 | 2.12 | -10.56 | -10.56 | 507.70 | -202.84 | 1.61 |
| 8150.00 | 76.930 | 2.15 | -10.19 | -10.19 | 505.34 | -209.78 | 1.63 |
| 8165.00 | 77.400 | 2.18 | -9.82 | -9.82 | 502.48 | -216.59 | 1.65 |
| 8180.00 | 77.850 | 2.21 | -9.46 | -9.46 | 499.15 | -223.22 | 1.67 |
| 8195.00 | 78.300 | 2.25 | -9.11 | -9.11 | 495.37 | -229.63 | 1.70 |
| 8210.00 | 78.730 | 2.29 | -8.77 | -8.77 | 491.16 | -235.79 | 1.72 |
| 8225.00 | 79.140 | 2.34 | -8.44 | -8.44 | 486.56 | -241.66 | 1.74 |
| 8240.00 | 79.550 | 2.39 | -8.13 | -8.13 | 481.60 | -247.22 | 1.77 |
| 8255.00 | 79.940 | 2.44 | -7.83 | -7.83 | 476.32 | -252.45 | 1.80 |
| 8270.00 | 80.310 | 2.49 | -7.55 | -7.55 | 470.75 | -257.32 | 1.83 |
| 8285.00 | 80.670 | 2.55 | -7.27 | -7.27 | 464.92 | -261.82 | 1.86 |
| 8300.00 | 81.010 | 2.60 | -7.02 | -7.02 | 458.88 | -265.94 | 1.88 |
| 8315.00 | 81.340 | 2.66 | -6.77 | -6.77 | 452.65 | -269.68 | 1.92 |
| 8330.00 | 81.650 | 2.72 | -6.54 | -6.54 | 446.28 | -273.04 | 1.95 |
| 8345.00 | 81.950 | 2.78 | -6.32 | -6.32 | 439.80 | -276.00 | 1.98 |
| 8360.00 | 82.230 | 2.84 | -6.11 | -6.11 | 433.24 | -278.58 | 2.01 |
| 8375.00 | 82.490 | 2.91 | -5.92 | -5.92 | 426.62 | -280.78 | 2.04 |
| 8390.00 | 82.740 | 2.97 | -5.73 | -5.73 | 419.98 | -282.62 | 2.07 |
| 8405.00 | 82.970 | 3.03 | -5.56 | -5.56 | 413.35 | -284.10 | 2.11 |
| 8420.00 | 83.190 | 3.09 | -5.39 | -5.39 | 406.74 | -285.23 | 2.14 |
| 8435.00 | 83.380 | 3.16 | -5.24 | -5.24 | 400.17 | -286.03 | 2.17 |
| 8450.00 | 83.560 | 3.22 | -5.09 | -5.09 | 393.68 | -286.51 | 2.20 |
| 8465.00 | 83.730 | 3.28 | -4.95 | -4.95 | 387.26 | -286.68 | 2.24 |
| 8480.00 | 83.870 | 3.34 | -4.82 | -4.82 | 380.95 | -286.57 | 2.27 |
| 8495.00 | 84.010 | 3.40 | -4.71 | -4.71 | 374.75 | -286.19 | 2.30 |
| 8510.00 | 84.120 | 3.46 | -4.59 | -4.59 | 368.68 | -285.54 | 2.34 |
| 8525.00 | 84.220 | 3.52 | -4.49 | -4.49 | 362.74 | -284.66 | 2.37 |
| 8540.00 | 84.300 | 3.58 | -4.39 | -4.39 | 356.94 | -283.55 | 2.40 |
| 8555.00 | 84.370 | 3.63 | -4.31 | -4.31 | 351.29 | -282.22 | 2.43 |
| 8570.00 | 84.420 | 3.68 | -4.22 | -4.22 | 345.80 | -280.70 | 2.46 |
| 8585.00 | 84.460 | 3.74 | -4.15 | -4.15 | 340.47 | -279.00 | 2.50 |
| 8600.00 | 84.480 | 3.79 | -4.08 | -4.08 | 335.30 | -277.13 | 2.53 |
| 8615.00 | 84.490 | 3.84 | -4.02 | -4.02 | 330.30 | -275.11 | 2.56 |
| 8630.00 | 84.480 | 3.88 | -3.97 | -3.97 | 325.46 | -272.94 | 2.59 |
| 8645.00 | 84.460 | 3.93 | -3.92 | -3.92 | 320.80 | -270.64 | 2.61 |
| 8660.00 | 84.430 | 3.97 | -3.88 | -3.88 | 316.31 | -268.23 | 2.64 |
| 8675.00 | 84.380 | 4.02 | -3.84 | -3.84 | 311.98 | -265.70 | 2.67 |
| 8690.00 | 84.320 | 4.06 | -3.81 | -3.81 | 307.83 | -263.08 | 2.70 |
| 8705.00 | 84.250 | 4.09 | -3.79 | -3.79 | 303.84 | -260.38 | 2.72 |
| 8720.00 | 84.160 | 4.13 | -3.77 | -3.77 | 300.02 | -257.59 | 2.75 |
| 8735.00 | 84.060 | 4.17 | -3.76 | -3.76 | 296.37 | -254.74 | 2.77 |
| 8750.00 | 83.950 | 4.20 | -3.75 | -3.75 | 292.87 | -251.82 | 2.80 |
| 8765.00 | 83.830 | 4.23 | -3.75 | -3.75 | 289.54 | -248.85 | 2.82 |
| 8780.00 | 83.700 | 4.25 | -3.75 | -3.75 | 286.36 | -245.84 | 2.84 |
| 8795.00 | 83.560 | 4.28 | -3.76 | -3.76 | 283.34 | -242.79 | 2.86 |
| 8810.00 | 83.410 | 4.30 | -3.77 | -3.77 | 280.48 | -239.70 | 2.88 |
| 8825.00 | 83.240 | 4.32 | -3.80 | -3.80 | 277.76 | -236.60 | 2.90 |
| 8840.00 | 83.070 | 4.34 | -3.82 | -3.82 | 275.19 | -233.47 | 2.92 |
| 8855.00 | 82.890 | 4.36 | -3.85 | -3.85 | 272.76 | -230.33 | 2.93 |
| 8870.00 | 82.710 | 4.37 | -3.89 | -3.89 | 270.47 | -227.18 | 2.95 |
| 8885.00 | 82.510 | 4.38 | -3.93 | -3.93 | 268.32 | -224.03 | 2.96 |
| 8900.00 | 82.310 | 4.39 | -3.97 | -3.97 | 266.31 | -220.87 | 2.98 |
| 8915.00 | 82.100 | 4.40 | -4.03 | -4.03 | 264.43 | -217.73 | 2.99 |
| 8930.00 | 81.880 | 4.41 | -4.08 | -4.08 | 262.67 | -214.59 | 3.00 |
| 8945.00 | 81.660 | 4.41 | -4.14 | -4.14 | 261.04 | -211.47 | 3.01 |
| 8960.00 | 81.440 | 4.41 | -4.21 | -4.21 | 259.54 | -208.37 | 3.02 |
| 8975.00 | 81.200 | 4.41 | -4.28 | -4.28 | 258.15 | -205.29 | 3.03 |
| 8990.00 | 80.970 | 4.40 | -4.36 | -4.36 | 256.88 | -202.23 | 3.04 |
| 9005.00 | 80.730 | 4.40 | -4.44 | -4.44 | 255.72 | -199.20 | 3.04 |
| 9020.00 | 80.480 | 4.39 | -4.53 | -4.53 | 254.68 | -196.20 | 3.05 |
| 9035.00 | 80.240 | 4.38 | -4.62 | -4.62 | 253.74 | -193.24 | 3.05 |
| 9050.00 | 79.980 | 4.36 | -4.72 | -4.72 | 252.90 | -190.32 | 3.05 |
| 9065.00 | 79.730 | 4.35 | -4.83 | -4.83 | 252.16 | -187.44 | 3.06 |
| 9080.00 | 79.480 | 4.33 | -4.93 | -4.93 | 251.53 | -184.60 | 3.06 |
| 9095.00 | 79.220 | 4.31 | -5.05 | -5.05 | 250.99 | -181.81 | 3.06 |
| 9110.00 | 78.960 | 4.28 | -5.17 | -5.17 | 250.54 | -179.06 | 3.06 |
| 9125.00 | 78.700 | 4.26 | -5.29 | -5.29 | 250.18 | -176.37 | 3.05 |
| 9140.00 | 78.440 | 4.23 | -5.42 | -5.42 | 249.90 | -173.73 | 3.05 |
| 9155.00 | 78.180 | 4.20 | -5.56 | -5.56 | 249.71 | -171.15 | 3.05 |
| 9170.00 | 77.920 | 4.17 | -5.70 | -5.70 | 249.60 | -168.62 | 3.04 |
| 9185.00 | 77.660 | 4.14 | -5.84 | -5.84 | 249.57 | -166.16 | 3.04 |
| 9200.00 | 77.410 | 4.10 | -5.99 | -5.99 | 249.61 | -163.75 | 3.03 |
| 9215.00 | 77.150 | 4.07 | -6.15 | -6.15 | 249.72 | -161.41 | 3.03 |
| 9230.00 | 76.890 | 4.03 | -6.31 | -6.31 | 249.89 | -159.14 | 3.02 |
| 9245.00 | 76.640 | 3.98 | -6.48 | -6.48 | 250.13 | -156.94 | 3.01 |
| 9260.00 | 76.380 | 3.94 | -6.65 | -6.65 | 250.43 | -154.80 | 3.01 |
| 9275.00 | 76.130 | 3.89 | -6.83 | -6.83 | 250.78 | -152.74 | 3.00 |
| 9290.00 | 75.880 | 3.85 | -7.02 | -7.02 | 251.19 | -150.74 | 2.99 |
| 9305.00 | 75.640 | 3.80 | -7.21 | -7.21 | 251.65 | -148.83 | 2.98 |
| 9320.00 | 75.390 | 3.74 | -7.40 | -7.40 | 252.15 | -146.98 | 2.97 |
| 9335.00 | 75.150 | 3.69 | -7.60 | -7.60 | 252.70 | -145.22 | 2.96 |
| 9350.00 | 74.920 | 3.63 | -7.81 | -7.81 | 253.28 | -143.53 | 2.95 |
| 9365.00 | 74.680 | 3.58 | -8.02 | -8.02 | 253.90 | -141.92 | 2.94 |
| 9380.00 | 74.450 | 3.52 | -8.24 | -8.24 | 254.55 | -140.39 | 2.93 |
| 9395.00 | 74.220 | 3.45 | -8.46 | -8.46 | 255.23 | -138.93 | 2.92 |
| 9410.00 | 73.990 | 3.39 | -8.69 | -8.69 | 255.92 | -137.56 | 2.91 |
| 9425.00 | 73.770 | 3.33 | -8.93 | -8.93 | 256.64 | -136.27 | 2.90 |
| 9440.00 | 73.550 | 3.26 | -9.16 | -9.16 | 257.37 | -135.06 | 2.89 |
| 9455.00 | 73.340 | 3.19 | -9.41 | -9.41 | 258.12 | -133.92 | 2.88 |
| 9470.00 | 73.130 | 3.12 | -9.65 | -9.65 | 258.86 | -132.87 | 2.87 |
| 9485.00 | 72.920 | 3.05 | -9.90 | -9.90 | 259.61 | -131.90 | 2.86 |
| 9500.00 | 72.710 | 2.98 | -10.16 | -10.16 | 260.36 | -131.00 | 2.85 |
| 9515.00 | 72.510 | 2.90 | -10.41 | -10.41 | 261.10 | -130.18 | 2.84 |
| 9530.00 | 72.320 | 2.83 | -10.67 | -10.67 | 261.83 | -129.44 | 2.83 |
| 9545.00 | 72.120 | 2.75 | -10.92 | -10.92 | 262.55 | -128.78 | 2.82 |
| 9560.00 | 71.930 | 2.67 | -11.18 | -11.18 | 263.25 | -128.18 | 2.82 |
| 9575.00 | 71.750 | 2.59 | -11.43 | -11.43 | 263.92 | -127.66 | 2.81 |
| 9590.00 | 71.560 | 2.50 | -11.68 | -11.68 | 264.58 | -127.20 | 2.80 |
| 9605.00 | 71.390 | 2.42 | -11.93 | -11.93 | 265.20 | -126.82 | 2.79 |
| 9620.00 | 71.210 | 2.34 | -12.17 | -12.17 | 265.78 | -126.49 | 2.79 |
| 9635.00 | 71.040 | 2.25 | -12.39 | -12.39 | 266.33 | -126.23 | 2.78 |
| 9650.00 | 70.870 | 2.16 | -12.61 | -12.61 | 266.84 | -126.02 | 2.78 |
| 9665.00 | 70.700 | 2.07 | -12.81 | -12.81 | 267.31 | -125.88 | 2.77 |
| 9680.00 | 70.540 | 1.98 | -13.00 | -13.00 | 267.73 | -125.78 | 2.77 |
| 9695.00 | 70.380 | 1.89 | -13.16 | -13.16 | 268.10 | -125.73 | 2.76 |
| 9710.00 | 70.230 | 1.80 | -13.30 | -13.30 | 268.42 | -125.72 | 2.76 |
| 9725.00 | 70.070 | 1.71 | -13.42 | -13.42 | 268.68 | -125.76 | 2.76 |
| 9740.00 | 69.920 | 1.61 | -13.52 | -13.52 | 268.89 | -125.82 | 2.75 |
| 9755.00 | 69.780 | 1.52 | -13.58 | -13.58 | 269.05 | -125.93 | 2.75 |
| 9770.00 | 69.630 | 1.42 | -13.62 | -13.62 | 269.14 | -126.06 | 2.75 |
| 9785.00 | 69.490 | 1.32 | -13.62 | -13.62 | 269.18 | -126.21 | 2.75 |
| 9800.00 | 69.350 | 1.23 | -13.60 | -13.60 | 269.15 | -126.38 | 2.75 |
| 9815.00 | 69.220 | 1.13 | -13.55 | -13.55 | 269.06 | -126.57 | 2.75 |
| 9830.00 | 69.090 | 1.03 | -13.48 | -13.48 | 268.91 | -126.77 | 2.76 |
| 9845.00 | 68.960 | 0.93 | -13.37 | -13.37 | 268.70 | -126.97 | 2.76 |
| 9860.00 | 68.830 | 0.83 | -13.25 | -13.25 | 268.43 | -127.17 | 2.76 |
| 9875.00 | 68.700 | 0.72 | -13.11 | -13.11 | 268.10 | -127.38 | 2.76 |
| 9890.00 | 68.580 | 0.62 | -12.95 | -12.95 | 267.70 | -127.57 | 2.77 |
| 9905.00 | 68.460 | 0.52 | -12.77 | -12.77 | 267.25 | -127.75 | 2.77 |
| 9920.00 | 68.340 | 0.41 | -12.58 | -12.58 | 266.74 | -127.92 | 2.78 |
| 9935.00 | 68.230 | 0.38 | -12.39 | -12.39 | 266.18 | -128.06 | 2.79 |
| 9950.00 | 68.120 | 0.40 | -12.18 | -12.18 | 265.56 | -128.18 | 2.79 |
| 9965.00 | 68.010 | 0.42 | -11.97 | -11.97 | 264.89 | -128.28 | 2.80 |
| 9980.00 | 67.900 | 0.43 | -11.75 | -11.75 | 264.17 | -128.35 | 2.81 |
| 9995.00 | 67.790 | 0.45 | -11.53 | -11.53 | 263.40 | -128.38 | 2.82 |
| 10010.00 | 67.690 | 0.46 | -11.31 | -11.31 | 262.59 | -128.37 | 2.82 |
| 10025.00 | 67.590 | 0.48 | -11.09 | -11.09 | 261.74 | -128.33 | 2.83 |
| 10040.00 | 67.490 | 0.49 | -10.86 | -10.86 | 260.85 | -128.24 | 2.84 |
| 10055.00 | 67.390 | 0.50 | -10.64 | -10.64 | 259.92 | -128.11 | 2.85 |
| 10070.00 | 67.300 | 0.51 | -10.43 | -10.43 | 258.96 | -127.93 | 2.86 |
| 10085.00 | 67.200 | 0.52 | -10.21 | -10.21 | 257.98 | -127.70 | 2.87 |
| 10100.00 | 67.110 | 0.53 | -10.00 | -10.00 | 256.96 | -127.43 | 2.88 |
| 10115.00 | 67.030 | 0.53 | -9.79 | -9.79 | 255.93 | -127.10 | 2.89 |
| 10130.00 | 66.940 | 0.54 | -9.58 | -9.58 | 254.88 | -126.71 | 2.91 |
| 10145.00 | 66.860 | 0.54 | -9.38 | -9.38 | 253.81 | -126.28 | 2.92 |
| 10160.00 | 66.780 | 0.54 | -9.18 | -9.18 | 252.73 | -125.78 | 2.93 |
| 10175.00 | 66.700 | 0.54 | -8.99 | -8.99 | 251.64 | -125.23 | 2.94 |
| 10190.00 | 66.630 | 0.54 | -8.80 | -8.80 | 250.54 | -124.63 | 2.95 |
| 10205.00 | 66.560 | 0.54 | -8.61 | -8.61 | 249.45 | -123.96 | 2.96 |
| 10220.00 | 66.490 | 0.53 | -8.43 | -8.43 | 248.35 | -123.24 | 2.98 |
| 10235.00 | 66.420 | 0.53 | -8.25 | -8.25 | 247.26 | -122.46 | 2.99 |
| 10250.00 | 66.360 | 0.52 | -8.08 | -8.08 | 246.18 | -121.63 | 3.00 |
| 10265.00 | 66.300 | 0.51 | -7.91 | -7.91 | 245.11 | -120.74 | 3.01 |
| 10280.00 | 66.240 | 0.51 | -7.74 | -7.74 | 244.06 | -119.79 | 3.02 |
| 10295.00 | 66.190 | 0.51 | -7.58 | -7.58 | 243.02 | -118.78 | 3.03 |
| 10310.00 | 66.140 | 0.51 | -7.42 | -7.42 | 242.01 | -117.72 | 3.04 |
| 10325.00 | 66.100 | 0.51 | -7.27 | -7.27 | 241.01 | -116.61 | 3.06 |
| 10340.00 | 66.050 | 0.50 | -7.12 | -7.12 | 240.05 | -115.44 | 3.07 |
| 10355.00 | 66.020 | 0.50 | -7.00 | -6.97 | 239.11 | -114.23 | 3.08 |
| 10370.00 | 65.980 | 0.49 | -7.21 | -6.83 | 238.21 | -112.96 | 3.09 |
| 10385.00 | 65.950 | 0.48 | -7.43 | -6.69 | 237.34 | -111.64 | 3.09 |
| 10400.00 | 65.930 | 0.47 | -7.65 | -6.55 | 236.50 | -110.27 | 3.10 |
| 10415.00 | 65.900 | 0.46 | -7.88 | -6.41 | 235.71 | -108.86 | 3.11 |
| 10430.00 | 65.890 | 0.44 | -8.11 | -6.28 | 234.96 | -107.39 | 3.12 |
| 10445.00 | 65.880 | 0.43 | -8.36 | -6.16 | 234.25 | -105.89 | 3.13 |
| 10460.00 | 65.870 | 0.41 | -8.61 | -6.03 | 233.59 | -104.34 | 3.13 |
| 10475.00 | 65.870 | 0.39 | -8.88 | -5.91 | 232.98 | -102.75 | 3.14 |
| 10490.00 | 65.870 | 0.37 | -9.15 | -5.79 | 232.42 | -101.13 | 3.14 |
| 10505.00 | 65.880 | 0.35 | -9.44 | -5.67 | 231.92 | -99.46 | 3.15 |
| 10520.00 | 65.890 | 0.33 | -9.73 | -5.56 | 231.46 | -97.76 | 3.15 |
| 10535.00 | 65.910 | 0.30 | -10.04 | -5.45 | 231.07 | -96.02 | 3.16 |
| 10550.00 | 65.930 | 0.27 | -10.36 | -5.34 | 230.73 | -94.26 | 3.16 |
| 10565.00 | 65.960 | 0.24 | -10.70 | -5.23 | 230.46 | -92.46 | 3.16 |
| 10580.00 | 66.000 | 0.21 | -11.05 | -5.13 | 230.25 | -90.63 | 3.16 |
| 10595.00 | 66.040 | 0.18 | -11.42 | -5.03 | 230.10 | -88.78 | 3.16 |
| 10610.00 | 66.090 | 0.15 | -11.80 | -4.93 | 230.02 | -86.90 | 3.16 |
| 10625.00 | 66.150 | 0.11 | -12.21 | -4.83 | 230.01 | -85.00 | 3.16 |
| 10640.00 | 66.210 | 0.07 | -12.63 | -4.73 | 230.07 | -83.08 | 3.15 |
| 10655.00 | 66.270 | 0.03 | -13.08 | -4.64 | 230.20 | -81.14 | 3.15 |
| 10670.00 | 66.350 | -0.01 | -13.56 | -4.55 | 230.40 | -79.18 | 3.15 |
| 10685.00 | 66.430 | -0.05 | -14.06 | -4.46 | 230.68 | -77.21 | 3.14 |
| 10700.00 | 66.520 | -0.09 | -14.60 | -4.37 | 231.04 | -75.23 | 3.13 |
| 10715.00 | 66.610 | -0.14 | -15.17 | -4.28 | 231.47 | -73.24 | 3.13 |
| 10730.00 | 66.720 | -0.19 | -15.78 | -4.20 | 231.99 | -71.25 | 3.12 |
| 10745.00 | 66.830 | -0.24 | -16.43 | -4.11 | 232.59 | -69.24 | 3.11 |
| 10760.00 | 66.940 | -0.29 | -17.12 | -4.03 | 233.27 | -67.24 | 3.10 |
| 10775.00 | 67.070 | -0.35 | -17.87 | -3.95 | 234.04 | -65.24 | 3.08 |
| 10790.00 | 67.200 | -0.40 | -18.67 | -3.87 | 234.90 | -63.24 | 3.07 |
| 10805.00 | 67.330 | -0.46 | -19.53 | -3.79 | 235.85 | -61.25 | 3.06 |
| 10820.00 | 67.480 | -0.52 | -20.52 | -3.72 | 236.89 | -59.27 | 3.04 |
| 10835.00 | 67.630 | -0.58 | -21.86 | -3.64 | 238.02 | -57.30 | 3.03 |
| 10850.00 | 67.790 | -0.65 | -23.46 | -3.57 | 239.24 | -55.35 | 3.01 |
| 10865.00 | 67.960 | -0.71 | -25.42 | -3.50 | 240.57 | -53.42 | 2.99 |
| 10880.00 | 68.130 | -0.78 | -27.95 | -3.42 | 241.99 | -51.51 | 2.97 |
| 10895.00 | 68.320 | -0.85 | -31.45 | -3.35 | 243.51 | -49.63 | 2.95 |
| 10910.00 | 68.510 | -0.92 | -36.66 | -3.29 | 245.14 | -47.79 | 2.93 |
| 10925.00 | 68.700 | -1.00 | -39.56 | -3.22 | 246.87 | -45.97 | 2.91 |
| 10940.00 | 68.910 | -1.07 | -33.91 | -3.15 | 248.70 | -44.20 | 2.89 |
| 10955.00 | 69.120 | -1.15 | -29.55 | -3.08 | 250.64 | -42.47 | 2.86 |
| 10970.00 | 69.330 | -1.23 | -26.52 | -3.02 | 252.69 | -40.80 | 2.84 |
| 10985.00 | 69.560 | -1.31 | -24.24 | -2.95 | 254.85 | -39.17 | 2.82 |
| 11000.00 | 69.790 | -1.40 | -23.62 | -2.89 | 257.13 | -37.61 | 2.79 |
| 11015.00 | 70.030 | -1.48 | -23.14 | -2.83 | 259.51 | -36.12 | 2.76 |
| 11030.00 | 70.270 | -1.57 | -23.89 | -2.76 | 262.01 | -34.69 | 2.74 |
| 11045.00 | 70.530 | -1.66 | -25.43 | -2.70 | 264.63 | -33.35 | 2.71 |
| 11060.00 | 70.780 | -1.75 | -27.29 | -2.64 | 267.36 | -32.09 | 2.68 |
| 11075.00 | 71.050 | -1.85 | -29.66 | -2.58 | 270.21 | -30.93 | 2.65 |
| 11090.00 | 71.320 | -1.95 | -32.91 | -2.52 | 273.18 | -29.87 | 2.62 |
| 11105.00 | 71.600 | -2.05 | -38.11 | -2.46 | 276.26 | -28.92 | 2.59 |
| 11120.00 | 71.880 | -2.15 | -50.84 | -2.40 | 279.46 | -28.08 | 2.56 |
| 11135.00 | 72.170 | -2.25 | -41.51 | -2.34 | 282.79 | -27.38 | 2.53 |
| 11150.00 | 72.470 | -2.29 | -34.70 | -2.29 | 286.22 | -26.81 | 2.50 |
| 11165.00 | 72.770 | -2.23 | -30.91 | -2.23 | 289.78 | -26.39 | 2.47 |
| 11180.00 | 73.080 | -2.17 | -28.28 | -2.17 | 293.44 | -26.12 | 2.44 |
| 11195.00 | 73.390 | -2.11 | -26.27 | -2.11 | 297.22 | -26.03 | 2.41 |
| 11210.00 | 73.710 | -2.06 | -24.65 | -2.06 | 301.10 | -26.11 | 2.38 |
| 11225.00 | 74.030 | -2.00 | -23.29 | -2.00 | 305.09 | -26.39 | 2.35 |
| 11240.00 | 74.360 | -1.94 | -22.12 | -1.94 | 309.17 | -26.87 | 2.32 |
| 11255.00 | 74.700 | -1.89 | -21.09 | -1.89 | 313.35 | -27.57 | 2.29 |
| 11270.00 | 75.040 | -1.83 | -20.17 | -1.83 | 317.61 | -28.50 | 2.26 |
| 11285.00 | 75.380 | -1.78 | -19.60 | -1.78 | 321.95 | -29.68 | 2.23 |
| 11300.00 | 75.730 | -1.72 | -19.73 | -1.72 | 326.36 | -31.12 | 2.20 |
| 11315.00 | 76.080 | -1.66 | -19.81 | -1.66 | 330.82 | -32.84 | 2.17 |
| 11330.00 | 76.440 | -1.61 | -19.84 | -1.61 | 335.33 | -34.84 | 2.14 |
| 11345.00 | 76.810 | -1.55 | -19.83 | -1.55 | 339.86 | -37.15 | 2.11 |
| 11360.00 | 77.170 | -1.50 | -19.77 | -1.50 | 344.41 | -39.78 | 2.08 |
| 11375.00 | 77.550 | -1.44 | -19.67 | -1.44 | 348.95 | -42.74 | 2.06 |
| 11390.00 | 77.920 | -1.38 | -19.54 | -1.38 | 353.46 | -46.04 | 2.03 |
| 11405.00 | 78.300 | -1.33 | -19.36 | -1.33 | 357.92 | -49.71 | 2.01 |
| 11420.00 | 78.690 | -1.27 | -19.17 | -1.27 | 362.30 | -53.75 | 1.99 |
| 11435.00 | 79.080 | -1.22 | -18.94 | -1.22 | 366.58 | -58.17 | 1.97 |
| 11450.00 | 79.480 | -1.16 | -18.70 | -1.16 | 370.72 | -62.98 | 1.95 |
| 11465.00 | 79.880 | -1.10 | -18.45 | -1.10 | 374.69 | -68.18 | 1.93 |
| 11480.00 | 80.280 | -1.04 | -18.18 | -1.04 | 378.46 | -73.79 | 1.92 |
| 11495.00 | 80.690 | -0.99 | -17.91 | -0.99 | 381.99 | -79.79 | 1.90 |
| 11510.00 | 81.100 | -0.93 | -17.64 | -0.93 | 385.24 | -86.20 | 1.89 |
| 11525.00 | 81.520 | -0.87 | -17.36 | -0.87 | 388.16 | -92.99 | 1.89 |
| 11540.00 | 81.940 | -0.81 | -17.09 | -0.81 | 390.72 | -100.16 | 1.88 |
| 11555.00 | 82.360 | -0.75 | -16.81 | -0.75 | 392.87 | -107.70 | 1.88 |
| 11570.00 | 82.790 | -0.69 | -16.55 | -0.69 | 394.56 | -115.57 | 1.88 |
| 11585.00 | 83.220 | -0.63 | -16.28 | -0.63 | 395.76 | -123.76 | 1.88 |
| 11600.00 | 83.660 | -0.57 | -16.02 | -0.57 | 396.42 | -132.22 | 1.89 |
| 11615.00 | 84.100 | -0.51 | -15.77 | -0.51 | 396.50 | -140.92 | 1.90 |
| 11630.00 | 84.550 | -0.45 | -15.52 | -0.45 | 395.96 | -149.81 | 1.92 |
| 11645.00 | 85.000 | -0.39 | -15.32 | -0.39 | 394.77 | -158.84 | 1.94 |
| 11660.00 | 85.450 | -0.33 | -16.06 | -0.33 | 392.90 | -167.96 | 1.96 |
| 11675.00 | 85.910 | -0.26 | -16.88 | -0.26 | 390.33 | -177.09 | 1.99 |
| 11690.00 | 86.370 | -0.20 | -17.81 | -0.20 | 387.05 | -186.17 | 2.02 |
| 11705.00 | 86.840 | -0.14 | -18.88 | -0.14 | 383.05 | -195.14 | 2.05 |
| 11720.00 | 87.310 | -0.07 | -20.12 | -0.07 | 378.34 | -203.91 | 2.10 |
| 11735.00 | 87.780 | -0.01 | -21.61 | -0.01 | 372.91 | -212.42 | 2.14 |
| 11750.00 | 88.250 | 0.05 | -23.45 | 0.05 | 366.80 | -220.60 | 2.19 |
| 11765.00 | 88.720 | 0.12 | -25.84 | 0.12 | 360.04 | -228.38 | 2.25 |
| 11780.00 | 89.200 | 0.18 | -29.24 | 0.18 | 352.66 | -235.68 | 2.31 |
| 11795.00 | 89.670 | 0.25 | -35.08 | 0.25 | 344.72 | -242.45 | 2.37 |
| 11810.00 | 90.150 | 0.31 | -57.98 | 0.31 | 336.26 | -248.64 | 2.44 |
| 11825.00 | 90.620 | 0.38 | -35.15 | 0.38 | 327.35 | -254.20 | 2.52 |
| 11840.00 | 91.090 | 0.44 | -28.99 | 0.44 | 318.06 | -259.08 | 2.60 |
| 11855.00 | 91.560 | 0.51 | -25.35 | 0.51 | 308.45 | -263.26 | 2.69 |
| 11870.00 | 92.020 | 0.57 | -22.74 | 0.57 | 298.60 | -266.73 | 2.79 |
| 11885.00 | 92.470 | 0.63 | -20.89 | 0.63 | 288.57 | -269.46 | 2.89 |
| 11900.00 | 92.910 | 0.69 | -22.17 | 0.69 | 278.46 | -271.45 | 2.99 |
| 11915.00 | 93.340 | 0.75 | -23.67 | 0.75 | 268.31 | -272.72 | 3.11 |
| 11930.00 | 93.760 | 0.81 | -25.45 | 0.81 | 258.21 | -273.28 | 3.22 |
| 11945.00 | 94.160 | 0.87 | -27.61 | 0.87 | 248.21 | -273.14 | 3.35 |
| 11960.00 | 94.540 | 0.93 | -30.26 | 0.93 | 238.38 | -272.34 | 3.48 |
| 11975.00 | 94.890 | 0.98 | -33.20 | 0.98 | 228.76 | -270.91 | 3.62 |
| 11990.00 | 95.220 | 1.03 | -34.73 | 1.03 | 219.41 | -268.89 | 3.76 |
| 12005.00 | 95.510 | 1.08 | -32.92 | 1.08 | 210.36 | -266.32 | 3.91 |
| 12020.00 | 95.770 | 1.13 | -30.03 | 1.13 | 201.65 | -263.25 | 4.06 |
| 12035.00 | 95.990 | 1.17 | -27.51 | 1.17 | 193.31 | -259.71 | 4.22 |
| 12050.00 | 96.160 | 1.20 | -25.47 | 1.20 | 185.37 | -255.75 | 4.38 |
| 12065.00 | 96.290 | 1.24 | -23.80 | 1.24 | 177.83 | -251.42 | 4.55 |
| 12080.00 | 96.370 | 1.26 | -22.41 | 1.26 | 170.72 | -246.76 | 4.71 |
| 12095.00 | 96.380 | 1.29 | -21.22 | 1.29 | 164.04 | -241.82 | 4.88 |
| 12110.00 | 96.340 | 1.30 | -20.20 | 1.30 | 157.80 | -236.64 | 5.05 |
| 12125.00 | 96.240 | 1.31 | -20.87 | 1.31 | 151.99 | -231.26 | 5.22 |
| 12140.00 | 96.070 | 1.32 | -21.91 | 1.32 | 146.61 | -225.71 | 5.38 |
| 12155.00 | 95.830 | 1.31 | -23.09 | 1.31 | 141.66 | -220.03 | 5.54 |
| 12170.00 | 95.520 | 1.30 | -24.44 | 1.30 | 137.13 | -214.25 | 5.70 |
| 12185.00 | 95.150 | 1.28 | -25.96 | 1.28 | 133.01 | -208.41 | 5.85 |
| 12200.00 | 94.710 | 1.26 | -27.65 | 1.26 | 129.29 | -202.53 | 5.99 |
| 12215.00 | 94.210 | 1.23 | -29.30 | 1.23 | 125.95 | -196.64 | 6.12 |
| 12230.00 | 93.640 | 1.18 | -30.45 | 1.18 | 122.98 | -190.76 | 6.24 |
| 12245.00 | 93.020 | 1.14 | -30.42 | 1.14 | 120.36 | -184.92 | 6.35 |
| 12260.00 | 92.340 | 1.08 | -29.25 | 1.08 | 118.09 | -179.12 | 6.44 |
| 12275.00 | 91.630 | 1.02 | -27.60 | 1.02 | 116.14 | -173.40 | 6.53 |
| 12290.00 | 90.870 | 0.95 | -25.94 | 0.95 | 114.49 | -167.77 | 6.60 |
| 12305.00 | 90.080 | 0.87 | -24.43 | 0.87 | 113.14 | -162.24 | 6.65 |
| 12320.00 | 89.260 | 0.79 | -23.73 | 0.79 | 112.06 | -156.82 | 6.69 |
| 12335.00 | 88.430 | 0.70 | -23.13 | 0.70 | 111.24 | -151.52 | 6.72 |
| 12350.00 | 87.590 | 0.61 | -22.50 | 0.61 | 110.66 | -146.35 | 6.74 |
| 12365.00 | 86.750 | 0.51 | -21.86 | 0.51 | 110.31 | -141.33 | 6.74 |
| 12380.00 | 85.910 | 0.41 | -21.25 | 0.41 | 110.17 | -136.45 | 6.73 |
| 12395.00 | 85.080 | 0.31 | -20.66 | 0.31 | 110.22 | -131.73 | 6.71 |
| 12410.00 | 84.270 | 0.21 | -20.11 | 0.21 | 110.46 | -127.17 | 6.68 |
| 12425.00 | 83.470 | 0.10 | -19.59 | 0.10 | 110.86 | -122.77 | 6.64 |
| 12440.00 | 82.700 | 0.00 | -19.13 | 0.00 | 111.42 | -118.54 | 6.60 |
| 12455.00 | 81.950 | -0.11 | -19.44 | -0.11 | 112.11 | -114.48 | 6.55 |
| 12470.00 | 81.240 | -0.22 | -19.69 | -0.22 | 112.93 | -110.59 | 6.49 |
| 12485.00 | 80.550 | -0.32 | -19.88 | -0.32 | 113.86 | -106.87 | 6.42 |
| 12500.00 | 79.900 | -0.43 | -20.01 | -0.43 | 114.89 | -103.32 | 6.36 |
| 12515.00 | 79.280 | -0.53 | -20.07 | -0.53 | 116.00 | -99.94 | 6.29 |
| 12530.00 | 78.690 | -0.63 | -20.06 | -0.63 | 117.20 | -96.73 | 6.22 |
| 12545.00 | 78.130 | -0.72 | -20.00 | -0.72 | 118.45 | -93.70 | 6.14 |
| 12560.00 | 77.610 | -0.82 | -20.22 | -0.82 | 119.76 | -90.83 | 6.07 |
| 12575.00 | 77.120 | -0.91 | -21.24 | -0.91 | 121.10 | -88.12 | 6.00 |
| 12590.00 | 76.660 | -1.00 | -22.28 | -1.00 | 122.48 | -85.57 | 5.92 |
| 12605.00 | 76.230 | -1.08 | -23.27 | -1.08 | 123.89 | -83.19 | 5.85 |
| 12620.00 | 75.830 | -1.16 | -24.15 | -1.16 | 125.30 | -80.95 | 5.78 |
| 12635.00 | 75.460 | -1.24 | -24.81 | -1.24 | 126.72 | -78.87 | 5.71 |
| 12650.00 | 75.110 | -1.31 | -25.14 | -1.31 | 128.13 | -76.93 | 5.65 |
| 12665.00 | 74.790 | -1.38 | -25.11 | -1.38 | 129.53 | -75.14 | 5.58 |
| 12680.00 | 74.490 | -1.45 | -25.18 | -1.45 | 130.91 | -73.48 | 5.52 |
| 12695.00 | 74.210 | -1.51 | -28.14 | -1.51 | 132.26 | -71.95 | 5.46 |
| 12710.00 | 73.950 | -1.57 | -32.46 | -1.57 | 133.57 | -70.54 | 5.41 |
| 12725.00 | 73.720 | -1.63 | -40.74 | -1.63 | 134.84 | -69.25 | 5.35 |
| 12740.00 | 73.500 | -1.68 | -43.71 | -1.68 | 136.07 | -68.07 | 5.30 |
| 12755.00 | 73.300 | -1.73 | -33.97 | -1.73 | 137.25 | -67.00 | 5.26 |
| 12770.00 | 73.120 | -1.77 | -29.51 | -1.77 | 138.36 | -66.02 | 5.21 |
| 12785.00 | 72.950 | -1.82 | -27.04 | -1.82 | 139.42 | -65.14 | 5.17 |
| 12800.00 | 72.790 | -1.86 | -28.33 | -1.86 | 140.41 | -64.35 | 5.14 |
| 12815.00 | 72.650 | -1.90 | -28.19 | -1.90 | 141.34 | -63.63 | 5.10 |
| 12830.00 | 72.530 | -1.94 | -26.84 | -1.94 | 142.19 | -62.98 | 5.07 |
| 12845.00 | 72.410 | -1.97 | -25.10 | -1.97 | 142.97 | -62.40 | 5.04 |
| 12860.00 | 72.310 | -2.00 | -24.28 | -2.00 | 143.68 | -61.88 | 5.02 |
| 12875.00 | 72.210 | -2.03 | -24.87 | -2.03 | 144.30 | -61.41 | 4.99 |
| 12890.00 | 72.130 | -2.06 | -25.47 | -2.06 | 144.85 | -60.98 | 4.97 |
| 12905.00 | 72.060 | -2.09 | -26.04 | -2.09 | 145.33 | -60.59 | 4.96 |
| 12920.00 | 71.990 | -2.11 | -26.58 | -2.11 | 145.72 | -60.23 | 4.94 |
| 12935.00 | 71.940 | -2.14 | -27.06 | -2.14 | 146.04 | -59.90 | 4.93 |
| 12950.00 | 71.890 | -2.16 | -27.80 | -2.16 | 146.28 | -59.58 | 4.92 |
| 12965.00 | 71.850 | -2.18 | -28.65 | -2.18 | 146.44 | -59.28 | 4.92 |
| 12980.00 | 71.820 | -2.20 | -29.25 | -2.20 | 146.52 | -58.99 | 4.91 |
| 12995.00 | 71.800 | -2.22 | -29.47 | -2.22 | 146.54 | -58.69 | 4.91 |
| 13010.00 | 71.790 | -2.24 | -29.30 | -2.24 | 146.48 | -58.39 | 4.92 |
| 13025.00 | 71.780 | -2.26 | -28.82 | -2.26 | 146.36 | -58.09 | 4.92 |
| 13040.00 | 71.780 | -2.27 | -28.13 | -2.27 | 146.16 | -57.77 | 4.93 |
| 13055.00 | 71.780 | -2.29 | -27.36 | -2.29 | 145.90 | -57.43 | 4.93 |
| 13070.00 | 71.800 | -2.31 | -26.56 | -2.31 | 145.58 | -57.07 | 4.94 |
| 13085.00 | 71.820 | -2.32 | -25.84 | -2.32 | 145.21 | -56.68 | 4.96 |
| 13100.00 | 71.840 | -2.34 | -25.34 | -2.34 | 144.78 | -56.27 | 4.97 |
| 13115.00 | 71.880 | -2.35 | -24.84 | -2.35 | 144.29 | -55.82 | 4.99 |
| 13130.00 | 71.920 | -2.37 | -24.35 | -2.37 | 143.76 | -55.33 | 5.00 |
| 13145.00 | 71.960 | -2.39 | -23.87 | -2.39 | 143.19 | -54.81 | 5.02 |
| 13160.00 | 72.020 | -2.40 | -23.40 | -2.40 | 142.57 | -54.24 | 5.05 |
| 13175.00 | 72.080 | -2.42 | -22.95 | -2.42 | 141.92 | -53.63 | 5.07 |
| 13190.00 | 72.150 | -2.43 | -22.52 | -2.43 | 141.23 | -52.97 | 5.09 |
| 13205.00 | 72.220 | -2.45 | -22.11 | -2.45 | 140.52 | -52.26 | 5.12 |
| 13220.00 | 72.300 | -2.47 | -21.71 | -2.47 | 139.77 | -51.51 | 5.14 |
| 13235.00 | 72.390 | -2.49 | -21.33 | -2.49 | 139.01 | -50.70 | 5.17 |
| 13250.00 | 72.480 | -2.51 | -20.97 | -2.51 | 138.23 | -49.85 | 5.20 |
| 13265.00 | 72.590 | -2.52 | -20.63 | -2.52 | 137.43 | -48.94 | 5.23 |
| 13280.00 | 72.700 | -2.54 | -20.29 | -2.54 | 136.62 | -47.98 | 5.26 |
| 13295.00 | 72.810 | -2.57 | -19.98 | -2.57 | 135.80 | -46.96 | 5.29 |
| 13310.00 | 72.940 | -2.59 | -19.67 | -2.59 | 134.98 | -45.90 | 5.32 |
| 13325.00 | 73.070 | -2.61 | -19.38 | -2.61 | 134.16 | -44.78 | 5.35 |
| 13340.00 | 73.210 | -2.63 | -19.10 | -2.63 | 133.34 | -43.60 | 5.38 |
| 13355.00 | 73.360 | -2.66 | -19.37 | -2.66 | 132.53 | -42.38 | 5.41 |
| 13370.00 | 73.510 | -2.68 | -19.80 | -2.68 | 131.73 | -41.10 | 5.45 |
| 13385.00 | 73.680 | -2.71 | -20.27 | -2.71 | 130.94 | -39.77 | 5.48 |
| 13400.00 | 73.850 | -2.74 | -20.77 | -2.74 | 130.16 | -38.39 | 5.51 |
| 13415.00 | 74.030 | -2.77 | -21.33 | -2.77 | 129.40 | -36.96 | 5.54 |
| 13430.00 | 74.210 | -2.80 | -21.93 | -2.80 | 128.66 | -35.48 | 5.57 |
| 13445.00 | 74.400 | -2.83 | -22.59 | -2.83 | 127.94 | -33.95 | 5.60 |
| 13460.00 | 74.610 | -2.87 | -23.32 | -2.87 | 127.26 | -32.38 | 5.63 |
| 13475.00 | 74.820 | -2.90 | -24.12 | -2.90 | 126.60 | -30.76 | 5.66 |
| 13490.00 | 75.030 | -2.94 | -25.03 | -2.94 | 125.96 | -29.10 | 5.69 |
| 13505.00 | 75.260 | -2.98 | -26.05 | -2.98 | 125.37 | -27.40 | 5.71 |
| 13520.00 | 75.490 | -3.02 | -27.22 | -3.02 | 124.80 | -25.65 | 5.74 |
| 13535.00 | 75.720 | -3.06 | -28.57 | -3.06 | 124.28 | -23.87 | 5.76 |
| 13550.00 | 75.970 | -3.11 | -30.16 | -3.11 | 123.79 | -22.05 | 5.78 |
| 13565.00 | 76.220 | -3.15 | -32.03 | -3.15 | 123.35 | -20.19 | 5.80 |
| 13580.00 | 76.480 | -3.20 | -34.22 | -3.20 | 122.94 | -18.30 | 5.82 |
| 13595.00 | 76.740 | -3.25 | -36.49 | -3.25 | 122.58 | -16.37 | 5.84 |
| 13610.00 | 77.010 | -3.30 | -37.75 | -3.30 | 122.27 | -14.42 | 5.85 |
| 13625.00 | 77.290 | -3.35 | -36.76 | -3.35 | 122.00 | -12.44 | 5.86 |
| 13640.00 | 77.570 | -3.41 | -34.48 | -3.41 | 121.79 | -10.43 | 5.87 |
| 13655.00 | 77.850 | -3.46 | -32.17 | -3.46 | 121.62 | -8.40 | 5.88 |
| 13670.00 | 78.140 | -3.52 | -30.16 | -3.52 | 121.51 | -6.34 | 5.88 |
| 13685.00 | 78.430 | -3.58 | -28.45 | -3.58 | 121.45 | -4.26 | 5.89 |
| 13700.00 | 78.730 | -3.65 | -26.99 | -3.65 | 121.44 | -2.16 | 5.89 |
| 13715.00 | 79.030 | -3.71 | -25.71 | -3.71 | 121.48 | -0.05 | 5.89 |
| 13730.00 | 79.330 | -3.78 | -24.59 | -3.78 | 121.59 | 2.08 | 5.88 |
| 13745.00 | 79.630 | -3.85 | -24.30 | -3.85 | 121.75 | 4.22 | 5.87 |
| 13760.00 | 79.930 | -3.92 | -24.11 | -3.92 | 121.96 | 6.37 | 5.86 |
| 13775.00 | 80.240 | -3.99 | -23.84 | -3.99 | 122.24 | 8.53 | 5.85 |
| 13790.00 | 80.540 | -4.06 | -23.51 | -4.06 | 122.58 | 10.69 | 5.83 |
| 13805.00 | 80.840 | -4.14 | -23.64 | -4.14 | 122.97 | 12.87 | 5.82 |
| 13820.00 | 81.150 | -4.22 | -24.25 | -4.22 | 123.43 | 15.04 | 5.80 |
| 13835.00 | 81.450 | -4.30 | -24.92 | -4.30 | 123.94 | 17.21 | 5.77 |
| 13850.00 | 81.750 | -4.38 | -25.65 | -4.38 | 124.52 | 19.38 | 5.75 |
| 13865.00 | 82.040 | -4.46 | -26.46 | -4.46 | 125.16 | 21.55 | 5.72 |
| 13880.00 | 82.330 | -4.55 | -27.34 | -4.55 | 125.87 | 23.71 | 5.69 |
| 13895.00 | 82.620 | -4.63 | -28.29 | -4.63 | 126.63 | 25.86 | 5.65 |
| 13910.00 | 82.910 | -4.72 | -29.31 | -4.72 | 127.46 | 27.99 | 5.62 |
| 13925.00 | 83.190 | -4.81 | -30.34 | -4.81 | 128.36 | 30.12 | 5.58 |
| 13940.00 | 83.460 | -4.90 | -31.31 | -4.90 | 129.31 | 32.23 | 5.54 |
| 13955.00 | 83.730 | -4.99 | -32.05 | -4.99 | 130.33 | 34.32 | 5.50 |
| 13970.00 | 83.990 | -5.09 | -32.37 | -5.09 | 131.42 | 36.39 | 5.46 |
| 13985.00 | 84.250 | -5.18 | -32.14 | -5.18 | 132.56 | 38.43 | 5.41 |
| 14000.00 | 84.500 | -5.28 | -31.41 | -5.28 | 133.77 | 40.45 | 5.36 |
| 14015.00 | 84.740 | -5.37 | -30.38 | -5.37 | 135.04 | 42.45 | 5.31 |
| 14030.00 | 84.970 | -5.47 | -29.22 | -5.47 | 136.38 | 44.41 | 5.26 |
| 14045.00 | 85.200 | -5.57 | -28.06 | -5.57 | 137.78 | 46.33 | 5.21 |
| 14060.00 | 85.420 | -5.67 | -26.94 | -5.67 | 139.24 | 48.23 | 5.16 |
| 14075.00 | 85.630 | -5.77 | -25.89 | -5.77 | 140.76 | 50.08 | 5.11 |
| 14090.00 | 85.830 | -5.79 | -24.92 | -5.87 | 142.34 | 51.89 | 5.05 |
| 14105.00 | 86.020 | -5.69 | -24.02 | -5.97 | 143.98 | 53.66 | 5.00 |
| 14120.00 | 86.210 | -5.59 | -23.18 | -6.07 | 145.68 | 55.38 | 4.94 |
| 14135.00 | 86.380 | -5.50 | -22.40 | -6.17 | 147.44 | 57.05 | 4.88 |
| 14150.00 | 86.540 | -5.41 | -21.68 | -6.27 | 149.26 | 58.67 | 4.82 |
| 14165.00 | 86.700 | -5.33 | -21.00 | -6.37 | 151.13 | 60.23 | 4.77 |
| 14180.00 | 86.850 | -5.25 | -20.36 | -6.47 | 153.06 | 61.74 | 4.71 |
| 14195.00 | 86.980 | -5.17 | -19.77 | -6.57 | 155.04 | 63.19 | 4.65 |
| 14210.00 | 87.110 | -5.10 | -19.20 | -6.67 | 157.07 | 64.57 | 4.59 |
| 14225.00 | 87.230 | -5.03 | -18.67 | -6.77 | 159.15 | 65.88 | 4.53 |
| 14240.00 | 87.340 | -4.96 | -18.45 | -6.87 | 161.28 | 67.13 | 4.47 |
| 14255.00 | 87.440 | -4.90 | -18.34 | -6.97 | 163.46 | 68.31 | 4.42 |
| 14270.00 | 87.530 | -4.84 | -18.20 | -7.07 | 165.68 | 69.41 | 4.36 |
| 14285.00 | 87.610 | -4.78 | -18.04 | -7.16 | 167.94 | 70.43 | 4.30 |
| 14300.00 | 87.680 | -4.73 | -17.86 | -7.26 | 170.24 | 71.38 | 4.24 |
| 14315.00 | 87.740 | -4.68 | -17.87 | -7.35 | 172.57 | 72.24 | 4.19 |
| 14330.00 | 87.790 | -4.63 | -18.18 | -7.45 | 174.94 | 73.02 | 4.13 |
| 14345.00 | 87.840 | -4.58 | -18.50 | -7.54 | 177.35 | 73.71 | 4.08 |
| 14360.00 | 87.870 | -4.54 | -18.82 | -7.63 | 179.78 | 74.30 | 4.02 |
| 14375.00 | 87.900 | -4.50 | -19.14 | -7.71 | 182.23 | 74.81 | 3.97 |
| 14390.00 | 87.910 | -4.46 | -19.46 | -7.80 | 184.70 | 75.22 | 3.92 |
| 14405.00 | 87.920 | -4.42 | -19.78 | -7.88 | 187.19 | 75.53 | 3.87 |
| 14420.00 | 87.920 | -4.39 | -20.09 | -7.96 | 189.70 | 75.75 | 3.81 |
| 14435.00 | 87.910 | -4.35 | -20.40 | -8.04 | 192.21 | 75.86 | 3.77 |
| 14450.00 | 87.890 | -4.32 | -20.69 | -8.11 | 194.72 | 75.87 | 3.72 |
| 14465.00 | 87.870 | -4.29 | -20.98 | -8.19 | 197.24 | 75.77 | 3.67 |
| 14480.00 | 87.840 | -4.27 | -21.25 | -8.25 | 199.75 | 75.57 | 3.62 |
| 14495.00 | 87.790 | -4.24 | -21.50 | -8.32 | 202.25 | 75.26 | 3.58 |
| 14510.00 | 87.740 | -4.22 | -21.74 | -8.38 | 204.74 | 74.84 | 3.53 |
| 14525.00 | 87.690 | -4.20 | -21.95 | -8.44 | 207.20 | 74.31 | 3.49 |
| 14540.00 | 87.620 | -4.18 | -22.14 | -8.50 | 209.64 | 73.67 | 3.45 |
| 14555.00 | 87.550 | -4.16 | -22.29 | -8.55 | 212.06 | 72.93 | 3.41 |
| 14570.00 | 87.470 | -4.14 | -22.42 | -8.59 | 214.43 | 72.07 | 3.37 |
| 14585.00 | 87.380 | -4.12 | -22.52 | -8.64 | 216.76 | 71.10 | 3.33 |
| 14600.00 | 87.280 | -4.11 | -22.58 | -8.67 | 219.04 | 70.02 | 3.30 |
| 14615.00 | 87.180 | -4.09 | -22.61 | -8.71 | 221.26 | 68.84 | 3.26 |
| 14630.00 | 87.070 | -4.08 | -22.61 | -8.74 | 223.43 | 67.55 | 3.23 |
| 14645.00 | 86.950 | -4.07 | -22.58 | -8.76 | 225.52 | 66.15 | 3.20 |
| 14660.00 | 86.830 | -4.06 | -22.53 | -8.78 | 227.55 | 64.65 | 3.17 |
| 14675.00 | 86.700 | -4.05 | -22.44 | -8.79 | 229.49 | 63.06 | 3.14 |
| 14690.00 | 86.560 | -4.04 | -22.33 | -8.80 | 231.35 | 61.36 | 3.12 |
| 14705.00 | 86.410 | -4.04 | -22.21 | -8.80 | 233.12 | 59.58 | 3.09 |
| 14720.00 | 86.260 | -4.03 | -22.06 | -8.80 | 234.79 | 57.70 | 3.07 |
| 14735.00 | 86.100 | -4.02 | -21.90 | -8.79 | 236.36 | 55.74 | 3.05 |
| 14750.00 | 85.930 | -4.02 | -21.73 | -8.78 | 237.82 | 53.70 | 3.03 |
| 14765.00 | 85.760 | -4.02 | -21.55 | -8.76 | 239.16 | 51.59 | 3.01 |
| 14780.00 | 85.580 | -4.02 | -21.35 | -8.74 | 240.39 | 49.41 | 2.99 |
| 14795.00 | 85.390 | -4.01 | -21.16 | -8.71 | 241.49 | 47.17 | 2.98 |
| 14810.00 | 85.190 | -4.01 | -20.96 | -8.67 | 242.46 | 44.87 | 2.96 |
| 14825.00 | 84.990 | -4.01 | -20.75 | -8.63 | 243.31 | 42.52 | 2.95 |
| 14840.00 | 84.780 | -4.02 | -20.55 | -8.59 | 244.01 | 40.13 | 2.94 |
| 14855.00 | 84.570 | -4.02 | -20.34 | -8.54 | 244.58 | 37.70 | 2.93 |
| 14870.00 | 84.350 | -4.02 | -20.13 | -8.48 | 245.00 | 35.25 | 2.93 |
| 14885.00 | 84.120 | -4.03 | -19.93 | -8.42 | 245.28 | 32.79 | 2.92 |
| 14900.00 | 83.880 | -4.03 | -19.72 | -8.35 | 245.42 | 30.31 | 2.92 |
| 14915.00 | 83.640 | -4.04 | -19.52 | -8.28 | 245.41 | 27.83 | 2.92 |
| 14930.00 | 83.390 | -4.04 | -19.32 | -8.20 | 245.25 | 25.36 | 2.92 |
| 14945.00 | 83.140 | -4.05 | -19.12 | -8.12 | 244.95 | 22.91 | 2.92 |
| 14960.00 | 82.870 | -4.06 | -18.92 | -8.04 | 244.50 | 20.48 | 2.93 |
| 14975.00 | 82.610 | -4.07 | -18.72 | -7.95 | 243.91 | 18.08 | 2.93 |
| 14990.00 | 82.330 | -4.08 | -18.53 | -7.86 | 243.17 | 15.73 | 2.94 |
| 15005.00 | 82.050 | -4.09 | -18.34 | -7.76 | 242.29 | 13.43 | 2.95 |
| 15020.00 | 81.770 | -4.10 | -18.15 | -7.66 | 241.28 | 11.19 | 2.96 |
| 15035.00 | 81.470 | -4.09 | -17.97 | -7.56 | 240.14 | 9.02 | 2.98 |
| 15050.00 | 81.170 | -4.02 | -17.79 | -7.45 | 238.87 | 6.93 | 2.99 |
| 15065.00 | 80.870 | -3.96 | -17.61 | -7.34 | 237.47 | 4.91 | 3.01 |
| 15080.00 | 80.560 | -3.89 | -17.43 | -7.23 | 235.96 | 2.99 | 3.03 |
| 15095.00 | 80.240 | -3.82 | -17.42 | -7.12 | 234.34 | 1.16 | 3.05 |
| 15110.00 | 79.920 | -3.75 | -17.64 | -7.00 | 232.61 | -0.56 | 3.07 |
| 15125.00 | 79.600 | -3.67 | -17.88 | -6.89 | 230.78 | -2.17 | 3.10 |
| 15140.00 | 79.270 | -3.59 | -18.12 | -6.77 | 228.86 | -3.67 | 3.12 |
| 15155.00 | 78.930 | -3.52 | -18.38 | -6.65 | 226.86 | -5.05 | 3.15 |
| 15170.00 | 78.590 | -3.44 | -18.65 | -6.52 | 224.79 | -6.30 | 3.18 |
| 15185.00 | 78.250 | -3.37 | -18.93 | -6.40 | 222.65 | -7.43 | 3.21 |
| 15200.00 | 77.910 | -3.29 | -19.23 | -6.28 | 220.44 | -8.43 | 3.24 |

# Appendix III

```
CM File: BESTBOWTIE.NEC
CM R-LOADED BOWTIE IN FREE SPACE WITH
CM Zo AS AN OPTIMIZATION PARAMETER.
CM Antenna in Y-Z plane.
CM Run ID: 07022011_221747
CM Fitness function:
CM [Min(Eff)+5*Min(Gmax)]/[|Zo-MaxRin|*(MaxVSWR-MinVSWR)*(MaxXin-MinXin)]
CM Arm Length = .051 meters
CM Bowtie HALF Angle = 39.4 degrees
CM Zo = 715 ohms
CM Rload = 166.93 ohms
CM NOTE: LOADING HAS BEEN REMOVED FOR FS AVG GAIN CALC
CM Loaded Seg # = 6/9
CM File ID 07032011225840
CM Nd = 5, p = 1, j = 35
```

FREE SPACE BOWTIE ***WITHOUT LOADING***: SUMMARY NEC DATA
======================================================

| F(MHz) | Rad Eff (%) | Fwd Gain (dBi) | Max Gain (dBi) | Min Gain (dBi) | Rin (ohms) | Xin (ohms) | VSWR//715 | Avg Pwr Gain** |
|---|---|---|---|---|---|---|---|---|
| 200.00 | 100.00 | 1.96 | 1.96 | 1.95 | 1.59 | -804.88 | 1019.18 | 1.044 |
| 215.00 | 100.00 | 1.96 | 1.96 | 1.95 | 1.84 | -741.76 | 805.42 | 1.044 |
| 230.00 | 100.00 | 1.97 | 1.97 | 1.95 | 2.12 | -686.38 | 649.43 | 1.044 |
| 245.00 | 100.00 | 1.97 | 1.97 | 1.95 | 2.41 | -637.33 | 532.82 | 1.044 |
| 260.00 | 100.00 | 1.97 | 1.97 | 1.95 | 2.72 | -593.50 | 443.79 | 1.044 |
| 275.00 | 100.00 | 1.97 | 1.97 | 1.95 | 3.06 | -554.03 | 374.56 | 1.044 |
| 290.00 | 100.00 | 1.98 | 1.98 | 1.95 | 3.41 | -518.25 | 319.81 | 1.044 |
| 305.00 | 100.00 | 1.98 | 1.98 | 1.95 | 3.79 | -485.62 | 275.88 | 1.044 |
| 320.00 | 100.00 | 1.98 | 1.98 | 1.95 | 4.19 | -455.67 | 240.16 | 1.044 |
| 335.00 | 100.00 | 1.99 | 1.99 | 1.95 | 4.61 | -428.06 | 210.76 | 1.044 |
| 350.00 | 100.00 | 1.99 | 1.99 | 1.95 | 5.05 | -402.47 | 186.32 | 1.044 |
| 365.00 | 100.00 | 1.99 | 1.99 | 1.95 | 5.52 | -378.66 | 165.78 | 1.044 |
| 380.00 | 100.00 | 2.00 | 2.00 | 1.94 | 6.02 | -356.41 | 148.38 | 1.044 |
| 395.00 | 100.00 | 2.00 | 2.00 | 1.94 | 6.53 | -335.54 | 133.52 | 1.044 |
| 410.00 | 100.00 | 2.01 | 2.01 | 1.94 | 7.08 | -315.91 | 120.72 | 1.044 |
| 425.00 | 100.00 | 2.01 | 2.01 | 1.94 | 7.65 | -297.36 | 109.64 | 1.044 |
| 440.00 | 100.00 | 2.02 | 2.02 | 1.94 | 8.25 | -279.78 | 99.97 | 1.044 |
| 455.00 | 100.00 | 2.02 | 2.02 | 1.94 | 8.87 | -263.08 | 91.49 | 1.044 |
| 470.00 | 100.00 | 2.03 | 2.03 | 1.94 | 9.53 | -247.18 | 84.01 | 1.044 |
| 485.00 | 100.00 | 2.03 | 2.03 | 1.94 | 10.21 | -231.98 | 77.38 | 1.044 |
| 500.00 | 100.00 | 2.04 | 2.04 | 1.94 | 10.93 | -217.43 | 71.48 | 1.044 |
| 515.00 | 100.00 | 2.04 | 2.04 | 1.94 | 11.68 | -203.47 | 66.20 | 1.044 |
| 530.00 | 100.00 | 2.05 | 2.05 | 1.94 | 12.46 | -190.04 | 61.46 | 1.044 |
| 545.00 | 100.00 | 2.05 | 2.05 | 1.94 | 13.27 | -177.09 | 57.19 | 1.045 |
| 560.00 | 100.00 | 2.06 | 2.06 | 1.94 | 14.12 | -164.59 | 53.33 | 1.045 |
| 575.00 | 100.00 | 2.07 | 2.07 | 1.94 | 15.00 | -152.49 | 49.83 | 1.045 |
| 590.00 | 100.00 | 2.07 | 2.07 | 1.94 | 15.92 | -140.76 | 46.64 | 1.045 |
| 605.00 | 100.00 | 2.08 | 2.08 | 1.94 | 16.89 | -129.36 | 43.73 | 1.045 |
| 620.00 | 100.00 | 2.09 | 2.09 | 1.94 | 17.89 | -118.28 | 41.07 | 1.045 |
| 635.00 | 100.00 | 2.09 | 2.09 | 1.94 | 18.93 | -107.49 | 38.62 | 1.045 |
| 650.00 | 100.00 | 2.10 | 2.10 | 1.94 | 20.02 | -96.95 | 36.37 | 1.045 |
| 665.00 | 100.00 | 2.11 | 2.11 | 1.94 | 21.15 | -86.65 | 34.30 | 1.045 |
| 680.00 | 100.00 | 2.12 | 2.12 | 1.94 | 22.33 | -76.57 | 32.39 | 1.045 |
| 695.00 | 100.00 | 2.13 | 2.13 | 1.93 | 23.56 | -66.69 | 30.62 | 1.045 |
| 710.00 | 100.00 | 2.13 | 2.13 | 1.93 | 24.84 | -56.99 | 28.97 | 1.045 |
| 725.00 | 100.00 | 2.14 | 2.14 | 1.93 | 26.17 | -47.46 | 27.44 | 1.045 |
| 740.00 | 100.00 | 2.15 | 2.15 | 1.93 | 27.55 | -38.08 | 26.02 | 1.045 |
| 755.00 | 100.00 | 2.16 | 2.16 | 1.93 | 29.00 | -28.85 | 24.70 | 1.045 |
| 770.00 | 100.00 | 2.17 | 2.17 | 1.93 | 30.50 | -19.74 | 23.46 | 1.045 |
| 785.00 | 100.00 | 2.18 | 2.18 | 1.93 | 32.07 | -10.74 | 22.30 | 1.045 |
| 800.00 | 100.00 | 2.19 | 2.19 | 1.93 | 33.70 | -1.85 | 21.22 | 1.045 |
| 815.00 | 100.00 | 2.20 | 2.20 | 1.93 | 35.39 | 6.96 | 20.20 | 1.045 |
| 830.00 | 100.00 | 2.21 | 2.21 | 1.93 | 37.16 | 15.67 | 19.25 | 1.045 |
| 845.00 | 100.00 | 2.22 | 2.22 | 1.93 | 39.01 | 24.30 | 18.35 | 1.045 |
| 860.00 | 100.00 | 2.23 | 2.23 | 1.92 | 40.93 | 32.87 | 17.50 | 1.045 |
| 875.00 | 100.00 | 2.25 | 2.25 | 1.92 | 42.94 | 41.39 | 16.71 | 1.045 |
| 890.00 | 100.00 | 2.26 | 2.26 | 1.92 | 45.03 | 49.85 | 15.96 | 1.045 |
| 905.00 | 100.00 | 2.27 | 2.27 | 1.92 | 47.21 | 58.27 | 15.25 | 1.045 |
| 920.00 | 100.00 | 2.28 | 2.28 | 1.92 | 49.48 | 66.66 | 14.58 | 1.045 |
| 935.00 | 100.00 | 2.29 | 2.29 | 1.92 | 51.85 | 75.02 | 13.94 | 1.046 |
| 950.00 | 100.00 | 2.31 | 2.31 | 1.92 | 54.33 | 83.35 | 13.34 | 1.046 |
| 965.00 | 100.00 | 2.32 | 2.32 | 1.92 | 56.92 | 91.67 | 12.77 | 1.046 |
| 980.00 | 100.00 | 2.33 | 2.33 | 1.91 | 59.62 | 99.99 | 12.23 | 1.046 |
| 995.00 | 100.00 | 2.35 | 2.35 | 1.91 | 62.44 | 108.30 | 11.72 | 1.046 |
| 1010.00 | 100.00 | 2.36 | 2.36 | 1.91 | 65.39 | 116.61 | 11.23 | 1.046 |
| 1025.00 | 100.00 | 2.38 | 2.38 | 1.91 | 68.47 | 124.93 | 10.76 | 1.046 |
| 1040.00 | 100.00 | 2.39 | 2.39 | 1.91 | 71.69 | 133.27 | 10.32 | 1.046 |
| 1055.00 | 100.00 | 2.41 | 2.41 | 1.91 | 75.06 | 141.62 | 9.90 | 1.046 |
| 1070.00 | 100.00 | 2.42 | 2.42 | 1.90 | 78.59 | 149.99 | 9.50 | 1.046 |
| 1085.00 | 100.00 | 2.44 | 2.44 | 1.90 | 82.29 | 158.40 | 9.12 | 1.046 |
| 1100.00 | 100.00 | 2.45 | 2.45 | 1.90 | 86.16 | 166.83 | 8.76 | 1.046 |
| 1115.00 | 100.00 | 2.47 | 2.47 | 1.90 | 90.21 | 175.30 | 8.41 | 1.046 |
| 1130.00 | 100.00 | 2.49 | 2.49 | 1.89 | 94.47 | 183.81 | 8.08 | 1.046 |
| 1145.00 | 100.00 | 2.51 | 2.51 | 1.89 | 98.93 | 192.37 | 7.76 | 1.046 |
| 1160.00 | 100.00 | 2.52 | 2.52 | 1.89 | 103.61 | 200.97 | 7.46 | 1.046 |
| 1175.00 | 100.00 | 2.54 | 2.54 | 1.89 | 108.52 | 209.62 | 7.17 | 1.047 |
| 1190.00 | 100.00 | 2.56 | 2.56 | 1.88 | 113.69 | 218.32 | 6.89 | 1.047 |
| 1205.00 | 100.00 | 2.58 | 2.58 | 1.88 | 119.12 | 227.08 | 6.62 | 1.047 |
| 1220.00 | 100.00 | 2.60 | 2.60 | 1.88 | 124.82 | 235.89 | 6.37 | 1.047 |
| 1235.00 | 100.00 | 2.62 | 2.62 | 1.88 | 130.85 | 244.77 | 6.12 | 1.047 |
| 1250.00 | 100.00 | 2.64 | 2.64 | 1.87 | 137.17 | 253.70 | 5.89 | 1.047 |
| 1265.00 | 100.00 | 2.66 | 2.66 | 1.87 | 143.84 | 262.68 | 5.67 | 1.047 |
| 1280.00 | 100.00 | 2.68 | 2.68 | 1.87 | 150.86 | 271.72 | 5.45 | 1.047 |
| 1295.00 | 100.00 | 2.71 | 2.71 | 1.86 | 158.28 | 280.81 | 5.24 | 1.047 |
| 1310.00 | 100.00 | 2.73 | 2.73 | 1.86 | 166.10 | 289.95 | 5.05 | 1.047 |
| 1325.00 | 100.00 | 2.75 | 2.75 | 1.85 | 174.36 | 299.13 | 4.86 | 1.047 |
| 1340.00 | 100.00 | 2.78 | 2.78 | 1.85 | 183.09 | 308.35 | 4.67 | 1.047 |
| 1355.00 | 100.00 | 2.80 | 2.80 | 1.85 | 192.32 | 317.60 | 4.50 | 1.048 |
| 1370.00 | 100.00 | 2.82 | 2.82 | 1.84 | 202.09 | 326.87 | 4.33 | 1.048 |
| 1385.00 | 100.00 | 2.85 | 2.85 | 1.84 | 212.44 | 336.14 | 4.17 | 1.048 |
| 1400.00 | 100.00 | 2.88 | 2.88 | 1.83 | 223.40 | 345.39 | 4.01 | 1.048 |
| 1415.00 | 100.00 | 2.90 | 2.90 | 1.83 | 235.01 | 354.62 | 3.86 | 1.048 |
| 1430.00 | 100.00 | 2.93 | 2.93 | 1.82 | 247.33 | 363.78 | 3.72 | 1.048 |
| 1445.00 | 100.00 | 2.96 | 2.96 | 1.82 | 260.40 | 372.85 | 3.58 | 1.048 |
| 1460.00 | 100.00 | 2.99 | 2.99 | 1.81 | 274.28 | 381.80 | 3.44 | 1.048 |
| 1475.00 | 100.00 | 3.02 | 3.02 | 1.81 | 289.03 | 390.59 | 3.31 | 1.049 |
| 1490.00 | 100.00 | 3.05 | 3.05 | 1.80 | 304.69 | 399.15 | 3.19 | 1.049 |
| 1505.00 | 100.00 | 3.08 | 3.08 | 1.79 | 321.34 | 407.44 | 3.07 | 1.049 |
| 1520.00 | 100.00 | 3.11 | 3.11 | 1.79 | 339.04 | 415.38 | 2.96 | 1.049 |
| 1535.00 | 100.00 | 3.14 | 3.14 | 1.78 | 357.85 | 422.90 | 2.85 | 1.049 |
| 1550.00 | 100.00 | 3.17 | 3.17 | 1.77 | 377.85 | 429.88 | 2.74 | 1.049 |
| 1565.00 | 100.00 | 3.21 | 3.21 | 1.77 | 399.10 | 436.23 | 2.64 | 1.049 |
| 1580.00 | 100.00 | 3.24 | 3.24 | 1.76 | 421.66 | 441.80 | 2.54 | 1.050 |
| 1595.00 | 100.00 | 3.28 | 3.28 | 1.75 | 445.60 | 446.45 | 2.44 | 1.050 |
| 1610.00 | 100.00 | 3.31 | 3.31 | 1.74 | 470.97 | 450.00 | 2.35 | 1.050 |
| 1625.00 | 100.00 | 3.35 | 3.35 | 1.73 | 497.82 | 452.24 | 2.27 | 1.050 |
| 1640.00 | 100.00 | 3.39 | 3.39 | 1.72 | 526.17 | 452.95 | 2.18 | 1.050 |
| 1655.00 | 100.00 | 3.43 | 3.43 | 1.71 | 556.02 | 451.86 | 2.10 | 1.051 |
| 1670.00 | 100.00 | 3.47 | 3.47 | 1.70 | 587.35 | 448.67 | 2.02 | 1.051 |
| 1685.00 | 100.00 | 3.51 | 3.51 | 1.69 | 620.08 | 443.06 | 1.95 | 1.051 |
| 1700.00 | 100.00 | 3.56 | 3.56 | 1.68 | 654.10 | 434.68 | 1.88 | 1.051 |
| 1715.00 | 100.00 | 3.60 | 3.60 | 1.67 | 689.22 | 423.14 | 1.81 | 1.051 |
| 1730.00 | 100.00 | 3.64 | 3.64 | 1.66 | 725.17 | 408.05 | 1.75 | 1.052 |
| 1745.00 | 100.00 | 3.69 | 3.69 | 1.65 | 761.60 | 389.02 | 1.69 | 1.052 |
| 1760.00 | 100.00 | 3.74 | 3.74 | 1.63 | 798.05 | 365.65 | 1.63 | 1.052 |
| 1775.00 | 100.00 | 3.79 | 3.79 | 1.62 | 833.93 | 337.62 | 1.58 | 1.053 |
| 1790.00 | 100.00 | 3.84 | 3.84 | 1.60 | 868.54 | 304.66 | 1.54 | 1.053 |
| 1805.00 | 100.00 | 3.89 | 3.89 | 1.59 | 901.06 | 266.62 | 1.50 | 1.053 |
| 1820.00 | 100.00 | 3.94 | 3.94 | 1.57 | 930.58 | 223.50 | 1.46 | 1.053 |
| 1835.00 | 100.00 | 4.00 | 4.00 | 1.56 | 956.12 | 175.51 | 1.43 | 1.054 |
| 1850.00 | 100.00 | 4.06 | 4.06 | 1.54 | 976.68 | 123.08 | 1.41 | 1.054 |
| 1865.00 | 100.00 | 4.12 | 4.12 | 1.52 | 991.29 | 66.89 | 1.40 | 1.054 |
| 1880.00 | 100.00 | 4.18 | 4.18 | 1.50 | 999.13 | 7.87 | 1.40 | 1.055 |
| 1895.00 | 100.00 | 4.24 | 4.24 | 1.48 | 999.55 | -52.82 | 1.41 | 1.055 |
| 1910.00 | 100.00 | 4.30 | 4.30 | 1.45 | 992.20 | -113.88 | 1.42 | 1.056 |
| 1925.00 | 100.00 | 4.37 | 4.37 | 1.43 | 977.00 | -173.91 | 1.45 | 1.056 |
| 1940.00 | 100.00 | 4.44 | 4.44 | 1.40 | 954.25 | -231.50 | 1.49 | 1.057 |
| 1955.00 | 100.00 | 4.51 | 4.51 | 1.38 | 924.54 | -285.38 | 1.54 | 1.057 |
| 1970.00 | 100.00 | 4.58 | 4.58 | 1.35 | 888.73 | -334.45 | 1.60 | 1.058 |
| 1985.00 | 100.00 | 4.66 | 4.66 | 1.32 | 847.89 | -377.88 | 1.66 | 1.058 |
| 2000.00 | 100.00 | 4.74 | 4.74 | 1.29 | 803.19 | -415.11 | 1.74 | 1.059 |
| 2015.00 | 100.00 | 4.82 | 4.82 | 1.25 | 755.82 | -445.79 | 1.82 | 1.059 |
| 2030.00 | 100.00 | 4.91 | 4.91 | 1.21 | 706.95 | -470.04 | 1.92 | 1.060 |
| 2045.00 | 100.00 | 5.00 | 5.00 | 1.17 | 657.60 | -488.10 | 2.02 | 1.061 |
| 2060.00 | 100.00 | 5.09 | 5.09 | 1.13 | 608.67 | -500.35 | 2.13 | 1.062 |
| 2075.00 | 100.00 | 5.19 | 5.19 | 1.08 | 560.87 | -507.29 | 2.26 | 1.063 |
| 2090.00 | 100.00 | 5.29 | 5.29 | 1.03 | 514.78 | -509.49 | 2.40 | 1.064 |
| 2105.00 | 100.00 | 5.39 | 5.39 | 0.98 | 470.79 | -507.54 | 2.55 | 1.065 |
| 2120.00 | 100.00 | 5.50 | 5.50 | 0.92 | 429.20 | -502.02 | 2.72 | 1.066 |
| 2135.00 | 100.00 | 5.61 | 5.61 | 0.86 | 390.16 | -493.48 | 2.91 | 1.067 |
| 2150.00 | 100.00 | 5.73 | 5.73 | 0.79 | 353.74 | -482.42 | 3.12 | 1.069 |
| 2165.00 | 100.00 | 5.85 | 5.85 | 0.71 | 319.95 | -469.31 | 3.35 | 1.070 |
| 2180.00 | 100.00 | 5.98 | 5.98 | 0.62 | 288.73 | -454.54 | 3.60 | 1.071 |
| 2195.00 | 100.00 | 6.11 | 6.11 | 0.53 | 260.01 | -438.46 | 3.89 | 1.073 |
| 2210.00 | 100.00 | 6.25 | 6.25 | 0.42 | 233.65 | -421.36 | 4.21 | 1.075 |
| 2225.00 | 100.00 | 6.39 | 6.39 | 0.30 | 209.55 | -403.50 | 4.57 | 1.076 |
| 2240.00 | 100.00 | 6.54 | 6.54 | 0.17 | 187.55 | -385.08 | 4.98 | 1.078 |
| 2255.00 | 100.00 | 6.70 | 6.70 | 0.02 | 167.53 | -366.28 | 5.44 | 1.080 |
| 2270.00 | 100.00 | 6.86 | 6.86 | -0.14 | 149.34 | -347.24 | 5.96 | 1.083 |
| 2285.00 | 100.00 | 7.03 | 7.03 | -0.34 | 132.86 | -328.06 | 6.55 | 1.085 |
| 2300.00 | 100.00 | 7.21 | 7.21 | -0.56 | 117.96 | -308.86 | 7.22 | 1.088 |
| 2315.00 | 100.00 | 7.39 | 7.39 | -0.81 | 104.52 | -289.69 | 7.98 | 1.091 |
| 2330.00 | 100.00 | 7.58 | 7.58 | -1.11 | 92.44 | -270.62 | 8.86 | 1.094 |
| 2345.00 | 100.00 | 7.77 | 7.77 | -1.45 | 81.61 | -251.69 | 9.86 | 1.097 |
| 2360.00 | 100.00 | 7.97 | 7.97 | -1.86 | 71.95 | -232.93 | 11.00 | 1.101 |
| 2375.00 | 100.00 | 8.16 | 8.16 | -2.35 | 63.38 | -214.38 | 12.30 | 1.105 |
| 2390.00 | 100.00 | 8.35 | 8.35 | -2.93 | 55.80 | -196.04 | 13.78 | 1.109 |
| 2405.00 | 100.00 | 8.53 | 8.53 | -3.64 | 49.17 | -177.92 | 15.45 | 1.113 |
| 2420.00 | 100.00 | 8.70 | 8.70 | -4.50 | 43.42 | -160.05 | 17.30 | 1.118 |
| 2435.00 | 100.00 | 8.85 | 8.85 | -5.54 | 38.48 | -142.42 | 19.32 | 1.122 |
| 2450.00 | 100.00 | 8.96 | 8.96 | -6.77 | 34.32 | -125.03 | 21.47 | 1.125 |
| 2465.00 | 100.00 | 9.02 | 9.02 | -8.12 | 30.89 | -107.88 | 23.67 | 1.128 |
| 2480.00 | 100.00 | 9.01 | 9.01 | -9.28 | 28.16 | -90.97 | 25.80 | 1.130 |
| 2495.00 | 100.00 | 8.93 | 8.93 | -9.95 | 26.08 | -74.30 | 27.71 | 1.130 |
| 2510.00 | 100.00 | 8.75 | 8.75 | -10.62 | 24.63 | -57.86 | 29.21 | 1.128 |
| 2525.00 | 100.00 | 8.46 | 8.46 | -11.63 | 23.79 | -41.65 | 30.16 | 1.124 |
| 2540.00 | 100.00 | 8.05 | 8.05 | -13.21 | 23.53 | -25.66 | 30.43 | 1.119 |
| 2555.00 | 100.00 | 7.52 | 7.52 | -15.02 | 23.83 | -9.89 | 30.01 | 1.111 |
| 2570.00 | 100.00 | 6.88 | 6.88 | -15.59 | 24.68 | 5.66 | 28.97 | 1.103 |
| 2585.00 | 100.00 | 6.13 | 6.13 | -20.71 | 26.06 | 21.00 | 27.46 | 1.095 |
| 2600.00 | 100.00 | 5.29 | 5.29 | -19.50 | 27.97 | 36.13 | 25.63 | 1.086 |
| 2615.00 | 100.00 | 4.37 | 4.37 | -31.38 | 30.39 | 51.05 | 23.65 | 1.079 |
| 2630.00 | 100.00 | 3.38 | 3.38 | -20.97 | 33.32 | 65.77 | 21.64 | 1.072 |
| 2645.00 | 100.00 | 2.33 | 2.33 | -25.03 | 36.75 | 80.28 | 19.70 | 1.066 |
| 2660.00 | 100.00 | 1.23 | 1.23 | -18.31 | 40.69 | 94.58 | 17.88 | 1.061 |
| 2675.00 | 100.00 | 0.09 | 1.45 | -19.34 | 45.12 | 108.66 | 16.21 | 1.057 |
| 2690.00 | 100.00 | -1.10 | 1.65 | -15.99 | 50.06 | 122.53 | 14.71 | 1.054 |
| 2705.00 | 100.00 | -2.33 | 1.80 | -16.35 | 55.49 | 136.18 | 13.35 | 1.051 |
| 2720.00 | 100.00 | -3.60 | 1.92 | -14.71 | 61.43 | 149.59 | 12.15 | 1.049 |
| 2735.00 | 100.00 | -4.90 | 2.02 | -14.49 | 67.88 | 162.76 | 11.08 | 1.047 |
| 2750.00 | 100.00 | -6.22 | 2.09 | -13.57 | 74.84 | 175.68 | 10.14 | 1.045 |
| 2765.00 | 100.00 | -7.53 | 2.15 | -13.26 | 82.31 | 188.34 | 9.30 | 1.044 |
| 2780.00 | 100.00 | -8.78 | 2.20 | -12.61 | 90.30 | 200.71 | 8.55 | 1.043 |
| 2795.00 | 100.00 | -9.88 | 2.24 | -12.37 | 98.81 | 212.78 | 7.89 | 1.043 |
| 2810.00 | 100.00 | -10.73 | 2.27 | -11.95 | 107.85 | 224.53 | 7.30 | 1.042 |
| 2825.00 | 100.00 | -11.23 | 2.29 | -11.65 | 117.41 | 235.94 | 6.77 | 1.042 |
| 2840.00 | 100.00 | -11.35 | 2.31 | -11.39 | 127.51 | 246.98 | 6.30 | 1.042 |
| 2855.00 | 100.00 | -11.13 | 2.32 | -11.13 | 138.14 | 257.63 | 5.87 | 1.042 |
| 2870.00 | 100.00 | -10.70 | 2.32 | -10.70 | 149.31 | 267.85 | 5.49 | 1.042 |
| 2885.00 | 100.00 | -10.14 | 2.33 | -10.14 | 161.00 | 277.62 | 5.14 | 1.042 |
| 2900.00 | 100.00 | -9.53 | 2.33 | -9.53 | 173.22 | 286.89 | 4.83 | 1.042 |
| 2915.00 | 100.00 | -8.93 | 2.33 | -8.93 | 185.96 | 295.63 | 4.54 | 1.042 |
| 2930.00 | 100.00 | -8.35 | 2.33 | -8.35 | 199.21 | 303.81 | 4.28 | 1.042 |
| 2945.00 | 100.00 | -7.81 | 2.32 | -7.81 | 212.96 | 311.38 | 4.04 | 1.042 |
| 2960.00 | 100.00 | -7.30 | 2.32 | -7.30 | 227.18 | 318.30 | 3.83 | 1.042 |
| 2975.00 | 100.00 | -6.83 | 2.31 | -6.83 | 241.86 | 324.52 | 3.63 | 1.042 |
| 2990.00 | 100.00 | -6.40 | 2.30 | -6.40 | 256.97 | 330.00 | 3.44 | 1.043 |
| 3005.00 | 100.00 | -6.01 | 2.29 | -6.01 | 272.48 | 334.69 | 3.27 | 1.043 |
| 3020.00 | 100.00 | -5.64 | 2.28 | -5.64 | 288.35 | 338.56 | 3.12 | 1.043 |
| 3035.00 | 100.00 | -5.30 | 2.27 | -5.30 | 304.54 | 341.55 | 2.97 | 1.043 |
| 3050.00 | 100.00 | -4.99 | 2.26 | -4.99 | 321.00 | 343.62 | 2.84 | 1.044 |
| 3065.00 | 100.00 | -4.69 | 2.25 | -4.69 | 337.67 | 344.73 | 2.71 | 1.044 |
| 3080.00 | 100.00 | -4.42 | 2.23 | -4.42 | 354.49 | 344.84 | 2.60 | 1.044 |
| 3095.00 | 100.00 | -4.17 | 2.22 | -4.17 | 371.40 | 343.93 | 2.49 | 1.045 |
| 3110.00 | 100.00 | -3.93 | 2.20 | -3.93 | 388.32 | 341.96 | 2.39 | 1.045 |
| 3125.00 | 100.00 | -3.71 | 2.19 | -3.71 | 405.16 | 338.90 | 2.29 | 1.045 |
| 3140.00 | 100.00 | -3.50 | 2.17 | -3.50 | 421.86 | 334.76 | 2.20 | 1.046 |
| 3155.00 | 100.00 | -3.30 | 2.16 | -3.30 | 438.31 | 329.52 | 2.12 | 1.046 |
| 3170.00 | 100.00 | -3.12 | 2.14 | -3.12 | 454.43 | 323.18 | 2.04 | 1.046 |
| 3185.00 | 100.00 | -2.94 | 2.12 | -2.94 | 470.12 | 315.76 | 1.97 | 1.047 |
| 3200.00 | 100.00 | -2.78 | 2.10 | -2.78 | 485.29 | 307.28 | 1.90 | 1.047 |
| 3215.00 | 100.00 | -2.62 | 2.09 | -2.62 | 499.84 | 297.78 | 1.83 | 1.047 |
| 3230.00 | 100.00 | -2.47 | 2.07 | -2.47 | 513.68 | 287.29 | 1.77 | 1.048 |
| 3245.00 | 100.00 | -2.33 | 2.05 | -2.33 | 526.73 | 275.87 | 1.71 | 1.048 |
| 3260.00 | 100.00 | -2.19 | 2.03 | -2.19 | 538.89 | 263.60 | 1.66 | 1.048 |
| 3275.00 | 100.00 | -2.06 | 2.01 | -2.06 | 550.10 | 250.54 | 1.61 | 1.049 |
| 3290.00 | 100.00 | -1.94 | 1.99 | -1.94 | 560.28 | 236.78 | 1.56 | 1.049 |
| 3305.00 | 100.00 | -1.82 | 1.96 | -1.82 | 569.38 | 222.42 | 1.51 | 1.049 |
| 3320.00 | 100.00 | -1.71 | 1.94 | -1.71 | 577.34 | 207.54 | 1.47 | 1.050 |
| 3335.00 | 100.00 | -1.60 | 1.92 | -1.60 | 584.13 | 192.26 | 1.43 | 1.050 |
| 3350.00 | 100.00 | -1.49 | 1.89 | -1.49 | 589.73 | 176.68 | 1.39 | 1.050 |
| 3365.00 | 100.00 | -1.39 | 1.87 | -1.39 | 594.12 | 160.92 | 1.36 | 1.051 |
| 3380.00 | 100.00 | -1.30 | 1.85 | -1.30 | 597.29 | 145.06 | 1.33 | 1.051 |
| 3395.00 | 100.00 | -1.20 | 1.82 | -1.20 | 599.27 | 129.24 | 1.30 | 1.051 |
| 3410.00 | 100.00 | -1.11 | 1.79 | -1.11 | 600.07 | 113.54 | 1.28 | 1.052 |
| 3425.00 | 100.00 | -1.03 | 1.77 | -1.03 | 599.72 | 98.07 | 1.26 | 1.052 |
| 3440.00 | 100.00 | -0.95 | 1.74 | -0.95 | 598.26 | 82.92 | 1.24 | 1.053 |
| 3455.00 | 100.00 | -0.87 | 1.71 | -0.87 | 595.76 | 68.17 | 1.23 | 1.053 |
| 3470.00 | 100.00 | -0.79 | 1.68 | -0.79 | 592.25 | 53.90 | 1.23 | 1.053 |
| 3485.00 | 100.00 | -0.72 | 1.65 | -0.72 | 587.81 | 40.19 | 1.23 | 1.054 |
| 3500.00 | 100.00 | -0.64 | 1.62 | -0.64 | 582.50 | 27.09 | 1.23 | 1.054 |
| 3515.00 | 100.00 | -0.58 | 1.58 | -0.58 | 576.40 | 14.65 | 1.24 | 1.054 |
| 3530.00 | 100.00 | -0.51 | 1.55 | -0.51 | 569.57 | 2.93 | 1.26 | 1.055 |
| 3545.00 | 100.00 | -0.45 | 1.52 | -0.45 | 562.10 | -8.06 | 1.27 | 1.055 |
| 3560.00 | 100.00 | -0.39 | 1.48 | -0.39 | 554.04 | -18.27 | 1.29 | 1.056 |
| 3575.00 | 100.00 | -0.33 | 1.44 | -0.33 | 545.49 | -27.69 | 1.32 | 1.056 |
| 3590.00 | 100.00 | -0.27 | 1.41 | -0.27 | 536.51 | -36.31 | 1.34 | 1.056 |
| 3605.00 | 100.00 | -0.22 | 1.37 | -0.22 | 527.17 | -44.13 | 1.37 | 1.057 |
| 3620.00 | 100.00 | -0.17 | 1.33 | -0.17 | 517.53 | -51.13 | 1.40 | 1.057 |
| 3635.00 | 100.00 | -0.12 | 1.29 | -0.12 | 507.66 | -57.33 | 1.43 | 1.058 |
| 3650.00 | 100.00 | -0.08 | 1.25 | -0.08 | 497.62 | -62.74 | 1.46 | 1.058 |
| 3665.00 | 100.00 | -0.03 | 1.21 | -0.03 | 487.46 | -67.37 | 1.49 | 1.058 |
| 3680.00 | 100.00 | 0.01 | 1.17 | 0.01 | 477.23 | -71.23 | 1.52 | 1.059 |
| 3695.00 | 100.00 | 0.04 | 1.13 | 0.04 | 466.99 | -74.35 | 1.56 | 1.059 |
| 3710.00 | 100.00 | 0.08 | 1.09 | 0.08 | 456.78 | -76.74 | 1.60 | 1.060 |
| 3725.00 | 100.00 | 0.11 | 1.05 | 0.11 | 446.63 | -78.44 | 1.63 | 1.060 |
| 3740.00 | 100.00 | 0.14 | 1.02 | 0.14 | 436.60 | -79.46 | 1.67 | 1.060 |
| 3755.00 | 100.00 | 0.16 | 0.98 | 0.16 | 426.70 | -79.82 | 1.71 | 1.061 |
| 3770.00 | 100.00 | 0.18 | 0.94 | 0.18 | 416.97 | -79.56 | 1.75 | 1.061 |
| 3785.00 | 100.00 | 0.20 | 0.90 | 0.20 | 407.45 | -78.70 | 1.79 | 1.061 |
| 3800.00 | 100.00 | 0.21 | 0.86 | 0.21 | 398.15 | -77.27 | 1.83 | 1.062 |
| 3815.00 | 100.00 | 0.22 | 0.82 | 0.22 | 389.10 | -75.28 | 1.87 | 1.062 |
| 3830.00 | 100.00 | 0.23 | 0.78 | 0.23 | 380.32 | -72.77 | 1.91 | 1.063 |
| 3845.00 | 100.00 | 0.23 | 0.74 | 0.23 | 371.82 | -69.76 | 1.95 | 1.063 |
| 3860.00 | 100.00 | 0.22 | 0.70 | 0.22 | 363.64 | -66.28 | 1.99 | 1.063 |
| 3875.00 | 100.00 | 0.22 | 0.65 | 0.22 | 355.77 | -62.34 | 2.03 | 1.063 |
| 3890.00 | 100.00 | 0.20 | 0.61 | 0.20 | 348.24 | -57.97 | 2.07 | 1.064 |
| 3905.00 | 100.00 | 0.18 | 0.58 | 0.18 | 341.05 | -53.19 | 2.11 | 1.064 |
| 3920.00 | 100.00 | 0.15 | 0.56 | 0.15 | 334.23 | -48.03 | 2.15 | 1.064 |
| 3935.00 | 100.00 | 0.12 | 0.54 | 0.12 | 327.78 | -42.50 | 2.19 | 1.064 |
| 3950.00 | 100.00 | 0.08 | 0.53 | 0.08 | 321.71 | -36.63 | 2.23 | 1.065 |
| 3965.00 | 100.00 | 0.04 | 0.53 | 0.04 | 316.03 | -30.43 | 2.27 | 1.065 |
| 3980.00 | 100.00 | -0.02 | 0.53 | -0.02 | 310.75 | -23.92 | 2.30 | 1.065 |
| 3995.00 | 100.00 | -0.08 | 0.54 | -0.08 | 305.89 | -17.13 | 2.34 | 1.065 |
| 4010.00 | 100.00 | -0.15 | 0.56 | -0.15 | 301.44 | -10.06 | 2.37 | 1.065 |
| 4025.00 | 100.00 | -0.23 | 0.59 | -0.24 | 297.43 | -2.74 | 2.40 | 1.065 |
| 4040.00 | 100.00 | -0.32 | 0.62 | -0.33 | 293.85 | 4.81 | 2.43 | 1.065 |
| 4055.00 | 100.00 | -0.42 | 0.67 | -0.44 | 290.72 | 12.57 | 2.46 | 1.065 |
| 4070.00 | 100.00 | -0.53 | 0.72 | -0.56 | 288.05 | 20.54 | 2.48 | 1.065 |
| 4085.00 | 100.00 | -0.65 | 0.79 | -0.69 | 285.84 | 28.68 | 2.51 | 1.065 |
| 4100.00 | 100.00 | -0.78 | 0.86 | -0.83 | 284.11 | 36.99 | 2.52 | 1.065 |
| 4115.00 | 100.00 | -0.93 | 0.94 | -0.97 | 282.86 | 45.44 | 2.54 | 1.064 |
| 4130.00 | 100.00 | -1.08 | 1.04 | -1.13 | 282.11 | 54.01 | 2.55 | 1.064 |
| 4145.00 | 100.00 | -1.25 | 1.14 | -1.30 | 281.86 | 62.69 | 2.56 | 1.064 |
| 4160.00 | 100.00 | -1.43 | 1.24 | -1.49 | 282.12 | 71.45 | 2.56 | 1.063 |
| 4175.00 | 100.00 | -1.63 | 1.36 | -1.69 | 282.90 | 80.27 | 2.57 | 1.063 |
| 4190.00 | 100.00 | -1.84 | 1.47 | -1.89 | 284.22 | 89.13 | 2.56 | 1.063 |
| 4205.00 | 100.00 | -2.06 | 1.60 | -2.11 | 286.08 | 98.00 | 2.55 | 1.062 |
| 4220.00 | 100.00 | -2.30 | 1.73 | -2.35 | 288.49 | 106.86 | 2.54 | 1.062 |
| 4235.00 | 100.00 | -2.55 | 1.86 | -2.59 | 291.47 | 115.68 | 2.53 | 1.061 |
| 4250.00 | 100.00 | -2.81 | 1.99 | -2.85 | 295.01 | 124.43 | 2.51 | 1.061 |
| 4265.00 | 100.00 | -3.09 | 2.12 | -3.13 | 299.13 | 133.09 | 2.49 | 1.060 |
| 4280.00 | 100.00 | -3.39 | 2.26 | -3.42 | 303.83 | 141.61 | 2.46 | 1.060 |
| 4295.00 | 100.00 | -3.70 | 2.39 | -3.72 | 309.12 | 149.98 | 2.44 | 1.059 |
| 4310.00 | 100.00 | -4.02 | 2.52 | -4.03 | 315.00 | 158.14 | 2.41 | 1.059 |
| 4325.00 | 100.00 | -4.36 | 2.65 | -4.37 | 321.48 | 166.06 | 2.37 | 1.058 |
| 4340.00 | 100.00 | -4.71 | 2.78 | -4.72 | 328.55 | 173.71 | 2.34 | 1.057 |
| 4355.00 | 100.00 | -5.08 | 2.91 | -5.08 | 336.21 | 181.03 | 2.30 | 1.057 |
| 4370.00 | 100.00 | -5.46 | 3.03 | -5.46 | 344.45 | 187.99 | 2.26 | 1.056 |
| 4385.00 | 100.00 | -5.85 | 3.15 | -5.85 | 353.28 | 194.54 | 2.22 | 1.056 |
| 4400.00 | 100.00 | -6.26 | 3.26 | -6.26 | 362.67 | 200.63 | 2.17 | 1.055 |
| 4415.00 | 100.00 | -6.67 | 3.37 | -6.67 | 372.60 | 206.21 | 2.13 | 1.055 |
| 4430.00 | 100.00 | -7.10 | 3.48 | -7.10 | 383.06 | 211.24 | 2.09 | 1.054 |
| 4445.00 | 100.00 | -7.53 | 3.58 | -7.53 | 394.02 | 215.66 | 2.04 | 1.054 |
| 4460.00 | 100.00 | -7.98 | 3.67 | -7.98 | 405.45 | 219.43 | 2.00 | 1.054 |
| 4475.00 | 100.00 | -8.42 | 3.77 | -8.42 | 417.31 | 222.49 | 1.95 | 1.053 |
| 4490.00 | 100.00 | -8.87 | 3.86 | -8.87 | 429.56 | 224.79 | 1.90 | 1.053 |
| 4505.00 | 100.00 | -9.32 | 3.94 | -9.32 | 442.15 | 226.30 | 1.86 | 1.053 |
| 4520.00 | 100.00 | -9.76 | 4.02 | -9.76 | 455.02 | 226.96 | 1.82 | 1.052 |
| 4535.00 | 100.00 | -10.19 | 4.10 | -10.19 | 468.11 | 226.74 | 1.77 | 1.052 |
| 4550.00 | 100.00 | -10.61 | 4.17 | -10.61 | 481.35 | 225.60 | 1.73 | 1.052 |
| 4565.00 | 100.00 | -11.01 | 4.24 | -11.01 | 494.68 | 223.52 | 1.69 | 1.052 |
| 4580.00 | 100.00 | -11.39 | 4.31 | -11.39 | 508.01 | 220.46 | 1.64 | 1.052 |
| 4595.00 | 100.00 | -11.73 | 4.37 | -11.73 | 521.27 | 216.43 | 1.60 | 1.052 |
| 4610.00 | 100.00 | -12.04 | 4.43 | -12.04 | 534.36 | 211.40 | 1.56 | 1.051 |
| 4625.00 | 100.00 | -12.30 | 4.49 | -12.30 | 547.20 | 205.39 | 1.52 | 1.051 |
| 4640.00 | 100.00 | -12.52 | 4.54 | -12.52 | 559.70 | 198.40 | 1.49 | 1.051 |
| 4655.00 | 100.00 | -12.68 | 4.59 | -12.68 | 571.76 | 190.46 | 1.45 | 1.051 |
| 4670.00 | 100.00 | -12.80 | 4.64 | -12.80 | 583.32 | 181.60 | 1.41 | 1.051 |
| 4685.00 | 100.00 | -12.87 | 4.69 | -12.87 | 594.26 | 171.86 | 1.38 | 1.051 |
| 4700.00 | 100.00 | -12.89 | 4.73 | -12.89 | 604.53 | 161.30 | 1.34 | 1.051 |
| 4715.00 | 100.00 | -12.87 | 4.78 | -12.87 | 614.04 | 149.97 | 1.31 | 1.051 |
| 4730.00 | 100.00 | -12.81 | 4.82 | -12.81 | 622.72 | 137.94 | 1.28 | 1.052 |
| 4745.00 | 100.00 | -12.72 | 4.86 | -12.72 | 630.51 | 125.30 | 1.25 | 1.052 |
| 4760.00 | 100.00 | -12.60 | 4.89 | -12.60 | 637.36 | 112.12 | 1.22 | 1.052 |
| 4775.00 | 100.00 | -12.47 | 4.93 | -12.47 | 643.24 | 98.50 | 1.20 | 1.052 |
| 4790.00 | 100.00 | -12.31 | 4.96 | -12.31 | 648.10 | 84.52 | 1.17 | 1.052 |
| 4805.00 | 100.00 | -12.15 | 4.99 | -12.15 | 651.94 | 70.28 | 1.15 | 1.052 |
| 4820.00 | 100.00 | -11.98 | 5.02 | -11.98 | 654.74 | 55.87 | 1.13 | 1.052 |
| 4835.00 | 100.00 | -11.80 | 5.05 | -11.80 | 656.50 | 41.40 | 1.11 | 1.053 |
| 4850.00 | 100.00 | -11.63 | 5.08 | -11.63 | 657.23 | 26.94 | 1.10 | 1.053 |
| 4865.00 | 100.00 | -11.45 | 5.10 | -11.45 | 656.96 | 12.58 | 1.09 | 1.053 |
| 4880.00 | 100.00 | -11.28 | 5.13 | -11.28 | 655.71 | -1.58 | 1.09 | 1.053 |
| 4895.00 | 100.00 | -11.11 | 5.15 | -11.11 | 653.52 | -15.49 | 1.10 | 1.054 |
| 4910.00 | 100.00 | -10.95 | 5.18 | -10.95 | 650.44 | -29.05 | 1.11 | 1.054 |
| 4925.00 | 100.00 | -10.79 | 5.20 | -10.79 | 646.51 | -42.22 | 1.13 | 1.054 |
| 4940.00 | 100.00 | -10.64 | 5.22 | -10.64 | 641.79 | -54.94 | 1.14 | 1.055 |
| 4955.00 | 100.00 | -10.49 | 5.24 | -10.49 | 636.33 | -67.15 | 1.17 | 1.055 |
| 4970.00 | 100.00 | -10.36 | 5.26 | -10.36 | 630.19 | -78.81 | 1.19 | 1.055 |
| 4985.00 | 100.00 | -10.23 | 5.28 | -10.23 | 623.43 | -89.90 | 1.21 | 1.056 |
| 5000.00 | 100.00 | -10.10 | 5.29 | -10.10 | 616.11 | -100.38 | 1.24 | 1.056 |
| 5015.00 | 100.00 | -9.99 | 5.31 | -9.99 | 608.30 | -110.23 | 1.26 | 1.056 |
| 5030.00 | 100.00 | -9.88 | 5.32 | -9.88 | 600.04 | -119.44 | 1.29 | 1.057 |
| 5045.00 | 100.00 | -9.78 | 5.34 | -9.78 | 591.41 | -128.00 | 1.31 | 1.057 |
| 5060.00 | 100.00 | -9.68 | 5.35 | -9.68 | 582.45 | -135.91 | 1.34 | 1.057 |
| 5075.00 | 100.00 | -9.60 | 5.36 | -9.60 | 573.21 | -143.16 | 1.37 | 1.058 |
| 5090.00 | 100.00 | -9.52 | 5.38 | -9.52 | 563.74 | -149.76 | 1.40 | 1.058 |
| 5105.00 | 100.00 | -9.44 | 5.39 | -9.44 | 554.10 | -155.72 | 1.42 | 1.059 |
| 5120.00 | 100.00 | -9.38 | 5.40 | -9.38 | 544.32 | -161.04 | 1.45 | 1.059 |
| 5135.00 | 100.00 | -9.32 | 5.41 | -9.32 | 534.45 | -165.76 | 1.48 | 1.059 |
| 5150.00 | 100.00 | -9.27 | 5.42 | -9.27 | 524.52 | -169.86 | 1.51 | 1.060 |
| 5165.00 | 100.00 | -9.22 | 5.42 | -9.22 | 514.57 | -173.39 | 1.54 | 1.060 |
| 5180.00 | 100.00 | -9.18 | 5.43 | -9.18 | 504.62 | -176.35 | 1.57 | 1.061 |
| 5195.00 | 100.00 | -9.15 | 5.43 | -9.15 | 494.71 | -178.76 | 1.60 | 1.061 |
| 5210.00 | 100.00 | -9.12 | 5.44 | -9.12 | 484.86 | -180.64 | 1.64 | 1.062 |
| 5225.00 | 100.00 | -9.10 | 5.44 | -9.10 | 475.10 | -182.02 | 1.67 | 1.062 |
| 5240.00 | 100.00 | -9.08 | 5.44 | -9.08 | 465.44 | -182.90 | 1.70 | 1.063 |
| 5255.00 | 100.00 | -9.07 | 5.45 | -9.07 | 455.90 | -183.32 | 1.73 | 1.063 |
| 5270.00 | 100.00 | -9.06 | 5.45 | -9.06 | 446.50 | -183.30 | 1.76 | 1.064 |
| 5285.00 | 100.00 | -9.06 | 5.44 | -9.06 | 437.26 | -182.84 | 1.80 | 1.064 |
| 5300.00 | 100.00 | -9.06 | 5.44 | -9.06 | 428.18 | -181.97 | 1.83 | 1.065 |
| 5315.00 | 100.00 | -9.07 | 5.44 | -9.07 | 419.27 | -180.72 | 1.86 | 1.065 |
| 5330.00 | 100.00 | -9.08 | 5.43 | -9.08 | 410.56 | -179.08 | 1.90 | 1.066 |
| 5345.00 | 100.00 | -9.10 | 5.43 | -9.10 | 402.03 | -177.10 | 1.93 | 1.066 |
| 5360.00 | 100.00 | -9.12 | 5.42 | -9.12 | 393.71 | -174.77 | 1.97 | 1.067 |
| 5375.00 | 100.00 | -9.14 | 5.41 | -9.14 | 385.60 | -172.11 | 2.00 | 1.067 |
| 5390.00 | 100.00 | -9.17 | 5.40 | -9.17 | 377.69 | -169.15 | 2.04 | 1.068 |
| 5405.00 | 100.00 | -9.20 | 5.39 | -9.20 | 370.01 | -165.89 | 2.07 | 1.068 |
| 5420.00 | 100.00 | -9.23 | 5.37 | -9.23 | 362.54 | -162.35 | 2.11 | 1.069 |
| 5435.00 | 100.00 | -9.26 | 5.36 | -9.26 | 355.30 | -158.53 | 2.14 | 1.069 |
| 5450.00 | 100.00 | -9.28 | 5.35 | -9.28 | 348.28 | -154.46 | 2.18 | 1.070 |
| 5465.00 | 100.00 | -9.31 | 5.33 | -9.31 | 341.49 | -150.14 | 2.21 | 1.070 |
| 5480.00 | 100.00 | -9.34 | 5.31 | -9.34 | 334.94 | -145.58 | 2.25 | 1.071 |
| 5495.00 | 100.00 | -9.36 | 5.29 | -9.36 | 328.61 | -140.79 | 2.28 | 1.071 |
| 5510.00 | 100.00 | -9.38 | 5.27 | -9.38 | 322.52 | -135.79 | 2.32 | 1.072 |
| 5525.00 | 100.00 | -9.39 | 5.25 | -9.39 | 316.66 | -130.57 | 2.35 | 1.072 |
| 5540.00 | 100.00 | -9.39 | 5.22 | -9.39 | 311.03 | -125.15 | 2.38 | 1.072 |
| 5555.00 | 100.00 | -9.38 | 5.20 | -9.38 | 305.65 | -119.54 | 2.42 | 1.073 |
| 5570.00 | 100.00 | -9.36 | 5.17 | -9.36 | 300.50 | -113.73 | 2.45 | 1.073 |
| 5585.00 | 100.00 | -9.33 | 5.13 | -9.33 | 295.60 | -107.74 | 2.48 | 1.074 |
| 5600.00 | 100.00 | -9.28 | 5.09 | -9.29 | 290.93 | -101.58 | 2.52 | 1.074 |
| 5615.00 | 100.00 | -9.22 | 5.06 | -9.28 | 286.51 | -95.25 | 2.55 | 1.075 |
| 5630.00 | 100.00 | -9.14 | 5.02 | -9.25 | 282.34 | -88.74 | 2.58 | 1.075 |
| 5645.00 | 100.00 | -9.04 | 4.98 | -9.21 | 278.41 | -82.08 | 2.61 | 1.075 |
| 5660.00 | 100.00 | -8.93 | 4.93 | -9.14 | 274.74 | -75.26 | 2.64 | 1.076 |
| 5675.00 | 100.00 | -8.79 | 4.88 | -9.16 | 271.31 | -68.29 | 2.66 | 1.076 |
| 5690.00 | 100.00 | -8.63 | 4.82 | -9.22 | 268.14 | -61.16 | 2.69 | 1.076 |
| 5705.00 | 100.00 | -8.46 | 4.76 | -9.27 | 265.24 | -53.89 | 2.71 | 1.076 |
| 5720.00 | 100.00 | -8.26 | 4.69 | -9.30 | 262.59 | -46.48 | 2.74 | 1.077 |
| 5735.00 | 100.00 | -8.05 | 4.62 | -9.31 | 260.22 | -38.92 | 2.76 | 1.077 |
| 5750.00 | 100.00 | -7.82 | 4.54 | -9.29 | 258.11 | -31.23 | 2.78 | 1.077 |
| 5765.00 | 100.00 | -7.57 | 4.46 | -9.25 | 256.29 | -23.40 | 2.79 | 1.077 |
| 5780.00 | 100.00 | -7.31 | 4.38 | -9.18 | 254.75 | -15.45 | 2.81 | 1.077 |
| 5795.00 | 100.00 | -7.04 | 4.29 | -9.24 | 253.50 | -7.36 | 2.82 | 1.077 |
| 5810.00 | 100.00 | -6.76 | 4.19 | -9.44 | 252.54 | 0.85 | 2.83 | 1.077 |
| 5825.00 | 100.00 | -6.47 | 4.09 | -9.63 | 251.90 | 9.18 | 2.84 | 1.077 |
| 5840.00 | 100.00 | -6.17 | 3.97 | -9.80 | 251.56 | 17.63 | 2.84 | 1.077 |
| 5855.00 | 100.00 | -5.87 | 3.85 | -9.94 | 251.55 | 26.19 | 2.85 | 1.077 |
| 5870.00 | 100.00 | -5.57 | 3.72 | -10.06 | 251.87 | 34.87 | 2.85 | 1.076 |
| 5885.00 | 100.00 | -5.26 | 3.58 | -10.14 | 252.54 | 43.65 | 2.84 | 1.076 |
| 5900.00 | 100.00 | -4.95 | 3.44 | -10.18 | 253.55 | 52.54 | 2.84 | 1.076 |
| 5915.00 | 100.00 | -4.65 | 3.28 | -10.18 | 254.93 | 61.52 | 2.83 | 1.076 |
| 5930.00 | 100.00 | -4.34 | 3.12 | -10.13 | 256.70 | 70.59 | 2.82 | 1.075 |
| 5945.00 | 100.00 | -4.05 | 2.95 | -10.03 | 258.85 | 79.75 | 2.80 | 1.075 |
| 5960.00 | 100.00 | -3.75 | 2.77 | -10.35 | 261.41 | 88.98 | 2.78 | 1.074 |
| 5975.00 | 100.00 | -3.46 | 2.59 | -10.78 | 264.40 | 98.28 | 2.76 | 1.074 |
| 5990.00 | 100.00 | -3.18 | 2.39 | -11.21 | 267.82 | 107.63 | 2.74 | 1.073 |
| 6005.00 | 100.00 | -2.90 | 2.18 | -11.63 | 271.70 | 117.03 | 2.71 | 1.073 |
| 6020.00 | 100.00 | -2.63 | 1.96 | -12.01 | 276.06 | 126.47 | 2.68 | 1.072 |
| 6035.00 | 100.00 | -2.36 | 1.73 | -12.34 | 280.91 | 135.92 | 2.65 | 1.071 |
| 6050.00 | 100.00 | -2.11 | 1.49 | -12.61 | 286.27 | 145.37 | 2.62 | 1.071 |
| 6065.00 | 100.00 | -1.86 | 1.23 | -12.80 | 292.17 | 154.81 | 2.58 | 1.070 |
| 6080.00 | 100.00 | -1.62 | 0.96 | -12.89 | 298.62 | 164.21 | 2.55 | 1.069 |
| 6095.00 | 100.00 | -1.38 | 0.68 | -12.88 | 305.65 | 173.56 | 2.51 | 1.068 |
| 6110.00 | 100.00 | -1.16 | 0.38 | -12.77 | 313.28 | 182.81 | 2.46 | 1.068 |
| 6125.00 | 100.00 | -0.94 | 0.07 | -12.57 | 321.52 | 191.94 | 2.42 | 1.067 |
| 6140.00 | 100.00 | -0.73 | -0.26 | -13.12 | 330.42 | 200.93 | 2.38 | 1.066 |
| 6155.00 | 100.00 | -0.53 | -0.53 | -14.01 | 339.98 | 209.73 | 2.33 | 1.065 |
| 6170.00 | 100.00 | -0.34 | -0.34 | -14.96 | 350.22 | 218.30 | 2.28 | 1.065 |
| 6185.00 | 100.00 | -0.15 | -0.15 | -15.93 | 361.17 | 226.60 | 2.24 | 1.064 |
| 6200.00 | 100.00 | 0.03 | 0.03 | -16.89 | 372.84 | 234.58 | 2.19 | 1.063 |
| 6215.00 | 100.00 | 0.20 | 0.20 | -17.77 | 385.26 | 242.17 | 2.14 | 1.062 |
| 6230.00 | 100.00 | 0.36 | 0.36 | -18.46 | 398.42 | 249.33 | 2.09 | 1.062 |
| 6245.00 | 100.00 | 0.52 | 0.52 | -18.83 | 412.35 | 255.99 | 2.04 | 1.061 |
| 6260.00 | 100.00 | 0.67 | 0.67 | -18.81 | 427.05 | 262.07 | 2.00 | 1.060 |
| 6275.00 | 100.00 | 0.81 | 0.81 | -18.41 | 442.50 | 267.50 | 1.95 | 1.060 |
| 6290.00 | 100.00 | 0.95 | 0.95 | -17.73 | 458.72 | 272.19 | 1.90 | 1.059 |
| 6305.00 | 100.00 | 1.08 | 1.08 | -16.91 | 475.68 | 276.07 | 1.85 | 1.058 |
| 6320.00 | 100.00 | 1.21 | 1.21 | -18.31 | 493.35 | 279.04 | 1.81 | 1.058 |
| 6335.00 | 100.00 | 1.33 | 1.33 | -20.33 | 511.71 | 281.00 | 1.76 | 1.057 |
| 6350.00 | 100.00 | 1.44 | 1.44 | -22.89 | 530.70 | 281.87 | 1.72 | 1.057 |
| 6365.00 | 100.00 | 1.55 | 1.55 | -26.39 | 550.26 | 281.53 | 1.67 | 1.056 |
| 6380.00 | 100.00 | 1.66 | 1.66 | -31.63 | 570.32 | 279.90 | 1.63 | 1.056 |
| 6395.00 | 100.00 | 1.76 | 1.76 | -35.87 | 590.80 | 276.88 | 1.59 | 1.056 |
| 6410.00 | 100.00 | 1.86 | 1.86 | -30.21 | 611.58 | 272.37 | 1.55 | 1.055 |
| 6425.00 | 100.00 | 1.95 | 1.95 | -25.67 | 632.54 | 266.30 | 1.51 | 1.055 |
| 6440.00 | 100.00 | 2.04 | 2.04 | -22.62 | 653.54 | 258.59 | 1.47 | 1.055 |
| 6455.00 | 100.00 | 2.12 | 2.12 | -22.00 | 674.45 | 249.19 | 1.44 | 1.054 |
| 6470.00 | 100.00 | 2.20 | 2.20 | -24.60 | 695.07 | 238.05 | 1.40 | 1.054 |
| 6485.00 | 100.00 | 2.28 | 2.28 | -27.33 | 715.25 | 225.14 | 1.37 | 1.054 |
| 6500.00 | 100.00 | 2.36 | 2.36 | -28.60 | 734.78 | 210.48 | 1.34 | 1.054 |
| 6515.00 | 100.00 | 2.43 | 2.43 | -27.00 | 753.47 | 194.07 | 1.31 | 1.054 |
| 6530.00 | 100.00 | 2.50 | 2.50 | -24.41 | 771.13 | 175.98 | 1.28 | 1.054 |
| 6545.00 | 100.00 | 2.56 | 2.56 | -27.44 | 787.56 | 156.29 | 1.26 | 1.054 |
| 6560.00 | 100.00 | 2.63 | 2.63 | -29.20 | 802.57 | 135.11 | 1.24 | 1.054 |
| 6575.00 | 100.00 | 2.69 | 2.69 | -34.59 | 815.98 | 112.57 | 1.22 | 1.054 |
| 6590.00 | 100.00 | 2.75 | 2.75 | -31.42 | 827.65 | 88.83 | 1.20 | 1.054 |
| 6605.00 | 100.00 | 2.80 | 2.80 | -26.26 | 837.44 | 64.09 | 1.20 | 1.054 |
| 6620.00 | 100.00 | 2.86 | 2.86 | -22.83 | 845.23 | 38.55 | 1.19 | 1.054 |
| 6635.00 | 100.00 | 2.91 | 2.91 | -20.36 | 850.95 | 12.43 | 1.19 | 1.054 |
| 6650.00 | 100.00 | 2.96 | 2.96 | -18.46 | 854.56 | -14.05 | 1.20 | 1.054 |
| 6665.00 | 100.00 | 3.01 | 3.01 | -16.92 | 856.03 | -40.66 | 1.21 | 1.054 |
| 6680.00 | 100.00 | 3.05 | 3.05 | -15.63 | 855.39 | -67.15 | 1.22 | 1.055 |
| 6695.00 | 100.00 | 3.10 | 3.10 | -14.53 | 852.69 | -93.33 | 1.24 | 1.055 |
| 6710.00 | 100.00 | 3.14 | 3.14 | -13.56 | 848.00 | -118.97 | 1.26 | 1.055 |
| 6725.00 | 100.00 | 3.19 | 3.19 | -12.71 | 841.43 | -143.89 | 1.28 | 1.056 |
| 6740.00 | 100.00 | 3.23 | 3.23 | -11.94 | 833.10 | -167.92 | 1.30 | 1.056 |
| 6755.00 | 100.00 | 3.27 | 3.27 | -11.25 | 823.14 | -190.92 | 1.33 | 1.056 |
| 6770.00 | 100.00 | 3.31 | 3.31 | -10.62 | 811.72 | -212.76 | 1.36 | 1.057 |
| 6785.00 | 100.00 | 3.34 | 3.34 | -10.05 | 798.98 | -233.34 | 1.39 | 1.057 |
| 6800.00 | 100.00 | 3.38 | 3.38 | -9.55 | 785.10 | -252.59 | 1.42 | 1.058 |
| 6815.00 | 100.00 | 3.42 | 3.42 | -9.10 | 770.23 | -270.47 | 1.45 | 1.058 |
| 6830.00 | 100.00 | 3.45 | 3.45 | -8.69 | 754.53 | -286.93 | 1.48 | 1.059 |
| 6845.00 | 100.00 | 3.48 | 3.48 | -8.30 | 738.16 | -301.97 | 1.51 | 1.059 |
| 6860.00 | 100.00 | 3.52 | 3.52 | -7.93 | 721.25 | -315.60 | 1.55 | 1.060 |
| 6875.00 | 100.00 | 3.55 | 3.55 | -7.58 | 703.95 | -327.84 | 1.58 | 1.061 |
| 6890.00 | 100.00 | 3.58 | 3.58 | -7.25 | 686.38 | -338.71 | 1.62 | 1.061 |
| 6905.00 | 100.00 | 3.61 | 3.61 | -6.94 | 668.64 | -348.27 | 1.65 | 1.062 |
| 6920.00 | 100.00 | 3.64 | 3.64 | -6.64 | 650.84 | -356.56 | 1.69 | 1.063 |
| 6935.00 | 100.00 | 3.66 | 3.66 | -6.36 | 633.08 | -363.63 | 1.73 | 1.063 |
| 6950.00 | 100.00 | 3.69 | 3.74 | -6.09 | 615.42 | -369.55 | 1.77 | 1.064 |
| 6965.00 | 100.00 | 3.72 | 3.85 | -5.84 | 597.94 | -374.38 | 1.81 | 1.065 |
| 6980.00 | 100.00 | 3.74 | 3.96 | -5.59 | 580.68 | -378.18 | 1.85 | 1.066 |
| 6995.00 | 100.00 | 3.77 | 4.07 | -5.36 | 563.71 | -381.01 | 1.89 | 1.067 |
| 7010.00 | 100.00 | 3.79 | 4.17 | -5.13 | 547.07 | -382.94 | 1.93 | 1.067 |
| 7025.00 | 100.00 | 3.81 | 4.28 | -4.92 | 530.78 | -384.01 | 1.97 | 1.068 |
| 7040.00 | 100.00 | 3.83 | 4.38 | -4.71 | 514.87 | -384.30 | 2.01 | 1.069 |
| 7055.00 | 100.00 | 3.85 | 4.48 | -4.51 | 499.36 | -383.86 | 2.06 | 1.070 |
| 7070.00 | 100.00 | 3.87 | 4.58 | -4.32 | 484.28 | -382.74 | 2.10 | 1.071 |
| 7085.00 | 100.00 | 3.89 | 4.67 | -4.13 | 469.63 | -380.99 | 2.15 | 1.072 |
| 7100.00 | 100.00 | 3.91 | 4.77 | -3.95 | 455.42 | -378.65 | 2.19 | 1.073 |
| 7115.00 | 100.00 | 3.93 | 4.86 | -3.78 | 441.65 | -375.77 | 2.24 | 1.075 |
| 7130.00 | 100.00 | 3.94 | 4.95 | -3.61 | 428.33 | -372.39 | 2.28 | 1.076 |
| 7145.00 | 100.00 | 3.95 | 5.04 | -3.45 | 415.45 | -368.55 | 2.33 | 1.077 |
| 7160.00 | 100.00 | 3.97 | 5.13 | -3.30 | 403.02 | -364.28 | 2.38 | 1.078 |
| 7175.00 | 100.00 | 3.98 | 5.21 | -3.15 | 391.03 | -359.61 | 2.43 | 1.079 |
| 7190.00 | 100.00 | 3.99 | 5.29 | -3.00 | 379.48 | -354.58 | 2.47 | 1.081 |
| 7205.00 | 100.00 | 3.99 | 5.37 | -2.86 | 368.35 | -349.20 | 2.52 | 1.082 |
| 7220.00 | 100.00 | 4.00 | 5.45 | -2.73 | 357.66 | -343.50 | 2.57 | 1.083 |
| 7235.00 | 100.00 | 4.00 | 5.52 | -2.62 | 347.38 | -337.52 | 2.62 | 1.085 |
| 7250.00 | 100.00 | 4.00 | 5.60 | -2.51 | 337.51 | -331.25 | 2.67 | 1.086 |
| 7265.00 | 100.00 | 4.00 | 5.67 | -2.41 | 328.06 | -324.73 | 2.72 | 1.088 |
| 7280.00 | 100.00 | 4.00 | 5.73 | -2.32 | 319.00 | -317.97 | 2.77 | 1.089 |
| 7295.00 | 100.00 | 3.99 | 5.80 | -2.23 | 310.34 | -310.99 | 2.82 | 1.091 |
| 7310.00 | 100.00 | 3.98 | 5.86 | -2.14 | 302.06 | -303.79 | 2.87 | 1.092 |
| 7325.00 | 100.00 | 3.97 | 5.92 | -2.06 | 294.17 | -296.39 | 2.92 | 1.094 |
| 7340.00 | 100.00 | 3.95 | 5.97 | -1.98 | 286.66 | -288.80 | 2.96 | 1.095 |
| 7355.00 | 100.00 | 3.93 | 6.03 | -1.90 | 279.52 | -281.03 | 3.01 | 1.097 |
| 7370.00 | 100.00 | 3.90 | 6.07 | -1.83 | 272.76 | -273.09 | 3.06 | 1.099 |
| 7385.00 | 100.00 | 3.87 | 6.12 | -1.77 | 266.37 | -264.98 | 3.10 | 1.100 |
| 7400.00 | 100.00 | 3.84 | 6.15 | -1.71 | 260.34 | -256.72 | 3.15 | 1.102 |
| 7415.00 | 100.00 | 3.79 | 6.19 | -1.66 | 254.68 | -248.31 | 3.19 | 1.104 |
| 7430.00 | 100.00 | 3.75 | 6.21 | -1.61 | 249.39 | -239.75 | 3.23 | 1.105 |
| 7445.00 | 100.00 | 3.69 | 6.24 | -1.56 | 244.46 | -231.05 | 3.27 | 1.107 |
| 7460.00 | 100.00 | 3.63 | 6.25 | -1.53 | 239.91 | -222.22 | 3.30 | 1.109 |
| 7475.00 | 100.00 | 3.56 | 6.26 | -1.50 | 235.73 | -213.26 | 3.33 | 1.110 |
| 7490.00 | 100.00 | 3.48 | 6.27 | -1.47 | 231.92 | -204.17 | 3.36 | 1.112 |
| 7505.00 | 100.00 | 3.40 | 6.26 | -1.45 | 228.50 | -194.97 | 3.39 | 1.114 |
| 7520.00 | 100.00 | 3.30 | 6.25 | -1.44 | 225.46 | -185.65 | 3.41 | 1.115 |
| 7535.00 | 100.00 | 3.20 | 6.23 | -1.44 | 222.82 | -176.22 | 3.42 | 1.116 |
| 7550.00 | 100.00 | 3.08 | 6.20 | -1.46 | 220.59 | -166.69 | 3.43 | 1.118 |
| 7565.00 | 100.00 | 2.95 | 6.16 | -1.50 | 218.77 | -157.06 | 3.44 | 1.119 |
| 7580.00 | 100.00 | 2.81 | 6.11 | -1.54 | 217.37 | -147.35 | 3.44 | 1.120 |
| 7595.00 | 100.00 | 2.65 | 6.05 | -1.60 | 216.41 | -137.55 | 3.44 | 1.121 |
| 7610.00 | 100.00 | 2.49 | 5.98 | -1.66 | 215.91 | -127.69 | 3.43 | 1.122 |
| 7625.00 | 100.00 | 2.30 | 5.89 | -1.73 | 215.87 | -117.76 | 3.41 | 1.122 |
| 7640.00 | 100.00 | 2.10 | 5.80 | -1.81 | 216.31 | -107.79 | 3.39 | 1.123 |
| 7655.00 | 100.00 | 1.89 | 5.69 | -1.90 | 217.24 | -97.79 | 3.36 | 1.123 |
| 7670.00 | 100.00 | 1.65 | 5.57 | -2.00 | 218.70 | -87.77 | 3.32 | 1.124 |
| 7685.00 | 100.00 | 1.40 | 5.43 | -2.10 | 220.68 | -77.76 | 3.28 | 1.123 |
| 7700.00 | 100.00 | 1.13 | 5.29 | -2.25 | 223.22 | -67.78 | 3.23 | 1.123 |
| 7715.00 | 100.00 | 0.84 | 5.13 | -2.42 | 226.33 | -57.86 | 3.18 | 1.123 |
| 7730.00 | 100.00 | 0.53 | 4.95 | -2.59 | 230.04 | -48.02 | 3.12 | 1.122 |
| 7745.00 | 100.00 | 0.20 | 4.76 | -2.77 | 234.35 | -38.30 | 3.06 | 1.122 |
| 7760.00 | 100.00 | -0.15 | 4.56 | -2.96 | 239.28 | -28.74 | 2.99 | 1.121 |
| 7775.00 | 100.00 | -0.52 | 4.34 | -3.15 | 244.86 | -19.37 | 2.92 | 1.119 |
| 7790.00 | 100.00 | -0.91 | 4.11 | -3.40 | 251.10 | -10.25 | 2.85 | 1.118 |
| 7805.00 | 100.00 | -1.33 | 3.87 | -3.67 | 258.00 | -1.44 | 2.77 | 1.117 |
| 7820.00 | 100.00 | -1.77 | 3.62 | -3.94 | 265.57 | 7.03 | 2.69 | 1.115 |
| 7835.00 | 100.00 | -2.23 | 3.35 | -4.21 | 273.81 | 15.08 | 2.61 | 1.114 |
| 7850.00 | 100.00 | -2.70 | 3.08 | -4.48 | 282.73 | 22.64 | 2.53 | 1.112 |
| 7865.00 | 100.00 | -3.20 | 2.80 | -4.76 | 292.30 | 29.65 | 2.45 | 1.110 |
| 7880.00 | 100.00 | -3.72 | 2.50 | -5.13 | 302.51 | 36.04 | 2.37 | 1.108 |
| 7895.00 | 100.00 | -4.26 | 2.21 | -5.49 | 313.32 | 41.72 | 2.29 | 1.106 |
| 7910.00 | 100.00 | -4.80 | 1.90 | -5.83 | 324.71 | 46.61 | 2.21 | 1.104 |
| 7925.00 | 100.00 | -5.36 | 1.60 | -6.16 | 336.60 | 50.65 | 2.14 | 1.103 |
| 7940.00 | 100.00 | -5.92 | 1.29 | -6.50 | 348.95 | 53.76 | 2.06 | 1.101 |
| 7955.00 | 100.00 | -6.47 | 0.99 | -6.92 | 361.67 | 55.86 | 1.99 | 1.099 |
| 7970.00 | 100.00 | -7.01 | 0.69 | -7.30 | 374.67 | 56.90 | 1.92 | 1.097 |
| 7985.00 | 100.00 | -7.53 | 0.39 | -7.64 | 387.85 | 56.81 | 1.86 | 1.095 |
| 8000.00 | 100.00 | -8.00 | 0.11 | -8.04 | 401.10 | 55.56 | 1.80 | 1.093 |
| 8015.00 | 100.00 | -8.41 | -0.16 | -8.41 | 414.29 | 53.11 | 1.74 | 1.092 |
| 8030.00 | 100.00 | -8.76 | -0.42 | -8.76 | 427.28 | 49.47 | 1.69 | 1.090 |
| 8045.00 | 100.00 | -9.02 | -0.65 | -9.02 | 439.96 | 44.62 | 1.64 | 1.088 |
| 8060.00 | 100.00 | -9.18 | -0.87 | -9.18 | 452.16 | 38.61 | 1.59 | 1.087 |
| 8075.00 | 100.00 | -9.26 | -1.06 | -9.26 | 463.78 | 31.48 | 1.55 | 1.085 |
| 8090.00 | 100.00 | -9.23 | -1.23 | -9.23 | 474.66 | 23.29 | 1.51 | 1.084 |
| 8105.00 | 100.00 | -9.13 | -1.37 | -9.13 | 484.69 | 14.15 | 1.48 | 1.083 |
| 8120.00 | 100.00 | -8.96 | -1.48 | -8.96 | 493.78 | 4.14 | 1.45 | 1.082 |
| 8135.00 | 100.00 | -8.73 | -1.57 | -8.73 | 501.82 | -6.60 | 1.43 | 1.080 |
| 8150.00 | 100.00 | -8.47 | -1.62 | -8.47 | 508.76 | -17.96 | 1.41 | 1.079 |
| 8165.00 | 100.00 | -8.18 | -1.65 | -8.18 | 514.54 | -29.78 | 1.39 | 1.078 |
| 8180.00 | 100.00 | -7.87 | -1.65 | -7.87 | 519.13 | -41.92 | 1.39 | 1.077 |
| 8195.00 | 100.00 | -7.55 | -1.63 | -7.55 | 522.53 | -54.26 | 1.39 | 1.077 |
| 8210.00 | 100.00 | -7.24 | -1.59 | -7.24 | 524.76 | -66.64 | 1.39 | 1.076 |
| 8225.00 | 100.00 | -6.92 | -1.53 | -6.92 | 525.83 | -78.93 | 1.39 | 1.075 |
| 8240.00 | 100.00 | -6.62 | -1.45 | -6.77 | 525.82 | -91.03 | 1.41 | 1.075 |
| 8255.00 | 100.00 | -6.32 | -1.36 | -6.80 | 524.76 | -102.81 | 1.42 | 1.074 |
| 8270.00 | 100.00 | -6.04 | -1.25 | -6.82 | 522.75 | -114.19 | 1.44 | 1.073 |
| 8285.00 | 100.00 | -5.76 | -1.14 | -6.84 | 519.85 | -125.09 | 1.46 | 1.073 |
| 8300.00 | 100.00 | -5.50 | -1.02 | -6.85 | 516.16 | -135.44 | 1.48 | 1.073 |
| 8315.00 | 100.00 | -5.25 | -0.89 | -6.86 | 511.77 | -145.18 | 1.51 | 1.072 |
| 8330.00 | 100.00 | -5.01 | -0.75 | -6.87 | 506.76 | -154.29 | 1.53 | 1.072 |
| 8345.00 | 100.00 | -4.79 | -0.62 | -6.87 | 501.22 | -162.73 | 1.56 | 1.072 |
| 8360.00 | 100.00 | -4.57 | -0.48 | -6.87 | 495.23 | -170.50 | 1.59 | 1.072 |
| 8375.00 | 100.00 | -4.37 | -0.34 | -6.86 | 488.88 | -177.58 | 1.62 | 1.071 |
| 8390.00 | 100.00 | -4.18 | -0.20 | -6.86 | 482.24 | -183.99 | 1.65 | 1.071 |
| 8405.00 | 100.00 | -4.00 | -0.06 | -6.86 | 475.37 | -189.73 | 1.68 | 1.071 |
| 8420.00 | 100.00 | -3.83 | 0.08 | -6.85 | 468.35 | -194.82 | 1.71 | 1.071 |
| 8435.00 | 100.00 | -3.66 | 0.21 | -6.85 | 461.23 | -199.28 | 1.74 | 1.071 |
| 8450.00 | 100.00 | -3.51 | 0.35 | -6.84 | 454.06 | -203.14 | 1.77 | 1.071 |
| 8465.00 | 100.00 | -3.36 | 0.48 | -6.84 | 446.88 | -206.42 | 1.80 | 1.071 |
| 8480.00 | 100.00 | -3.23 | 0.61 | -6.84 | 439.74 | -209.15 | 1.84 | 1.071 |
| 8495.00 | 100.00 | -3.10 | 0.73 | -6.84 | 432.67 | -211.36 | 1.87 | 1.071 |
| 8510.00 | 100.00 | -2.98 | 0.85 | -6.84 | 425.70 | -213.08 | 1.90 | 1.071 |
| 8525.00 | 100.00 | -2.86 | 0.97 | -6.84 | 418.86 | -214.35 | 1.93 | 1.072 |
| 8540.00 | 100.00 | -2.76 | 1.08 | -6.84 | 412.17 | -215.18 | 1.96 | 1.072 |
| 8555.00 | 100.00 | -2.66 | 1.19 | -6.84 | 405.64 | -215.62 | 1.99 | 1.072 |
| 8570.00 | 100.00 | -2.56 | 1.30 | -6.85 | 399.29 | -215.69 | 2.02 | 1.072 |
| 8585.00 | 100.00 | -2.48 | 1.40 | -6.86 | 393.14 | -215.41 | 2.04 | 1.072 |
| 8600.00 | 100.00 | -2.39 | 1.50 | -6.87 | 387.20 | -214.81 | 2.07 | 1.073 |
| 8615.00 | 100.00 | -2.32 | 1.59 | -6.89 | 381.46 | -213.91 | 2.10 | 1.073 |
| 8630.00 | 100.00 | -2.25 | 1.68 | -6.90 | 375.95 | -212.75 | 2.13 | 1.073 |
| 8645.00 | 100.00 | -2.19 | 1.77 | -6.92 | 370.66 | -211.33 | 2.15 | 1.073 |
| 8660.00 | 100.00 | -2.13 | 1.85 | -6.94 | 365.59 | -209.69 | 2.18 | 1.074 |
| 8675.00 | 100.00 | -2.07 | 1.93 | -6.96 | 360.76 | -207.83 | 2.20 | 1.074 |
| 8690.00 | 100.00 | -2.03 | 2.01 | -6.99 | 356.16 | -205.79 | 2.22 | 1.074 |
| 8705.00 | 100.00 | -1.98 | 2.08 | -7.01 | 351.78 | -203.57 | 2.24 | 1.074 |
| 8720.00 | 100.00 | -1.95 | 2.14 | -7.04 | 347.65 | -201.19 | 2.26 | 1.075 |
| 8735.00 | 100.00 | -1.91 | 2.21 | -7.07 | 343.74 | -198.67 | 2.28 | 1.075 |
| 8750.00 | 100.00 | -1.88 | 2.27 | -7.10 | 340.06 | -196.03 | 2.30 | 1.075 |
| 8765.00 | 100.00 | -1.86 | 2.32 | -7.14 | 336.61 | -193.26 | 2.32 | 1.075 |
| 8780.00 | 100.00 | -1.84 | 2.37 | -7.17 | 333.39 | -190.40 | 2.33 | 1.076 |
| 8795.00 | 100.00 | -1.83 | 2.42 | -7.21 | 330.40 | -187.45 | 2.35 | 1.076 |
| 8810.00 | 100.00 | -1.82 | 2.46 | -7.25 | 327.63 | -184.42 | 2.36 | 1.076 |
| 8825.00 | 100.00 | -1.82 | 2.50 | -7.29 | 325.08 | -181.32 | 2.37 | 1.076 |
| 8840.00 | 100.00 | -1.82 | 2.54 | -7.33 | 322.76 | -178.16 | 2.38 | 1.077 |
| 8855.00 | 100.00 | -1.82 | 2.57 | -7.37 | 320.65 | -174.96 | 2.39 | 1.077 |
| 8870.00 | 100.00 | -1.83 | 2.60 | -7.41 | 318.76 | -171.72 | 2.40 | 1.077 |
| 8885.00 | 100.00 | -1.84 | 2.63 | -7.45 | 317.08 | -168.45 | 2.41 | 1.077 |
| 8900.00 | 100.00 | -1.86 | 2.65 | -7.49 | 315.62 | -165.16 | 2.41 | 1.077 |
| 8915.00 | 100.00 | -1.89 | 2.66 | -7.54 | 314.37 | -161.86 | 2.42 | 1.078 |
| 8930.00 | 100.00 | -1.91 | 2.68 | -7.63 | 313.33 | -158.56 | 2.42 | 1.078 |
| 8945.00 | 100.00 | -1.95 | 2.69 | -7.73 | 312.50 | -155.26 | 2.42 | 1.078 |
| 8960.00 | 100.00 | -1.98 | 2.69 | -7.83 | 311.87 | -151.98 | 2.42 | 1.078 |
| 8975.00 | 100.00 | -2.02 | 2.69 | -7.93 | 311.45 | -148.72 | 2.42 | 1.078 |
| 8990.00 | 100.00 | -2.07 | 2.69 | -8.04 | 311.23 | -145.49 | 2.41 | 1.078 |
| 9005.00 | 100.00 | -2.12 | 2.68 | -8.15 | 311.21 | -142.30 | 2.41 | 1.078 |
| 9020.00 | 100.00 | -2.17 | 2.67 | -8.27 | 311.39 | -139.16 | 2.40 | 1.078 |
| 9035.00 | 100.00 | -2.23 | 2.66 | -8.39 | 311.77 | -136.07 | 2.39 | 1.078 |
| 9050.00 | 100.00 | -2.30 | 2.64 | -8.51 | 312.34 | -133.05 | 2.39 | 1.078 |
| 9065.00 | 100.00 | -2.36 | 2.62 | -8.63 | 313.11 | -130.10 | 2.38 | 1.078 |
| 9080.00 | 100.00 | -2.44 | 2.60 | -8.76 | 314.07 | -127.24 | 2.37 | 1.078 |
| 9095.00 | 100.00 | -2.51 | 2.57 | -8.89 | 315.22 | -124.46 | 2.35 | 1.078 |
| 9110.00 | 100.00 | -2.60 | 2.54 | -9.01 | 316.55 | -121.79 | 2.34 | 1.078 |
| 9125.00 | 100.00 | -2.68 | 2.50 | -9.14 | 318.06 | -119.22 | 2.33 | 1.078 |
| 9140.00 | 100.00 | -2.77 | 2.46 | -9.27 | 319.76 | -116.78 | 2.31 | 1.078 |
| 9155.00 | 100.00 | -2.87 | 2.41 | -9.40 | 321.63 | -114.47 | 2.29 | 1.078 |
| 9170.00 | 100.00 | -2.97 | 2.37 | -9.52 | 323.68 | -112.30 | 2.28 | 1.077 |
| 9185.00 | 100.00 | -3.07 | 2.31 | -9.65 | 325.88 | -110.28 | 2.26 | 1.077 |
| 9200.00 | 100.00 | -3.18 | 2.26 | -9.77 | 328.26 | -108.43 | 2.24 | 1.077 |
| 9215.00 | 100.00 | -3.30 | 2.20 | -9.88 | 330.78 | -106.75 | 2.22 | 1.077 |
| 9230.00 | 100.00 | -3.41 | 2.14 | -9.99 | 333.46 | -105.25 | 2.20 | 1.077 |
| 9245.00 | 100.00 | -3.54 | 2.07 | -10.10 | 336.27 | -103.95 | 2.18 | 1.076 |
| 9260.00 | 100.00 | -3.66 | 2.00 | -10.19 | 339.22 | -102.86 | 2.16 | 1.076 |
| 9275.00 | 100.00 | -3.80 | 1.93 | -10.28 | 342.29 | -101.99 | 2.14 | 1.076 |
| 9290.00 | 100.00 | -3.93 | 1.85 | -10.37 | 345.47 | -101.35 | 2.12 | 1.076 |
| 9305.00 | 100.00 | -4.07 | 1.77 | -10.44 | 348.76 | -100.95 | 2.10 | 1.075 |
| 9320.00 | 100.00 | -4.22 | 1.68 | -10.50 | 352.12 | -100.81 | 2.08 | 1.075 |
| 9335.00 | 100.00 | -4.37 | 1.59 | -10.55 | 355.57 | -100.93 | 2.06 | 1.075 |
| 9350.00 | 100.00 | -4.52 | 1.50 | -10.59 | 359.07 | -101.32 | 2.04 | 1.074 |
| 9365.00 | 100.00 | -4.68 | 1.40 | -10.76 | 362.61 | -102.00 | 2.03 | 1.074 |
| 9380.00 | 100.00 | -4.84 | 1.30 | -11.05 | 366.17 | -102.97 | 2.01 | 1.074 |
| 9395.00 | 100.00 | -5.01 | 1.20 | -11.35 | 369.74 | -104.24 | 1.99 | 1.073 |
| 9410.00 | 100.00 | -5.18 | 1.09 | -11.66 | 373.29 | -105.81 | 1.97 | 1.073 |
| 9425.00 | 100.00 | -5.36 | 0.98 | -11.98 | 376.80 | -107.70 | 1.96 | 1.073 |
| 9440.00 | 100.00 | -5.54 | 0.87 | -12.31 | 380.25 | -109.89 | 1.94 | 1.072 |
| 9455.00 | 100.00 | -5.72 | 0.75 | -12.65 | 383.62 | -112.41 | 1.93 | 1.072 |
| 9470.00 | 100.00 | -5.91 | 0.63 | -13.00 | 386.87 | -115.24 | 1.92 | 1.072 |
| 9485.00 | 100.00 | -6.10 | 0.50 | -13.36 | 389.99 | -118.38 | 1.90 | 1.071 |
| 9500.00 | 100.00 | -6.29 | 0.38 | -13.73 | 392.95 | -121.83 | 1.89 | 1.071 |
| 9515.00 | 100.00 | -6.49 | 0.24 | -14.10 | 395.72 | -125.58 | 1.89 | 1.071 |
| 9530.00 | 100.00 | -6.69 | 0.11 | -14.49 | 398.27 | -129.61 | 1.88 | 1.070 |
| 9545.00 | 100.00 | -6.90 | -0.03 | -14.87 | 400.59 | -133.93 | 1.87 | 1.070 |
| 9560.00 | 100.00 | -7.11 | -0.17 | -15.26 | 402.64 | -138.51 | 1.87 | 1.070 |
| 9575.00 | 100.00 | -7.31 | -0.29 | -15.66 | 404.40 | -143.33 | 1.87 | 1.070 |
| 9590.00 | 100.00 | -7.53 | -0.18 | -16.05 | 405.85 | -148.37 | 1.87 | 1.069 |
| 9605.00 | 100.00 | -7.74 | -0.08 | -16.43 | 406.96 | -153.62 | 1.87 | 1.069 |
| 9620.00 | 100.00 | -7.96 | 0.02 | -16.80 | 407.73 | -159.03 | 1.88 | 1.069 |
| 9635.00 | 100.00 | -8.17 | 0.12 | -17.16 | 408.12 | -164.60 | 1.89 | 1.069 |
| 9650.00 | 100.00 | -8.39 | 0.22 | -17.50 | 408.14 | -170.28 | 1.89 | 1.068 |
| 9665.00 | 100.00 | -8.61 | 0.32 | -17.80 | 407.76 | -176.05 | 1.91 | 1.068 |
| 9680.00 | 100.00 | -8.83 | 0.42 | -18.07 | 406.98 | -181.87 | 1.92 | 1.068 |
| 9695.00 | 100.00 | -9.04 | 0.51 | -18.30 | 405.79 | -187.71 | 1.93 | 1.068 |
| 9710.00 | 100.00 | -9.26 | 0.61 | -18.47 | 404.20 | -193.54 | 1.95 | 1.068 |
| 9725.00 | 100.00 | -9.47 | 0.70 | -18.59 | 402.20 | -199.33 | 1.97 | 1.068 |
| 9740.00 | 100.00 | -9.68 | 0.79 | -18.65 | 399.81 | -205.03 | 1.99 | 1.067 |
| 9755.00 | 100.00 | -9.88 | 0.88 | -18.65 | 397.02 | -210.63 | 2.02 | 1.067 |
| 9770.00 | 100.00 | -10.08 | 0.97 | -18.59 | 393.85 | -216.08 | 2.04 | 1.067 |
| 9785.00 | 100.00 | -10.28 | 1.06 | -18.47 | 390.32 | -221.37 | 2.07 | 1.067 |
| 9800.00 | 100.00 | -10.46 | 1.15 | -18.31 | 386.44 | -226.45 | 2.10 | 1.067 |
| 9815.00 | 100.00 | -10.64 | 1.23 | -18.11 | 382.24 | -231.32 | 2.13 | 1.067 |
| 9830.00 | 100.00 | -10.81 | 1.31 | -17.87 | 377.73 | -235.94 | 2.17 | 1.067 |
| 9845.00 | 100.00 | -10.97 | 1.40 | -18.35 | 372.93 | -240.29 | 2.20 | 1.067 |
| 9860.00 | 100.00 | -11.12 | 1.48 | -19.04 | 367.88 | -244.37 | 2.24 | 1.067 |
| 9875.00 | 100.00 | -11.26 | 1.55 | -19.77 | 362.60 | -248.14 | 2.28 | 1.068 |
| 9890.00 | 100.00 | -11.39 | 1.63 | -20.52 | 357.10 | -251.62 | 2.32 | 1.068 |
| 9905.00 | 100.00 | -11.50 | 1.71 | -21.28 | 351.43 | -254.77 | 2.36 | 1.068 |
| 9920.00 | 100.00 | -11.60 | 1.78 | -22.04 | 345.60 | -257.60 | 2.40 | 1.068 |
| 9935.00 | 100.00 | -11.69 | 1.85 | -22.76 | 339.65 | -260.11 | 2.45 | 1.068 |
| 9950.00 | 100.00 | -11.76 | 1.93 | -23.39 | 333.59 | -262.28 | 2.50 | 1.069 |
| 9965.00 | 100.00 | -11.82 | 2.00 | -23.88 | 327.45 | -264.14 | 2.55 | 1.069 |
| 9980.00 | 100.00 | -11.86 | 2.06 | -24.17 | 321.25 | -265.66 | 2.60 | 1.069 |
| 9995.00 | 100.00 | -11.90 | 2.13 | -24.22 | 315.02 | -266.87 | 2.65 | 1.070 |
| 10010.00 | 100.00 | -11.91 | 2.20 | -24.01 | 308.78 | -267.76 | 2.70 | 1.070 |
| 10025.00 | 100.00 | -11.92 | 2.26 | -23.59 | 302.54 | -268.35 | 2.76 | 1.070 |
| 10040.00 | 100.00 | -11.91 | 2.32 | -23.01 | 296.32 | -268.64 | 2.81 | 1.071 |
| 10055.00 | 100.00 | -11.89 | 2.38 | -22.34 | 290.15 | -268.64 | 2.87 | 1.071 |
| 10070.00 | 100.00 | -11.86 | 2.44 | -21.62 | 284.03 | -268.36 | 2.93 | 1.072 |
| 10085.00 | 100.00 | -11.82 | 2.50 | -20.88 | 277.98 | -267.82 | 2.99 | 1.072 |
| 10100.00 | 100.00 | -11.77 | 2.56 | -20.16 | 272.01 | -267.02 | 3.05 | 1.073 |
| 10115.00 | 100.00 | -11.71 | 2.61 | -19.46 | 266.13 | -265.97 | 3.11 | 1.073 |
| 10130.00 | 100.00 | -11.65 | 2.67 | -18.79 | 260.36 | -264.69 | 3.17 | 1.074 |
| 10145.00 | 100.00 | -11.58 | 2.72 | -18.15 | 254.70 | -263.19 | 3.23 | 1.075 |
| 10160.00 | 100.00 | -11.50 | 2.77 | -17.54 | 249.15 | -261.48 | 3.30 | 1.075 |
| 10175.00 | 100.00 | -11.43 | 2.82 | -16.96 | 243.73 | -259.57 | 3.36 | 1.076 |
| 10190.00 | 100.00 | -11.34 | 2.87 | -16.62 | 238.44 | -257.47 | 3.43 | 1.077 |
| 10205.00 | 100.00 | -11.26 | 2.91 | -16.59 | 233.28 | -255.20 | 3.50 | 1.077 |
| 10220.00 | 100.00 | -11.17 | 2.96 | -16.51 | 228.26 | -252.76 | 3.56 | 1.078 |
| 10235.00 | 100.00 | -11.08 | 3.00 | -16.55 | 223.38 | -250.16 | 3.63 | 1.079 |
| 10250.00 | 100.00 | -10.99 | 3.04 | -17.01 | 218.64 | -247.42 | 3.70 | 1.080 |
| 10265.00 | 100.00 | -10.91 | 3.08 | -17.46 | 214.04 | -244.54 | 3.76 | 1.081 |
| 10280.00 | 100.00 | -10.82 | 3.12 | -17.90 | 209.59 | -241.54 | 3.83 | 1.082 |
| 10295.00 | 100.00 | -10.73 | 3.15 | -18.32 | 205.29 | -238.41 | 3.90 | 1.083 |
| 10310.00 | 100.00 | -10.64 | 3.19 | -18.70 | 201.14 | -235.17 | 3.97 | 1.084 |
| 10325.00 | 100.00 | -10.55 | 3.22 | -19.04 | 197.12 | -231.83 | 4.04 | 1.085 |
| 10340.00 | 100.00 | -10.47 | 3.25 | -19.33 | 193.26 | -228.39 | 4.10 | 1.086 |
| 10355.00 | 100.00 | -10.38 | 3.28 | -20.03 | 189.54 | -224.86 | 4.17 | 1.087 |
| 10370.00 | 100.00 | -10.30 | 3.31 | -20.86 | 185.96 | -221.25 | 4.24 | 1.088 |
| 10385.00 | 100.00 | -10.22 | 3.34 | -21.72 | 182.53 | -217.56 | 4.30 | 1.089 |
| 10400.00 | 100.00 | -10.14 | 3.36 | -22.61 | 179.24 | -213.80 | 4.37 | 1.090 |
| 10415.00 | 100.00 | -10.06 | 3.38 | -23.49 | 176.09 | -209.97 | 4.43 | 1.091 |
| 10430.00 | 100.00 | -9.99 | 3.40 | -24.34 | 173.08 | -206.08 | 4.49 | 1.092 |
| 10445.00 | 100.00 | -9.92 | 3.42 | -25.10 | 170.21 | -202.14 | 4.56 | 1.093 |
| 10460.00 | 100.00 | -9.85 | 3.43 | -25.68 | 167.47 | -198.13 | 4.61 | 1.094 |
| 10475.00 | 100.00 | -9.78 | 3.44 | -26.02 | 164.87 | -194.08 | 4.67 | 1.096 |
| 10490.00 | 100.00 | -9.71 | 3.46 | -26.08 | 162.40 | -189.98 | 4.73 | 1.097 |
| 10505.00 | 100.00 | -9.65 | 3.46 | -25.84 | 160.07 | -185.84 | 4.78 | 1.098 |
| 10520.00 | 100.00 | -9.58 | 3.47 | -25.38 | 157.87 | -181.66 | 4.84 | 1.099 |
| 10535.00 | 100.00 | -9.52 | 3.47 | -24.77 | 155.79 | -177.45 | 4.89 | 1.101 |
| 10550.00 | 100.00 | -9.46 | 3.47 | -24.09 | 153.85 | -173.20 | 4.93 | 1.102 |
| 10565.00 | 100.00 | -9.40 | 3.47 | -23.38 | 152.03 | -168.91 | 4.98 | 1.103 |
| 10580.00 | 100.00 | -9.34 | 3.47 | -22.68 | 150.35 | -164.60 | 5.02 | 1.104 |
| 10595.00 | 100.00 | -9.28 | 3.46 | -22.00 | 148.78 | -160.26 | 5.06 | 1.105 |
| 10610.00 | 100.00 | -9.23 | 3.45 | -21.37 | 147.35 | -155.90 | 5.09 | 1.107 |
| 10625.00 | 100.00 | -9.17 | 3.44 | -20.77 | 146.04 | -151.51 | 5.12 | 1.108 |
| 10640.00 | 100.00 | -9.11 | 3.42 | -20.21 | 144.85 | -147.10 | 5.15 | 1.109 |
| 10655.00 | 100.00 | -9.06 | 3.40 | -19.69 | 143.79 | -142.67 | 5.18 | 1.110 |
| 10670.00 | 100.00 | -9.00 | 3.38 | -19.21 | 142.85 | -138.22 | 5.20 | 1.112 |
| 10685.00 | 100.00 | -8.94 | 3.35 | -18.77 | 142.04 | -133.76 | 5.22 | 1.113 |
| 10700.00 | 100.00 | -8.88 | 3.33 | -18.36 | 141.36 | -129.27 | 5.23 | 1.114 |
| 10715.00 | 100.00 | -8.82 | 3.30 | -17.97 | 140.79 | -124.78 | 5.24 | 1.115 |
| 10730.00 | 100.00 | -8.76 | 3.26 | -17.62 | 140.36 | -120.27 | 5.24 | 1.116 |
| 10745.00 | 100.00 | -8.70 | 3.22 | -17.29 | 140.05 | -115.75 | 5.24 | 1.117 |
| 10760.00 | 100.00 | -8.63 | 3.18 | -16.98 | 139.87 | -111.21 | 5.24 | 1.118 |
| 10775.00 | 100.00 | -8.56 | 3.14 | -16.70 | 139.82 | -106.67 | 5.23 | 1.119 |
| 10790.00 | 100.00 | -8.49 | 3.09 | -16.44 | 139.89 | -102.12 | 5.22 | 1.120 |
| 10805.00 | 100.00 | -8.42 | 3.04 | -16.20 | 140.10 | -97.57 | 5.20 | 1.121 |
| 10820.00 | 100.00 | -8.34 | 2.98 | -15.98 | 140.44 | -93.01 | 5.18 | 1.122 |
| 10835.00 | 100.00 | -8.27 | 2.92 | -15.77 | 140.92 | -88.44 | 5.15 | 1.123 |
| 10850.00 | 100.00 | -8.19 | 2.86 | -15.59 | 141.54 | -83.87 | 5.12 | 1.124 |
| 10865.00 | 100.00 | -8.10 | 2.79 | -15.41 | 142.29 | -79.31 | 5.09 | 1.124 |
| 10880.00 | 100.00 | -8.01 | 2.73 | -15.25 | 143.19 | -74.74 | 5.05 | 1.125 |
| 10895.00 | 100.00 | -7.92 | 2.65 | -15.11 | 144.24 | -70.18 | 5.01 | 1.126 |
| 10910.00 | 100.00 | -7.83 | 2.58 | -14.97 | 145.43 | -65.62 | 4.96 | 1.126 |
| 10925.00 | 100.00 | -7.73 | 2.50 | -14.85 | 146.77 | -61.07 | 4.91 | 1.127 |
| 10940.00 | 100.00 | -7.63 | 2.41 | -14.75 | 148.27 | -56.53 | 4.85 | 1.127 |
| 10955.00 | 100.00 | -7.53 | 2.32 | -14.65 | 149.94 | -52.00 | 4.80 | 1.128 |
| 10970.00 | 100.00 | -7.42 | 2.23 | -14.56 | 151.76 | -47.49 | 4.73 | 1.128 |
| 10985.00 | 100.00 | -7.31 | 2.14 | -14.48 | 153.75 | -43.00 | 4.67 | 1.129 |
| 11000.00 | 100.00 | -7.20 | 2.04 | -14.41 | 155.92 | -38.53 | 4.60 | 1.129 |
| 11015.00 | 100.00 | -7.08 | 1.94 | -14.35 | 158.26 | -34.09 | 4.53 | 1.129 |
| 11030.00 | 100.00 | -6.97 | 1.83 | -14.30 | 160.79 | -29.69 | 4.45 | 1.130 |
| 11045.00 | 100.00 | -6.85 | 1.72 | -14.26 | 163.51 | -25.32 | 4.38 | 1.130 |
| 11060.00 | 100.00 | -6.73 | 1.61 | -14.22 | 166.42 | -21.00 | 4.30 | 1.130 |
| 11075.00 | 100.00 | -6.60 | 1.50 | -14.19 | 169.54 | -16.73 | 4.22 | 1.130 |
| 11090.00 | 100.00 | -6.48 | 1.38 | -14.17 | 172.86 | -12.51 | 4.14 | 1.130 |
| 11105.00 | 100.00 | -6.35 | 1.25 | -14.15 | 176.40 | -8.36 | 4.05 | 1.130 |
| 11120.00 | 100.00 | -6.22 | 1.13 | -14.13 | 180.16 | -4.29 | 3.97 | 1.130 |
| 11135.00 | 100.00 | -6.10 | 1.00 | -14.13 | 184.14 | -0.30 | 3.88 | 1.130 |
| 11150.00 | 100.00 | -5.97 | 0.87 | -14.12 | 188.36 | 3.59 | 3.80 | 1.130 |
| 11165.00 | 100.00 | -5.84 | 0.73 | -14.12 | 192.83 | 7.37 | 3.71 | 1.130 |
| 11180.00 | 100.00 | -5.71 | 0.59 | -14.12 | 197.54 | 11.03 | 3.62 | 1.130 |
| 11195.00 | 100.00 | -5.57 | 0.45 | -14.13 | 202.51 | 14.56 | 3.53 | 1.130 |
| 11210.00 | 100.00 | -5.44 | 0.31 | -14.14 | 207.74 | 17.92 | 3.44 | 1.129 |
| 11225.00 | 100.00 | -5.31 | 0.16 | -14.15 | 213.23 | 21.12 | 3.36 | 1.129 |
| 11240.00 | 100.00 | -5.18 | 0.01 | -14.16 | 219.00 | 24.13 | 3.27 | 1.129 |
| 11255.00 | 100.00 | -5.05 | -0.15 | -14.18 | 225.06 | 26.92 | 3.18 | 1.129 |
| 11270.00 | 100.00 | -4.92 | -0.31 | -14.19 | 231.39 | 29.47 | 3.10 | 1.128 |
| 11285.00 | 100.00 | -4.79 | -0.47 | -14.21 | 238.01 | 31.76 | 3.01 | 1.128 |
| 11300.00 | 100.00 | -4.66 | -0.49 | -14.22 | 244.91 | 33.76 | 2.93 | 1.128 |
| 11315.00 | 100.00 | -4.53 | -0.41 | -14.24 | 252.10 | 35.44 | 2.84 | 1.128 |
| 11330.00 | 100.00 | -4.40 | -0.33 | -14.25 | 259.58 | 36.76 | 2.76 | 1.127 |
| 11345.00 | 100.00 | -4.27 | -0.24 | -14.26 | 267.33 | 37.69 | 2.68 | 1.127 |
| 11360.00 | 100.00 | -4.14 | -0.16 | -14.28 | 275.34 | 38.19 | 2.61 | 1.127 |
| 11375.00 | 100.00 | -4.02 | -0.08 | -14.29 | 283.61 | 38.22 | 2.53 | 1.127 |
| 11390.00 | 100.00 | -3.89 | 0.00 | -14.29 | 292.12 | 37.74 | 2.46 | 1.126 |
| 11405.00 | 100.00 | -3.77 | 0.07 | -14.30 | 300.83 | 36.70 | 2.38 | 1.126 |
| 11420.00 | 100.00 | -3.64 | 0.15 | -14.30 | 309.73 | 35.07 | 2.32 | 1.126 |
| 11435.00 | 100.00 | -3.52 | 0.23 | -14.30 | 318.77 | 32.79 | 2.25 | 1.126 |
| 11450.00 | 100.00 | -3.40 | 0.30 | -14.30 | 327.92 | 29.82 | 2.19 | 1.125 |
| 11465.00 | 100.00 | -3.28 | 0.38 | -14.43 | 337.12 | 26.11 | 2.12 | 1.125 |
| 11480.00 | 100.00 | -3.16 | 0.45 | -14.54 | 346.32 | 21.63 | 2.07 | 1.125 |
| 11495.00 | 100.00 | -3.04 | 0.52 | -14.60 | 355.43 | 16.34 | 2.01 | 1.125 |
| 11510.00 | 100.00 | -2.92 | 0.60 | -14.62 | 364.40 | 10.21 | 1.96 | 1.125 |
| 11525.00 | 100.00 | -2.80 | 0.67 | -14.60 | 373.12 | 3.21 | 1.92 | 1.125 |
| 11540.00 | 100.00 | -2.68 | 0.74 | -14.54 | 381.50 | -4.68 | 1.87 | 1.125 |
| 11555.00 | 100.00 | -2.57 | 0.81 | -14.43 | 389.45 | -13.45 | 1.84 | 1.125 |
| 11570.00 | 100.00 | -2.45 | 0.88 | -14.29 | 396.84 | -23.09 | 1.80 | 1.125 |
| 11585.00 | 100.00 | -2.34 | 0.95 | -14.12 | 403.58 | -33.58 | 1.78 | 1.125 |
| 11600.00 | 100.00 | -2.23 | 1.02 | -14.09 | 409.54 | -44.87 | 1.76 | 1.125 |
| 11615.00 | 100.00 | -2.12 | 1.09 | -14.05 | 414.61 | -56.91 | 1.74 | 1.125 |
| 11630.00 | 100.00 | -2.01 | 1.15 | -14.01 | 418.68 | -69.60 | 1.73 | 1.125 |
| 11645.00 | 100.00 | -1.90 | 1.22 | -14.07 | 421.65 | -82.85 | 1.73 | 1.125 |
| 11660.00 | 100.00 | -1.79 | 1.28 | -14.26 | 423.44 | -96.53 | 1.74 | 1.125 |
| 11675.00 | 100.00 | -1.68 | 1.35 | -14.44 | 423.98 | -110.52 | 1.75 | 1.126 |
| 11690.00 | 100.00 | -1.57 | 1.41 | -14.63 | 423.21 | -124.67 | 1.77 | 1.126 |
| 11705.00 | 100.00 | -1.47 | 1.48 | -14.82 | 421.12 | -138.81 | 1.79 | 1.126 |
| 11720.00 | 100.00 | -1.36 | 1.54 | -15.01 | 417.69 | -152.79 | 1.83 | 1.127 |
| 11735.00 | 100.00 | -1.26 | 1.60 | -15.19 | 412.96 | -166.45 | 1.87 | 1.127 |
| 11750.00 | 100.00 | -1.16 | 1.66 | -15.38 | 406.97 | -179.62 | 1.91 | 1.128 |
| 11765.00 | 100.00 | -1.06 | 1.72 | -15.57 | 399.80 | -192.17 | 1.97 | 1.129 |
| 11780.00 | 100.00 | -0.96 | 1.78 | -15.76 | 391.53 | -203.96 | 2.03 | 1.129 |
| 11795.00 | 100.00 | -0.86 | 1.83 | -15.95 | 382.29 | -214.88 | 2.10 | 1.130 |
| 11810.00 | 100.00 | -0.76 | 1.89 | -16.13 | 372.19 | -224.84 | 2.17 | 1.131 |
| 11825.00 | 100.00 | -0.67 | 1.94 | -16.31 | 361.37 | -233.76 | 2.25 | 1.132 |
| 11840.00 | 100.00 | -0.57 | 1.99 | -16.49 | 349.97 | -241.60 | 2.34 | 1.133 |
| 11855.00 | 100.00 | -0.48 | 2.04 | -16.98 | 338.12 | -248.33 | 2.43 | 1.134 |
| 11870.00 | 100.00 | -0.39 | 2.09 | -17.98 | 325.97 | -253.94 | 2.53 | 1.135 |
| 11885.00 | 100.00 | -0.30 | 2.13 | -19.07 | 313.63 | -258.46 | 2.64 | 1.136 |
| 11900.00 | 100.00 | -0.21 | 2.18 | -20.28 | 301.23 | -261.90 | 2.75 | 1.137 |
| 11915.00 | 100.00 | -0.13 | 2.22 | -21.60 | 288.88 | -264.31 | 2.87 | 1.139 |
| 11930.00 | 100.00 | -0.05 | 2.25 | -22.98 | 276.66 | -265.74 | 2.99 | 1.140 |
| 11945.00 | 100.00 | 0.03 | 2.29 | -24.32 | 264.67 | -266.25 | 3.13 | 1.141 |
| 11960.00 | 100.00 | 0.11 | 2.32 | -25.35 | 252.98 | -265.91 | 3.26 | 1.143 |
| 11975.00 | 100.00 | 0.18 | 2.35 | -25.67 | 241.63 | -264.78 | 3.41 | 1.144 |
| 11990.00 | 100.00 | 0.25 | 2.37 | -25.15 | 230.69 | -262.94 | 3.56 | 1.146 |
| 12005.00 | 100.00 | 0.32 | 2.39 | -24.02 | 220.17 | -260.45 | 3.72 | 1.148 |
| 12020.00 | 100.00 | 0.38 | 2.40 | -22.65 | 210.12 | -257.38 | 3.88 | 1.150 |
| 12035.00 | 100.00 | 0.44 | 2.41 | -21.27 | 200.55 | -253.79 | 4.05 | 1.151 |
| 12050.00 | 100.00 | 0.49 | 2.41 | -19.98 | 191.47 | -249.75 | 4.22 | 1.153 |
| 12065.00 | 100.00 | 0.54 | 2.41 | -18.81 | 182.88 | -245.31 | 4.40 | 1.155 |
| 12080.00 | 100.00 | 0.59 | 2.40 | -18.45 | 174.79 | -240.53 | 4.58 | 1.157 |
| 12095.00 | 100.00 | 0.62 | 2.38 | -18.48 | 167.18 | -235.45 | 4.76 | 1.159 |
| 12110.00 | 100.00 | 0.66 | 2.38 | -18.51 | 160.06 | -230.13 | 4.95 | 1.161 |
| 12125.00 | 100.00 | 0.68 | 2.37 | -18.52 | 153.41 | -224.59 | 5.14 | 1.163 |
| 12140.00 | 100.00 | 0.70 | 2.35 | -18.53 | 147.22 | -218.88 | 5.33 | 1.165 |
| 12155.00 | 100.00 | 0.71 | 2.32 | -18.52 | 141.48 | -213.03 | 5.52 | 1.166 |
| 12170.00 | 100.00 | 0.72 | 2.28 | -18.50 | 136.16 | -207.07 | 5.71 | 1.168 |
| 12185.00 | 100.00 | 0.72 | 2.24 | -18.47 | 131.26 | -201.03 | 5.89 | 1.170 |
| 12200.00 | 100.00 | 0.70 | 2.18 | -18.44 | 126.76 | -194.92 | 6.07 | 1.171 |
| 12215.00 | 100.00 | 0.68 | 2.11 | -18.39 | 122.65 | -188.78 | 6.25 | 1.173 |
| 12230.00 | 100.00 | 0.65 | 2.03 | -18.34 | 118.90 | -182.61 | 6.42 | 1.174 |
| 12245.00 | 100.00 | 0.62 | 1.94 | -18.28 | 115.50 | -176.44 | 6.58 | 1.176 |
| 12260.00 | 100.00 | 0.57 | 1.84 | -18.21 | 112.45 | -170.28 | 6.73 | 1.177 |
| 12275.00 | 100.00 | 0.51 | 1.72 | -18.13 | 109.72 | -164.14 | 6.87 | 1.178 |
| 12290.00 | 100.00 | 0.44 | 1.60 | -18.05 | 107.29 | -158.04 | 7.00 | 1.178 |
| 12305.00 | 100.00 | 0.37 | 1.46 | -17.97 | 105.17 | -151.97 | 7.11 | 1.179 |
| 12320.00 | 100.00 | 0.28 | 1.30 | -17.88 | 103.33 | -145.96 | 7.21 | 1.179 |
| 12335.00 | 100.00 | 0.18 | 1.13 | -17.78 | 101.76 | -140.01 | 7.30 | 1.179 |
| 12350.00 | 100.00 | 0.07 | 0.95 | -17.69 | 100.45 | -134.12 | 7.37 | 1.179 |
| 12365.00 | 100.00 | -0.05 | 0.75 | -17.58 | 99.40 | -128.31 | 7.43 | 1.179 |
| 12380.00 | 100.00 | -0.17 | 0.54 | -17.48 | 98.58 | -122.58 | 7.47 | 1.179 |
| 12395.00 | 100.00 | -0.31 | 0.32 | -17.37 | 98.00 | -116.93 | 7.49 | 1.178 |
| 12410.00 | 100.00 | -0.46 | 0.08 | -17.26 | 97.64 | -111.36 | 7.50 | 1.177 |
| 12425.00 | 100.00 | -0.61 | -0.18 | -17.15 | 97.50 | -105.89 | 7.50 | 1.176 |
| 12440.00 | 100.00 | -0.78 | -0.44 | -17.03 | 97.56 | -100.52 | 7.48 | 1.175 |
| 12455.00 | 100.00 | -0.95 | -0.73 | -16.91 | 97.83 | -95.25 | 7.44 | 1.174 |
| 12470.00 | 100.00 | -1.14 | -1.03 | -16.79 | 98.28 | -90.07 | 7.39 | 1.172 |
| 12485.00 | 100.00 | -1.32 | -1.32 | -16.80 | 98.92 | -85.01 | 7.33 | 1.170 |
| 12500.00 | 100.00 | -1.52 | -1.52 | -16.89 | 99.75 | -80.06 | 7.26 | 1.169 |
| 12515.00 | 100.00 | -1.72 | -1.72 | -16.96 | 100.75 | -75.22 | 7.18 | 1.167 |
| 12530.00 | 100.00 | -1.93 | -1.93 | -16.99 | 101.91 | -70.50 | 7.09 | 1.165 |
| 12545.00 | 100.00 | -2.15 | -2.15 | -16.99 | 103.24 | -65.90 | 6.99 | 1.163 |
| 12560.00 | 100.00 | -2.37 | -2.37 | -16.96 | 104.73 | -61.43 | 6.88 | 1.161 |
| 12575.00 | 100.00 | -2.59 | -2.59 | -16.90 | 106.38 | -57.09 | 6.77 | 1.159 |
| 12590.00 | 100.00 | -2.82 | -2.82 | -17.05 | 108.17 | -52.88 | 6.65 | 1.157 |
| 12605.00 | 100.00 | -3.05 | -3.05 | -17.48 | 110.10 | -48.81 | 6.52 | 1.154 |
| 12620.00 | 100.00 | -3.28 | -3.28 | -17.84 | 112.17 | -44.88 | 6.40 | 1.152 |
| 12635.00 | 100.00 | -3.51 | -3.51 | -18.09 | 114.38 | -41.10 | 6.27 | 1.150 |
| 12650.00 | 100.00 | -3.75 | -3.75 | -18.24 | 116.71 | -37.47 | 6.14 | 1.148 |
| 12665.00 | 100.00 | -3.98 | -3.98 | -18.28 | 119.17 | -33.99 | 6.01 | 1.146 |
| 12680.00 | 100.00 | -4.22 | -4.22 | -19.38 | 121.74 | -30.68 | 5.88 | 1.144 |
| 12695.00 | 100.00 | -4.45 | -4.45 | -20.58 | 124.42 | -27.53 | 5.76 | 1.142 |
| 12710.00 | 100.00 | -4.68 | -4.68 | -21.66 | 127.20 | -24.56 | 5.63 | 1.140 |
| 12725.00 | 100.00 | -4.91 | -4.91 | -22.47 | 130.08 | -21.76 | 5.50 | 1.138 |
| 12740.00 | 100.00 | -5.14 | -5.14 | -22.82 | 133.05 | -19.14 | 5.38 | 1.136 |
| 12755.00 | 100.00 | -5.36 | -5.36 | -22.63 | 136.11 | -16.71 | 5.26 | 1.134 |
| 12770.00 | 100.00 | -5.58 | -5.53 | -24.84 | 139.23 | -14.48 | 5.14 | 1.132 |
| 12785.00 | 100.00 | -5.80 | -5.69 | -29.11 | 142.42 | -12.43 | 5.02 | 1.130 |
| 12800.00 | 100.00 | -6.00 | -5.84 | -36.48 | 145.66 | -10.60 | 4.91 | 1.128 |
| 12815.00 | 100.00 | -6.21 | -6.01 | -37.21 | 148.95 | -8.97 | 4.80 | 1.127 |
| 12830.00 | 100.00 | -6.40 | -6.18 | -29.80 | 152.27 | -7.54 | 4.70 | 1.125 |
| 12845.00 | 100.00 | -6.59 | -6.35 | -25.65 | 155.62 | -6.34 | 4.60 | 1.124 |
| 12860.00 | 100.00 | -6.77 | -6.53 | -22.89 | 158.97 | -5.35 | 4.50 | 1.122 |
| 12875.00 | 100.00 | -6.94 | -6.71 | -20.84 | 162.32 | -4.58 | 4.41 | 1.121 |
| 12890.00 | 100.00 | -7.11 | -6.90 | -19.22 | 165.65 | -4.03 | 4.32 | 1.119 |
| 12905.00 | 100.00 | -7.26 | -7.10 | -17.89 | 168.95 | -3.71 | 4.23 | 1.118 |
| 12920.00 | 100.00 | -7.40 | -7.30 | -16.78 | 172.20 | -3.60 | 4.15 | 1.117 |
| 12935.00 | 100.00 | -7.54 | -7.49 | -16.03 | 175.40 | -3.72 | 4.08 | 1.116 |
| 12950.00 | 100.00 | -7.66 | -7.66 | -15.28 | 178.51 | -4.05 | 4.01 | 1.114 |
| 12965.00 | 100.00 | -7.78 | -7.78 | -14.56 | 181.54 | -4.60 | 3.94 | 1.113 |
| 12980.00 | 100.00 | -7.88 | -7.88 | -13.88 | 184.45 | -5.35 | 3.88 | 1.112 |
| 12995.00 | 100.00 | -7.97 | -7.97 | -13.24 | 187.24 | -6.30 | 3.82 | 1.111 |
| 13010.00 | 100.00 | -8.06 | -8.06 | -12.65 | 189.90 | -7.45 | 3.77 | 1.110 |
| 13025.00 | 100.00 | -8.13 | -8.13 | -12.09 | 192.40 | -8.78 | 3.72 | 1.109 |
| 13040.00 | 100.00 | -8.20 | -8.20 | -11.58 | 194.73 | -10.27 | 3.67 | 1.109 |
| 13055.00 | 100.00 | -8.25 | -8.25 | -11.35 | 196.88 | -11.93 | 3.63 | 1.108 |
| 13070.00 | 100.00 | -8.30 | -8.30 | -11.21 | 198.84 | -13.73 | 3.60 | 1.107 |
| 13085.00 | 100.00 | -8.33 | -8.33 | -11.06 | 200.59 | -15.65 | 3.57 | 1.106 |
| 13100.00 | 100.00 | -8.36 | -8.36 | -10.92 | 202.13 | -17.68 | 3.54 | 1.106 |
| 13115.00 | 100.00 | -8.39 | -8.32 | -10.77 | 203.44 | -19.81 | 3.52 | 1.105 |
| 13130.00 | 100.00 | -8.40 | -8.17 | -10.92 | 204.53 | -22.01 | 3.50 | 1.105 |
| 13145.00 | 100.00 | -8.41 | -7.99 | -11.25 | 205.38 | -24.26 | 3.49 | 1.104 |
| 13160.00 | 100.00 | -8.41 | -7.81 | -11.61 | 206.00 | -26.55 | 3.48 | 1.104 |
| 13175.00 | 100.00 | -8.41 | -7.64 | -11.98 | 206.38 | -28.85 | 3.47 | 1.103 |
| 13190.00 | 100.00 | -8.40 | -7.46 | -12.37 | 206.53 | -31.14 | 3.47 | 1.103 |
| 13205.00 | 100.00 | -8.39 | -7.29 | -12.77 | 206.44 | -33.41 | 3.47 | 1.102 |
| 13220.00 | 100.00 | -8.37 | -7.13 | -13.20 | 206.13 | -35.63 | 3.48 | 1.102 |
| 13235.00 | 100.00 | -8.35 | -6.92 | -13.65 | 205.60 | -37.78 | 3.49 | 1.102 |
| 13250.00 | 100.00 | -8.33 | -6.71 | -14.13 | 204.86 | -39.86 | 3.50 | 1.101 |
| 13265.00 | 100.00 | -8.31 | -6.51 | -14.63 | 203.91 | -41.83 | 3.52 | 1.101 |
| 13280.00 | 100.00 | -8.28 | -6.32 | -15.17 | 202.78 | -43.70 | 3.54 | 1.101 |
| 13295.00 | 100.00 | -8.25 | -6.15 | -15.74 | 201.47 | -45.43 | 3.56 | 1.101 |
| 13310.00 | 100.00 | -8.22 | -5.98 | -16.34 | 200.00 | -47.03 | 3.59 | 1.100 |
| 13325.00 | 100.00 | -8.20 | -5.82 | -16.98 | 198.39 | -48.48 | 3.62 | 1.100 |
| 13340.00 | 100.00 | -8.17 | -5.66 | -17.65 | 196.64 | -49.76 | 3.66 | 1.100 |
| 13355.00 | 100.00 | -8.14 | -5.52 | -18.37 | 194.78 | -50.89 | 3.69 | 1.100 |
| 13370.00 | 100.00 | -8.11 | -5.38 | -19.12 | 192.82 | -51.84 | 3.73 | 1.100 |
| 13385.00 | 100.00 | -8.08 | -5.25 | -19.90 | 190.77 | -52.62 | 3.77 | 1.100 |
| 13400.00 | 100.00 | -8.05 | -5.13 | -20.69 | 188.65 | -53.22 | 3.81 | 1.100 |
| 13415.00 | 100.00 | -8.03 | -5.02 | -21.45 | 186.48 | -53.64 | 3.86 | 1.100 |
| 13430.00 | 100.00 | -8.01 | -4.91 | -22.15 | 184.28 | -53.88 | 3.90 | 1.099 |
| 13445.00 | 100.00 | -7.99 | -4.82 | -22.70 | 182.04 | -53.95 | 3.95 | 1.099 |
| 13460.00 | 100.00 | -7.97 | -4.72 | -23.02 | 179.80 | -53.84 | 4.00 | 1.099 |
| 13475.00 | 100.00 | -7.95 | -4.64 | -23.07 | 177.56 | -53.56 | 4.05 | 1.099 |
| 13490.00 | 100.00 | -7.94 | -4.56 | -22.81 | 175.34 | -53.12 | 4.10 | 1.099 |
| 13505.00 | 100.00 | -7.93 | -4.49 | -22.42 | 173.14 | -52.52 | 4.15 | 1.099 |
| 13520.00 | 100.00 | -7.92 | -4.43 | -23.63 | 170.98 | -51.76 | 4.20 | 1.099 |
| 13535.00 | 100.00 | -7.91 | -4.37 | -25.03 | 168.87 | -50.85 | 4.26 | 1.099 |
| 13550.00 | 100.00 | -7.91 | -4.32 | -26.72 | 166.81 | -49.80 | 4.31 | 1.099 |
| 13565.00 | 100.00 | -7.91 | -4.27 | -28.80 | 164.81 | -48.62 | 4.36 | 1.098 |
| 13580.00 | 100.00 | -7.92 | -4.23 | -31.51 | 162.88 | -47.31 | 4.41 | 1.098 |
| 13595.00 | 100.00 | -7.93 | -4.20 | -35.33 | 161.04 | -45.88 | 4.46 | 1.098 |
| 13610.00 | 100.00 | -7.94 | -4.17 | -40.92 | 159.27 | -44.34 | 4.51 | 1.098 |
| 13625.00 | 100.00 | -7.96 | -4.15 | -41.57 | 157.59 | -42.70 | 4.55 | 1.098 |
| 13640.00 | 100.00 | -7.99 | -4.14 | -35.87 | 156.00 | -40.96 | 4.60 | 1.097 |
| 13655.00 | 100.00 | -8.01 | -4.13 | -31.89 | 154.50 | -39.13 | 4.64 | 1.097 |
| 13670.00 | 100.00 | -8.04 | -4.13 | -29.10 | 153.11 | -37.21 | 4.68 | 1.097 |
| 13685.00 | 100.00 | -8.08 | -4.13 | -26.97 | 151.81 | -35.23 | 4.72 | 1.097 |
| 13700.00 | 100.00 | -8.12 | -4.14 | -25.26 | 150.62 | -33.17 | 4.76 | 1.096 |
| 13715.00 | 100.00 | -8.16 | -4.16 | -23.84 | 149.53 | -31.06 | 4.79 | 1.096 |
| 13730.00 | 100.00 | -8.21 | -4.18 | -22.63 | 148.55 | -28.89 | 4.82 | 1.095 |
| 13745.00 | 100.00 | -8.27 | -4.20 | -21.57 | 147.68 | -26.68 | 4.85 | 1.095 |
| 13760.00 | 100.00 | -8.32 | -4.23 | -20.64 | 146.91 | -24.42 | 4.87 | 1.095 |
| 13775.00 | 100.00 | -8.39 | -4.22 | -19.80 | 146.25 | -22.13 | 4.89 | 1.094 |
| 13790.00 | 100.00 | -8.45 | -4.17 | -19.05 | 145.70 | -19.81 | 4.91 | 1.094 |
| 13805.00 | 100.00 | -8.53 | -4.12 | -18.37 | 145.25 | -17.47 | 4.93 | 1.093 |
| 13820.00 | 100.00 | -8.60 | -4.08 | -17.74 | 144.91 | -15.11 | 4.94 | 1.092 |
| 13835.00 | 100.00 | -8.68 | -4.04 | -17.17 | 144.68 | -12.73 | 4.94 | 1.092 |
| 13850.00 | 100.00 | -8.77 | -4.00 | -16.64 | 144.56 | -10.35 | 4.95 | 1.091 |
| 13865.00 | 100.00 | -8.86 | -3.95 | -16.30 | 144.54 | -7.97 | 4.95 | 1.090 |
| 13880.00 | 100.00 | -8.95 | -3.90 | -16.55 | 144.62 | -5.59 | 4.94 | 1.090 |
| 13895.00 | 100.00 | -9.05 | -3.85 | -16.80 | 144.81 | -3.22 | 4.94 | 1.089 |
| 13910.00 | 100.00 | -9.16 | -3.81 | -17.05 | 145.09 | -0.86 | 4.93 | 1.088 |
| 13925.00 | 100.00 | -9.26 | -3.77 | -17.29 | 145.48 | 1.49 | 4.91 | 1.087 |
| 13940.00 | 100.00 | -9.37 | -3.73 | -17.53 | 145.97 | 3.82 | 4.90 | 1.087 |
| 13955.00 | 100.00 | -9.49 | -3.67 | -17.77 | 146.56 | 6.12 | 4.88 | 1.086 |
| 13970.00 | 100.00 | -9.61 | -3.62 | -18.00 | 147.24 | 8.39 | 4.86 | 1.085 |
| 13985.00 | 100.00 | -9.73 | -3.57 | -18.21 | 148.02 | 10.63 | 4.83 | 1.084 |
| 14000.00 | 100.00 | -9.85 | -3.52 | -18.42 | 148.89 | 12.83 | 4.80 | 1.083 |
| 14015.00 | 100.00 | -9.98 | -3.48 | -18.61 | 149.86 | 15.00 | 4.77 | 1.082 |
| 14030.00 | 100.00 | -10.11 | -3.43 | -18.79 | 150.91 | 17.12 | 4.74 | 1.082 |
| 14045.00 | 100.00 | -10.24 | -3.39 | -18.95 | 152.05 | 19.19 | 4.71 | 1.081 |
| 14060.00 | 100.00 | -10.38 | -3.36 | -19.10 | 153.28 | 21.21 | 4.67 | 1.080 |
| 14075.00 | 100.00 | -10.52 | -3.32 | -19.22 | 154.60 | 23.17 | 4.63 | 1.079 |
| 14090.00 | 100.00 | -10.66 | -3.28 | -19.33 | 155.99 | 25.08 | 4.59 | 1.078 |
| 14105.00 | 100.00 | -10.80 | -3.25 | -19.41 | 157.47 | 26.93 | 4.55 | 1.077 |
| 14120.00 | 100.00 | -10.94 | -3.22 | -19.46 | 159.02 | 28.71 | 4.50 | 1.076 |
| 14135.00 | 100.00 | -11.09 | -3.19 | -19.50 | 160.66 | 30.42 | 4.46 | 1.075 |
| 14150.00 | 100.00 | -11.23 | -3.16 | -19.51 | 162.36 | 32.05 | 4.41 | 1.074 |
| 14165.00 | 100.00 | -11.37 | -3.12 | -19.50 | 164.14 | 33.61 | 4.37 | 1.073 |
| 14180.00 | 100.00 | -11.52 | -3.07 | -19.47 | 165.99 | 35.09 | 4.32 | 1.072 |
| 14195.00 | 100.00 | -11.66 | -3.03 | -19.42 | 167.90 | 36.48 | 4.27 | 1.072 |
| 14210.00 | 100.00 | -11.80 | -2.98 | -19.35 | 169.87 | 37.79 | 4.22 | 1.071 |
| 14225.00 | 100.00 | -11.94 | -2.94 | -19.26 | 171.91 | 39.01 | 4.17 | 1.070 |
| 14240.00 | 100.00 | -12.07 | -2.90 | -19.15 | 174.00 | 40.13 | 4.12 | 1.069 |
| 14255.00 | 100.00 | -12.20 | -2.86 | -19.03 | 176.15 | 41.15 | 4.07 | 1.068 |
| 14270.00 | 100.00 | -12.33 | -2.82 | -18.90 | 178.34 | 42.07 | 4.02 | 1.067 |
| 14285.00 | 100.00 | -12.45 | -2.78 | -18.75 | 180.58 | 42.89 | 3.97 | 1.066 |
| 14300.00 | 100.00 | -12.56 | -2.75 | -18.59 | 182.86 | 43.59 | 3.93 | 1.065 |
| 14315.00 | 100.00 | -12.67 | -2.71 | -18.48 | 185.19 | 44.18 | 3.88 | 1.064 |
| 14330.00 | 100.00 | -12.77 | -2.68 | -18.87 | 187.54 | 44.66 | 3.83 | 1.063 |
| 14345.00 | 100.00 | -12.86 | -2.64 | -19.27 | 189.92 | 45.02 | 3.78 | 1.063 |
| 14360.00 | 100.00 | -12.94 | -2.61 | -19.68 | 192.32 | 45.26 | 3.73 | 1.062 |
| 14375.00 | 100.00 | -13.01 | -2.57 | -20.09 | 194.75 | 45.36 | 3.69 | 1.061 |
| 14390.00 | 100.00 | -13.06 | -2.54 | -20.50 | 197.18 | 45.35 | 3.64 | 1.060 |
| 14405.00 | 100.00 | -13.11 | -2.51 | -20.91 | 199.62 | 45.20 | 3.60 | 1.059 |
| 14420.00 | 100.00 | -13.14 | -2.48 | -21.31 | 202.06 | 44.91 | 3.55 | 1.059 |
| 14435.00 | 100.00 | -13.16 | -2.45 | -21.70 | 204.49 | 44.49 | 3.51 | 1.058 |
| 14450.00 | 100.00 | -13.16 | -2.42 | -22.05 | 206.90 | 43.94 | 3.47 | 1.057 |
| 14465.00 | 100.00 | -13.15 | -2.38 | -22.37 | 209.29 | 43.24 | 3.43 | 1.056 |
| 14480.00 | 100.00 | -13.13 | -2.35 | -22.63 | 211.66 | 42.40 | 3.39 | 1.056 |
| 14495.00 | 100.00 | -13.09 | -2.32 | -22.83 | 213.98 | 41.42 | 3.35 | 1.055 |
| 14510.00 | 100.00 | -13.03 | -2.30 | -22.96 | 216.26 | 40.30 | 3.32 | 1.054 |
| 14525.00 | 100.00 | -12.96 | -2.27 | -22.99 | 218.49 | 39.03 | 3.28 | 1.053 |
| 14540.00 | 100.00 | -12.87 | -2.24 | -22.94 | 220.65 | 37.63 | 3.25 | 1.053 |
| 14555.00 | 100.00 | -12.77 | -2.21 | -22.79 | 222.74 | 36.08 | 3.22 | 1.052 |
| 14570.00 | 100.00 | -12.66 | -2.18 | -22.57 | 224.75 | 34.39 | 3.19 | 1.051 |
| 14585.00 | 100.00 | -12.53 | -2.14 | -22.26 | 226.66 | 32.57 | 3.16 | 1.051 |
| 14600.00 | 100.00 | -12.39 | -2.08 | -21.90 | 228.48 | 30.61 | 3.14 | 1.050 |
| 14615.00 | 100.00 | -12.24 | -2.03 | -21.49 | 230.18 | 28.52 | 3.11 | 1.049 |
| 14630.00 | 100.00 | -12.08 | -1.98 | -21.05 | 231.77 | 26.31 | 3.09 | 1.049 |
| 14645.00 | 100.00 | -11.91 | -1.93 | -20.58 | 233.22 | 23.97 | 3.07 | 1.048 |
| 14660.00 | 100.00 | -11.72 | -1.88 | -20.10 | 234.54 | 21.52 | 3.05 | 1.048 |
| 14675.00 | 100.00 | -11.53 | -1.83 | -19.60 | 235.70 | 18.96 | 3.04 | 1.047 |
| 14690.00 | 100.00 | -11.34 | -1.77 | -19.11 | 236.71 | 16.30 | 3.02 | 1.046 |
| 14705.00 | 100.00 | -11.13 | -1.72 | -18.62 | 237.55 | 13.55 | 3.01 | 1.046 |
| 14720.00 | 100.00 | -10.92 | -1.67 | -18.14 | 238.22 | 10.72 | 3.00 | 1.045 |
| 14735.00 | 100.00 | -10.71 | -1.62 | -17.66 | 238.71 | 7.81 | 3.00 | 1.045 |
| 14750.00 | 100.00 | -10.49 | -1.57 | -17.19 | 239.00 | 4.84 | 2.99 | 1.044 |
| 14765.00 | 100.00 | -10.27 | -1.52 | -16.73 | 239.10 | 1.83 | 2.99 | 1.044 |
| 14780.00 | 100.00 | -10.04 | -1.47 | -16.28 | 239.00 | -1.23 | 2.99 | 1.044 |
| 14795.00 | 100.00 | -9.81 | -1.41 | -15.84 | 238.70 | -4.31 | 3.00 | 1.043 |
| 14810.00 | 100.00 | -9.58 | -1.36 | -15.41 | 238.18 | -7.40 | 3.00 | 1.043 |
| 14825.00 | 100.00 | -9.35 | -1.31 | -14.99 | 237.46 | -10.49 | 3.01 | 1.042 |
| 14840.00 | 100.00 | -9.11 | -1.26 | -14.58 | 236.52 | -13.56 | 3.02 | 1.042 |
| 14855.00 | 100.00 | -8.87 | -1.21 | -14.18 | 235.36 | -16.60 | 3.04 | 1.042 |
| 14870.00 | 100.00 | -8.64 | -1.16 | -13.79 | 234.00 | -19.59 | 3.06 | 1.041 |
| 14885.00 | 100.00 | -8.40 | -1.11 | -13.41 | 232.42 | -22.52 | 3.08 | 1.041 |
| 14900.00 | 100.00 | -8.16 | -1.06 | -13.04 | 230.63 | -25.38 | 3.10 | 1.041 |
| 14915.00 | 100.00 | -7.92 | -1.01 | -12.67 | 228.64 | -28.14 | 3.13 | 1.040 |
| 14930.00 | 100.00 | -7.69 | -0.96 | -12.31 | 226.46 | -30.80 | 3.16 | 1.040 |
| 14945.00 | 100.00 | -7.45 | -0.91 | -11.96 | 224.08 | -33.34 | 3.20 | 1.040 |
| 14960.00 | 100.00 | -7.21 | -0.86 | -11.62 | 221.52 | -35.75 | 3.24 | 1.040 |
| 14975.00 | 100.00 | -6.97 | -0.81 | -11.28 | 218.78 | -38.01 | 3.28 | 1.040 |
| 14990.00 | 100.00 | -6.74 | -0.76 | -10.95 | 215.89 | -40.12 | 3.32 | 1.040 |
| 15005.00 | 100.00 | -6.50 | -0.71 | -10.65 | 212.84 | -42.06 | 3.37 | 1.040 |
| 15020.00 | 100.00 | -6.27 | -0.67 | -10.64 | 209.65 | -43.83 | 3.42 | 1.040 |
| 15035.00 | 100.00 | -6.04 | -0.62 | -10.63 | 206.33 | -45.41 | 3.48 | 1.040 |
| 15050.00 | 100.00 | -5.81 | -0.57 | -10.61 | 202.90 | -46.79 | 3.54 | 1.040 |
| 15065.00 | 100.00 | -5.58 | -0.53 | -10.60 | 199.36 | -47.98 | 3.60 | 1.040 |
| 15080.00 | 100.00 | -5.35 | -0.48 | -10.58 | 195.74 | -48.96 | 3.67 | 1.040 |
| 15095.00 | 100.00 | -5.13 | -0.44 | -10.57 | 192.04 | -49.74 | 3.74 | 1.040 |
| 15110.00 | 100.00 | -4.90 | -0.40 | -10.55 | 188.29 | -50.30 | 3.82 | 1.040 |
| 15125.00 | 100.00 | -4.68 | -0.36 | -10.53 | 184.49 | -50.65 | 3.90 | 1.040 |
| 15140.00 | 100.00 | -4.46 | -0.32 | -10.52 | 180.66 | -50.78 | 3.98 | 1.041 |
| 15155.00 | 100.00 | -4.24 | -0.28 | -10.50 | 176.81 | -50.70 | 4.07 | 1.041 |
| 15170.00 | 100.00 | -4.03 | -0.24 | -10.48 | 172.96 | -50.41 | 4.16 | 1.041 |
| 15185.00 | 100.00 | -3.81 | -0.21 | -10.46 | 169.12 | -49.91 | 4.25 | 1.042 |
| 15200.00 | 100.00 | -3.60 | -0.17 | -10.43 | 165.30 | -49.20 | 4.35 | 1.042 |

**IMPORTANT NOTE: AVG PWR GAIN _MUST_ BE IN THE RANGE 0.8-1.2 FOR A VALID NEC MODEL (IDEALLY = 1)!
VALUES OUTSIDE THIS RANGE INDICATE A FLAWED MODEL, USUALLY A RESULT OF INCORRECT SEGMENTATION.
SEE DISCUSSION, p.101, NEC 4.1 USER'S MANUAL, LLNL UCRL-MA-109338 Pt. I (Gerald J. Burke), 1992.